\documentclass[twocolumn,aps,pra,10pt,longbibliography]{revtex4-1}

    \usepackage{appendix}
	\usepackage[usenames,dvipsnames]{xcolor}
	\usepackage{amsmath}
	\usepackage{amsfonts}
	\usepackage{amssymb}
	\usepackage{mathtools}
	\usepackage[colorlinks=true,citecolor=blue,linkcolor=red]{hyperref}
	\usepackage{graphicx}
	\usepackage{bbold}					
	\usepackage[makeroom]{cancel}		
	\usepackage{multirow}				
	\usepackage[normalem]{ulem}        
	\usepackage{array} 
	\usepackage{mathrsfs} 	
    \usepackage[section]{placeins}
    \usepackage{setspace}

\newcolumntype{L}{>{$}l<{$}}
\newcolumntype{R}{>{$}r<{$}}
\newcolumntype{C}{>{$}c<{$}}

	\newcolumntype{x}[1]{>{\centering\let\newline\\\arraybackslash\hspace{0pt}}p{#1}}
		
	\DeclareMathOperator{\sign}{sign}  	

	\DeclareMathAlphabet{\mathbbold}{U}{bbold}{m}{n}
	\DeclareMathAlphabet{\mathpzc}{OT1}{pzc}{m}{it}
	
	\def\imi{\mathrm{i}}				

	\newcounter{subeqn} %
	\makeatletter
	\@addtoreset{subeqn}{equation}
	\makeatother

\newenvironment{rsmallmatrix}
  {\begin{psmallmatrix*}[r]}
  {\end{psmallmatrix*}}

\newcommand{\ie}{{\it i.e.}\;}
\newcommand{\eg}{{\it e.g.}\;}

\begin{document}

\title{Multi-gap topological conversion of Euler class via band-node braiding: minimal models, $PT$-linked nodal rings, and chiral heirs}
\author{Adrien Bouhon$^{1,2}$}
\thanks{Refer for correspondence to \href{mailto:ab2859@cam.ac.uk}{ab2859@cam.ac.uk} and \href{rjs269@cam.ac.uk}{rjs269@cam.ac.uk}.}
\author{Robert-Jan Slager$^{1}$}
\affiliation{\vspace{0.3em}$^{1}$TCM Group, Cavendish Laboratory, University of Cambridge, J. J. Thomson Avenue, Cambridge CB3 0HE, United Kingdom}
\affiliation{$^{2}$Nordita, Stockholm University and KTH Royal Institute of Technology, Hannes Alfv{\'e}ns v{\"a}g 12, SE-106 91 Stockholm, Sweden}

\begin{abstract}
The past few years have seen rapid progress in characterizing topological band structures using symmetry eigenvalue indicated methods.  Recently, however, there has been increasing theoretical and experimental interest in multi-gap dependent topological phases that cannot be captured by this paradigm. These topologies arise by braiding band degeneracies that reside between different bands and carry non-Abelian charges due 
to the presence of either $C_2T$ or $PT$ symmetry, culminating in different invariants such as $\mathbb{Z}$-valued Euler class. Here, we present a universal formulation for Euler phases motivated by their homotopy classification that is related to the Skyrmion-profile of a single unit-vector in three-level systems, and that of two unit-vectors in four-level systems. In addition, upon employing the strategy of systematically building 3D models from a pair of sub-dimensional Euler phases, we show that phase transitions between any two inequivalent Euler phases are mediated by the presence of adjacent (in-gap) nodal rings linked with sub-gap nodal lines, forming trajectories corresponding to the braiding or debraiding of nodal points. The stability of the linked adjacent nodal rings is furthermore demonstrated to be indicated by an Euler class monopole charge matching with its $\mathbb{Z}$-valued linking numbers. We finally also systematically address the conversion of  Euler phases into descendant Chern phases upon breaking the $C_2T$  or $PT$ symmetry. All the topological phases discussed in this work are corroborated with explicit minimal lattice models. These models can  themselves directly serve as an extra impetus for experimental searches or be employed for theoretical studies, thereby underpinning the upcoming of this nascent pursuit.
\end{abstract}

\maketitle
\section{Introduction}
Topological materials \cite{Rmp1,Rmp2,Weylrmp} entail an active field in condensed matter, encompassing studies that range from theoretical pursuits to material science impetuses. These intensive efforts have resulted in a wide charted field of phases and a plethora of topological characterizations \cite{volovik2003universe,Clas3,Wi2,InvTIBernevig, Clas1, InvTIVish, Clas2, probes_2D, Shiozaki14, Codefects2,SchnyderClass, Wi1,  Wi3, UnsupMach, ShiozakiSatoGomiK, Clas4, ran2009one, UnifiedBBc, teodef,Clas5,Codefects1,Bouhon_HHL, BbcWeyl,alex2019crystallographic,Mode2, Chenprb2012,Unal2019, Cornfeld_2021}. Much of this progress has been rooted in symmetry eigenvalue analyses. Using the information of representations at high symmetry points in the Brillouin zone~\cite{Clas3,Wi2} a significant fraction of topological phases in momentum space can be efficiently characterized and, upon comparing which of these combinations have an atomic limit, versatile classification schemes have been formulated~\cite{Clas4,Clas5}.

These endeavours have also transpired in the retrieval of different topological phases exhibiting different features. First, within the symmetry eigenvalue setting, it was found that certain topologies phases can be fragile~\cite{Ft1}. Such phases have led to new explorations into their properties  \cite{bouhon2018wilson,Bradlyn_fragile,Hwang_inversion_fragile,Song794, paulmbo_metricw, SubD_Gunnar,Wieder_axion} and also resulted in experimental signatures \cite{Peri797}. 
More recently, however, a new class of topological phases, that depend on multi-gap conditions, have increasingly been gaining interest. 
A prominent example in this regard occur in phases enjoying a real Hamiltonian representation by virtue of $C_2T$ symmetry or $PT$ symmetry. Band degeneracies between different bands (which we will refer to as `gaps') can then carry non-Abelian frame charges~\cite{Wu1273, Tiwari:2019,BJY_nielsen, bouhon2019nonabelian}, akin to $\pi$-disclination defects in bi-axial nematics~\cite{Kamienrmp, Genqcs2016,volovik2018investigation, Beekman20171}, and braiding them around in momentum space leads to similarly-valued band touchings within a certain gap. The resulting obstruction to annihilate 
these band touchings is directly related to a multi-gap topological invariant, known as Euler class~\cite{BJY_linking,bouhon2019nonabelian,BJY_nielsen,bouhon2020geometric,Unal_quenched_Euler}. This invariant corresponds 
to a  characteristic form being the real counterpart of the complex variant that underlies Chern numbers. 

We have recently shown that Euler phases can generally be understood as arising from refined partition schemes and classified by specific homotopy characterizations, which in turn can also be used reversely as general strategy to construct models having desired Euler class \cite{bouhon2020geometric}. Such Euler class models are increasingly becoming of importance and have for example been proposed to induce monopole-antimonople generation in quench setups~\cite{Unal_quenched_Euler}, while the observation of this physical observable has just been reported in trapped-ion experiments~\cite{zhao2022observation}. In addition, the braiding and emerging of such non-Abelian charges and its relation to Euler class have also been inspiring pursuits in other experimental contexts that range from from phononic systems~\cite{park2021,Lange2022,Peng2021,peng2022multi} and electronic systems~\cite{chen2021manipulation, magnetic, bouhon2019nonabelian, Koneye2021, SubD_Gunnar, Eulersc,biao2020landau,guan2021landau} to acoustic, photonic and electric circuit metamaterials~\cite{Guo1Dexp,park2022nodal,Jiang2021,qiu2022minimal,ezawa2021euler}. 

Given this interest, we here wish to further underpin these developments by introducing simple models that exhibit non-trivial Euler topologies and controlled band node formation. In particular this allows us to further examine braiding processes~\cite{bouhon2019nonabelian}  and their interplay with symmetries as well as relation to homotopy perspectives~\cite{bouhon2020geometric}. More precisely, we retrieve a simple formulation of tight-binding models with Euler topology in term of Skyrmion winding numbers of a single \cite{BzduSigristRobust} or double unit vector for the three and four band cases, respectively. Remarkably, this allows us to formulate a great variety of Euler phases by exploiting the phase diagram of a single two-band Chern lattice model parametrized to produce a Chern number ranging from $-2$ to $2$ \cite{sticlet2012engineering}. Extending the above intrinsic considerations, we find that the transition between inequivalent Euler phases, while preserving the reality condition, are generically mediated by the presence of ``adjacent'' nodal rings linked with ``sub-gap'' nodal lines, the former of which appearing within the gap of the Euler phases, and the later being formed by the band crossings of the connected two-band subspaces of the Euler phases. Running through the transition, the nodal points extend into nodal braids forming trajectories that correspond to the braiding or debraiding of nodal points \cite{Tiwari:2019}. The stability of the adjacent nodal rings is moreover found to be indicated by specific monopole charges dictated by the ``difference'' of Euler classes and corresponding to the linking numbers \cite{BJY_linking} of the nodal rings. This point of view thus culminates in the systematic building of 3D $PT$-symmetric models, obtained through the embedding of pairs of 2D Euler phases within the 3D Brillouin zone, that host linked adjacent nodal rings. As an other extension, we address the systematic conversion of Euler phases into descendant Chern phases upon breaking the $C_2T$ or $PT$ symmetry. Then, similarly to the $PT$-symmetric case, we build 3D $C_2T$-symmetric chiral phases, obtained from pairs of Euler phases, that trap a number of Weyl nodes that is again dictated by the same ``difference'' of Euler classes. From a practical point of view, all these models can directly be implemented by experimental and modeling pursuits with the hope of further advancing this new field.

This paper is organized as follows. In Sec. \ref{sec:modeling} we begin our discussion by reviewing the generic homotopy-induced strategy to model Euler phases~\cite{bouhon2020geometric} for both three-level and four-level systems, where the latter are specified by the relative balance between the Euler class of the two two-band subspaces. In particular, we introduce the parametrization of three-band and four-band models through the Skyrmion winding number of one unit vector, and that of two independent unit vectors, respectively. Readers interested in the concrete models may directly skip to Sec. \ref{sec_min_models}, where we formulate the models of interest hosting the Euler topology of orientable phases. Given our generic framework that can generate any kind of model, these examples are on purpose taken as simple as possible. This however may generically induce additional symmetries, which is the topic that we address systematically in Sec.~\ref{sec_sym}. In Sec.~\ref{sec_3D_NL} we then expose the relation between inequivalent 2D Euler phases and their relation to 3D $PT$ symmetric nodal lines structures when the 2D phases are seen as planar cuts of a 3D embedding and discuss the quantification in terms of monopole charges and linking numbers. This general point of view is then again made concrete with readily implementable models in Sec.~\ref{sec_3D_NL_numerics}, for everyone of which we present the linked nodal ring structures obtained numerically. Finally, we discuss how breaking symmetries can lead to descendant topologies, such as 2D Chern phases, and the notion of 3D chiral phases in Sec.~\ref{sec_chiral}, before concluding in Sec.~\ref{sec:conclusions}.

\section{Geometric and homotopic modeling of orientable Euler phases}\label{sec:modeling}
In this section we review the geometry and homotopy frameworks that motivate the derivation of explicit models with Euler class topology. In particular, we obtain that flattened and two-by-two Euler Hamiltonians are fully parametrized in terms of three-component unit vectors winding on a sphere. These homotopy representative Hamiltonians are then used in the next Section to derive explicit minimal tight-binding Hamiltonians for a variety of Euler phases. In the whole work we assume that the system has a $C_2T$ symmetry (spinful or spinless), with $C_2$ the $\pi$ rotation axis perpendicular to the system's basal plane and $T$ is time reversal, with $[C_2T]^2=+\mathbb{1}$. (Equivalently, the system can host a $PT$ symmetry, with $P$ the inversion symmetry, still with $[PT]^2=+\mathbb{1}$. In that case however, the system must be spinless.) Because the anti-unitary symmetry squares to the identity, it can be shown (through the Takagi factorization of the symmetric unitary matrix that represents $C_2T$ in the Bloch orbital basis, see below) that there exists a special basis for which the Bloch Hamiltonian matrix is real and symmetric~\cite{bouhon2019nonabelian, chen2021manipulation}. In the following we call it the reality condition of Euler phases. We again note that this section is more technical in nature and can be skipped by readers interested in the minimal models for direct implementation that are presented in the subsequent Sections.

\subsection{Homotopy classification of two-dimensional orientable Euler phases}

We review the homotopy classification and modeling of Euler phases obtained in Ref.\;\cite{bouhon2020geometric}. Let us consider the Bloch Hamiltonian operator 
\begin{equation}
    \mathcal{H} = \sum\limits_{\boldsymbol{k}\in \mathrm{BZ}}\sum\limits_{ab} \vert \phi_{a}, \boldsymbol{k} \rangle H_{ab}(\boldsymbol{k}) \langle \phi_{b},\boldsymbol{k} \vert,  
\end{equation}
where the wave-vector $\boldsymbol{k}$ of the two-dimensional system is a point of the Brillouin zone $\mathbb{T}^2=\{(k_1,k_2)\vert k_1\in [-\pi,\pi),k_2\in[-\pi,\pi)\}$, and where $\{\vert \phi_{a}, \boldsymbol{k} \rangle \}_{a}$ is assumed to be a Bloch orbital basis obtained from the Fourier transform of a localized Wannier basis. Our starting point is the spectral decomposition of the real and symmetric Bloch Hamiltonian matrix, \ie 
\begin{equation}
    H(\boldsymbol{k}) = R(\boldsymbol{k}) \cdot D(\boldsymbol{k}) \cdot R(\boldsymbol{k})^T,    
\end{equation}
with the diagonal matrix of energy ordered eigenvalues $D(\boldsymbol{k})=\mathrm{diag}[E_1(\boldsymbol{k}),\dots,E_N(\boldsymbol{k})]$, such that $E_{n}(\boldsymbol{k})\leq E_{n+1}(\boldsymbol{k})$ for $n=1,\dots,N-1$, the matrix of column eigenvectors $R(\boldsymbol{k}) = \left[ u_1(\boldsymbol{k})~\cdots~u_N(\boldsymbol{k}) \right]\in \mathsf{O}(N)$, and where $N$ is the total number of bands.

Assuming that the first $p$ bands are separated from the higher $(N-p)$ bands by an energy gap, the classifying space of the Hamiltonian takes the form of a real Grassmannian, $\mathsf{Gr}^{\mathbb{R}}_{p,N} = \mathsf{O}(N)/[\mathsf{O}(p)\times \mathsf{O}(N-p)]$. In the following, we will use the flattened Hamiltonian 
\begin{equation}
    Q(\boldsymbol{k}) = R(\boldsymbol{k})\cdot (-\mathbb{1}_{p}\oplus \mathbb{1}_{N-p})\cdot R(\boldsymbol{k})^T, 
\end{equation}
as the homotopy representative of the dispersive Hamiltonian $H(\boldsymbol{k})$. 

We start the homotopy characterization by noting that two-dimensional systems can host nontrivial one-dimensional topologies as indicated by the nontrivial first homotopy group, $\pi^{(l)}_1[\mathsf{Gr}^{\mathbb{R}}_{p,N}]  = \mathbb{Z}_2$, over one non-contractible direction of the two-dimensional Brillouin zone torus, \ie $l \in \{l^{(k_2)}_1,l^{(k_1)}_2\}$ where $l^{(k_2)}_{1} = \{(k_1,k_2) \vert k_1\in [-\pi,\pi)\}$, \ie the path crossing the Brillouin zone at a fixed $k_2$, and similarly for $l^{(k_1)}_2$. The one-dimensional topologies are indicated by the quantized Berry phase $\gamma_B[l] \in \{0,\pi\}\;\mathrm{mod}\; 2\pi$ or, equivalently, by the first Stiefel-Whitney class \cite{BJY_linking}, which characterize the {\it orientability} of the phase, \ie whether the frame of eigenvector $R(\boldsymbol{k})$ can be chosen to be continuous and periodic across the Brillouin zone \cite{bouhon2020geometric,BJY_linking}. In the following we write the homotopy classes of one-dimensional cuts as $\alpha_{1(2)} = [l^{(k_{2(1)})}_{1(2)}] \in \pi_1[\mathsf{Gr}^{\mathbb{R}}_{2,3}]$.

In Ref.\;\cite{bouhon2020geometric}, we have derived the general homotopy classification of Euler phases with multiple energy gaps, in which case the classifying space takes the form of a generalized real flag manifold. For this work, it is sufficient to consider the topological classification of the two-dimensional Euler phases with a single principal gap. We nevertheless relate these phases to the braiding of multi-gap nodes and the conversion of their non-Abelian homotopy charges. Indeed, we show in Section \ref{sec_3D_NL} that the mapping of an Euler phase to another requires the {\it braiding} of nodes from an adjacent gap. In the following, we label the occupied (unoccupied) band-subspace with the roman letter $I$ ($II$), \eg we write their respective Euler classes $(\chi_I,\chi_{II})$.   

In this work, we only consider {\it orientable} phases, \ie with $R(\boldsymbol{k})$ periodic. The homotopy classification and modeling of {\it orientable} phases (\ie with trivial one-dimensional topology) is most conveniently obtained from the classification of two-dimensional {\it oriented} vector bundles, \ie with the oriented Grassmannian as the classifying space, $\widetilde{\mathsf{Gr}}^{\mathbb{R}}_{p,N} = \mathsf{SO}(N)/[\mathsf{SO}(p)\times \mathsf{SO}(N-p)]$. Indeed, we have for the orientable phases \cite{bouhon2020geometric}
\begin{equation}
\label{eq_homotopy}
\begin{aligned}
    [\mathbb{T}^2,\mathsf{Gr}^{\mathbb{R}}_{p,N}]^{(\alpha_1=0,\alpha_2=0)} 
    &= [\mathbb{S}^2,\mathsf{Gr}^{\mathbb{R}}_{p,N}] \\
    &= \pi_2[\mathsf{Gr}^{\mathbb{R}}_{p,N}]/{\sim} \\
    &= \pi_2[\widetilde{\mathsf{Gr}}^{\mathbb{R}}_{p,N}]/{\sim} ,
\end{aligned}
\end{equation}
where the equivalence relation $\sim$ corresponds to the Euler class reversal map
\begin{equation}
    (\chi_I,\chi_{II}) \sim (-\chi_I,-\chi_{II}).
\end{equation}
The important point here is that $\pi_2[\widetilde{\mathsf{Gr}}^{\mathbb{R}}_{p,N}]$ is known. (It can be computed through the long exact sequence of homotopy groups associated to fiber bundles \cite{Hatcher_1}.) The reduction of the homotopy classification through the equivalence relation ${\sim}$ for the orientable phases, as compared to the oriented phases classified by $\pi_2[\widetilde{\mathsf{Gr}}^{\mathbb{R}}_{p,N}]$, is due to the absence of a fixed base point in the definition of the homotopy classes $[\mathbb{T}^2,\mathsf{Gr}^{\mathbb{R}}_{p,N}]$ that capture the topology of Hamiltonians (contrary to homotopy groups that are defined assuming a fixed base point). The absence of a fixed base point permits the nontrivial action of a generator of the first homotopy group (\ie the deformation of the Hamiltonian along one non-contractible loop of the classifying space) on the elements of the second homotopy group. More precisely, this action defines an automorphism between distinct elements of the second homotopy group while remaining within the same homotopy class \cite{bouhon2020geometric,wojcik2020homotopy,Hatcher_1}. 

We also use the oriented Grassmannian $\widetilde{\mathsf{Gr}}_{p,N}^{\mathbb{R}}$ for the modelling of Euler phases. Indeed, starting from the representative $R\in \mathsf{SO}(N)$ of a point of the oriented Grassmannian (here defined as a coset) $[R] = \{R\cdot (O_I\oplus O_{II})\vert  O_I\in \mathsf{SO}(p), O_{II}\in \mathsf{SO}(N-p) \}$, the flattened Hamiltonian $Q=R\cdot (-\mathbb{1} \oplus \mathbb{1}) \cdot R^T$ inherits the equivalence relation $\sim$ defined above. This directly follows from the higher gauge freedom of the Hamiltonian form as compared to the coset element $[R]$. (Explicitly, the transformation $R\rightarrow R \cdot (O_I\oplus O_{II})$ with $O_{I}\in \mathsf{O}(p)$ and $O_{II}\in \mathsf{O}(N-p)$, leaves $Q$ invariant, while it maps to a frame that is not necessarily represented by the coset $[R]$.)

In the following, we concentrate on the three-band and four-band Euler phases, \ie for $(p,N)=(2,3)$ and $(p,N)=(2,4)$. 

\subsection{Three-band Euler phases}
The homotopy classification of (two-dimensional) orientable three-band Euler phases that split into $2+1$-band-subspaces is given by
\begin{equation}
\begin{aligned}
    \pi_2[\widetilde{\mathsf{Gr}}^{\mathbb{R}}_{2,3}]/{\sim} &= \pi_2[\mathbb{S}^2]/{\sim} \\
   & = \{a \in 2\mathbb{Z}\vert a \sim -a\}= 2\mathbb{N} \ni \beta(H^{(2+1)}),
\end{aligned}
\end{equation}
where we have used the identities $\widetilde{\mathsf{Gr}}^{\mathbb{R}}_{2,3}= \mathsf{SO}(3)/\mathsf{SO}(2) = \mathbb{S}^2$, and $\beta(H^{(2+1)})$ represents the homotopy class of the Bloch Hamiltonian $H^{(2+1)}(\boldsymbol{k})$ with eigenvalues that split as $E_1(\boldsymbol{k}) \leq E_2(\boldsymbol{k}) < E_3(\boldsymbol{k})$ for all $\boldsymbol{k}\in \mathbb{T}^2$. The factor two in the classifying set $2\mathbb{N}$ will become clear below. We define the Euler class of the two-band subspace, \ie \cite{bouhon2019nonabelian,BJY_linking,BJY_nielsen,Zhao_PT}
\begin{subequations}
\begin{equation}
\begin{aligned}
    \chi_I[\{u_1,u_2\}] &= \dfrac{1}{2\pi} \int_{\mathrm{BZ}} \mathsf{Eu}(\boldsymbol{k}) \in \mathbb{Z},
\end{aligned}
\end{equation}
where the Euler form $\mathsf{Eu}$ is obtained from the connection $\mathsf{a} = u_1^T\cdot du_2$ through $\mathsf{Eu}=d\mathsf{a}$, leading to
\begin{equation}
    \mathsf{Eu}(\boldsymbol{k}) = \left(\partial_{k_1} u_1^T\cdot \partial_{k_2} u_2-\partial_{k_2} u_1^T\cdot \partial_{k_1} u_2\right) dk_1\wedge dk_2  .
\end{equation}
\end{subequations}
The homotopy invariant is then readily given as an equivalence class
\begin{equation}
    \begin{aligned}
    \beta\left( H^{(2+1)} \right)  &= \left[ \, \chi_I[\{u_1,u_2\}] \,\right] \,,\\
    & = \left\{ \vert\, \chi_I[\{u_1,u_2\}]\, \vert \,, - \vert\,\chi_I[\{u_1,u_2\}] \,\vert \right\}\,.
    \end{aligned}
\end{equation}
We note that while we could simply take $\vert \chi_I \vert \in 2\mathbb{N}$ as a the number representative of the equivalence class $[\chi_I]$, we will see in Section \ref{sec_3D_NL} that the equivalence class must be used to predict the correct expression of the monopole charge and linking number of $PT$-symmetry protected linked nodal rings. Importantly, the above definition of the Euler class holds for any orientable two-band subspace $\{u_{n},u_{n+1}\}$ isolated from all the other bands, \ie $\chi_{\nu}[\{u_n,u_{n+1}\}]$ ($\nu=I,II,\dots$) is well defined whenever $E_{n-1}(\boldsymbol{k}) < E_n(\boldsymbol{k}) \leq E_{n+1}(\boldsymbol{k}) < E_{n+2}(\boldsymbol{k})$ for all $\boldsymbol{k}\in \mathbb{T}^2$.

The modeling of three-band Euler phases with two occupied bands can then be readily obtained from a representative $R\in \mathsf{SO}(3)$ of the coset $[R] \in \mathsf{SO}(3)/\mathsf{SO}(2) = \mathbb{S}^2$. The spherical frame readily satisfies this condition, \ie $R(\phi,\theta) = (u_1~u_2~u_3) = (e_{\theta}~e_{\phi}~e_{r})$, with $e_{\theta} = (\cos\phi\cos\theta,\sin\phi,\cos\theta,-\sin\theta)$, $e_{\phi}=(-\sin\phi,\cos\phi,0)$ and $e_{r} = (\cos\phi\sin\theta,\sin\phi\sin\theta,\cos\theta)$, from which we get \cite{BzduSigristRobust,Wu1273}
\begin{equation}
\label{eq_flat_21}
    Q^{(2+1)}[\boldsymbol{n}(\phi,\theta)] = 2 \boldsymbol{n}(\phi,\theta)\cdot \boldsymbol{n}(\phi,\theta)^T - \mathbb{1}_3,
\end{equation}
with the unit vector $\boldsymbol{n}(\phi,\theta)=e_r$ (the superscript `$2+1$' refers to the spectral decomposition into one two-band subspace and one single band). Since for orientable phases we can simplify the Brillouin zone to a sphere, see Eq.\;(\ref{eq_homotopy}), let us represent a point of the Brillouin zone by the angles $(\phi_{\boldsymbol{k}},\theta_{\boldsymbol{k}})\in \mathbb{S}^2_{\mathrm{BZ}}$. More concretely, this follows \eg by choosing
\begin{equation}
\label{eq_BZ_mapping}
\begin{aligned}
	\phi_{\boldsymbol{k}} &=  \mathrm{arg}\,(k_1 + \mathrm{i} k_2)\, ,\\
	\theta_{\boldsymbol{k}} &=  \max  \left(\vert k_1 \vert , \vert k_2 \vert \right)  \,.
\end{aligned}
\end{equation} 
We can then define the Euler phases through the ansatz
\begin{equation}
\label{eq_ansatz_angles}
\begin{aligned}
    \phi_q(\phi_{\boldsymbol{k}},\theta_{\boldsymbol{k}}) &= q \phi_{\boldsymbol{k}}\, ,\\
    \theta_q(\phi_{\boldsymbol{k}},\theta_{\boldsymbol{k}}) &= (1-\delta_{0,q}) \theta_{\boldsymbol{k}}\, ,
\end{aligned}
\end{equation}
where the integer $q\in \mathbb{Z}$ fixes the number of times the mapping $(\phi_q,\theta_q)$ wraps the sphere as we cover the base sphere $\mathbb{S}^2_{\mathrm{BZ}}$ one time, which is computed by the Skyrmion winding number~\cite{bouhon2019nonabelian,Unal_quenched_Euler}
\begin{equation}
\label{eq_skyrmion}
\begin{aligned}
W[\boldsymbol{n}_q]
    &=  \dfrac{1}{4\pi} \int_{\mathbb{S}^2_{\mathrm{BZ}}} \boldsymbol{n}_q 
        \cdot \left(\partial_{\phi_{\boldsymbol{k}}} \boldsymbol{n}_q \times \partial_{\theta_{\boldsymbol{k}} } \boldsymbol{n}_q \right) d\phi_{\boldsymbol{k}} \wedge d\theta_{\boldsymbol{k}},\\
        &  = q  \in \mathbb{Z} \, ,
\end{aligned}
\end{equation}
with $\boldsymbol{n}_q = \boldsymbol{n}(\phi_q,\theta_q)$. Substituting the above ansatz in the expression for the connection $\mathsf{a}(\phi_q,\theta_q) = e_{\theta}^T\cdot de_{\phi}$, we get the Euler form
\begin{subequations}\label{eq_Euler_form_sphere}
\begin{equation}
\begin{aligned}
    \mathsf{Eu}(\phi_q,\theta_q) &=  \left(\partial_{\phi_{\boldsymbol{k}}} u_1^T\cdot \partial_{\theta_{\boldsymbol{k}}} u_2-\partial_{\theta_{\boldsymbol{k}}} u_1^T\cdot \partial_{\phi_{\boldsymbol{k}}} u_2\right) d\phi_{\boldsymbol{k}}\wedge d\theta_{\boldsymbol{k}},\\
    &= -q (1-\delta_{0,q}) \sin\left[(1-\delta_{0,q}) \theta_{\boldsymbol{k}}\right] d\phi_{\boldsymbol{k}}\wedge d\theta_{\boldsymbol{k}},
\end{aligned}
\end{equation}
and then the Euler class
\begin{equation}
\label{eq_Euler_class_3B}
\begin{aligned}
    \chi_I[\{e_{\theta},e_{\phi}\}] &= \dfrac{1}{2\pi} \int_{\mathbb{S}_{\mathrm{BZ}}^2}  \mathsf{Eu}(\phi_q,\theta_q),\\
    &= q \left(-1+\cos [ (1-\delta_{0,q})\pi ] \right) = -2q,
\end{aligned}
\end{equation}
\end{subequations}
from which we see that the Euler class is doubled, \ie only even values of the Euler class are permitted. This is a direct consequence of the fact that the Hamiltonian $Q^{(2+1)}$ is given by the ``square'' of the winding unit vector $\boldsymbol{n}$. Taking the equivalence relation $\chi_I\sim -\chi_I$ in account, the homotopy classes of three-band Euler phases, splitting into $2+1$-band-subspaces, are thus classified by one even number through
\begin{equation}
    \beta(H^{(2+1)}) = [ \chi_I ] = [  2q]\,,~\text{with}~ \vert 2q\vert\in 2\mathbb{N},    
\end{equation}
such that the corresponding Euler phases are all represented up-to-homotopy by the flattened Hamiltonian $Q^{(2+1)}(\boldsymbol{n}_q)$, \ie by Eq.\;(\ref{eq_flat_21}) with the ansazt Eq.\;(\ref{eq_ansatz_angles}) for $q\in \mathbb{N}$. We give minimal tight-binding models for the phases $\chi_I \in \{2,4\}$ in Section \ref{sec_min_3B}. 

For completeness, we give an example of an Euler class reversal map \cite{bouhon2020geometric,wojcik2020homotopy}. For this we first define a representation of the nontrivial element of $\pi_1[\mathbb{R}P^2]=\mathbb{Z}_2$ through the deformation of the flattened Hamiltonian (noting $\mathsf{Gr}^{\mathbb{R}}_{2,3}=\mathbb{R}P^2$) \cite{bouhon2020geometric,wojcik2020homotopy}
\begin{equation}
\label{eq_euler_reversal}
\begin{aligned}
    \ell_{\boldsymbol{n}} &: [0,1]\rightarrow \mathbb{R}P^2
    \\
    &: t\mapsto \ell_{\boldsymbol{n}}(t)=Q^{(2+1)}[S(t)\cdot \boldsymbol{n}(\phi,\theta)], \\
    \text{with}~S(t) &= \left(
        \begin{array}{ccc}
            \cos \pi t & 0 & -\sin \pi t \\
            0 & 1 & 0\\
            \sin \pi t & 0 & \cos \pi t
        \end{array}
    \right).
\end{aligned}
\end{equation}
The transformation acts non-trivially on all the points of the classifying space, except at $\boldsymbol{n}(\pi/2,\pi/2)=(0,1,0)$ since $S(t)\cdot (0,1,0) = (0,1,0)$. In particular, the deformation starting at $\boldsymbol{n}_0=\boldsymbol{n}(\phi=0,\theta=0)=(0,0,1)$ defines a closed loop in $\mathbb{R}P^2$ since $\ell_{\boldsymbol{n}_0}(1) = Q^{(2+1)}[S(1)\cdot\boldsymbol{n}_0] = Q^{(2+1)}[-\boldsymbol{n}_0] = Q^{(2+1)}[\boldsymbol{n}_0] = \ell_{\boldsymbol{n}_0}(0)$. Noting that $Q^{(2+1)}[S(t)\cdot\boldsymbol{n}]=S(t)\cdot Q^{(2+1)}[\boldsymbol{n}]\cdot S(t)^T $, the transformation of the frame at the reference base point $(\phi,\theta)=(0,0)$, is $R_{t=1}(0,0)=(u_{1}(0,0)~u_{2}(0,0)~u_{3}(0,0))_{t=1} = (-u_{1}(0,0)~u_{2}(0,0)~-u_{3}(0,0))_{t=0}$, such that the Berry phase factors over the loop $\ell_{\boldsymbol{n}_0}$ for the two band subspaces are $\mathrm{e}^{-i\gamma_B[\ell_{\boldsymbol{n}_0};\{u_3\}]} = -1$ and $\mathrm{e}^{-i\gamma_B[\ell_{\boldsymbol{n}_0};\{u_1,u_2\}]} = -1$. We then conclude that the homotopy class $[[0,1],\ell_{\boldsymbol{n}_0}]$, with $\ell_{\boldsymbol{n}_0}(1)=\ell_{\boldsymbol{n}_0}(0)$, represents the generator of $\pi_1[\mathbb{R}P^2]=\mathbb{Z}_2$, as indicated by the $\pi$-Berry phase. We now want to compare the Euler class of the Hamiltonians before, $Q^{(2+1)}[\boldsymbol{n}(\phi,\theta)]$, and after the transformation, $Q^{(2+1)}[S(1)\cdot\boldsymbol{n}(\phi,\theta)]$. The Euler classes can only be compared if we chose the same gauge with respect to the same chosen reference point for both Hamiltonians. This reference point must be taken as the base point $\boldsymbol{n}_0$ that generates the above nontrivial loop. Keeping $\boldsymbol{n}_0$ fixed for the evaluation of the Euler class, we must thus compare the winding number of $\boldsymbol{n}$ in $Q^{(2+1)}[\boldsymbol{n}]$, with the winding number of $-S(t)\cdot \boldsymbol{n}$ in $Q^{(2+1)}[-S(t)\cdot \boldsymbol{n}](=Q^{(2+1)}[S(t)\cdot \boldsymbol{n}])$, since $-S(1)\cdot \boldsymbol{n}_0 = \boldsymbol{n}_0$. We conclude that the transformation reverses the signed Euler class from $\chi_I=2W[\boldsymbol{n}]=2q$ to $\chi_I = 2W[-S(1)\cdot\boldsymbol{n}] = -2q$.

We end this part by noting that $3$ is the minimal number of bands permitting a nontrivial Euler phase. Indeed, in the case of a (orientable) two-band system, the frame of eigenvectors $R\in \mathsf{SO}(2)$ can be written 
\begin{equation}
    R(f) = \left(u_1(f)~u_2(f) \right)= \left(
        \begin{array}{cc}
            \cos f & -\sin f \\
            \sin f & \cos f 
        \end{array}
    \right),
\end{equation} 
with $f\in \mathbb{S}^1$. Representing again the points of the Brillouin zone by the points of a sphere $\mathbb{S}^2_{\mathrm{BZ}}\ni (\phi_{\boldsymbol{k}},\theta_{\boldsymbol{k}})$ (assuming the orientability of the phase), we find that the Euler form Eq.\;(\ref{eq_Euler_form_sphere}) is identically zero.

\subsection{Four-band Euler phases}
The homotopy classification of the two-dimensional orientable four-band Euler phases that split into $2+2$-band subspaces is given by
\begin{equation}
\begin{aligned}
    \pi_2[\widetilde{\mathsf{Gr}}^{\mathbb{R}}_{2,4} ]/{\sim}  &= \pi_2[\mathbb{S}^2\times \mathbb{S}^2 ]/{\sim},\\
    &= \{(a,b)\in \mathbb{Z}^2\vert (a,b){\sim} (-a,-b)\}\ni \beta(H^{(2+2)}),
\end{aligned}
\end{equation}
where we have used the diffeomorphism $\widetilde{\mathsf{Gr}}^{\mathbb{R}}_{2,4} \approx \mathbb{S}^2\times \mathbb{S}^2$, and $\beta(H^{(2+2)})$ represents the homotopy class of the Bloch Hamiltonian $H^{(2+2)}(\boldsymbol{k})$ with eigenvalues that split as $E_1(\boldsymbol{k}) \leq E_2(\boldsymbol{k}) < E_3(\boldsymbol{k})\leq E_4(\boldsymbol{k})$ for all $\boldsymbol{k}\in\mathbb{T}^2$. The homotopy invariants are computed through the Euler classes of the two-band occupied and unoccupied subspaces, $(\chi_I , \chi_{II})$, modulo the homotopy equivalence $(\chi_I , \chi_{II}) \sim (-\chi_I , -\chi_{II})$, which we show is a consequence of the existence of an adiabatic deformation of the Hamiltonian reversing both Euler classes at the same time. (This is a consequence of the facts that $(i)$ while the Euler class is a homotopy invariant of an {\it oriented} vector bundle, the real Bloch Hamiltonians are only {\it orientable}, and $(ii)$ the topology of Bloch Hamiltonians are captured by homotopy classes (\ie with no base point) rather than by homotopy groups (\ie with a fixed base point) \cite{bouhon2020geometric}.) In the following, we write the homotopy invariant in terms of an equivalence class of Euler classes, \ie
\begin{equation}
\begin{aligned}
     \beta\left( H^{(2+2)} \right) &= [\chi_I,\chi_{II}] = [\boldsymbol{\chi}]   \\
     &= \{(\chi_I,\chi_{II}),(-\chi_I,-\chi_{II})\}\,.
\end{aligned}
\end{equation}

The modeling of the four-band Euler phases is obtained from a representative $R\in \mathsf{SO}(4)$ of the coset $[R]\in \mathsf{SO}(4)/[\mathsf{SO}(2)\times \mathsf{SO}(2)] \cong \mathbb{S}^2_+\times \mathbb{S}^2_-$. Using the Pl{\"u}cker embedding, we find (see Ref.\;\cite{bouhon2020geometric,abouhon_EulerClassTightBinding})
\begin{equation}
    R(\phi_+,\theta_+,\phi_-,\theta_-) = [\text{Eq.\;(\ref{eq_R_full})\;in\;Appendix}].
\end{equation}
Then, setting 
\begin{equation}
    \begin{aligned}
        (\phi_+, \theta_+)&=(\phi,\theta),\\
        (\phi_-, \theta_-)& =(\phi'+\pi/2,\theta'+\pi/2),
    \end{aligned}    
\end{equation}
we obtain the twofold degenerated Hamiltonian,
\begin{subequations}
\begin{equation}
\label{eq_H_imb_gen}
\begin{aligned}
     &H[\boldsymbol{n},\boldsymbol{n}';\epsilon_1,\epsilon_2] \\
     &= R(\phi,\theta,\phi',\theta') \cdot
            \left( \epsilon_1 \mathbb{1}_2
                \oplus 
                \epsilon_2 \mathbb{1}_2
            \right) \cdot R(\phi,\theta,\phi',\theta')^T, \\
    &= \dfrac{1}{2} \left\{ 
    (\epsilon_1+\epsilon_2)\Gamma_{00}  
    + (-\epsilon_1+\epsilon_2) Q^{(2+2)}[\boldsymbol{n},\boldsymbol{n}'] \right\},
\end{aligned}
\end{equation}
such that the gap condition reads $\epsilon_{1}<\epsilon_{2}$, with the flattened Hamiltonian (\ie for $\epsilon_2=-\epsilon_1=+1$) given by
\begin{equation}
\label{eq_flat_4B}
\begin{aligned}
    Q^{(2+2)}[\boldsymbol{n},
    \boldsymbol{n}']
    &=H[\boldsymbol{n},\boldsymbol{n}';-1,1] \\
    &=
      n_1' 
   \left(
        -n_1 \Gamma_{33} + n_2 \Gamma_{31} + n_3 \Gamma_{10}
    \right) \\
     &- n_2' 
    \left(
        +n_1 \Gamma_{13} - n_2 \Gamma_{11} + n_3 \Gamma_{30}
    \right)\\
     &+ n_3' 
    \left(
        +n_1 \Gamma_{01} + n_2 \Gamma_{03} - n_3 \Gamma_{22}
    \right),\\
    &= \boldsymbol{n}^T\cdot \underline{\Gamma} \cdot \boldsymbol{n}',
\end{aligned}
\end{equation}
which is determined by two unit vectors $\boldsymbol{n}(\phi,\theta) = (\cos\phi\sin\theta,\sin\phi,\sin\theta,\cos\theta)$ and $\boldsymbol{n}' = \boldsymbol{n}(\phi',\theta')$, with
\begin{equation}
\label{eq_gamma_matrices}
     \underline{\Gamma}= 
        \left(
            \begin{array}{rrr}
                -\Gamma_{33} & -\Gamma_{13} &
                \Gamma_{01} \\
                \Gamma_{31} &
                \Gamma_{11} &
                \Gamma_{03} \\
                \Gamma_{10} &
                -\Gamma_{30} &
                -\Gamma_{22}
            \end{array}
        \right)
     \;,
\end{equation}
\end{subequations}
where the Dirac matrices $\Gamma_{ij} = \sigma_i \otimes \sigma_j$ are defined from the Pauli matrices $\sigma_1 = \begin{rsmallmatrix}0&1\\1&0\end{rsmallmatrix}$, $\sigma_2 = \begin{rsmallmatrix}0&-i\\i&0\end{rsmallmatrix}$, $\sigma_3 = \begin{rsmallmatrix}1&0\\0&-1\end{rsmallmatrix}$, and $\sigma_0 = \mathbb{1}_2$. Inversely, inserting two generic vectors $\boldsymbol{h}$ and $\boldsymbol{h}'$ (\ie non unit vectors), we get
\begin{equation}
    Q^{(2+2)}[\boldsymbol{h},
    \boldsymbol{h}'] = H\left[
        \dfrac{\boldsymbol{h}}{\vert \boldsymbol{h}\vert},
        \dfrac{\boldsymbol{h}'}{\vert \boldsymbol{h}'\vert}
    ; \vert \boldsymbol{h}\vert, \vert \boldsymbol{h}'\vert\right] \;.
\end{equation}

Simplifying the Brillouin zone to the sphere $\mathbb{S}_{\mathrm{BZ}}^2$ (without loss of generality for the orientable phases), we can define all Euler phases in terms of two integers $(q,q')\in \mathbb{Z}^2$ through the ansatz
\begin{equation}
\label{eq_4B_para}
    \begin{aligned}
        (\phi_{q},\theta_q)  &= (q \phi_{\boldsymbol{k}},[1-\delta_{0,q}] \theta_{\boldsymbol{k}}),\\
    (\phi'_{q'},\theta'_{q'})  &= (q' \phi_{\boldsymbol{k}},[1-\delta_{0,q'}] \theta_{\boldsymbol{k}}).
    \end{aligned}
\end{equation}
We note that this parametrization readily implies that the Euler phases are characterized by the two winding numbers [Eq.\;(\ref{eq_skyrmion})].
\begin{equation}
        q = W[\boldsymbol{n}_{q}],~
        q' = W[\boldsymbol{n}'_{q'}].
\end{equation}
Substituting Eq.\;(\ref{eq_4B_para}), we obtain the Euler forms
\begin{equation*}
        \mathsf{Eu}_I  =  \dfrac{\sin \theta_0}{2} \left( q  - q' 
        \right),~
        \mathsf{Eu}_{II}  = \dfrac{\sin \theta_0}{2} \left( q  + q' \right),
\end{equation*}
leading to the Euler classes
\begin{equation}
\label{eq_winding_to_Euler}
        \chi_I  =  q-q',~
        \chi_{II}  = q+q'.
\end{equation}
We thus conclude that $Q^{(2+2)}[\boldsymbol{n}_q,\boldsymbol{n}'_{q'}]$ in Eq.\;(\ref{eq_flat_4B}) represents all the homotopy classes $[\chi_I,\chi_{II}]$ of four-band Euler phases, that is a pair of Euler classes modulo the equivalence relation $(\chi_I,\chi_{II}) \sim (-\chi_I,-\chi_{II})$, or, in terms of the winding numbers, $(q,q')\sim (-q,-q')$. As discussed above, this reduction comes from the existence of an adiabatic mapping that reverses both Euler classes, see Section \ref{sec_reversal_map_4B} below. It is important to note that, as a consequence, we can only keep track of the {\it relative} signs of the Euler classes in Eq.\;(\ref{eq_winding_to_Euler}). 

Importantly, we find the sum rule
\begin{equation}
    \chi_I + \chi_{II} = 0\mod 2,
\end{equation}
which guarantees the global cancellation of the second Stiefel Whitney class, \ie from the definition $w_{2,I(II)} = \chi_{I(II)} \mod 2$, we get $w_{2,I}+w_{2,II} = 0 \mod 2$. This is actually a requirement for any total oriented real vector bundle, \ie here taking all $N$ bands of the Bloch Hamiltonian. (Here, we implicitly assume that all the elements $\{H_{ab}(\boldsymbol{k})\}_{a,b=1,\dots,N}$ of the Bloch Hamiltonian are analytic functions of the momentum, or in words, that these are given by finite Fourier series.) 

\subsection{Balanced vs imbalanced four-band phases}

When one winding number is zero, we obtain $\vert \chi_I \vert = \vert \chi_{II} \vert$, \ie the absolute Euler classes are equal across the energy gap. We call these phases the {\it balanced} Euler phases. Whenever both winding numbers are nonzero, \ie $\vert q\vert ,\vert q'\vert>0$, we get unequal absolute Euler classes across the energy gap, \ie $\vert \chi_{I}\vert \neq \vert \chi_{II}\vert$. We call these phases the {\it imbalanced} Euler phases. The different responses under an external magnetic field between the balanced and the imbalanced Euler phases via their Hofstadter spectrum have been systematically studied in Ref.\;\cite{guan2021landau}. 

Fixing one constant unit vector, say $\boldsymbol{n}'_{q'=0} = \boldsymbol{n}(\phi'_c,\theta'_c)$, we importantly note the topological non-equivalence of the two balanced phases
\begin{subequations}
\begin{equation}
        H_a = H[\boldsymbol{n}_q,\boldsymbol{n}'_0;\epsilon_1,\epsilon_2] \not\simeq  H[\boldsymbol{n}'_0,\boldsymbol{n}_q;\epsilon_1,\epsilon_2] = H_b ,
\end{equation}
as indicated by the two inequivalent homotopy invariants, \ie
\begin{equation}
    \begin{aligned}
    \beta(H_a) &= [\boldsymbol{\chi}_a] = \{(q,q),(-q,-q)\} \\
    \neq \beta(H_b) &= [\boldsymbol{\chi}_b] = \{(-q,q),(-q,q)\}\,.
    \end{aligned}
\end{equation}
\end{subequations}
We discuss in detail in Section \ref{sec_3D_NL} an indicator that distinguishes the phases $[\chi_I,\chi_{II}]$ and $[\chi_I,-\chi_{II}]$.

\subsection{Mirror Chern number of the balanced degenerate Euler phases}\label{subsec_basal_mirror}
We show in Appendix \ref{ap_mirror}, see also Ref.\;\cite{guan2021landau}, that all the balanced and degenerate Euler phases possess an effective spinful basal mirror symmetry (\ie $\sigma_h H_{\text{bal}}^{\text{deg}} \sigma_h = H_{\text{bal}}^{\text{deg}}$ with $\sigma_h^2=-1$) which permits the definition of a mirror Chern number. Fixing $\boldsymbol{n}'=(0,0,1)$, we obtain in Appendix \ref{ap_mirror} 
\begin{equation}
\begin{aligned}
    C^{(-\mathrm{i})}_{I} &= -C^{(\mathrm{i})}_{I} =
    C^{(\mathrm{i})}_{II}=
    -C^{(-\mathrm{i})}_{II} ,\\
    &= W[\boldsymbol{n}] = q = \chi_I = \chi_{II}.
\end{aligned}
\end{equation}

The reverse is true, namely all Euler phases that are not twofold degenerate have no basal mirror symmetry and the mirror Chern number is not defined, leaving the Euler class as the unique fundamental topological invariant (\ie discarding crystalline topologies characterized by symmetry-indicators \cite{Clas5,Clas4,Khalaf_sum_indicators, Clas2, Clas3}, see also Section \ref{sec_sym}).

\subsection{Stable nodal points and nodal lines in 3D}
Let us label every block of isolated bands ordered in energy from below with a roman number $\nu = I,II,III,\dots$, \ie $E_{\nu,m_{\nu}} < E_{\nu+I,1}$, where $m_{\nu}$ is the number of bands in the $\nu$-th block of bands with the eigenenergies $\{E_{\nu,1}\leq \dots \leq E_{\nu,m_{\nu}}\}$. Given that the Euler class is only well defined for orientable two-band subspaces \cite{bouhon2019nonabelian,BJY_linking}, the most striking observable of the nontrivial Euler topology of a two-band subspace is the presence of stable nodal points that cannot be annihilated as long as the two bands remain separated from all the other bands (while preserving $C_2T$ symmetry). More precisely, given a two-band subspace, say the $\nu$-th block, with the Bloch eigenenergies $E_{\nu,1}\leq E_{\nu,2}$ and the Bloch eigenvectors $\{u_{\nu,1},u_{\nu,2}\}$, it must host a number $2\vert\chi_{\nu}\vert \in 2\mathbb{N}$ of stable nodal points determined by its Euler class $\chi_{\nu}[\{u_{\nu,1},u_{\nu,2}\}]$. This has for instance the consequence that any nontrivial two-band subspace with flat energy levels must necessarily be twofold degenerate. Non-trivial flat bands thus host mirror Chern numbers, and pairs of anti-propagating chiral branches must appear on each edge of the system. (We note that the situation is more subtle in the case of the coexistence of flat bands with dispersive bands, \eg as in twisted bilayer Graphene.)

\subsection{Nodal line continuations of nodal points in 3D}
Upon the adiabatic deformation of any Euler phase, say by a term of the real Hamiltonian scaling with the parameter $\lambda \in \mathbb{R}$, the nodal points extend into nodal lines within the three-dimensional parameter space $(\boldsymbol{k},\lambda)\in\mathbb{T}^2\times \mathbb{R}$. We show in Section \ref{sec_3D_NL}, and in Section \ref{sec_3D_NL_numerics} with concrete models, how this allows us to systematically generate 3D tight-binding Hamiltonian with linked nodal rings characterized by (1D) non-Abelian frame charges and (2D) linking numbers. 

\subsection{Euler class reversal map}\label{sec_reversal_map_4B}

For completeness, we here elaborate on the Euler class reversing map in the four-band case \cite{bouhon2020geometric}. Using the compact notation $\boldsymbol{p}=(\boldsymbol{n},\boldsymbol{n}')$, and taking $\boldsymbol{p}_0 = (\boldsymbol{n}_0 , \boldsymbol{n}'_0)$, with $\boldsymbol{n}_0=\boldsymbol{n}'_0=\boldsymbol{n}(\phi=0,\theta=0)  = (0,0,1)$, as a reference base point, the transformation
\begin{equation}
\begin{aligned}
    \ell_{\boldsymbol{p}} &: [0,1]\rightarrow \mathsf{Gr}^{\mathbb{R}}_{2,4}
    \\
    &: t\mapsto \ell_{\boldsymbol{p}}(t)=Q^{(2+2)}[S(t)\cdot \boldsymbol{p}]\,.
\end{aligned}
\end{equation}
(the unit interval here, $[0,1]$, should not be confused with one equivalence class), with $S(t)$ given in Eq.\;(\ref{eq_euler_reversal}), induces a non-trivial closed loop in the Hamiltonian space at $ \boldsymbol{p}_0$, since $\ell_{\boldsymbol{p}_0}(1) = Q^{(2+2)}[S(1)\cdot \boldsymbol{p}_0] = Q^{(2+2)}[- \boldsymbol{p}_0] = Q^{(2+2)}[ \boldsymbol{p}_0] = \ell_{\boldsymbol{p}_0}(0)$. The gapped spectrum of the Hamiltonian remains constant ($-\epsilon_1 = \epsilon_2=1$) through the whole transformation, since $\vert S(t)\cdot\boldsymbol{n} \vert=\vert S(t)\cdot\boldsymbol{n}' \vert =1$ [by Eq.\;(\ref{eq_flat_4B}) and Eq.\;(\ref{eq_H_imb_gen})]. We thus conclude that there exists a similitude relation $Q^{(2+2)}[ S(t)\cdot \boldsymbol{p}_0] = O(t)\cdot Q^{(2+2)}[ \boldsymbol{p}_0] \cdot O(t)^T$ with $O(t)\in \mathsf{SO}(4)$, from which we get the action on the eigen-frame, $O(t)\cdot R[\boldsymbol{p}_0] = (u_1(\boldsymbol{p}_0)~\cdots~u_4(\boldsymbol{p}_0))_{t}$. This leads to the Berry phase factors $\mathrm{e}^{-\gamma_B[\ell_{\boldsymbol{p}_0};\{u_1,u_2\}]} = -1$ and $\mathrm{e}^{-\gamma_B[\ell_{\boldsymbol{p}_0};\{u_3,u_4\}]} = -1$, indicating that the homotopy class $[[0,1],\ell_{\boldsymbol{p}_0}]$ represents the generator of $\pi_1[\mathsf{Gr}^{\mathbb{R}}_{2,4}]=\mathbb{Z}_2$. As in the three-band case, the Euler classes before and after the transformation must be evaluated with respect to the same gauge at the fixed base point $\boldsymbol{p}_0 = (\boldsymbol{n}_0,\boldsymbol{n}'_0)$. Since $S(1)\cdot \boldsymbol{p}_0 = -\boldsymbol{p}_0$, we compare the winding numbers $(q,q')$ for $Q^{(2+2)}[ \boldsymbol{p}]$ with those for $Q^{(2+2)}[ -S(1)\cdot \boldsymbol{p}]$ (\ie fixing the same reference point with the same gauge). We conclude that the transformation reverses the winding numbers $(q,q') = (W[\boldsymbol{n}],W[\boldsymbol{n}'])$ to $(W[-S(1)\cdot\boldsymbol{n}],W[-S(1)\cdot\boldsymbol{n}']) = (-q,-q')$, and thus reverses the Euler classes from $(\chi_I,\chi_{II})=(q-q',q+q')$ to $(\chi_I,\chi_{II})=(-q+q',-q-q')$.

\section{Minimal models with Euler class topology}\label{sec_min_models}
In this section we formulate the models of interest hosting the Euler topology of orientable phases. In particular we address concrete forms of several three and four band models from lower to higher Euler classes.

As reviewed in the previous Section, while the three-band case is characterized by a single even Euler class, the four-band case is classified by two Euler classes, thus permitting a greater variety of inequivalent topological phases, as we will detail in the following. Remarkably, we formulate explicit tight-binding models for a great variety of Euler phases by simply combining the generic forms derived in the previous Section together with a minimal parametrization of a two-band Chern model with the Chern number ranging from $-2$ to $2$. An alternative approach consists in truncating the inverse Fourier transform (\ie from the reciprocal space to the direct lattice space) of the degenerate Bloch Hamiltonians given in the previous Section, see \cite{bouhon2020geometric,guan2021landau} for more detail and Ref.\;\cite{abouhon_EulerClassTightBinding} for the implementation in a \texttt{Mathematica} notebook. 

\subsection{3-band case}\label{sec_min_3B}
The generic 3-band real Hermitian Hamiltonian can be written in terms of the five real Gell-Mann matrices
\begin{equation}
\begin{array}{l}
    \begin{array}{lll}
        \Lambda_1 \,=\, 
        \begin{rsmallmatrix}
            0 & 1 & 0 \\ 1 & 0 & 0 \\ 0 & 0 & 0
        \end{rsmallmatrix},&
            \Lambda_3 \,=\, \begin{rsmallmatrix}
            1 & 0 & 0 \\ 0 & -1 & 0 \\ 0 & 0 & 0
            \end{rsmallmatrix},&
            \Lambda_4 \,=\, \begin{rsmallmatrix}
            0 & 0 & 1 \\ 0 & 0 & 0 \\ 1 & 0 & 0
            \end{rsmallmatrix}, 
        \end{array}\\
        \begin{array}{ll}
             \Lambda_6 \,=\, \begin{rsmallmatrix}
            0 & 0 & 0 \\ 0 & 0 & 1 \\ 0 & 1 & 0
            \end{rsmallmatrix},&
            \Lambda_8 \,=\, \frac{1}{\sqrt{3}}\begin{rsmallmatrix}
            1 & 0 & 0 \\ 0 & 1 & 0 \\ 0 & 0 & -2
            \end{rsmallmatrix},
    \end{array}
    \end{array}
\end{equation}
to which we add the identity matrix $\Lambda_0=\mathbb{1}_3$, as
\begin{equation}
\begin{aligned}
    H^{[\chi_{I}]}_{3\text{B}}(\boldsymbol{k}) &= \boldsymbol{f}^{[\chi_{I}]}(\boldsymbol{k}) \cdot \boldsymbol{\Lambda} \\
    &= \sum\limits_{i= 0,1,3,4,6,8} f^{[\chi_{I}]}_i(\boldsymbol{k})\; \Lambda_i,
\end{aligned}
\end{equation}
where $\vert\chi_I\vert$ is the maximum Euler class reachable for the given ansatz (see below). 

We give here minimal tight-binding models for the topological Euler phases $\chi_I=2,4$. While the generalization to an arbitrary high Euler class is straightforward, the distance in the hopping processes required in order to achieve the nontrivial topology increases with the Euler class, making the experimental realization of higher Euler classes more involving.     

\begin{figure}[t!]
\centering
\begin{tabular}{c}
	\includegraphics[width=0.8\linewidth]{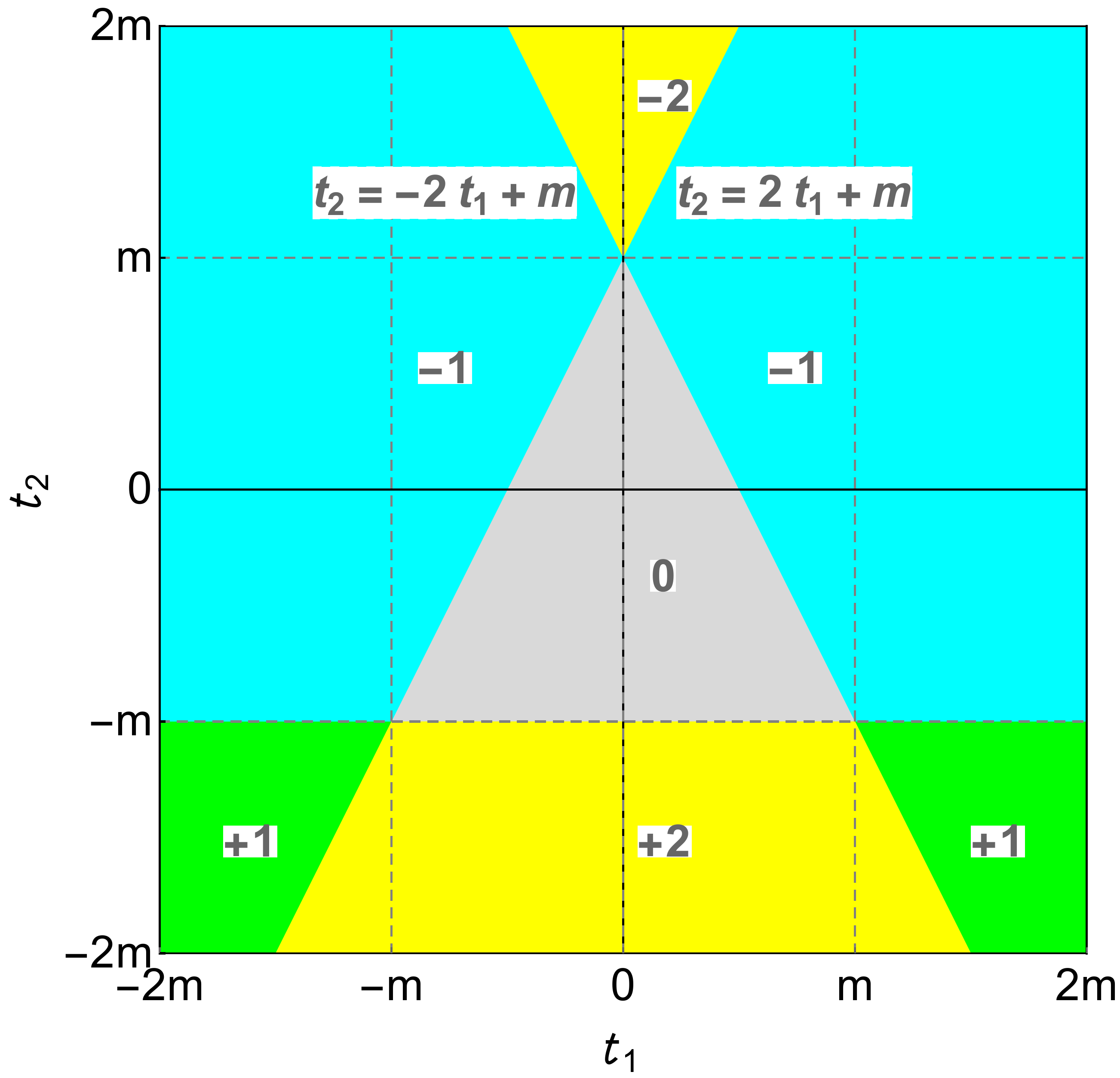}
\end{tabular}
\caption{\label{fig_phase_diag} Phase diagram of the Chern number (winding number) $c_1[\boldsymbol{h}]$ given by Eq.\;(\ref{eq_chern_3band}), adapted from Ref.\;\cite{sticlet2012engineering}.
}
\end{figure}

Taking advantage of the specific form of the flattened Hamiltonian Eq.\;(\ref{eq_flat_21}), we can readily use the ansatz of a two-band Chern model $H_{\text{2B}} = \boldsymbol{h}^{\alpha}\cdot \boldsymbol{\sigma}$, for which the Chern number is given by the winding of the vector $\boldsymbol{h}^{\alpha}(\boldsymbol{k}) =(h^{\alpha}_1(\boldsymbol{k}),h^{\alpha}_2(\boldsymbol{k}),h^{\alpha}_3(\boldsymbol{k}))$, \ie by $W[\boldsymbol{h}^{\alpha}/\vert \boldsymbol{h}^{\alpha}\vert]$ in Eq.\;(\ref{eq_skyrmion}). Defining the functions 
\begin{equation}
\label{eq_para_chern}
\begin{aligned}
    a(\boldsymbol{k})&=\lambda \sin k_1,~
    b(\boldsymbol{k})= \lambda \sin k_2\,,\\
    c(\boldsymbol{k}) &= m 
        - t_1 (\cos k_1  + \cos k_2) 
        - t_2 \cos(k_1+k_2) \,,
\end{aligned}
\end{equation}
we set $\boldsymbol{h}^{\alpha}$ ($\alpha=A,B,C$) to be one of the cyclic permutations of $(a,b,c)$, \ie
\begin{equation}
    \boldsymbol{h}^{A} = (a,b,c)\,,~
    \boldsymbol{h}^{B} = (b,c,a)\,,~
    \boldsymbol{h}^{C} = (c,a,b)\,.
\end{equation}
Substituting this ansatz within the two-band Chern model $\boldsymbol{h}^{\alpha}\cdot \boldsymbol{\sigma}$, the Chern number takes the generic form (assuming $\vert\lambda\vert>0$) \cite{sticlet2012engineering} 
\begin{multline}
\label{eq_chern_3band}
    c_1[\boldsymbol{h}^{\alpha}] = W\left[\dfrac{\boldsymbol{h}^{\alpha}(\boldsymbol{k})}{\vert \boldsymbol{h}^{\alpha}(\boldsymbol{k})\vert}\right] =
     \sign[-m -  t_2] \\
     + \dfrac{1}{2}\left(\sign[m + 2 t_1 -  t_2 ] + \sign[m - 2 t_1 -  t_2 ]\right) \,,
\end{multline}
which is independent to the cyclic form $\alpha=A,B,C$. In the following we set $\lambda=1$, without loos of generality. We show the generic phase diagram for $c_1[\boldsymbol{h}]$ for a fixed parameter $m$ in Figure \ref{fig_phase_diag} (which is slightly adapted from Ref.\,\cite{sticlet2012engineering}). 
We readily conclude that the Chern number is bounded as $-2\leq  c_1[\boldsymbol{h}] \leq 2$.

Making the substitution $\boldsymbol{n}\rightarrow \boldsymbol{h}^{\alpha}(\boldsymbol{k})$ in Eq.\;(\ref{eq_flat_21}), we obtain the following degenerate Bloch Hamiltonian (non-flattened because $\boldsymbol{h}^{\alpha}$ is not a unit vector)
\begin{subequations}
\begin{equation}
\label{eq_3B_nonflat}
\begin{aligned}
    H^{\alpha}_{\text{3B,deg}}[m,t_1,t_2,\lambda](\boldsymbol{k}) &= 
     \boldsymbol{h}^{\alpha}(\boldsymbol{k})\cdot \boldsymbol{h}^{\alpha}(\boldsymbol{k})^T \,\\
     &= \boldsymbol{f}^{[\chi_I]}(\boldsymbol{k})\cdot \boldsymbol{\Lambda},   
\end{aligned}
\end{equation}
where we have discarded the factor $2$ and the term $-\mathbb{1}_3$ used in Eq.\;(\ref{eq_flat_21}) that scale and shift the whole spectrum without changing the topology, with
\begin{equation}
\label{eq_3B_nonflat_b}
\begin{array}{rcl rcl}
    f^{[\chi_I]}_0 &=& \dfrac{\vert \boldsymbol{h}^{\alpha}\vert^2}{3} \,, & 
    f^{[\chi_I]}_1 &=& h_1^{\alpha} h_2^{\alpha} \,, \\
    f^{[\chi_I]}_3 &=& \dfrac{1}{2} (h_1^{\alpha}-h_2^{\alpha})(h^{\alpha}_1+h^{\alpha}_2) \,, &
    f^{[\chi_I]}_4 &=& h^{\alpha}_1 h^{\alpha}_3 \,,\\
    f^{[\chi_I]}_8 &=& \dfrac{\vert \boldsymbol{h}^{\alpha}\vert^2-3 {h^{\alpha}_3}^2}{2\sqrt{3}} \,, &
    f^{[\chi_I]}_6 &=& h^{\alpha}_2 h^{\alpha}_3 \,.
\end{array}
\end{equation}
\end{subequations}
We readily observe that any three-band model of a phase with nontrivial Euler topology requires a winding vector $\boldsymbol{h}^{\alpha}(\boldsymbol{k})$, such that the three components $\{h^{\alpha}_i(\boldsymbol{k})\}_{i=1,2,3}$ cannot be identically zero, which itself implies that all real Gell-Mann matrices must be present in Eq.\;(\ref{eq_3B_nonflat}). 

The analytical eigenvalues are now
\begin{equation}
\label{eq_spec_3B}
\begin{aligned}
    E_1(\boldsymbol{k}) &= E_2(\boldsymbol{k}) = 0 ,\\
    E_3(\boldsymbol{k}) &=  \boldsymbol{h}^{\alpha} \cdot \boldsymbol{h}^{\alpha} = a(\boldsymbol{k})^2 + b(\boldsymbol{k})^2 +c(\boldsymbol{k})^2\,,
\end{aligned}
\end{equation}
where the two-band subspace is still flat on top of being degenerate. Whenever the phase is gapped, combining $\vert \chi_I \vert= \vert 2q \vert $ [Eq.\;(\ref{eq_Euler_class_3B})] with $q=c_1[\boldsymbol{h}^{\alpha}]$ [Eq.\;(\ref{eq_chern_3band})], we find that the topology of the gapped Euler phase is given by 
\begin{equation}
\label{eq_chern_Euler_3B}
    [ \chi_I ] = [\, 2 c_1[\boldsymbol{h}^{\alpha}] \,] \,,
\end{equation}
if $E_3(\boldsymbol{k}) >  0$ for all $\boldsymbol{k}\in \text{BZ}$.

\subsubsection{$ 
[\chi_I]=[2]$}

Setting $(t_2,\lambda)=(0,1)$, we deduce from Eq.\;(\ref{eq_chern_3band}) and Eq.\;(\ref{eq_chern_Euler_3B}) that the maximum Euler class is $\chi_I=2$. From Eq.\;(\ref{eq_spec_3B}), we readily find the energy gap \cite{ezawa2021euler}
\begin{equation}
\begin{aligned}
    \Delta_{(t_2=0,\lambda=1)} &= \min\limits_{\boldsymbol{k},m,t_1} \left\{ E_3(\boldsymbol{k}) - E_2(\boldsymbol{k}) \right\},\\
    &= \min \left\{m^2, (m\pm 2t_1)^2 \right\}.
\end{aligned}
\end{equation}
We thus conclude, given Eq.\;(\ref{eq_chern_3band}) and (\ref{eq_chern_Euler_3B}), that the gapped phases and their topologies are given by ($\lambda=1,t_2=0$)
\begin{equation}
\begin{aligned}
    \vert \chi_I \vert &= 2~\text{for}~ \vert t_1 \vert > \dfrac{\vert m \vert }{2}~\text{and}~\vert m\vert>0,\\
    \text{and}~\vert \chi_I \vert &= 0~\text{for}~ \vert t_1 \vert < \dfrac{\vert m \vert }{2}~\text{and}~\vert m\vert>0\,.
\end{aligned}
\end{equation}

\begin{figure}[t!]
\centering
\begin{tabular}{ll}
    (a) & (b) \\
	\includegraphics[width=0.5\linewidth]{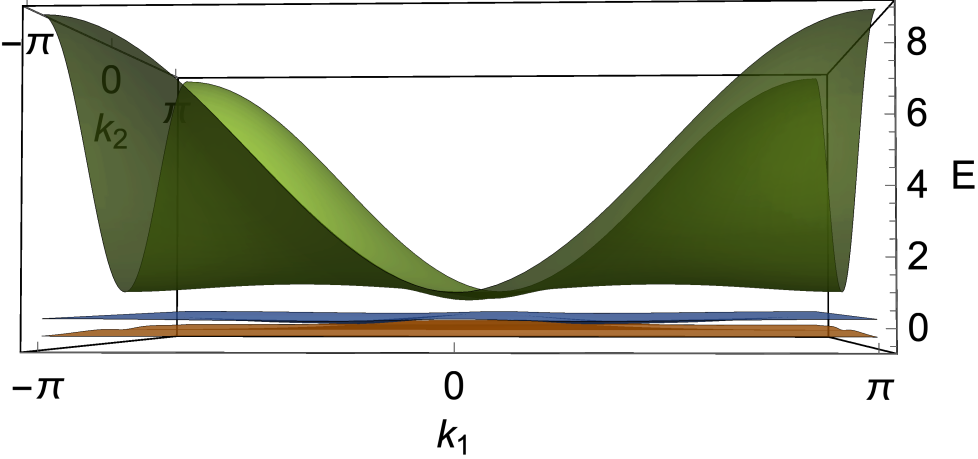} &
	\includegraphics[width=0.5\linewidth]{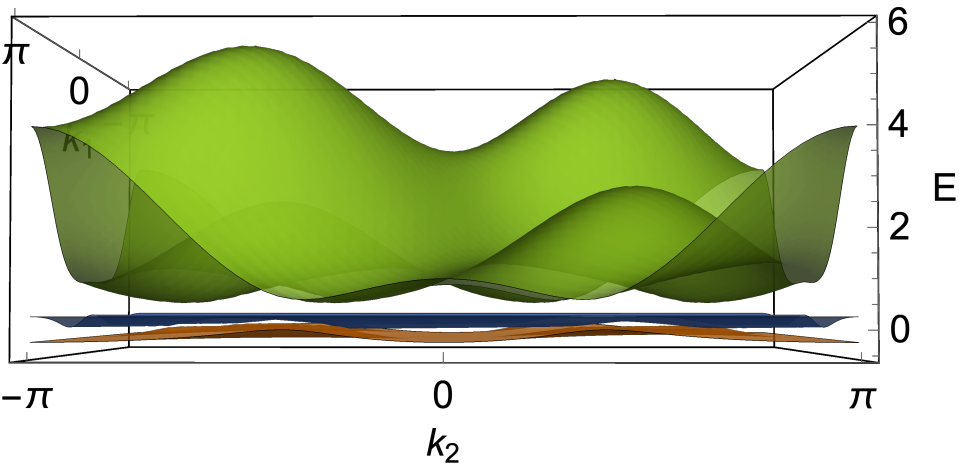}    \\
	(c) & (d) \\
	\includegraphics[width=0.5\linewidth]{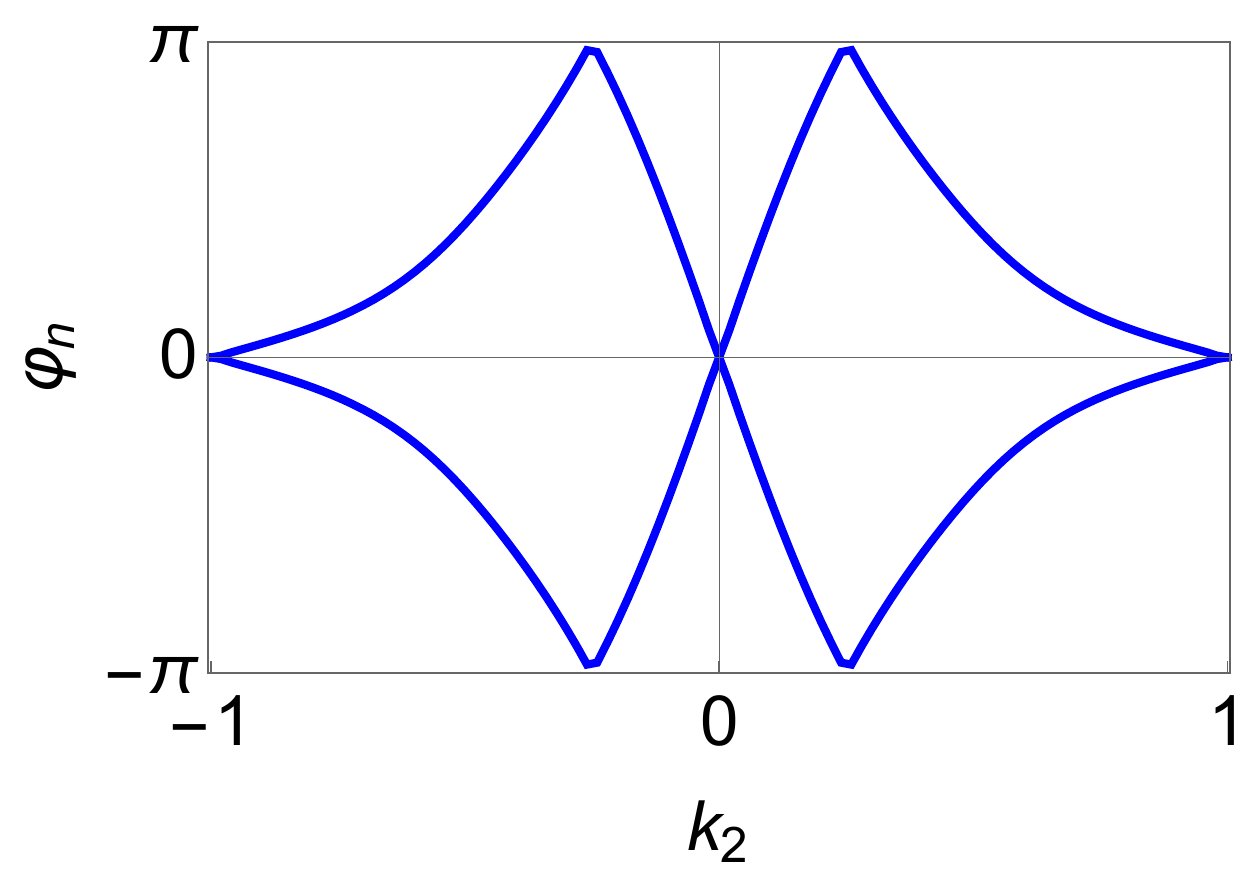}  &
	\includegraphics[width=0.5\linewidth]{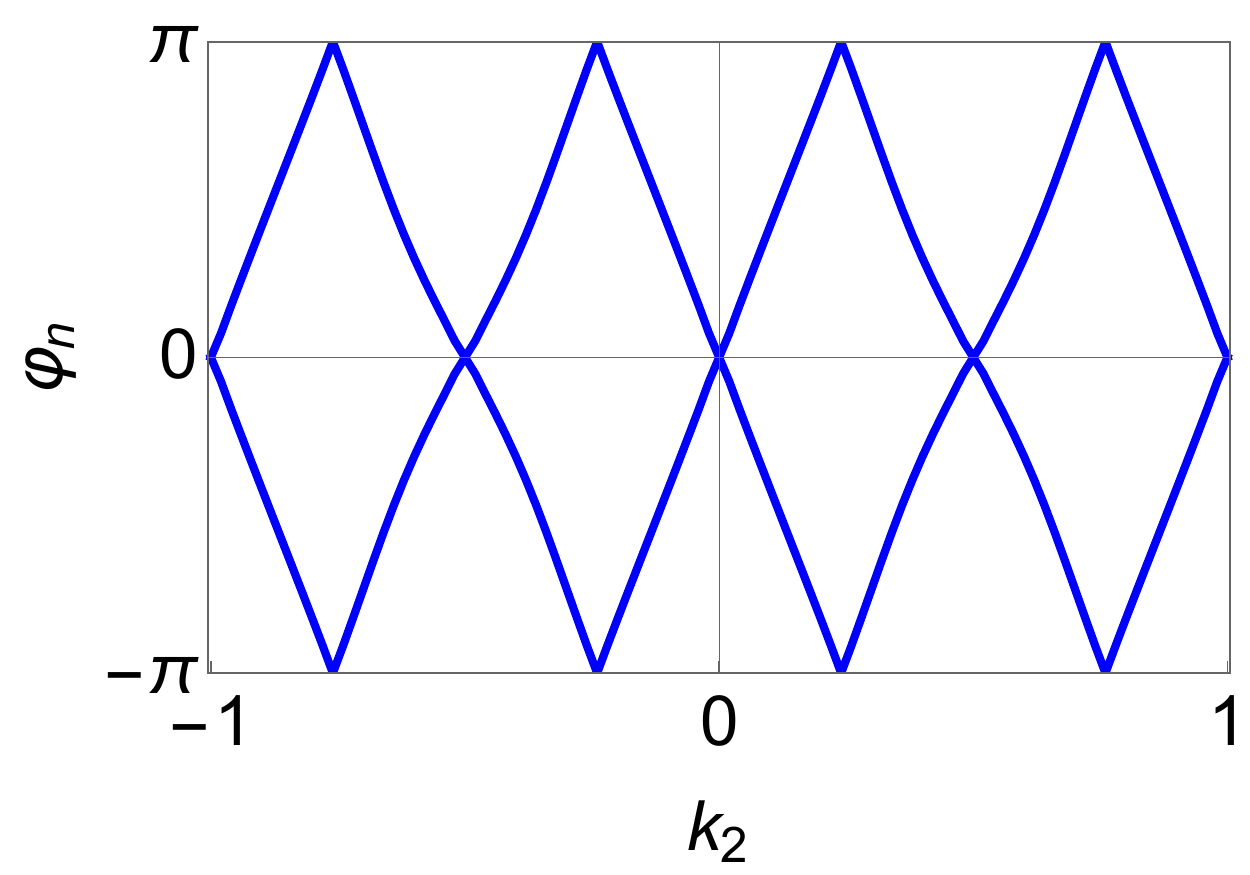} 
\end{tabular}
\caption{\label{fig_3B} (a,b) Band structures and (c,d) Wilson loop of the gapped (orientable) Euler phases in three-level systems, obtained from the minimal models (a,c) $H^{[2]}_{\text{3B}}$ for $(m,t_1,\delta)=(1,1,1/4)$ [Eq.\;(\ref{eq_min_H_E2})] for the Euler class $\chi_I =2 $, and (b,d) $H^{[4]}_{\text{3B}}$ for $(m,t_1,t_2,\delta)=(1/2,0,-3/2,1/4)$ [Eq.\;(\ref{eq_min_H_E4})] for the Euler class $\chi_I =4 $. We have taken $k_2$ in units of $\pi$.
}
\end{figure}

By choosing $\boldsymbol{h}^{A}$ and adding a small constant term that splits the degeneracy of the lower two-band, we obtain the minimal model
\begin{equation}
\label{eq_min_H_E2}
    H^{[2]}_{\text{3B}}(\boldsymbol{k})  =  H^A_{\text{3B,deg}}[m,t_1,0,1](\boldsymbol{k}) + \delta\, \Lambda_3\,,
\end{equation}
where we choose $\vert\delta\vert > 0$ such that the band gap remains open. For instance, setting $m=t_1=1$, the gap remains open for $\vert \delta\vert<1/2$. We show the band structure of $H^{[2]}_{\text{3B}}(\boldsymbol{k})$ for $(m,t_1,\delta)=(1,1,1/4)$ in Figure \ref{fig_3B}(a), and the winding of Wilson loop of the lower two-band subspace in Figure \ref{fig_3B}(c) indicating an Euler class $\vert\chi_I\vert =2$. 

\subsubsection{$[\chi_I]=[4]$}

Including the term in $t_2$ in Eq.\;(\ref{eq_para_chern}), we find the maximum Euler class $\chi_I=4$ [from Eq.\;(\ref{eq_chern_3band}) and Eq.\;(\ref{eq_chern_Euler_3B})]. The values of the parameters for which the phase is gapless are readily defined at the jumps of the step function Eq.\;(\ref{eq_chern_3band}) (since these correspond to a transition between different Chern phases, which requires the closing of the gap) \cite{sticlet2012engineering}, \ie ($\lambda=1$)
\begin{multline}
    \Delta = \min \left\{E_3 - E_2 \right\} =  0 \\
        \Leftrightarrow 
     (t_2=-m) ~\text{or}~ (t_2=\pm2 t_1 +m) \,,
\end{multline}
From the relation $\vert \chi_I\vert = \vert 2c_1[\boldsymbol{h}]\vert$ and $\lambda=1$, we find the following conditions for the gapped phase of maximum Euler class, assuming $m\geq 0$, (see Figure \ref{fig_phase_diag}) 
\begin{multline}
    \vert\chi_I \vert= 4 \Leftrightarrow \\
\begin{array}{cl}    
    & \left( t_2> 2t_1+m ~\text{and}~ t_2>-2t_1+m\right) \\
    \text{or} & \left( t_2< -m ~\text{and}~ t_2<2t_1+m~\text{and}~ t_2<-2t_2+m\right)\,.
\end{array}
\end{multline}
 
 Choosing $\boldsymbol{h}^A$, we then define the minimal Bloch Hamiltonian as ($\lambda=1$)
\begin{equation}
\label{eq_min_H_E4}
    H^{[4]}_{\text{3B}}(\boldsymbol{k})  = 
    H^A_{\text{3B,deg}}[m,t_1,t_2,1](\boldsymbol{k}) +\delta\, \Lambda_3\,,
\end{equation}
where $\vert \delta\vert >0$ lifts the degeneracy of the two-band subspace, leaving eight stable nodal points connecting the bands 1 and 2. The parameter $\delta$ must be chosen as a function of $(m,t_1,t_2)$ under the condition that the energy gap remains open. For instance, setting $(m,t_1,t_2)=(1/2,0,-3/2)$, the gap remains open for $\vert \delta \vert < 3/7$. We show the band structure of $H^{[4]}_{\text{3B}}$ for $(m,t_1,t_2,\delta)=(1/2,0,-3/2,1/4)$ in Figure \ref{fig_3B}(a), and the winding of Wilson loop of the lower two-band subspace in Figure \ref{fig_3B}(c) indicating an Euler class $\vert\chi_I\vert =4$.

\subsection{4-band case}\label{sec_4B_models}
We now turn to four-band models with a $2+2$-band splitting. We closely follow the same strategy, as for the three-band models, of using the minimal parametrization of a two-band Chern model Eq.\;(\ref{eq_para_chern}) and with the Euler topology inferred from the phase diagram in Figure \ref{fig_phase_diag} [Eq.\;(\ref{eq_chern_3band}) \cite{sticlet2012engineering}]. Writing the Euler classes of the two subspaces as the vector $\boldsymbol{\chi} = (\chi_I,\chi_{II})$ and corresponding Homotopy class $[\boldsymbol{\chi}]=[\chi_I,\chi_{II}]$, the generic 4-band real Hermitian Hamiltonian can be written in terms of nine real gamma matrices as
\begin{equation}
\label{eq_generic_4B}
\begin{aligned}
    H^{[\boldsymbol{\chi}]}_{4\text{B}}(\boldsymbol{k}) 
    &= \sum\limits_{ \substack{ ij \in 01,03,10,30, \\ 11,13,22,31,33 } } g^{[\boldsymbol{\chi}]}_{ij}(\boldsymbol{k}) \Gamma_{ij} \,,\\
    &= \boldsymbol{g}^{[\boldsymbol{\chi}]}(\boldsymbol{k}) \cdot \boldsymbol{\Gamma} \,,
\end{aligned}
\end{equation}
and where we have used the vector notation $\boldsymbol{\Gamma}=(\Gamma_{01},\Gamma_{03},\Gamma_{10},\Gamma_{30},\Gamma_{11},\Gamma_{13},\Gamma_{22},\Gamma_{31},\Gamma_{33})$ [see below Eq.\;(\ref{eq_gamma_matrices})] and similarly for $\boldsymbol{g}^{[\boldsymbol{\chi}]}$.

We make the substitution $\boldsymbol{n}\rightarrow \boldsymbol{h}^{\alpha}$ and $\boldsymbol{n}'\rightarrow \boldsymbol{h}'^{\beta}$ in Eq.\;(\ref{eq_flat_4B}), defining the vectors $\boldsymbol{h}^{\alpha}$ and $\boldsymbol{h}'^{\beta}$, for $\alpha,\beta=A,B,C$, as one of the cyclic permutations
\begin{equation}
\label{eq_permutations}
\begin{array}{rcl rcl rcl}
    \boldsymbol{h}^{A} &=& (a,b,c),
    &\boldsymbol{h}^{B} &=& (b,c,a),
    &\boldsymbol{h}^{C} &=& (c,a,b),\\
    \boldsymbol{h}'^{A} &=& (a',b',c'),
    &\boldsymbol{h}'^{B} &=& (b',c',a'),
    &\boldsymbol{h}'^{C} &=& (c',a',b'),
\end{array}
\end{equation}
with $a(\boldsymbol{k})$, $b(\boldsymbol{k})$ and $c(\boldsymbol{k})$ defined in Eq.\;(\ref{eq_para_chern}), and similarly for $a'(\boldsymbol{k})$, $b'(\boldsymbol{k})$ and $c'(\boldsymbol{k})$ with the substitution $(m,t_1,t_2,\lambda)\rightarrow (m',t_1',t_2',\lambda')$. This gives the degenerate (non-flattened) Bloch Hamiltonian
\begin{subequations}
\begin{equation}
\label{eq_4B_deg}
\begin{aligned}
    H^{\alpha\beta}_{\text{4B,deg}}
    \left[
    \begin{smallmatrix}
        m & t_1 & t_2 & \lambda \\
        m' & t_1' & t_2' & \lambda'
    \end{smallmatrix}
    \right]
    (\boldsymbol{k}) &= 
     \boldsymbol{h}^{\alpha}(\boldsymbol{k})^T\cdot
     \underline{\Gamma}\cdot \boldsymbol{h}'^{\beta}(\boldsymbol{k})  \\
     &= \boldsymbol{g}^{[\boldsymbol{\chi}]}(\boldsymbol{k})\cdot \boldsymbol{\Gamma},    
\end{aligned}
\end{equation}
with
\begin{equation}
\begin{array}{rcr rcr rcr}
    g^{[\boldsymbol{\chi}]}_{01} &=&  h_1^{\alpha} h_3'^{\beta}\,, & 
    g^{[\boldsymbol{\chi}]}_{03} &=&  h_2^{\alpha} h_3'^{\beta} \,, &
    g^{[\boldsymbol{\chi}]}_{10} &=& h_3^{\alpha} h_1'^{\beta} \,,\\
    g^{[\boldsymbol{\chi}]}_{30} &=& -h_3^{\alpha} h_2'^{\beta} \,, &
    g^{[\boldsymbol{\chi}]}_{11} &=& h_2^{\alpha} h_2'^{\beta} \,,&
    g^{[\boldsymbol{\chi}]}_{13} &=& -h_1^{\alpha} h_2'^{\beta} \,, \\ g^{[\boldsymbol{\chi}]}_{22} &=& -h_3^{\alpha} h_3'^{\beta} \,, & 
    g^{[\boldsymbol{\chi}]}_{31} &=& h_2^{\alpha} h_1'^{\beta} \,, &
    g^{[\boldsymbol{\chi}]}_{33} &=& -h_1^{\alpha} h_1'^{\beta} \,.
\end{array}
\end{equation}
\end{subequations}
The eigenvalues are now
\begin{equation}
\label{eq_4band_eigenvalues}
\begin{aligned}
    &E_1(\boldsymbol{k}) = E_2(\boldsymbol{k}) = -\epsilon(\boldsymbol{k}) \,,~E_3(\boldsymbol{k}) = E_4(\boldsymbol{k}) = \epsilon(\boldsymbol{k})  \,,\\
    &\epsilon(\boldsymbol{k}) = \vert \boldsymbol{h}(\boldsymbol{k}) \vert \vert \boldsymbol{h}'(\boldsymbol{k}) \vert \,,
\end{aligned}
\end{equation}
such that the spectrum is gapped whenever $\epsilon(\boldsymbol{k}) >0$ for all $\boldsymbol{k} \in \text{BZ}$. 

From the identifications
\begin{equation}
\label{eq_4band_chern_euler}
    c_1[\boldsymbol{h}^{\alpha}] = q\,,~c_1[\boldsymbol{h}'^{\beta}] = q'\,,
\end{equation}  
and the phase diagram of Eq.\;(\ref{eq_chern_3band}), together with Eq.\;(\ref{eq_winding_to_Euler}), we remarkably obtain that the ansatz of the two-band Chern model Eq.\;(\ref{eq_para_chern}) is sufficient to generate all the following four-band ($2+2$)-Euler phases
\begin{equation}
\begin{aligned}
    [\boldsymbol{\chi}] &\in \left\{
        \begin{array}{l}
            (q-q',q+q') \\
            \sim (-q+q',-q-q')
        \end{array}
        \left\vert
        \begin{array}{r}
         -2\leq q \leq 2,\\
         -2\leq q' \leq 2
        \end{array} \right.\right\} \,,\\ 
    & \in \left\{ 
       [0,0], [1,1],[1,-1],[2,2],[2,-2],\right.\\
    & \quad~~\, [2,0],[4,0],[3,1],[3,-1],  \\
    & \quad~\; \left.[0,2],[0,4],[1,3],[1,-3]
    \right\}\,.
\end{aligned}
\end{equation}
In the following, we only consider $\{[2,0],[4,0],[3,1],[3,-1]\}$ among the imbalanced phases, since the imbalanced phases $\{[0,2],[0,4],[1,3],[1,-3]\}$ can readily be obtained from the former through the transformation $H\rightarrow -H$ of the Hamiltonian.

\subsubsection{Balanced phase $[\boldsymbol{\chi}] =[1,1]$}

\begin{figure}[t!]
\centering
\begin{tabular}{ll}
    (a) & (b) \\
	\includegraphics[width=0.5\linewidth]{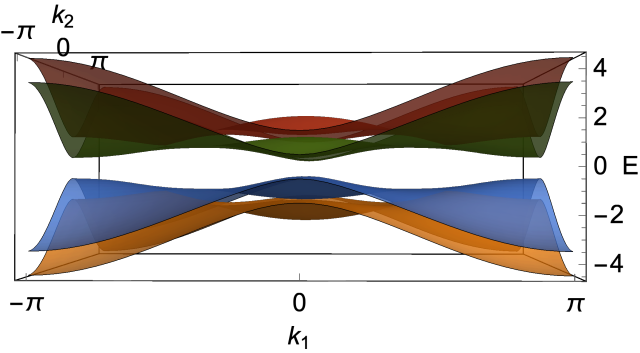} &
	\includegraphics[width=0.5\linewidth]{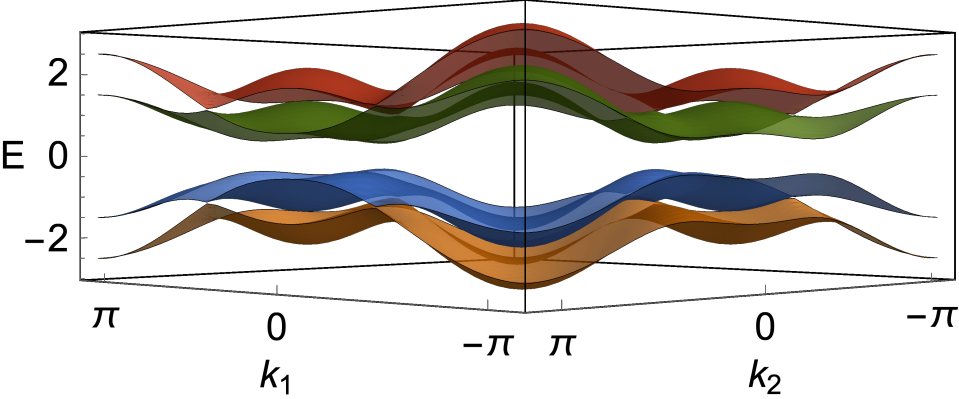}    \\
	(c) & (d) \\
	\includegraphics[width=0.5\linewidth]{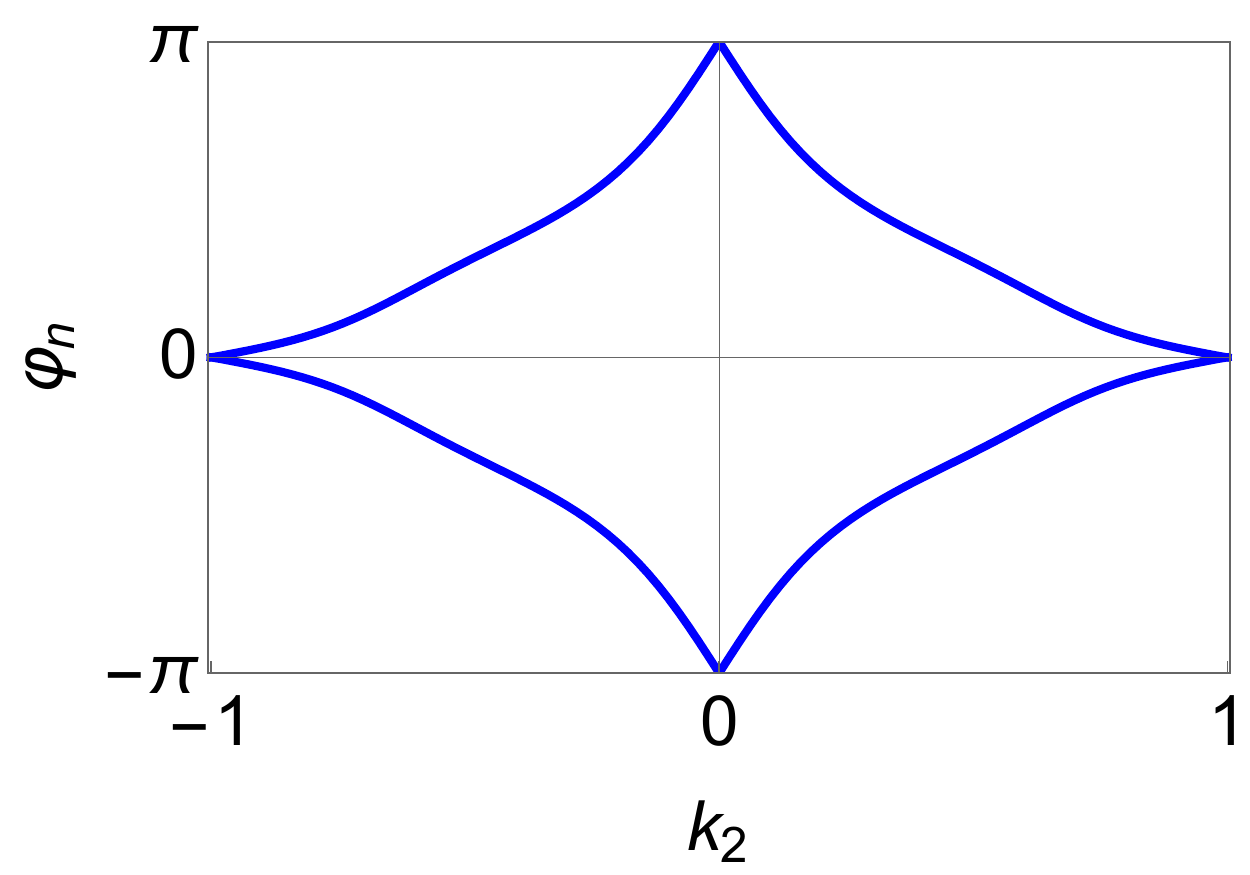}  &
	\includegraphics[width=0.5\linewidth]{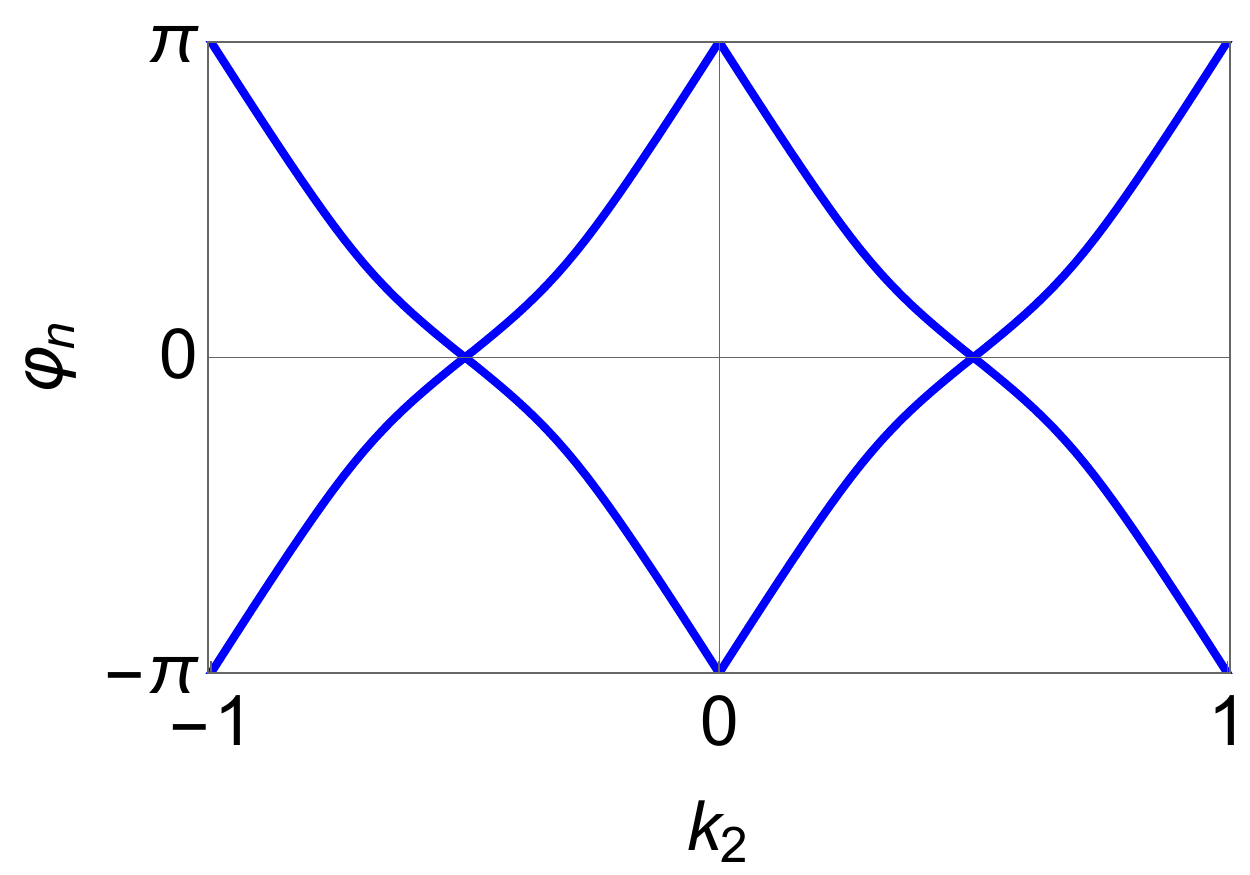} 
\end{tabular}
\caption{\label{fig_4B} (a,b) Band structures and (c,d) Wilson loop of the gapped (orientable) four-band balanced Euler phases belonging to the homotopy classes $[\boldsymbol{\chi}]=[1,1]$ and $[2,2]$, obtained for the minimal models (a,c) $H^{[1,1]}_{\text{4B}}$ with $(m,t_1,
\delta)=(1,3/2,1/2)$ in Eq.\;(\ref{eq_min_H_E2}), and (b,d) $H^{[2,2]}_{\text{4B}}$ with $(m,t_1,t_2,\delta)=(1/2,0,-3/2,1/2)$ in Eq.\;(\ref{eq_min_H_E4}). We have taken $k_2$ in units of $\pi$. By the symmetry of Eq.\;(\ref{eq_4band_eigenvalues}) under the permutations Eq.\;(\ref{eq_permutations}), (a,c) remain the same for $H^{[1,-1]}_{\text{4B}}$ with $(m',t_1',\delta)=(1,3/2,1/2)$, and (b,d) for $H^{[2,-2]}_{\text{4B}}$ with $(m',t_1',t_2',\delta)=(1/2,0,-3/2,1/2)$.
}
\end{figure}

We start with the balanced model for the homotopy class $[\boldsymbol{\chi}] = [\chi_I,\chi_{II}] = [1,1] = \{(1,1),(-1,-1)\} $ obtained for $q=c_1[\boldsymbol{h}]=-1$ and $q'=c_1[\boldsymbol{h}']=0$ [by Eq.\;(\ref{eq_winding_to_Euler}) and Eq.\;(\ref{eq_4band_chern_euler})]. Choosing $\boldsymbol{h}^A$ and $\boldsymbol{h}'^A$, and setting $(t_2,\lambda)=(0,1)$ and $(m',t_1',t_2',\lambda') = (1,0,0,0)$
(leading to $\boldsymbol{h}'^A=(0,0,1)$ and $c_1[\boldsymbol{h}'^A]=0$), we define from Eq.\;(\ref{eq_4B_deg}) the minimal model
\begin{equation}
\label{eq_min_H_E11}
\begin{aligned}
    H^{[1,1]}_{\text{4B}}(\boldsymbol{k})  &=  
     H^{AA}_{\text{4B,deg}}
    \left[\begin{smallmatrix}
        m & t_1 & 0 & 1 \\
        1 & 0 & 0 & 0
    \end{smallmatrix}\right]
    (\boldsymbol{k}) 
    + \delta\, \Gamma_{13}\,,\\
    & = \sin k_1 \Gamma_{01} + \sin k_2 \Gamma_{03} + \delta\, \Gamma_{13} \\
    &  - [m-t_1(\cos k_1+\cos k_2)] \Gamma_{22} \,,
\end{aligned}
\end{equation}
where the parameter $\vert \delta\vert>0$ is taken in order to lift the degeneracy of the two two-band subspaces, while keeping the band gap open. For instance, setting $(m,t_1)=(1,1)$, the gap remains open for $\vert \delta \vert < 1$. We show in Figure \ref{fig_4B}(a) the band structure and (c) the Wilson loop for the model $H^{[1,1]}_{\text{4B}}(\boldsymbol{k})$ for $(m,t_1,\delta)=(3/2,1,1/2)$. (Note that the choice $(m,t_1)=(3/2,1)$ implies the gap condition $\vert \delta \vert < 2/\sqrt{5}$.) The winding of Wilson loop of the two-band subspaces indicates the Euler classes $[\boldsymbol{\chi}]=[1,1]$ or $[\boldsymbol{\chi}]=[1,-1]$. Indeed, we cannot read the relative signs of the Euler classes $(\chi_I,\chi_{II})$ from the Wilson loop only. We assign the homotopy indicator $[1,1]$ from the fact that we know the winding numbers $q$ and $q'$ by construction, \ie here $(q,q')=(-1,0)$. In the next Section we present an indicator that differentiates the homotopy classes $[1,1]$ and $[1,-1]$, in the form of a linked nodal ring with nontrivial monopole Euler charge (equivalently a non-Abelian frame charge) at the interface of the two phases.

\subsubsection{Balanced phase $[\boldsymbol{\chi}] =[1,-1]$}

The phase $[\boldsymbol{\chi}] = [1,-1]$ is obtained for $q=c_1[\boldsymbol{h}]=0$ and $q'=c_1[\boldsymbol{h}']=-1$. Choosing $\boldsymbol{h}^B$ and $\boldsymbol{h}'^A$, and setting the parameters $(t_2',\lambda')=(0,1)$ and $(m,t_1,t_2,\lambda) = (1,0,0,0)$ (that gives $\boldsymbol{h}^B=(b,c,a)=(0,0,1)$ and $c_1[\boldsymbol{h}^B]=0$), we define the minimal model
\begin{equation}
\label{eq_min_H_E11m}
\begin{aligned}
    H^{[1,-1]}_{\text{4B}}(\boldsymbol{k})  &=  
     H_{\text{4B,deg}}^{BA}
    \left[\begin{smallmatrix}
        1 & 0 & 0 & 0 \\
        m' & t_1' & 1 & 0
    \end{smallmatrix}\right]
    (\boldsymbol{k}) 
    + \delta\, \Gamma_{13}\,,\\
    & = \sin k_1 \Gamma_{31} + \sin k_2 \Gamma_{11} + \delta\, \Gamma_{13} \\
    &  + [m' - t_1'(\cos k_1+\cos k_2)] \Gamma_{03} \,.
\end{aligned}
\end{equation}
Noting the symmetry of Eq.\;(\ref{eq_4band_eigenvalues}) under the permutations of $\{a,b,c\}$ and $\{a',b',c'\}$ Eq.\;(\ref{eq_permutations}), the band structure for $H^{[1,-1]}_{\text{4B}}$ is identical to the one of $H^{[1,1]}_{\text{4B}}$, upon exchanging non-primed to primed parameters. The gap condition is thus the same as in the previous case, \eg $\vert \delta \vert < 1$ if we set $(m',t_1')=(1,1)$, or $\vert \delta \vert < 2/\sqrt{5}$ if we set $(m',t_1')=(1,3/2)$. For $(m',t_1',\delta')=(1,3/2,1/2)$, the band structure is again given by Figure \ref{fig_4B}(a). Moreover, since the Wilson loop does not capture the relative sign of the Euler classes it is also the same as in Figure \ref{fig_4B}(c).

\subsubsection{Balanced phase $[\boldsymbol{\chi}] =[2,2]$}

For the Euler phase $[\boldsymbol{\chi}]=[2,2]$, we take $q=c_1[\boldsymbol{h}]=2$ and $q'=c_1[\boldsymbol{h}']=0$, and define the minimal model
\begin{equation}
\label{eq_min_H_E22}
\begin{aligned}
    H^{[2,2]}_{\text{4B}} (\boldsymbol{k})  &=  
     H^{AA}_{\text{4B,deg}}
    \left[\begin{smallmatrix}
        m & t_1 & t_2 & 1 \\
        1 & 0 & 0 & 0
    \end{smallmatrix}\right]
    (\boldsymbol{k}) 
    + \delta\, \Gamma_{13}\,,\\
    & = \sin k_1 \Gamma_{01} + \sin k_2 \Gamma_{03} + \delta\, \Gamma_{13}  \\
    &- [m-t_1(\cos k_1+\cos k_2)- t_2 \cos(k_1+k_2)] \Gamma_{22}   \,.
\end{aligned}
\end{equation}
Setting $(m,t_1,t_2,\delta)=(1/2,0,-3/2,1/2)$, we plot the band structure and the Wilson loop in Figure \ref{fig_4B}(b,d).

\subsubsection{Balanced phase $[\boldsymbol{\chi}] =[2,-2]$}
For the Euler phase $[2,-2]$ we take $q=c_1[\boldsymbol{h}]=0$ and $q'=c_1[\boldsymbol{h}']=2$, which, similarly to the previous case, leads to the minimal model
\begin{equation}
\label{eq_min_H_E22m}
\begin{aligned}
    H^{[2,-2]}_{\text{4B}}(\boldsymbol{k})  &=  
     H^{BA}_{\text{4B,deg}}
    \left[\begin{smallmatrix}
        1 & 0 & 0 & 0 \\
        m' & t_1' & 1 & t_2'
    \end{smallmatrix}\right]
    (\boldsymbol{k}) 
    + \delta\, \Gamma_{13}\,,\\
    & = \sin k_1 \Gamma_{31} + \sin k_2 \Gamma_{11} + \delta\, \Gamma_{13}  \\
    &+ [m'-t_1'(\cos k_1+\cos k_2)- t_2' \cos(k_1+k_2)] \Gamma_{03}   \,.
\end{aligned}
\end{equation}
Setting $(m',t_1',t_2',\delta)=(1/2,0,-3/2,1/2)$, again by the symmetry of Eq.\;(\ref{eq_4band_eigenvalues}) we obtain the same band structure obtained for $H^{[2,-2]}_{\text{4B}}(\boldsymbol{k})$ in \ref{fig_4B}(b). The Wilson loop is also the same as for the $[2,2]$-phase shown in Figure \ref{fig_4B}(d).

\subsubsection{Imbalanced phase $[\boldsymbol{\chi}] =[2,0]$}

\begin{figure}[t!]
\centering
\begin{tabular}{ll}
    (a) & (b) \\
	\includegraphics[width=0.5\linewidth]{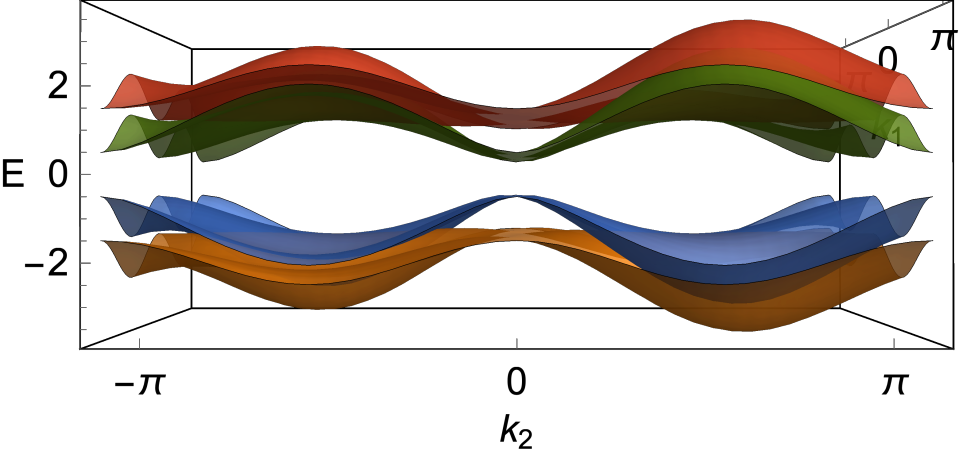} &
	\includegraphics[width=0.5\linewidth]{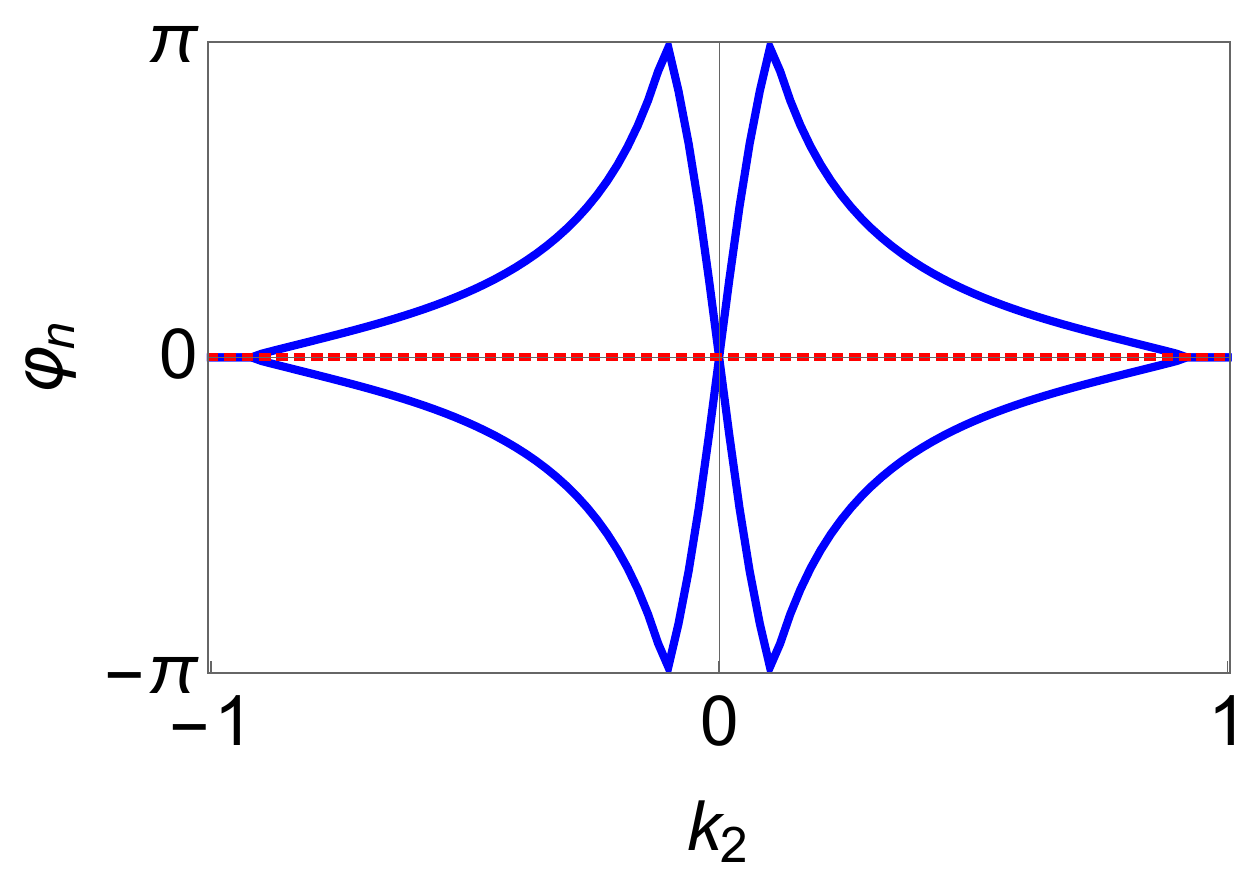}    \\
	(c) & (d) \\
	\includegraphics[width=0.5\linewidth]{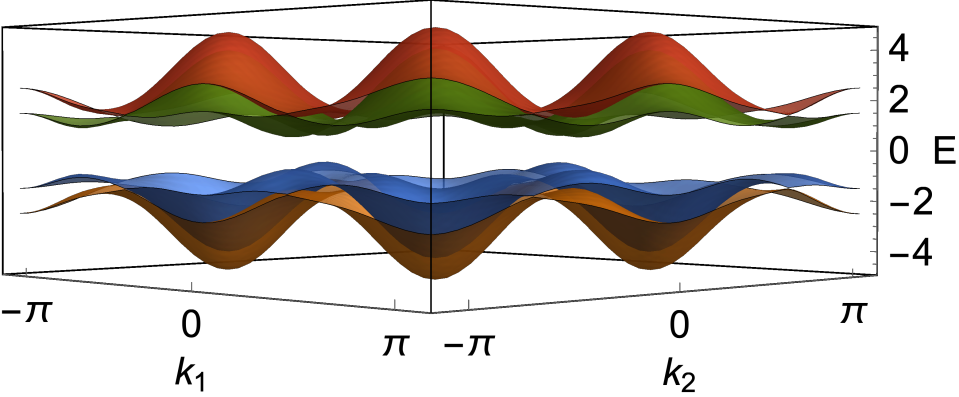}  &
	\includegraphics[width=0.5\linewidth]{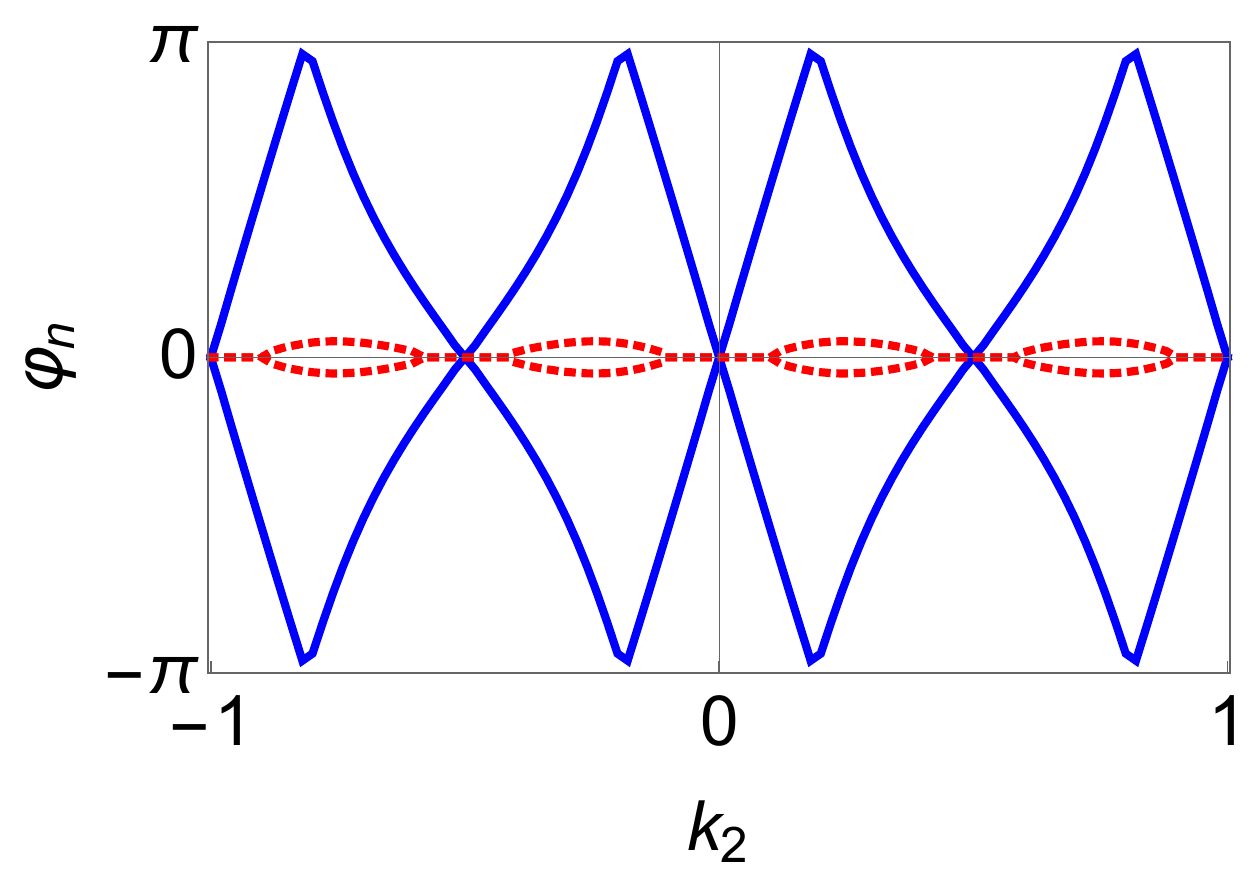} \\
	(e) & (f) \\
	\includegraphics[width=0.5\linewidth]{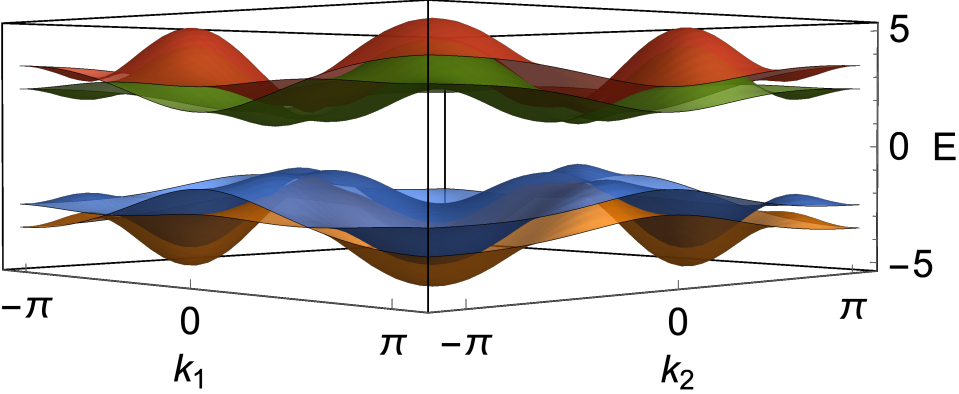}  &
	\includegraphics[width=0.5\linewidth]{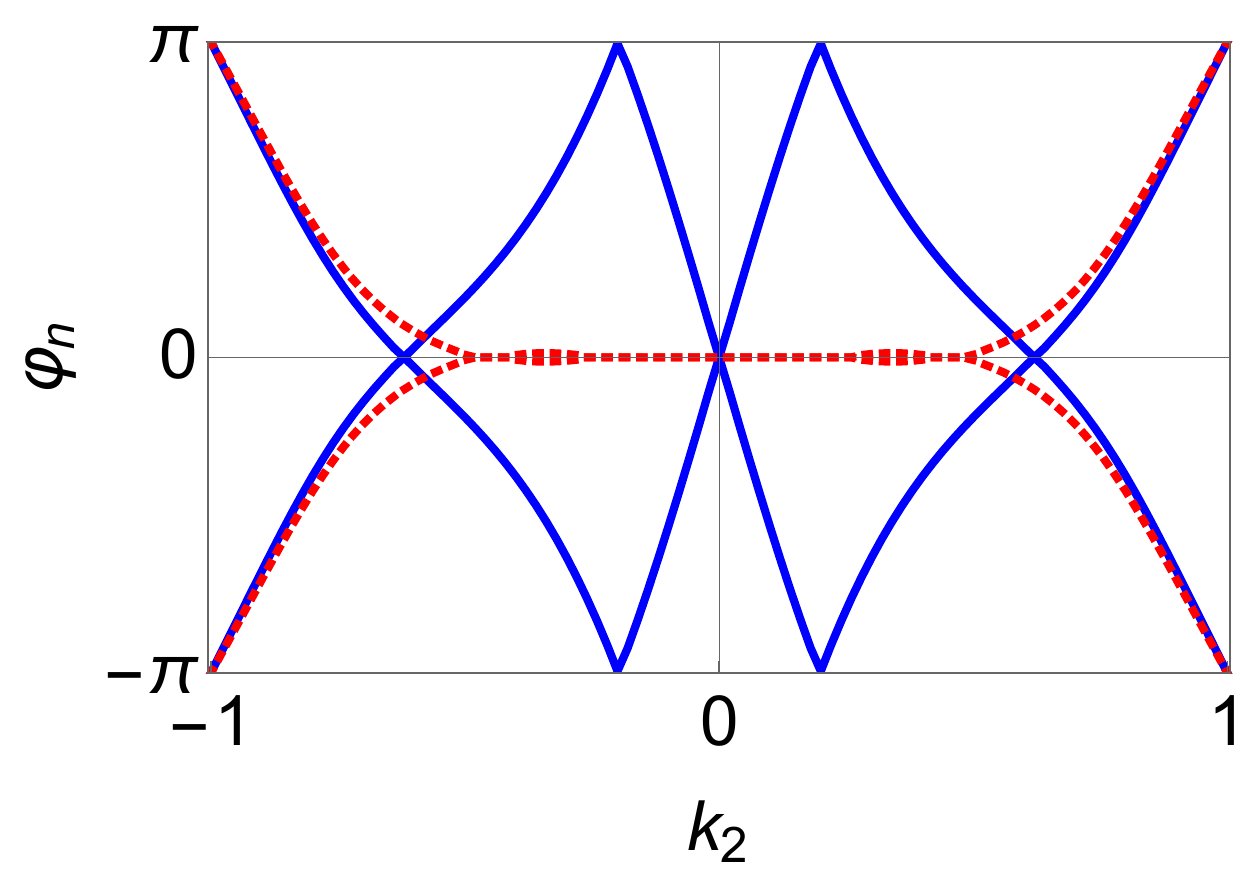} 
\end{tabular}
\caption{\label{fig_4B_imbalanced} Band structure and Wilson loop of the imbalanced gapped (orientable) Euler phases in four-level systems. The Wilson loop of the lower (higher) two-band subspace is the blue full line (red dashed line). (a,b) Imbalanced phase $[\boldsymbol{\chi}]=[2,0]$ defined by the model $H^{[2,0]}_{\text{4B}}$ Eq.\;(\ref{eq_min_H_E20}). (c,d) Imbalanced phase $[\boldsymbol{\chi}]=[4,0]$ defined by the model $H^{[4,0]}_{\text{4B}}$ Eq.\;(\ref{eq_min_H_E40}). (e,f) Imbalanced phase $[\boldsymbol{\chi}]=[3,1]$ defined by the model $H^{[3,1]}_{\text{4B}}$ Eq.\;(\ref{eq_min_H_E31}). 
}
\end{figure}

For the Euler phase $[2,0]$, we take $q=c_1[\boldsymbol{h}]=-1$ and $q'=c_1[\boldsymbol{h}']=1$, that is compatible with the minimal model
\begin{equation}
\label{eq_min_H_E20}
\begin{aligned}
    H^{[2,0]}_{\text{4B}}(\boldsymbol{k})  &=  
     \dfrac{1}{4} H^{AA}_{\text{4B,deg}}
    \left[\begin{smallmatrix}
        1 & 1 & 1 & 2 \\
        -1 & -1 & -1 & 2
    \end{smallmatrix}\right]
    (\boldsymbol{k}) 
    + \dfrac{1}{2}\, \Gamma_{11}\,,\\
    & = 
    \left(
        \begin{array}{c}
        \bar{a} \\
        \bar{b}\\
        \bar{c} 
    \end{array}\right)^T \cdot \underline{\Gamma} \cdot 
    \left(
        \begin{array}{c}
        \bar{a} \\
        \bar{b}\\
        -\bar{c} 
    \end{array}\right)\,,
\end{aligned}
\end{equation}
with
\begin{equation}
\begin{aligned}
    \bar{a}& =\sin k_1\,,~\bar{b}=\sin k_2\,,\\
    \bar{c}&=\frac{1}{2}(1-\cos k_1-\cos k_2-\cos(k_1+k_2))\,,
\end{aligned}
\end{equation}
where we have set $(m,t_1,t_2,\lambda)=(-m',-t_1',-t_2',\lambda')=(1/2,1/2,1/2,1)$, and with $\underline{\Gamma}$ defined in Eq.\;(\ref{eq_flat_4B}). We show the band structure and the Wilson loop (blue for the lower two-band subspace, and dashed red for the higher two-band subspace) in Figure \ref{fig_4B_imbalanced}(a,b).

\subsubsection{Imbalanced phase $[\boldsymbol{\chi}] =[4,0]$}

For the Euler phase $[4,0]$, we take $q=c_1[\boldsymbol{h}]=2$ and $q'=c_1[\boldsymbol{h}']=-2$, that is compatible with the minimal model
\begin{equation}
\label{eq_min_H_E40}
\begin{aligned}
    H^{[4,0]}_{\text{4B}}(\boldsymbol{k})  &=  
     \dfrac{1}{4} H^{AA}_{\text{4B,deg}}
    \left[\begin{smallmatrix}
        1 & 0 & -3 & 2 \\
        1 & 0 & 3 & 2
    \end{smallmatrix}\right]
    (\boldsymbol{k}) 
    + \dfrac{1}{2}\, \Gamma_{13}\,,\\
    & = 
    \left(
        \begin{array}{c}
        \bar{a} \\
        \bar{b}\\
        \bar{c} 
    \end{array}\right)^T \cdot \underline{\Gamma} \cdot 
    \left(
        \begin{array}{c}
        -\bar{a} \\
        -\bar{b}\\
        \bar{c}' 
    \end{array}\right)\,,
\end{aligned}
\end{equation}
with
\begin{equation}
\begin{aligned}
    \bar{a}& =\sin k_1\,,~
    \bar{c}=\frac{1}{2}(1+3 \cos(k_1+k_2))\,,\\
    \bar{b}&=\sin k_2\,,~
    \bar{c}'=\frac{1}{2}(1-3 \cos(k_1+k_2))\,,
\end{aligned}
\end{equation}
where we have taken $(m,t_1,t_2,\lambda)=(m',t_1',-t_2',-\lambda')=(1/2,0,-3/2,1)$. We show the band structure and the Wilson loop in Figure \ref{fig_4B_imbalanced}(c,d).

\subsubsection{Imbalanced phase $[\boldsymbol{\chi}] =[3,1]$}

For the Euler phase $[3,1]$, we take $q=c_1[\boldsymbol{h}]=2$ and $q'=c_1[\boldsymbol{h}']=-1$, that is compatible with the minimal model
\begin{equation}
\label{eq_min_H_E31}
\begin{aligned}
    H^{[3,1]}_{\text{4B}}(\boldsymbol{k})  &=  
     \dfrac{1}{4} H^{AA}_{\text{4B,deg}}
    \left[\begin{smallmatrix}
        1 & -1 & -3 & 2 \\
        3 & 2 & 1 & 2
    \end{smallmatrix}\right]
    (\boldsymbol{k}) 
    + \dfrac{1}{2}\, \Gamma_{13}\,,\\
    & = 
    \left(
        \begin{array}{c}
        \bar{a} \\
        \bar{b}\\
        \bar{c} 
    \end{array}\right)^T \cdot \underline{\Gamma} \cdot 
    \left(
        \begin{array}{c}
        \bar{a} \\
        \bar{b}\\
        \bar{c}' 
    \end{array}\right)\,,
\end{aligned}
\end{equation}
with
\begin{equation}
\begin{aligned}
    \bar{a}& =\sin k_1\,,~\bar{b}=\sin k_2\,,\\
    \bar{c}&=\frac{1}{2}(1+1 \cos(k_1+k_2)+3 \cos(k_1+k_2))\,,\\
    \bar{c}'&=\frac{1}{2}(3-2 \cos(k_1+k_2)-\cos(k_1+k_2))\,,
\end{aligned}
\end{equation}
where we have taken $(m,t_1,t_2,\lambda)=(1/2,-1/2,-3/2,1)$ and $(m',t_1',t_2',\lambda')=(3/2,1,1/2,1)$. We show the band structure and the Wilson loop in Figure \ref{fig_4B_imbalanced}(e,f).

\subsubsection{Imbalanced phase $[\boldsymbol{\chi}] =[3,-1]$}

For the Euler phase $[3,-1]$, we take $q=c_1[\boldsymbol{h}]=1$ and $q'=c_1[\boldsymbol{h}']=-2$, that is compatible with the minimal model
\begin{equation}
\label{eq_min_H_E31m}
\begin{aligned}
    H^{[3,-1]}_{\text{4B}}(\boldsymbol{k})  &=  
     \dfrac{1}{4} H^{AA}_{\text{4B,deg}}
    \left[\begin{smallmatrix}
        -3 & -2 & -1 & 2 \\
        -1 & 1 & 3 & 2
    \end{smallmatrix}\right]
    (\boldsymbol{k}) 
    + \dfrac{1}{2}\, \Gamma_{13}\,,\\
    & = 
    \left(
        \begin{array}{c}
        \bar{a} \\
        \bar{b}\\
        \bar{c} 
    \end{array}\right)^T \cdot \underline{\Gamma} \cdot 
    \left(
        \begin{array}{c}
        \bar{a} \\
        \bar{b}\\
        \bar{c}' 
    \end{array}\right)\,,
\end{aligned}
\end{equation}
with
\begin{equation}
\begin{aligned}
    \bar{a}& =\sin k_1\,,~\bar{b}=\sin k_2\,,\\
    \bar{c}&=\frac{1}{2}(-3+2 (\cos k_1+\cos k_2) +  \cos(k_1+k_2))\,,\\
    \bar{c}'&=-\frac{1}{2}(1+ \cos k_1+\cos k_2 +3 \cos(k_1+k_2))\,,
\end{aligned}
\end{equation}
where we have taken $(m,t_1,t_2,\lambda)=(-3/2,-1,-1/2,1)$ and $(m',t_1',t_2',\lambda')=(-1/2,1/2,3/2,1)$. We obtain the same band structure and the Wilson loop as for the phase $[3,1]$.

\section{Symmetries and symmetry-breaking terms of the four-level systems}\label{sec_sym}

The purpose of defining minimal models is to simplify their realization in experiments. While our strategy to systematically generate simple models is solely conditioned by the targeted Euler topology, requiring the reality condition (from a $C_2T$ or a $PT$ symmetry that squares to the identity), the simplicity of the four-band models itself make them symmetric under additional symmetries. While the presence of extra symmetries does not affect the Euler topology, these bring their own phenomenology which should not be confused with the manifestations of the Euler topology {\it per se}. We therefore identify all the additional symmetries for each model and define the minimal terms that break them. 

The ansatz of the three-band models Eq.\;(\ref{eq_3B_nonflat}) is general enough, by spanning all the real Gell-Mann matrices [see the remark below Eq.\;(\ref{eq_3B_nonflat_b})], to break all symmetries at the exception of the reality condition \footnote{One caveat here comes from the presence of some accidental symmetries in the three-band models and in the imbalanced four-band models that merely come from the specific ansatz in Eq.\;(\ref{eq_3B_nonflat}). The effect of these accidental symmetries is further discussed in Section \ref{sec_chiral}.}.

\begin{table*}[t!]
\begin{equation*}
\arraycolsep=2.4pt\def\arraystretch{1.4}
\begin{array}{ | l || c | c | c | c | c | c | c | c |}
\hline
	g(T) & [1,1] &
	[1,-1] & [2,2] & [2,-2] & [2,0] & [4,0] & [3,1] & [3,-1] \\
	\hline
    C_{2z}T & \mathbb{1}_4\mathcal{K} & \mathbb{1}_4\mathcal{K} & \mathbb{1}_4\mathcal{K} & \mathbb{1}_4\mathcal{K} & \mathbb{1}_4\mathcal{K} & \mathbb{1}_4\mathcal{K} & \mathbb{1}_4\mathcal{K} & \mathbb{1}_4\mathcal{K} \\
    C_{2z} & \imi\,  \Gamma_{22} & \imi\, \Gamma_{03} & \imi\,  \Gamma_{22} & \imi\, \Gamma_{03} & \imi\,  \Gamma_{22} & \imi\,  \Gamma_{22} & \imi\,  \Gamma_{22} & \imi\,  \Gamma_{22} \\
    T & -\imi \,\Gamma_{22}\mathcal{K} & -\imi\, \Gamma_{03}\mathcal{K} & -\imi \,\Gamma_{22}\mathcal{K} & -\imi\, \Gamma_{03}\mathcal{K} & -\imi\,  \Gamma_{22}\mathcal{K} & -\imi\,  \Gamma_{22}\mathcal{K} & -\imi\,  \Gamma_{22}\mathcal{K} & -\imi\,  \Gamma_{22}\mathcal{K} \\
    m_{y} & \Gamma_{31} & \Gamma_{31} & - & - & - & - & - & - \\
    m_{x} & \Gamma_{13} & \Gamma_{10} & - & - & - & - & - & - \\
    m_{y}T & -\imi\,\Gamma_{13}\mathcal{K} & -\imi\,\Gamma_{10}\mathcal{K} & - & - & - & - & - & - \\
    m_{x}T & -\imi\,\Gamma_{31}\mathcal{K} & -\imi\,\Gamma_{13}\mathcal{K} & - & - & - & - & - & - \\
    \hline
    Sg(T) & S=\Gamma_{12} & S=\Gamma_{02} & S=\Gamma_{12} & S=\Gamma_{02} &  &  &  & \\
    \hline
    SC_{2z}T & \Gamma_{12}\mathcal{K} & \Gamma_{02}\mathcal{K} & \Gamma_{12}\mathcal{K} & \Gamma_{02}\mathcal{K} & - & - & - & - \\
    SC_{2z} & -  \Gamma_{30} & -\Gamma_{01} & -\Gamma_{30} & -\Gamma_{01} & - & - & - & - \\
    ST & \Gamma_{30}\mathcal{K} & \Gamma_{01}\mathcal{K}  & \Gamma_{30}\mathcal{K} & \Gamma_{01}\mathcal{K} & - & - & - & - \\
    Sm_{y} & -\Gamma_{23} & -\imi\,\Gamma_{33} & - & - & - & - & - & - \\
    Sm_{x} & \imi\,\Gamma_{01} & \Gamma_{12} & - & - & - & - & - & - \\
    Sm_{y}T & \Gamma_{01} \mathcal{K} & -\imi\,\Gamma_{12}\mathcal{K}  & - & - & - & - & - & - \\
    Sm_{x}T & \imi\,\Gamma_{23}\mathcal{K} & \Gamma_{11}\mathcal{K} & - & - & - & - & - & - \\
    \hline
\end{array}
\end{equation*}
\caption{Extra symmetries $(S)g(T)$ with their representation in the Bloch orbital basis, $U_{(S)g(T)}(\mathcal{K})$, for the minimal four-band models $H^{[\boldsymbol{\chi}]}_{\text{4B}}$ of Section \ref{sec_min_models}, listed for each Euler phase $[\boldsymbol{\chi}] = [\chi_I, \chi_{II}]$. The symmetries are composed of the crystalline symmetries $g\in\{E,C_{2z},m_y,m_x\}$ ($E$ is the identity), the spinless time reversal $T$ ($T^2=+1$), with $\mathcal{K}$ the complex conjugation, and the chiral symmetry $S$ ($S^2=+1$). We have imposed $U_{Sg(T)} = S U_{g(T)}$. The mirror symmetries hold when we take $(x_1,x_2)=(x,y)$ and $(k_1,k_2)=(k_x,k_y)$.
}
\label{tab_1}
\end{table*}

In our context, the symmetries of the system take the form a constraint to be satisfied by the Bloch Hamiltonian. Symmetries can be of four types, depending on whether it is unitary (\ie no complex conjugation) or anti-unitary (\ie with complex conjugation), or if it is a ``symmetry'' (\ie commuting with the Hamiltonian) or an ``anti-symmetry'' (\ie anti-commuting with the Hamiltonian) \cite{Shiozaki14}, see the examples for each case below. Since there are only twofold (and symmorphic) crystalline symmetries in our models, all the topological classes protected by these symmetries, when combined with the time reversal, chiral, and particle-hole symmetries, have been classified in Ref.\;\cite{Shiozaki14}. 

The symmetries of any (hermitian) four-band models, built from the Dirac matrices $\Gamma_{ij} = \sigma_{i}\otimes \sigma_{j}$ ($i,j=0,1,2,3$), are easily determined from the anti-commutation of the Pauli matrices. Indeed, we have that $\Gamma_{ij}$ commutes with
\begin{equation}
        \Gamma_{ij}, \Gamma_{i0},\Gamma_{0j},\text{\;and\;}
        \left(\Gamma_{kl}\right)_{k\neq i,l\neq j}\,,
\end{equation}
and anti-commutes with
\begin{equation}
        \left(\Gamma_{k0}\right)_{k\neq i},
        \left(\Gamma_{0l}\right)_{l\neq j},
        \left(\Gamma_{kj}\right)_{k\neq i},
     \left(\Gamma_{il}\right)_{l\neq j}\,,
\end{equation}
where take $\sigma_0 =\mathbb{1}_2$. Taking the generic four-band model $H_{\text{4B}}(\boldsymbol{k}) = \boldsymbol{g}(\boldsymbol{k})\cdot \boldsymbol{\Gamma}$ in Eq.\;(\ref{eq_generic_4B}), the symmetries and anti-symmetries are then readily obtained from whether $g_{ij}(\boldsymbol{k})$ is purely real or imaginary, and whether it is even or odd under the independent flip of the momentum coordinates, \ie $k_1\rightarrow -k_1$ and $k_2\rightarrow -k_2$. In the following we assume that the momentum coordinates match with the coordinates of an orthorhombic Bravais lattice (\ie a rectangular lattice) and take $\boldsymbol{k}=(k_1,k_2)=(k_x,k_y)$. 

We list in Table~\ref{tab_1}, for each minimal four-band model, all the symmetries and anti-symmetries with their representation in the orbital basis of the Bloch Hamiltonian, where $T$ is the spinless time-reversal ($T^2=\mathbb{1}_4$), and $S$ is the operator of chiral symmetry ($S^2=\mathbb{1}_4$). An example of unitary symmetry is $C_{2z}$ ($\pi$ rotation around $\hat{z}$) acting on the Bloch orbital basis as
\begin{equation}
        {}^{C_{2z}}\vert\boldsymbol{\phi},\boldsymbol{k} \rangle 
        = \vert\boldsymbol{\phi},C_{2z}\boldsymbol{k} \rangle \cdot U_{C_{2z}}\,,
\end{equation}
with $C_{2z}(k_x,k_y) = (-k_x,-k_y)$ and $\boldsymbol{\phi}=(\phi_1,\phi_2,\phi_3,\phi_4)$, leading to the constraint of the Bloch Hamiltonian
\begin{equation}
\label{eq_sym_us}
    U_{C_{2z}} \cdot H^{[\boldsymbol{\chi}]}_{\text{4B}}(-\boldsymbol{k}) \cdot U^{\dagger}_{C_{2z}} = H^{[\boldsymbol{\chi}]}_{\text{4B}}(\boldsymbol{k})\,. 
\end{equation}
where the unitary representation $U_{C_{2z}}$ is listed in Table \ref{tab_1} for each Euler phase $[\boldsymbol{\chi}]$. An example of anti-unitary symmetry is time reversal $T$, acting on the Bloch orbital basis as
\begin{equation}
        {}^{T}\vert\boldsymbol{\phi},\boldsymbol{k} \rangle
        = \vert\boldsymbol{\phi},-\boldsymbol{k} \rangle \cdot U_{T} \mathcal{K}\,,
\end{equation}
and leading to the constraint 
\begin{equation}
\label{eq_sym_aus}
    U_{T} \cdot H^*(-\boldsymbol{k}) \cdot U^{\dagger}_{T} = H(\boldsymbol{k})\,,
\end{equation}
where $\mathcal{K}$ is complex conjugation. An example of unitary anti-symmetry is $Sm_y$, with $S$ the chiral symmetry and the mirror symmetry $m_y(k_x,k_y)=(k_x,-k_y)$, acting as 
\begin{equation}
        {}^{Sm_y}\vert\boldsymbol{\phi},\boldsymbol{k} \rangle
        = \vert\boldsymbol{\phi},m_y\boldsymbol{k} \rangle \cdot U_{Sm_y}\,,
\end{equation}
and leading to the constraint 
\begin{equation}
\label{eq_sym_uas}
    U_{Sm_y} \cdot H(m_y\boldsymbol{k}) \cdot U^{\dagger}_{Sm_y} = - H(\boldsymbol{k})\,.
\end{equation}
Note that we have imposed for all unitary anti-symmetries $U_{Sg} = SU_{g}$. Then, an example of anti-unitary anti-symmetry is $SC_{2z}T$, acting as 
\begin{equation}
        {}^{Sm_yT}\vert\boldsymbol{\phi},\boldsymbol{k} \rangle
        = \vert\boldsymbol{\phi},-m_y\boldsymbol{k} \rangle \cdot U_{Sm_yT} \mathcal{K}\,,
\end{equation}
and leading to the constraint 
\begin{equation}
\label{eq_sym_auas}
    U_{Sm_yT} \cdot H^*(-m_y\boldsymbol{k}) \cdot U^{\dagger}_{Sm_yT} = - H(\boldsymbol{k})\,.
\end{equation}
Note that we have again imposed for all anti-unitary anti-symmetries that $U_{SgT} = SU_{gT}$.

It is now straightforward to break any symmetry $g(T)$, or anti-symmetry $Sg(T)$ with $g\in \{E,C_{2z},m_y,m_x\}$, by adding a term that does not commute, or anti-commute, with the representation of $g(T)$, or $Sg(T)$, in the sense of Eq.\;(\ref{eq_sym_us}) and Eq.\;(\ref{eq_sym_aus}), or of Eq.\;(\ref{eq_sym_uas}) and Eq.\;(\ref{eq_sym_auas}), respectively. 

For instance, the model $H^{[1,1]}_{\text{4B}}(\boldsymbol{k})$ in Eq.\;(\ref{eq_min_H_E11}) has the mirror symmetry $m_y$ (taking $(k_1,k_2)=(k_x,k_y)$), represented by $\Gamma_{31}$. The only Dirac matrix in Eq.\;(\ref{eq_min_H_E11}) that does not commute with $\Gamma_{31}$ is $\Gamma_{03}$, which comes with the factor $\sin k_y$ that is odd under $m_yk_y\rightarrow -k_y$. The other terms of Eq.\;(\ref{eq_min_H_E11}) all have a Dirac matrix that commutes with $\Gamma_{31}$ (\ie these are $\{\Gamma_{01},\Gamma_{13},\Gamma_{22}\}$) and each with a factor that is even under the reversal of $k_2=k_y$. Then, the breaking of $m_y$ is achieved by adding any one of the following terms
\begin{equation}
        \widetilde{g}_e(\boldsymbol{k}) \left\{
            \Gamma_{02},\Gamma_{03},\Gamma_{10},\Gamma_{11},\Gamma_{20},\Gamma_{21},\Gamma_{32},\Gamma_{33}
        \right\}\,,
\end{equation}
where $\widetilde{g}_e(\boldsymbol{k})$ is even under $k_y\rightarrow -k_y$, or
\begin{equation}
    \widetilde{g}_o(\boldsymbol{k}) \left\{
            \Gamma_{00},\Gamma_{01},\Gamma_{12},\Gamma_{13},\Gamma_{22},\Gamma_{23},\Gamma_{30},\Gamma_{31}
        \right\}\,,
\end{equation}
where $\widetilde{g}_o(\boldsymbol{k})$ is odd under $k_y\rightarrow -k_y$. This analysis for the symmetry $m_y$ can be straightforwardly extended to all other symmetries. We note that any term with a complex Dirac matrix breaks $C_{2z}T$ symmetry, which we discuss in more detail in Section \ref{sec_chiral}. 

While the additional symmetries do not play a special role for the intrinsic manifestations of the Euler class topology, \eg see the next Section where we systematically generate linked nodal rings in 3D from 2D Euler phases, they must be considered in Section \ref{sec_chiral} where we address the conversion of the Euler phases into Chern phases (\eg a remaining mirror symmetry enforces zero Chern numbers).

We conclude this section by noting that the spinfull basal mirror symmetry and spinfull $PT$ symmetri ($[PT]^2=+1$), discussed in Section \ref{subsec_basal_mirror} and in Appendix \ref{ap_mirror}, are broken in all the models of Section \ref{sec_min_models} since the two-band subspaces are not degenerate. 

\section{From 2D Euler phases to $PT$-protected adjacent linked nodal rings in 3D}\label{sec_3D_NL}
In this section we address the intricate interplay of non-Abelian multi-gap topology and Euler class. In particular, we find that these notions directly tie to linked nodal structures protected by $PT$-symmetry (\ie inversion and time reversal) in three-dimension, providing for a rather rich topological underpinning. This section focuses on the conceptual aspects of the linked nodal structures obtained from pairs of inequivalent 2D Euler phases, that is, the explicit tight-binding models and their linked nodal structures obtained numerically are discussed in the next section. In particular, we address the relation between the non-Abelian frame charges \cite{Wu1273,Tiwari:2019} and the refined patch Euler class characterization \cite{bouhon2019nonabelian,Jiang2021,Peng2021}. We then introduce the homotopy invariant of the linked nodal structure that is invariant under all changes of gauge and all Euler class reversal maps. We finally introduce linking numbers defined as the Euler class-valued monopole charges of linked nodal rings, and show their relation with the homotopy invariant. We note that this section has overlap with the earlier works Ref.\;\cite{BJY_linking}, which introduced the Euler class for nodal rings while focusing on its $\mathbb{Z}_2$ reduction (to the second Stiefel-Whitney class) for occupied subspaces with more than two bands, and with Ref.\;\cite{Wu1273,Tiwari:2019}, which introduced the non-Abelian frame charges for nodal rings. This section goes beyond these works by $(i)$ fully exploiting the $\mathbb{Z}$ Euler class classification of two-band subspaces separated by two energy gaps (from above and from below), which leads to many topological configurations that have not been considered before, $(ii)$ by providing a systematic method for the building of arbitrary linked nodal structures obtained as a transition between two inequivalent Euler phases, and $(iii)$ by clarifying the effect of the adiabatic Euler class reversal maps in terms of the relation between the homotopy invariant of a linked nodal structure and its linking numbers assuming a fixed gauge choice.

\subsection{General 3D ansatz}

We begin by noting that our construction of minimal two-dimensional Euler models allows us to then systematically deform one topological Euler phase ($H^{[\chi^0]}$) into any other neighboring topological Euler phase ($H^{[\chi^1]}$). Assuming that such topological deformations are controlled by a single parameter, say $\lambda_{PT}\in [0,1]$, that does not break the reality condition, we can generically model the transition from one phase to another through a linear combination of Hamiltonians, \ie
\begin{multline}
     H^{\Delta[\chi]}(\boldsymbol{k},\lambda_{PT}) = \cos \left( \dfrac{\lambda_{PT}\pi}{2}\right) H^{[\chi^{0}]}(\boldsymbol{k})+ \\
     \sin \left( \dfrac{\lambda_{PT}\pi}{2}\right) H^{[\chi^{1}]}(\boldsymbol{k}) \;.
\end{multline}
In the above we noted the ``difference'' of equivalence classes
\begin{equation}
    \Delta[\chi] = [\chi^1]-[\chi^0] \,,
\end{equation} 
which we define more precisely and evaluate below, with the Euler classes $\chi^{0,1} = \chi(\lambda_{PT}=1,0)$. The nodal points of the initial 2D phases extend to nodal lines, or braids, in the enlarged parameter space $(\boldsymbol{k},\lambda_{PT})\in \mathbb{T}^2\times [0,1]$. As an alternative approach, which is the one we take below, one may also embed a pair of 2D Euler Bloch Hamiltonians into one real 3D Bloch Hamiltonian through
\begin{multline}
\label{eq_3D_PT_nodal}
    H^{\Delta[\chi]}_{PT}(\boldsymbol{k}_{\parallel},k_z) = 
    \dfrac{1+\cos k_z}{2} H^{[\chi^{0}]}(\boldsymbol{k}_{\parallel}) +\\
    \dfrac{1-\cos k_z}{2} H^{[\chi^{\pi}]}(\boldsymbol{k}_{\parallel}),
\end{multline}
with $\Delta[\chi] = [\chi^{\pi}] - [\chi^{0}]$ and $\chi^{0,\pi}=\chi(k_z=0,\pi)$, such that the section of the 3D model at $k_z=0$ is given by $H^{[\chi^{0}]}(\boldsymbol{k}_{\parallel})$, and the section at $k_z=\pi$ is $H^{[\chi^{\pi}]}(\boldsymbol{k}_{\parallel})$, where $k_z$ is the momentum perpendicular to the $\boldsymbol{k}_{\parallel}=(k_1,k_2)$-plane. Evidently, $k_z$ then acts as the deformation parameter between the two Euler phases. By retaining the reality condition, the resulting 3D Hamiltonian preserves an effective spinless $PT$ symmetry, \ie with $[PT]^2=+1$, which supports stable band crossings in the form of nodal lines \cite{volovik2003universe,burkov2011semimetal,FuC2T,BBS_nodal_lines}. We find that all such deformations produce nodal braids from adjacent energy gaps to be linked together. While these nodal braids are characterized by complementary non-Abelian charges \cite{Wu1273, Tiwari:2019,BJY_nielsen,bouhon2019nonabelian}, we show in the subsequent that the knowledge of the initial ($k_z=0$) and final ($k_z=\pi$) Euler classes of the deformation provides a greatly refined characterization of linked nodal rings.

The rationale for the presence of linked nodal lines directly relates to the multi-gap nature of the Euler class. Indeed, a change in the Euler class necessitates the creation or the removal of pairs of stable nodal points of a two-band subspace, which can only happen via their braiding around nodes present in one of the two adjacent gaps, \ie the gaps above and below in energy, while the reality condition (protected by $C_2T$ or $PT$) is maintained \cite{BJY_nielsen, Wu1273, Tiwari:2019,bouhon2019nonabelian}. Considering the total trajectory of the nodes through a braiding, or through the path, we obtain nodal braids that form linked nodal rings. Although nontrivial linked nodal rings and their non-Abelian charges have been detailed conceptually~\cite{Tiwari:2019}, no explicit models have so far been formulated in the general context of Euler topology. Furthermore, the present characterization of linked nodal rings in terms of the $\mathbb{Z}$ Euler class for two-band subspaces \cite{BzduSigristRobust,bouhon2018wilson} (see below) constitutes a substantial refinement compared to the finite non-Abelian group of loop charges of the frame of eigenvectors~\cite{Tiwari:2019}, \ie computed over base loops encircling the nodal rings (which provides an effective $\mathbb{Z}_4$ counting in each gap, see details below), and of the $\mathbb{Z}_2$ monopole charge \cite{FuC2T,BJY_linking} when more than two bands must be considered (corresponding to the reduction of the Euler class to the $\mathbb{Z}_2$ second Stiefel-Whitney class). This work thus fills these gaps by providing concrete minimal tight-binding models that can be readily used as a guide for the design of acoustic metamaterials \cite{Jiang2021,Guo1Dexp}, photonic crystals \cite{park2022nodal}, electronic circuits \cite{ezawa2021euler}, and optical traps for cold atoms \cite{Unal_quenched_Euler, zhao2022observation}.

\subsection{Non-Abelian frame charge of nodal braids}

We here discuss in more detail the braiding of nodes taking place at the transition between inequivalent Euler phases. For this we introduce the non-Abelian frame charge of nodal braids that complements the Euler classes. 

To this end, let us first consider the special case of changing Euler class in the $I$-th two-band subspace from some finite value at $k_z=0$ to zero at $k_z=\pi$, while the Euler class of the $II$-th block of bands remains unchanged at zero, \ie we have $\vert \chi^{0}_I \vert \neq \vert\chi^{\pi}_I\vert = 0$ and $\chi^{\pi}_{II} = \chi^{0}_{II} = 0$. That is, a transition from the homotopy class $[\chi^{0}_I,0]=[\chi^{0}_I]$ to $[0,0]=[0]$. Then, by varying $k_z$ from $0$ to $\pi$, the number of stable nodal points must change from $2\vert \chi^{0}_I \vert$ to zero, implying that a number $\vert \chi^{0}_I\vert$ of nodes of the $I$-th two-band subspace must be braided with some adjacent nodes located in the energy gap between the $I$-th and the $II$-th blocks. We explicitly illustrate this result with the example below.  

In the following, we refer to nodes  formed by the crossing of the energy levels $E_{I,1}$ and $E_{I,2}$ as {\it $I$-th nodes}, and denote the adjacent nodes in the gap between the $I$-th and $II$-th blocks (\ie formed by the crossing of the energy levels $E_{I,2}$ and $E_{II,1}$) as {\it $(I,II)$-nodes}. Since by construction the $(I,II)$-energy gap is open (\ie $E_{II,1}-E_{I,1}>0$) at the initial ($k_z=0$) and at the final ($k_z=\pi$) phases, the complete trajectory of the intermediary adjacent nodes must form closed rings, which we refer to as $\mathcal{L}_{(I,II),i} = \{\boldsymbol{k}\vert E_{I,2}(\boldsymbol{k}) = E_{II,1}(\boldsymbol{k})\}_i$ where $i=1,2,\dots$ lists all the connected components (in the example below we take a single adjacent nodal ring). We then denote the $I$-th braids by $\mathcal{L}_{I,i} = \{\boldsymbol{k}\vert E_{I,1}(\boldsymbol{k}) = E_{I,2}(\boldsymbol{k})\}_i$, where $i=1,2,\dots$ again lists the distinct connected components.

\begin{figure}[t!]
\centering
\begin{tabular}{l}
	\includegraphics[width=0.8\linewidth]{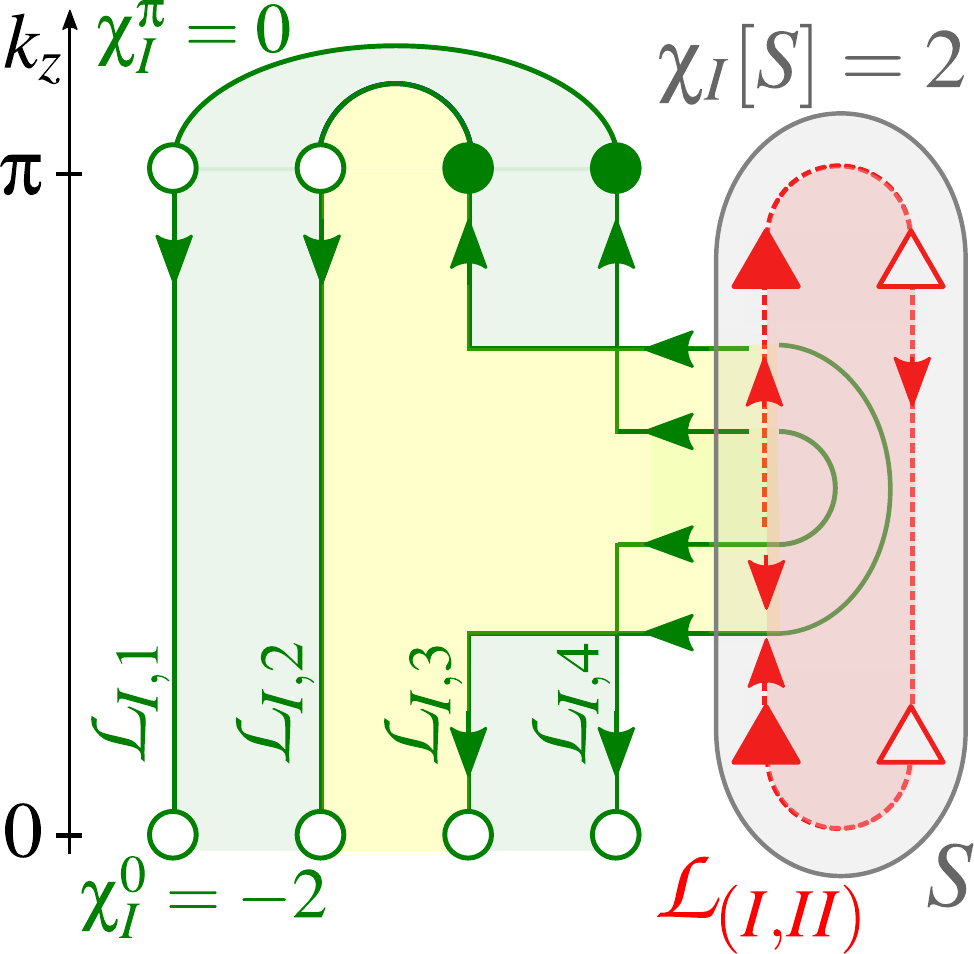}
\end{tabular}
\caption{\label{fig_braiding_monopole} Linking nodal structure protected by $PT$ symmetry for a transition of Euler phases from $[\chi^{0}_I,\chi^{0}_{II}] = [2,0]$ at $k_z=0$ to $[\chi^{\pi}_I,\chi^{\pi}_{II}] = [0,0]$ at $k_z=\pi$. The green circles and braids correspond to the trajectories of the $I$-th nodal points (see text), with their signed non-Abelian frame charges (see text) represented by the open and full circles at the $k_{z}=0,\pi$-planes, and by arrows along the braids. The conversion of Euler class is mediated by the presence of a linked nodal ring $\mathcal{L}_{(I,II)}$ (red dashed ring) formed by the trajectory of the adjacent $(I,II)$-nodes, with their signed non-Abelian frame charges represented by open and full triangles at $k_z=0,\pi$, and by arrows along the braid. The non-Abelian frame charge of a braid is flipped whenever it runs {\it below} one adjacent braid \cite{Tiwari:2019,Jiang2021,Peng2021}, in which case it must cross one adjacent Dirac sheet (see text), here represented by the greenish, yellowish, and reddish surfaces between pairs of braids. The linking numbers of the adjacent nodal ring (red dashed) linked with the $I$-th nodal braids is defined by its $I$-th monopole Euler class, \ie $\mathsf{Ln}_I[\mathcal{L}_{(I,II)}] = \chi_I[S] = \Delta \chi_I = +2$ (see text), with $S$ the wrapping envelope (gray).  
}
\end{figure}

We illustrate the intuitive picture for $[\chi^0_I]=[2]$ and $[\chi^{\pi}_I]=[0]$ in Figure \ref{fig_braiding_monopole}. Taking a plane at a fixed $k_z$-value in the Figure, every nodal braid is crossed at a point, to which we attribute one (loop) non-Abelian frame charge. We do this for every gap that hosts a nodal line. For instance, in the three-band case, the non-Abelian frame charges are given by the elements of the quaternion group, \ie $\pm i$ for the $I$-th nodes and $\pm j$ for the adjacent $(I,II)$-nodes \cite{Wu1273,Tiwari:2019,bouhon2019nonabelian}. We represent the sign of the non-Abelian nodal charges with open and full symbols, \ie open or full circles for the $I$-nodes [green in Fig.\;\ref{fig_braiding_monopole}], and open and full triangles for the adjacent $(I,II)$-nodes [red in Fig.\;\ref{fig_braiding_monopole}]. While these signs are gauge dependent, we assume that a reference point has been chosen with a fixed choice of gauge phases of the (real) eigenvectors. The presence of nodal points constitutes an obstruction to assign a smooth choice of gauge over the whole 2D cut. Indeed, each single node hosts a $\pi$-Berry phase disinclination of the pair of eigenstates forming the node. This obstruction to define a globally smooth gauge sign can be conveniently represented through a Dirac string \cite{BJY_nielsen} connecting every pair of nodes, \ie the two eigenvectors forming the nodes undergo sign-flip across the Dirac string. See \cite{Jiang2021,Peng2021} for a systematic method for the consistent global attribution of non-Abelian frame charges in 2D systems.

Since we can repeat the above analysis for any fixed $k_z$ value, we can thus attribute a sign-dependent non-Abelian frame charge to each nodal braid. Following \cite{Tiwari:2019}, we represent these charges by an oriented arrow on each braid, see Figure \ref{fig_braiding_monopole}, and use different symbols to represent nodal braids from different gaps, \ie green full lines for the $I$-th nodal braids, $\{\mathcal{L}_{I,i}\}_{i=1,\dots,4}$, and red dashed lines for the adjacent nodal braid, $\mathcal{L}_{(I,II)}$. Finally, we note that each Dirac sting of a $k_z$-plane cut extends for varying $k_z$-values into a {\it Dirac sheet} connecting pair of nodal braids, which we have represented in light colors (greenish, yellowish and reddish) in Figure \ref{fig_braiding_monopole}. Very importantly, whenever one nodal braid, say $\mathcal{L}_{I,3}$, runs below an adjacent nodal braid ($\mathcal{L}_{(I,II)}$) it must cross the adjacent Dirac sheet (reddish) and its non-Abelian frame charge (green arrow) must be flipped \cite{Tiwari:2019,BJY_nielsen}.

We finish this part with a few comments on the strict relation between the Euler class and the non-Abelian frame charges. Strictly speaking, the set of non-Abelian frame charges available to characterize the nodes in each gap is $\mathbb{Z}_4= \{1,q_{(n,n+1)},-q_{(n,n+1)},-1\}$, where $1$ is the trivial frame charge, $\pm q_{(n,n+1)}$ are the charge for an odd number of nodes in the gap between the bands $n$ and $n+1$, and $q_{(n,n+1)}^2=-1$ is the frame charge for an even number modulo 4 of nodes with the same charge. The Euler class of two-band subspace (a two-dimensional invariant) thus provides a generalization of the counting of stable nodes from $\mathbb{Z}_4$ to an arbitrary number in $\mathbb{Z}$. While we have assumed that there is no adjacent nodes at $k_{z}=0$ and $k_{z}=\pi$ (where we have the {\it gapped} Euler phases), we can readily generalize the $\mathbb{Z}$-counting even in the presence of adjacent nodes. This is done using the patch Euler class \cite{BJY_nielsen,bouhon2019nonabelian,Jiang2021,Peng2021}
\begin{equation}
    \chi[D] = \dfrac{1}{2\pi} \left(\displaystyle \int_{D} \mathsf{Eu} - \oint_{\partial D} \mathsf{a}\right) \in \mathbb{Z} \,,
\end{equation}
where $D$ is a disk, in the $(k_1,k_2)$-plane at a fixed $k_z$, covering one nodal point, and $\partial D$ is the oriented boundary of $D$. Assuming that each patch Euler class is evaluated with respect to the same global choice of gauge at a fixed reference base point, the patch Euler classes of all the nodal points of a two-band subspace can be added together, leading to the counting of arbitrary many stable nodal braids within each gap. We note that, similarly, to the non-Abelian charge, the patch Euler class of one nodal braid changes sign whenever the braid passes below one adjacent nodal braid, namely when it crosses one adjacent Dirac sheet \cite{BJY_nielsen,Jiang2021,Peng2021}.

\subsection{Homotopy invariant of the linked nodal structures} 

The assignment of {\it signed} non-Abelian frame charges to all the nodal braids relies on a choice of gauge at a reference base point. Yet, the Euler class reversal maps of the Euler phases at $k_z=0$ and $k_z=\pi$, inducing the equivalence $(\chi^{k_z}_I,\chi^{k_z}_{II}) \sim (-\chi^{k_z}_I,-\chi^{k_z}_{II}) $, would flip the signs of the charges at $k_z=0$ and $k_z=\pi$, independently. These maps are adiabatic, in the sense that they don't require the closing of the $(I,II)$-gaps of the initial and final Euler phases. As a consequence, the nodal structures must eventually be classified up-to-homotopy by an equivalence class that does not depend on a specific choice of gauge. Through the detail discussion of two simple examples in the next section, we motivate that the homotopy invariant of Euler-generated linked nodal structures can be defined by
\begin{equation}
\label{eq_diff_equivalence}
\begin{aligned}
    \Delta[\boldsymbol{\chi}] &= [\chi^{\pi}_I,\chi^{\pi}_{II}] - [\chi^{0}_I,\chi^{0}_{II}] \\
    &= \left\{\,
        [\chi^{\pi}_I-\chi^{0}_I,\chi^{\pi}_{II} -\chi^{0}_{II}] \sim
        [\chi^{\pi}_I+\chi^{0}_I,\chi^{\pi}_{II} +\chi^{0}_{II}]
    \,\right\} \,,
\end{aligned}
\end{equation}
\ie the difference of equivalence classes is now explicitly valued as an equivalence class of equivalence classes. While we used the ``sign-forgetful'' function by considering the gauge invariant homotopy classes, \ie
\begin{equation}
    G^{-1} : (\chi^{k_z}_I,\chi^{k_z}_{II}) \mapsto [\chi^{k_z}_I,\chi^{k_z}_{II}]\,,
\end{equation}
it will be convenient to also have a ``gauge-fixing'' map
\begin{equation}
    G:[\chi^{k_z}_I,\chi^{k_z}_{II}] \rightarrow (\chi^{k_z}_I,\chi^{k_z}_{II}),
\end{equation}
that represents the assignment of signed frame charges with respect to a chosen gauge at a fixed base point, \eg in the example of Figure \ref{fig_braiding_monopole} we have taken $G([\chi^{k_z,1}_I]) = G([2]) = -2$. We remark that contrary to the function $G^{-1}$ which is surjective, the gauge-fixing map is multi-valued, corresponding to all the possible choices of global gauges and locations of the Dirac sheets. 

In particular, the gauge fixing of a 3D $PT$-symmetric phase built from the transition between two Euler phases provides signed differences of Euler classes, \ie 
\begin{equation}
    G(\Delta[\boldsymbol{\chi}]) = 
    G([\boldsymbol{\chi}^{\pi}])-G([\boldsymbol{\chi}^{0}]) = (\Delta\chi_I,\Delta\chi_{II})\,,
\end{equation}
which we will use below to introduce the signed linking numbers of the adjacent nodal rings.

\subsection{Euler-class valued linking numbers, or the monopole charges of nodal rings} 

In this section we motivate that the linked nodal ring mediating a transition of Euler phases can be generally characterized by two linking numbers which we obtain by gauge fixing through
\begin{equation}
\begin{aligned}
    G(\Delta[\boldsymbol{\chi}]) &=
    G([\boldsymbol{\chi}^{\pi}])-G([\boldsymbol{\chi}^{0}])  \,,\\
    &= \left(\Delta\chi_I , \Delta\chi_{II} \right)\,,\\
    &= \left(\chi_I[S] , \chi_{II}[S] \right)\,,\\
    & = \left(                          \mathsf{Ln}_{I}[\mathcal{L}_{(I,II)}],\mathsf{Ln}_{II}[\mathcal{L}_{(I,II)}]    
    \right)\,,\\
    &= \boldsymbol{\mathsf{Ln}}[\mathcal{L}_{(I,II)}] \,,
\end{aligned}
\end{equation}
where $\chi_{I,II}[S]$ are the Euler class-valued monopole charges of the adjacent nodal ring $\mathcal{L}_{(I,II)}$ wrapped by the surface $S$. We motive this definition with two examples below.

\subsubsection{Single Euler class transition}
Returning to the example of Figure \ref{fig_braiding_monopole}, \ie a single Euler class transition with $[\chi^0_I]=[2]$, $[\chi^{\pi}_I]=[0]$ and $[\chi^0_{II}]=[\chi^{\pi}_{II}]=[0]$, the signed difference of Euler class is $G(\Delta [\chi_I]) = G([0]-[2]) = 0-(-2) = +2$, given the gauge choice of Figure \ref{fig_braiding_monopole}. We have noted above that the cancellation of Euler class at $k_z=\pi$ requires that a number $\vert G(\Delta [\chi_I])\vert =2$ of $I$-th nodes be braided with an adjacent node. As a consequence, the adjacent nodal ring $\mathcal{L}_{(I,II)}$ mediating the change of Euler class of the $I$-th subspace must be linked with two $I$-th nodal braids, those are $\{\mathcal{L}_{I,i}\}_{i=3,4}$ in Figure \ref{fig_braiding_monopole}. 

This motivates the definition of the linking number \cite{BJY_linking} of the adjacent nodal ring with the $I$-th nodal braids by the $I$-th {\it monopole} Euler classes, \ie
\begin{equation}
\begin{aligned}
    \mathsf{Ln}_I [\mathcal{L}_{(I,II)}] &= \chi_I[S]\,\\
    &= G(\Delta [\chi_I]) = +2 \;,
\end{aligned}
\end{equation}
where $S$ is an oriented envelope wrapping $\mathcal{L}_{(I,II)}$, see Fig.\;\ref{fig_braiding_monopole} in gray. We note that the monopole Euler class $\chi_I[S]$ is readily given by half the number of oriented $I$-th braids crossing the surface $S$, for a fixed choice of gauge. From Fig.\;\ref{fig_braiding_monopole}, we readily get $\chi_I[S]  = (1+1+1+1)/2 = 2$, \ie we count $+1$ for each $I$-th nodal braid oriented outwards the surface $S$. The linking number corresponds to the number of {\it stable} linking among the nodal braids, that is, not counting possible accidental linking that can be removed adiabatically (\ie while preserving the $(I,II)$-gap of the initial and final Euler phases). 

\begin{figure}[t!]
\centering
\begin{tabular}{l}
    (a) \\
    \includegraphics[width=0.7\linewidth]{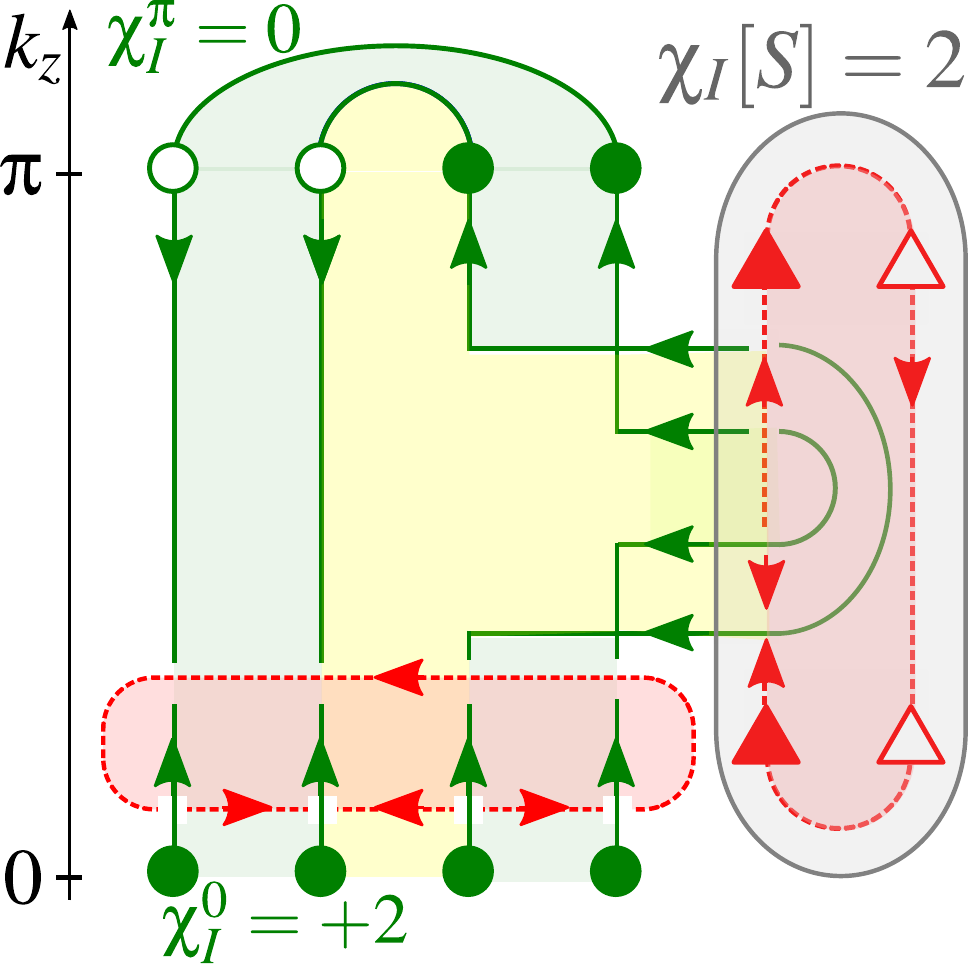} \\
    (b) \\
	\includegraphics[width=0.7\linewidth]{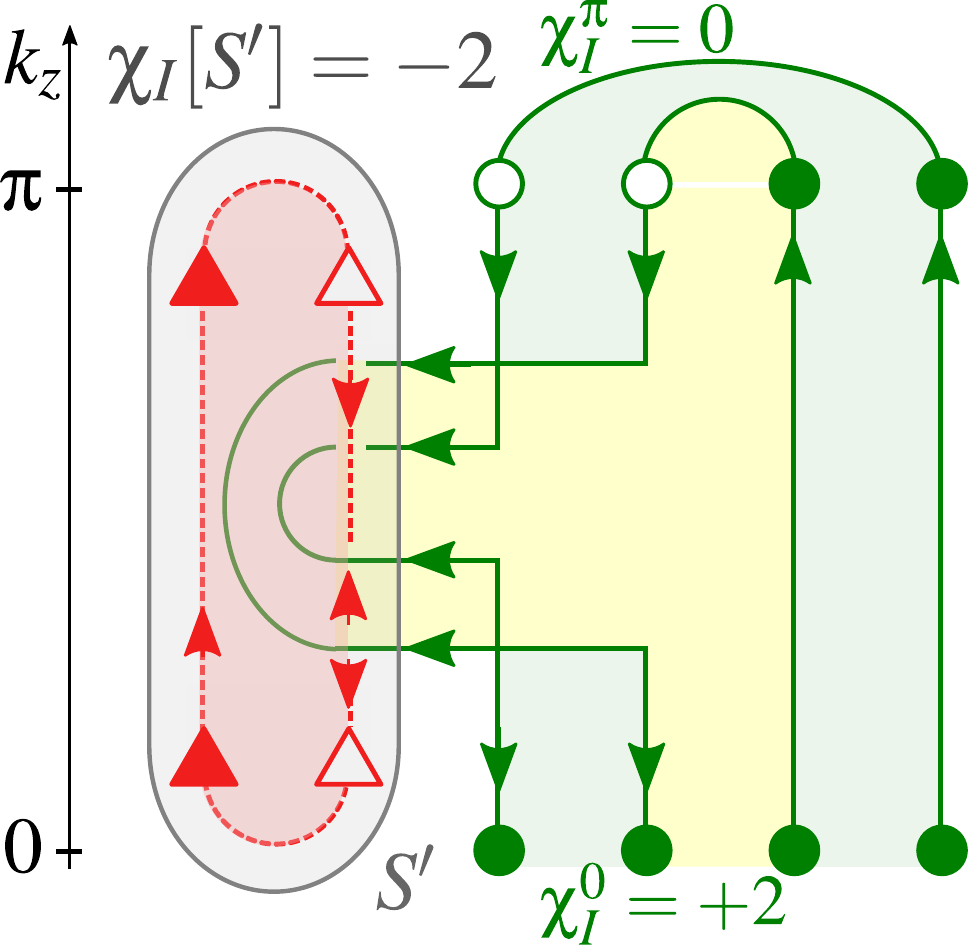}
\end{tabular}
\caption{\label{fig_braiding_reversed} The same linked nodal structure as in Fig.\;\ref{fig_braiding_monopole} (where it has the linking numbers $\mathsf{Ln}[\mathcal{L}_{(I,II)}]=(+2,0)$) upon the adiabatic reversal of the Euler class at $k_z=0$. (a) The Euler class reversal at $k_z=0$ has the effect of introducing one extra adjacent nodal ring (dashed line red) encircling all $I$-th nodal braids (full line green). (b) The resulting linked nodal structure after combining the two adjacent nodal rings, giving the linking numbers $\mathsf{Ln}[\mathcal{L}_{(I,II)}]=(-2,0)$. 
}
\end{figure}
\begin{figure*}[t!]
\centering
\begin{tabular}{lll}
    (a) & (b) & (c) \\
	\includegraphics[width=0.3\linewidth]{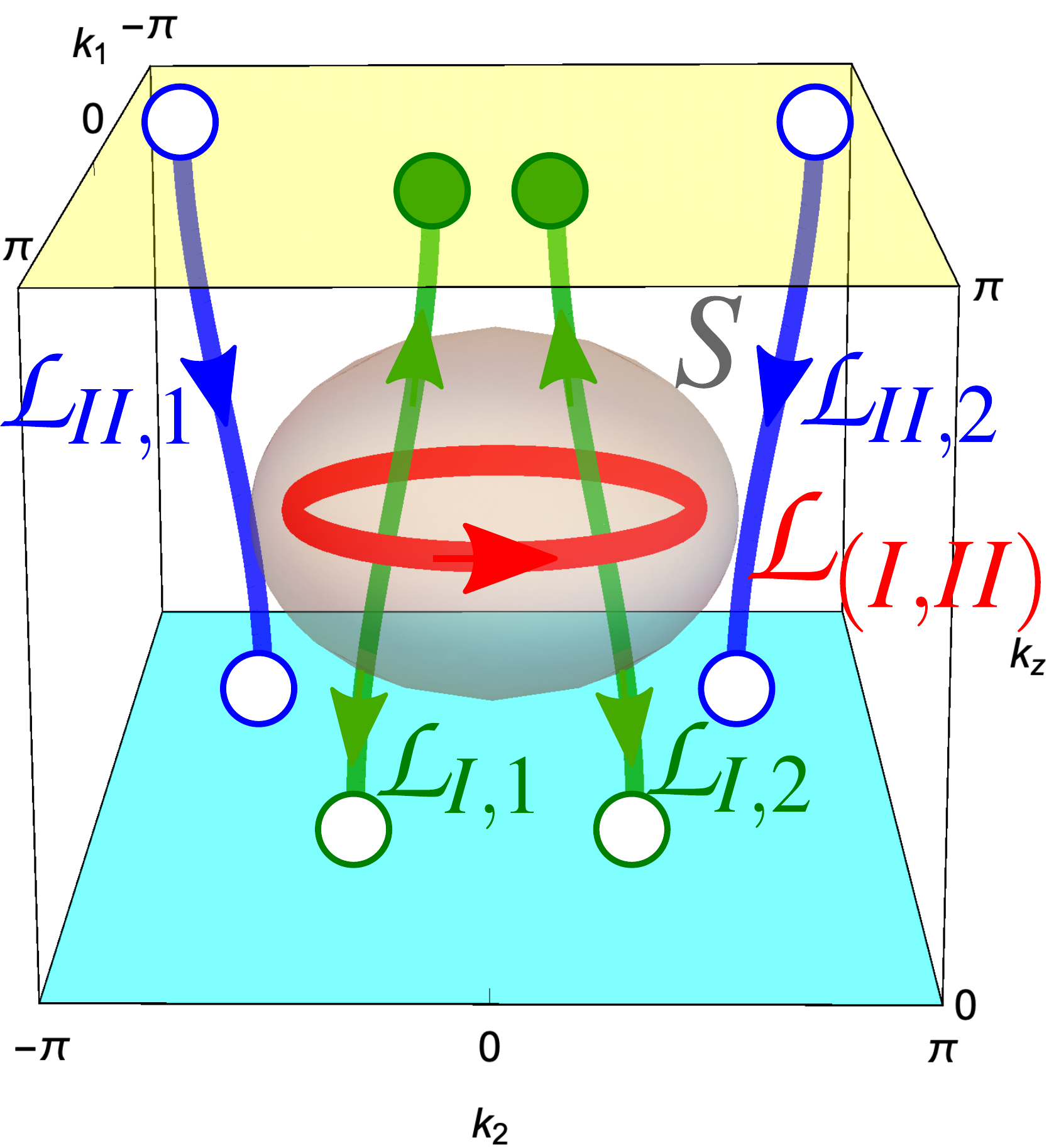} & 
	\includegraphics[width=0.3\linewidth]{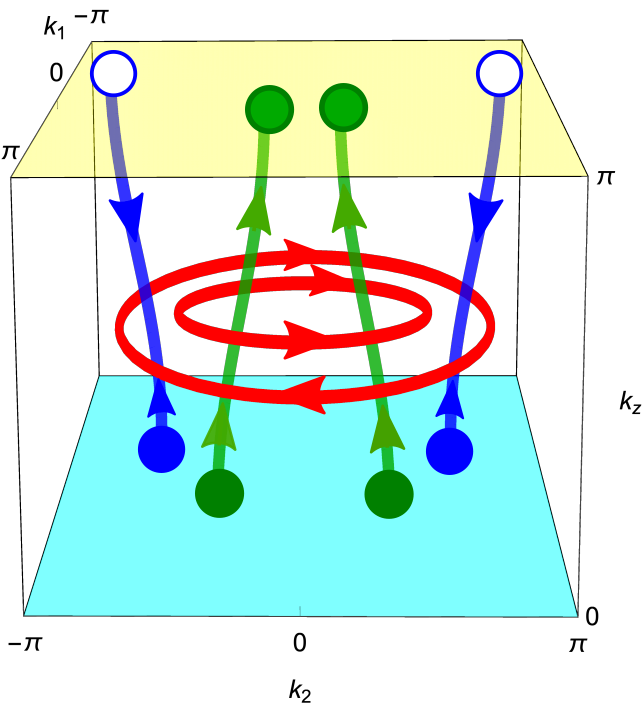} &
	\includegraphics[width=0.3\linewidth]{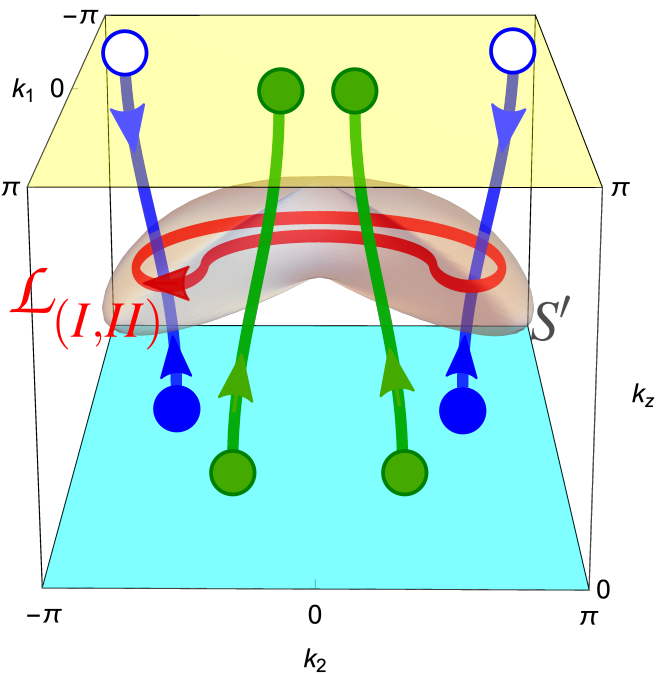}
\end{tabular}
\caption{\label{fig_linking_equivalence} Nodal braids and linked adjacent nodal ring obtained for the topological transition between the Euler phases $[\chi^0_{I},\chi^0_{II}]=[1,1]$ at $k_z=0$ and $[\chi^{\pi}_{I},\chi^{\pi}_{II}]=[1,-1]$ at $k_z=\pi$. By fixing the gauge globally, we allocate consistent non-Abelian frame charges to every nodal braid, here representing the charges through open/full circles at $k_z=0,\pi$ and through arrows \cite{Tiwari:2019} at intermediary $k_z$ (see text). We write the gauge fixing as $G([\boldsymbol{\chi}^0])=(-1,-1)$ and $G([\boldsymbol{\chi}^{\pi}])=(+1,-1)$, with $G$ the ``gauge-fixing'' map. From (a) to (c), we show the effect of the adiabatic reversal of Euler class at $k_z=0$ on the linking numbers that are computed in terms of Euler monopole charges. First, in (a) the adjacent nodal ring encircles the $I$-th nodal braids (green). Then, the Euler class reversal at $k_z=0$ induces the creation of an extra adjacent nodal ring in (b), which, after combining with the preexistent adjacent nodal ring, gives rise to an adjacent nodal ring that now encircles the $II$-th nodal braids. We find different linking numbers in (a), $\boldsymbol{\mathsf{Ln}}(\mathcal{L}_{(I,II)}) = (\chi_I[S],\chi_{II}[S]) = (+2,0)$, and in (c), $\boldsymbol{\mathsf{Ln}}(\mathcal{L}_{(I,II)}) = (\chi_I[S'],\chi_{II}[S']) = (0,-2)$, where $S$, and $S'$ are the surfaces wrapping the adjacent nodal ring in (a), and in (c), respectively. From the homotopy equivalence of (a) and (c), the linked nodal structure is hence characterized by the unique homotopy invariant $\Delta[\boldsymbol{\chi}] = [1,1] - [1,-1] =  \{[2,0],[0,2]\}$ (see text).
}
\end{figure*}

Combining the above result with the fact that there is no linking with the $II$-th nodal braids, we conclude that the nodal structure is characterized by the linking numbers
\begin{equation}
\begin{aligned}
    \boldsymbol{\mathsf{Ln}}[\mathcal{L}_{(I,II)}] &= \left(\chi_I[S] , \chi_{II}[S] \right)\,,\\
    & = G([\boldsymbol{\chi}^{\pi}])-G([\boldsymbol{\chi}^{0}]) \,,\\
    &= (0,0)-(-2,0) = (+2,0)\,.
\end{aligned}
\end{equation}

Let us now address the effect of the (adiabatic) Euler class reversal map on the linking numbers. On one hand, we note that the chosen gauge can be flipped globally, in which case the linking numbers become $(-2, 0)$. On the other hand, we can act with the Euler class reversal map on the Euler phases at $k_z = 0$ and at $k_z = \pi$, independently. We show in Figure \ref{fig_braiding_reversed} the effect of reversing the Euler classes at $k_z=0$. First, an extra adjacent nodal ring (dashed red) is introduced, which encircles all the nodal braids [Fig.\;\ref{fig_braiding_reversed}(a)]. After recombining the two adjacent nodal rings [Fig.\;\ref{fig_braiding_reversed}(b)], we get the linking numbers $\mathsf{Ln}[\mathcal{L}_{(I,II)}] = \chi_I[S']=(-2,0)$.

While the effects of the Euler class reversal and that of a global change of gauge appear to be the same, with all the homotopy equivalent phases captured by $[\,\mathsf{Ln}[\mathcal{L}_{(I,II)}]\,]$, we show below with an other example that the Euler reversals allows more possibilities leading to a larger homotopy equivalence class given by Eq.\;(\ref{eq_diff_equivalence}).

\subsubsection{Double Euler class transition}

We now generalize the above results to the cases when the Euler classes of both subspaces (the $I$-th and $II$-th) change. For this, we must again address all the consequences of the homotopy equivalence $(\chi_I,\chi_{II}) \sim (-\chi_I,-\chi_{II})$ on the linking numbers (\ie the Euler class-valued monopole charges) of the adjacent nodal ring. To this end, we consider the example of the transition from an Euler phase $[\chi^0_I,\chi^0_{II}]=[1,1]$ to $[\chi^{\pi}_I,\chi^{\pi}_{II}]=[1,-1]$, shown schematically in Figure \ref{fig_linking_equivalence}(a) with the choice of gauge in which $G(\boldsymbol{\chi}^0) = (-1,-1)$ and $G(\boldsymbol{\chi}^{\pi})=(+1,-1)$.

Taking into account that the initial and final Euler phases are only defined up-to-homotopy by an equivalence class, we deduce that the linking nodal structure is in principle characterized by all the combinatorial differences of Euler classes $(\chi^{0}_{I},\chi^{0}_{II}) - (\chi^{\pi}_{I},\chi^{\pi}_{II})$, \ie
\begin{alignat*}{2}
    (+1,-1) &- (-1,-1) && = (+2,0)\,,\\
    (+1,-1) &- (+1,+1) && = (0,-2)\,,\\
    (-1,+1) &- (-1,-1) && = (0,+2)\,,\\
    (-1,+1) &- (+1,+1) && = (-2,0)\,.
\end{alignat*}
Combining the pairs $(\pm2,0)$ and $(0,\pm2)$ into equivalence classes, \ie $[2,0]$ and $[0,2]$, we are left with showing the homotopy equivalence $[2,0] \sim [0,2]$, which is the meaning of the definition of the difference of two equivalence classes $[1,1] - [1,-1]$ as a pair of equivalence classes, \ie $\Delta[\boldsymbol{\chi}] = [1,1] - [1,-1] = \left\{ [2,0] , [0,2]\right\}$.

Using Figure \ref{fig_linking_equivalence}, we find that the above algebraic expression gives the right homotopy invariant of the linked nodal structures obtained from the transition of Euler phases. Starting from panel (a) obtained for a fixed global choice of gauge, we see that there is a change of (signed) Euler class of the $I$-th subspace from $\chi^0_I=1$ to $\chi^0_I=-1$, which is mediated by the presence of an adjacent $(I,II)$-nodal ring, $\mathcal{L}_{(I,II)}$ (red), linked to the $I$-th nodal braids (green). Wrapping $\mathcal{L}_{(I,II)}$ with the grey sphere in Fig.\;\ref{fig_linking_equivalence}(a), we obtain the linking numbers of the adjacent nodal ring in terms of signed Euler monopole charges  
\begin{equation}
\begin{aligned}
    \boldsymbol{\mathsf{Ln}}[\mathcal{L}_{(I,II)}]&=(\chi_I[S],\chi_{II}[S])=G\left(\Delta[\boldsymbol{\chi}] \right)=(+2,0)\,,
\end{aligned}
\end{equation}
where the map $G$ emphasizes that this holds upon the fixing of the gauge with respect to a unique base point.

We now proceed with the homotopy equivalence between the {\it a priori} different linking nodal structures corresponding to the linking numbers $(\pm2,0)$ and $(0,\pm2)$. We start with Fig.\;\ref{fig_linking_equivalence}(a), where one adjacent nodal ring encircles the two $I$-th nodal braids (green), which mediates the charge conversion of the latter from the plane $k_z=0$ to $k_z=\pi$. Fig.\;\ref{fig_linking_equivalence}(b,c) shows the effect of the adiabatic reversal of Euler class of the phase on the $k_z=0$-plane (similarly to Fig.\;\ref{fig_braiding_reversed}). The Euler class-reversal induces the creation of a new adjacent nodal ring that encircles all the nodal braids, both from the $I$-th and $II$-th gaps, in Fig.\;\ref{fig_linking_equivalence}(b) (note the reversed non-Abelian frame charges of the nodal braids at $k_z=0$). After combining the two adjacent nodal rings, we get Fig.\;\ref{fig_linking_equivalence}(c), where the resulting adjacent nodal ring now encircles the two $II$-th nodal braids (blue). We say that the deformation is adiabatic because we do not need to close the ($I,II$)-band gaps at the $k_z=0,\pi$-planes, which is the rule of the game here. After the adiabatic deformation, we define the new oriented wrapping surface $S'$ in Fig.\;\ref{fig_linking_equivalence}(c), with respect to which we obtain the linking numbers in terms of the signed Euler monopole charges, \ie 
\begin{equation}
    \boldsymbol{\mathsf{Ln}}[\mathcal{L}_{(I,II)}]
    =(\chi_I[S'],\chi_{II}[S']) = (0,-2)\,,
\end{equation}
where we assumed the same global choice of gauge as Fig.\;\ref{fig_linking_equivalence}(a). 

Since the two configurations Fig.\;\ref{fig_linking_equivalence}(a) and Fig.\;\ref{fig_linking_equivalence}(c) are homotopy equivalent, the nodal structure resulting from the Euler phase transition $[1,1]\leftrightarrow[1,-1]$ is allowed to realize all the linking numbers $\boldsymbol{\mathsf{Ln}}[\mathcal{L}_{(I,II)}]\in \{(2,0),(-2,0),(0,2),(0,-2)\}$, where the transition from one pair of linking numbers (given a fixed choice of gauge) to another is obtained through the adiabatic reversal of Euler classes at $k_z=0,\pi$. We thus conclude that the linked nodal structure is characterized by the single homotopy invariant 
\begin{equation}
    \Delta[\boldsymbol{\chi}] = [1,1] - [1,-1] = \left\{ [2,0] , [0,2]\right\}\,.
\end{equation}

\section{Linked nodal structures in 3D}\label{sec_3D_NL_numerics}

In this section, we specify the 3D models hosting nodal line structures for every pair of 2D Euler models presented in Section \ref{sec_min_models}. We show that the transition from one Euler phase to an other, which is homotopically inequivalent, must be mediated by an {\it adjacent} nodal ring, that is linked with the nodal braids of the $I$-th or $II$-th subspaces. We show the nodal structures obtained numerically and we characterize them with the concepts introduced in Section \ref{sec_3D_NL}. 
 
Within this section, we use the general ansatz Eq.\;(\ref{eq_3D_PT_nodal}) for $H_{PT}^{\Delta[\boldsymbol{\chi}]}(\boldsymbol{k}_{\parallel},k_z)$, and hence we refer to the two 2D building blocks $H^{[\boldsymbol{\chi}^0]}(\boldsymbol{k}_{\parallel})$ and $H^{[\boldsymbol{\chi}^{\pi}]}(\boldsymbol{k}_{\parallel})$ chosen among the models of Section \ref{sec_min_models}, where we take $(k_1,k_2) = \boldsymbol{k}_{\parallel} $.  

\begin{figure*}[t!]
\centering
\begin{tabular}{lll}
    (a) ~  $\Delta[\chi_I]=[0]-[2]$ &
    (b)~ $\Delta[\chi_I]=[2]-[4]$  & 
    (c) ~  $\Delta[\chi_I]=[0]-[4]$\\
	\includegraphics[width=0.31\linewidth]{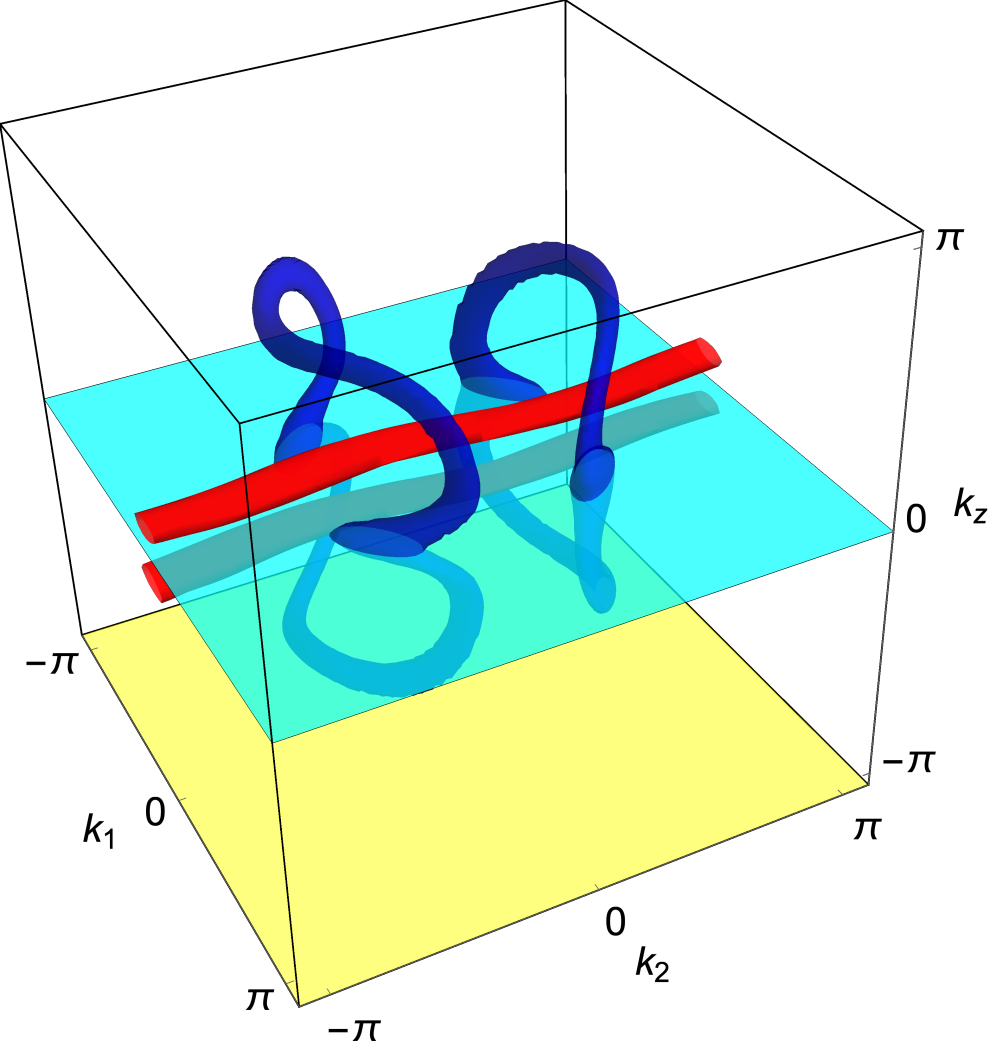}&
	\includegraphics[width=0.31\linewidth]{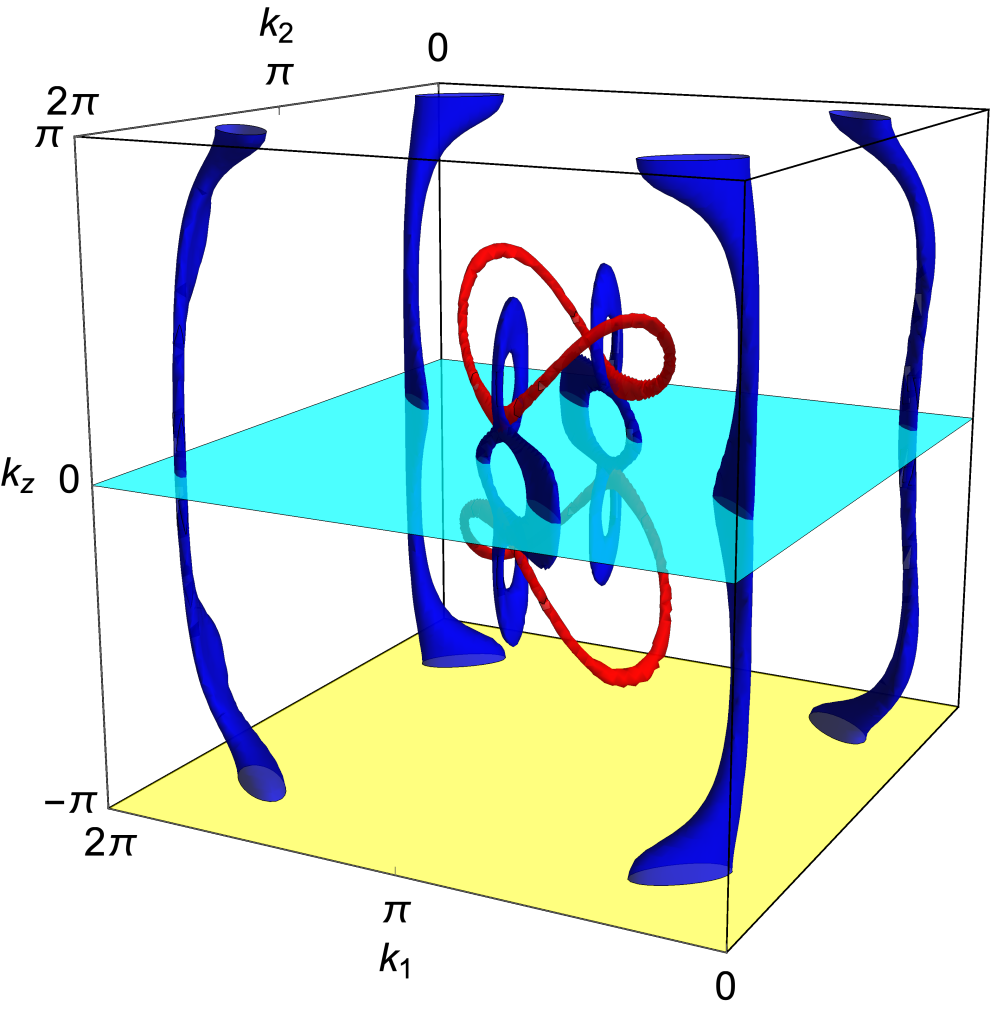} &
	\includegraphics[width=0.31\linewidth]{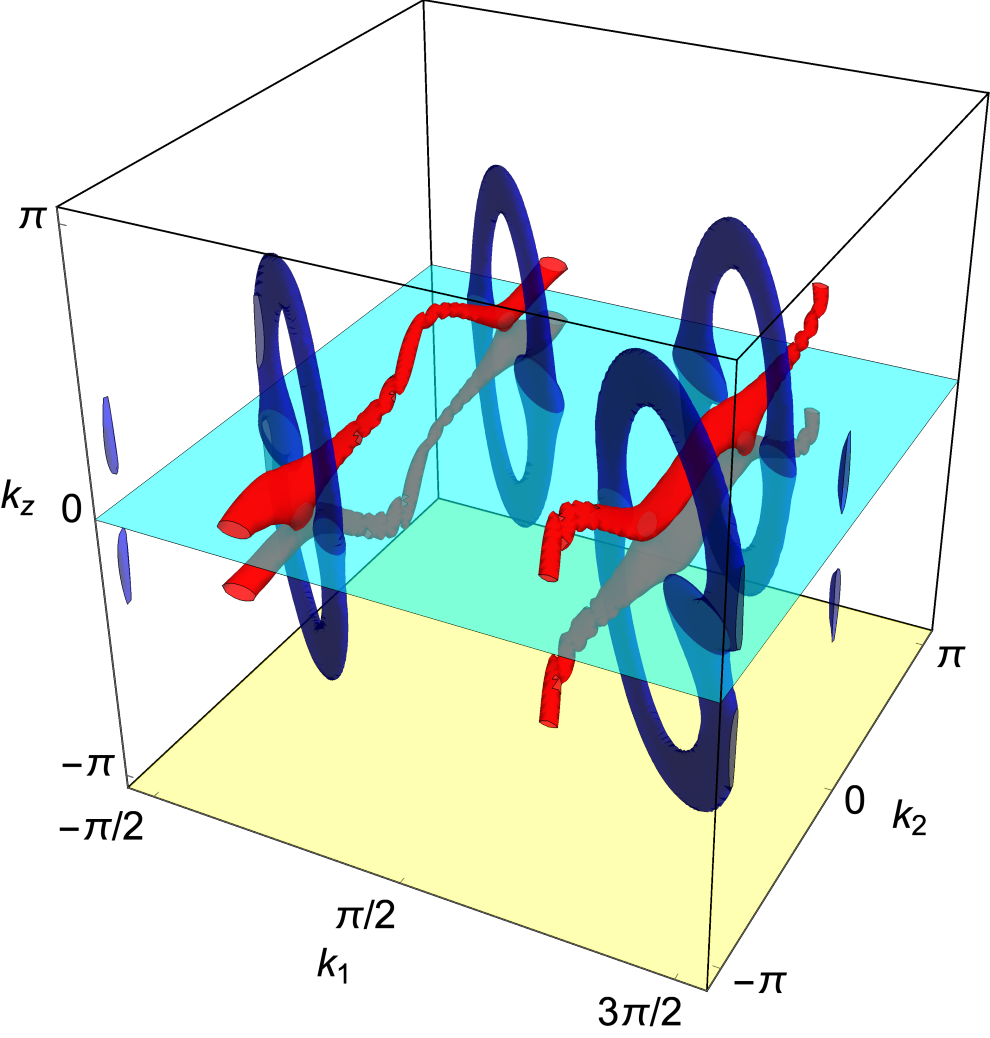}
\end{tabular}
\caption{\label{fig_3B_NL} $PT$-symmetry protected linked nodal structures in 3D generated by a pair of 2D Euler phases, located at $k_z=0$ (cyan plane) and $\vert k_z\vert=\pi$ (yellow plane), in the three-band case. The nodal braids belonging to the $I$-th two-band subspace are colored in blue, and the adjacent nodal rings, belonging to the $(I,II)$-gap are in red. The linking of an adjacent nodal rings $\mathcal{L}_{(I,II)}$ with the $I$-th nodal braids are required to mediate the transition between any two inequivalent Euler phases, with an Euler class-valued monopole charge corresponding its linking number with respect to the $I$-th subspaces $\mathsf{Ln}_I \in 2\mathbb{Z}$ (in the three-band and orientable case, only even Euler class can be formed). Note that the {\it signed} linking number requires the global fixing of the gauge (see Section \ref{sec_3D_NL}). (a) Linked nodal structure for the Euler phases $[\chi^0_I]=[2]$ and $[\chi^{\pi}_I]=[0]$. The linking number of the adjacent nodal ring is $\vert \mathcal{Ln}\vert = 2$, matching with the monopole Euler class computed over a wrapping cylinder (since $\mathcal{L}_{(I,II)}$ is winding through one non-contractible loop of the Brillouin zone). (b) Linked nodal structure for the Euler phases $[\chi^0_I]=[4]$ and $[\chi^{\pi}_I]=[2]$, inducing an adjacent nodal ring with the linking number $\vert\mathsf{Ln}_I\vert= 2$. (c) Linked nodal ring with a linking number $\vert\mathsf{Ln}_I\vert= 4$, induced by the Euler phases $[\chi^0_I]=[4]$ and $[\chi^{\pi}_I]=[0]$.     
}
\end{figure*}

\subsection{Three-band $PT$-models and their linked nodal structure}\label{sec_PT_3B}

\subsubsection{$\Delta[\chi_I] = [0] - [2]$} 

We take 
\begin{equation}
        H^{[\chi^0_I]} (\boldsymbol{k}) = H^{[2]}_{3B}(\boldsymbol{k})\,,~
        H^{[\chi^{\pi}_I]} (\boldsymbol{k}) = H^{[0]}_{3B}(\boldsymbol{k})\,,
\end{equation}
with $(m,t_1,t_2,\lambda,\delta)=(1,1,0,1,1/4)$ in Eq.\;(\ref{eq_min_H_E2}) for the phase $\chi_I=[2]$, and for the trivial phase, we take
\begin{equation}
        H^{[0]}_{3B}(\boldsymbol{k}) =  H^A_{3B,\text{deg}}[5,0,0,1](\boldsymbol{k}) - \dfrac{1}{4} \text{diag}(0,1,-1)\,,
\end{equation}
\ie that is Eq.\;(\ref{eq_3B_nonflat}) with $(m,t_1,t_2,\lambda)=(5,0,0,1)$, and an additional mass term. We show the linked nodal structure generated by the 3D model in Fig.\;\ref{fig_3B_NL}(a). The blue lines correspond to the $I$-th nodal braids, and the red line is the adjacent $(I,II)$-nodal ring, $\mathcal{L}_{(I,II)}$. The cyan plane at $k_z=0$ locates the Euler phase $[\chi^0_I]=[2]$ with four stable nodal points,\ie at the intersections of the blue nodal braids, and the yellow plane at $k_z=\pi$ locates the trivial phase with no node. The adjacent nodal ring is characterized by a linking number $\vert \mathsf{Ln}_I[\mathcal{L}_{(I,II)}] \vert = 2$. We note that $\mathcal{L}_{(I,II)}$ is also winding through the $k_2$-axis of the Brillouin zone.

\subsubsection{$\Delta[\chi_I] = [2] - [4]$} 

We take 
\begin{equation}
        H^{[\chi^0_I]} (\boldsymbol{k}) = H^{[4]}_{3B}(\boldsymbol{k})\,,~
        H^{[\chi^{\pi}_I]} (\boldsymbol{k}) = H^{[2]}_{3B}(\boldsymbol{k})\,,
\end{equation}
where $H^{[2]}_{3B}(\boldsymbol{k})$ is the same as above. For the phase $[4]$, we take Eq.\;(\ref{eq_min_H_E4}) with $\boldsymbol{h}^B$ and $(m,t_1,t_2,\lambda)=(1/2,0,5/2,1/4)$, \ie
\begin{equation}
        H^{[4]}_{3B}(\boldsymbol{k}) =  H^B_{3B,\text{deg}}[1/2,0,5/2,1](\boldsymbol{k}) - \dfrac{1}{4} \text{diag}(1,0,-1)\,.
\end{equation}
We show the linked nodal structure in Fig.\;\ref{fig_3B_NL}(b) [note the shift of the axes, when comparing the $k_z=\pi$ plane in (b) with the $k_z=0$ plane in (a)], which is now characterized by eight stable nodal points (blue) on the $k_z=0$ plane, and an adjacent nodal ring (red) with a linking number $\vert\mathsf{Ln}_I(\mathcal{L}_{(I,II)})\vert = 2$.

\subsubsection{$\Delta[\chi_I] = [0] - [4]$} 

We take the same model $H^{[0]}_{3B}(\boldsymbol{k})$ for the trivial phase, and for the $[4]$ phase we take, 
\begin{equation}
        H^{[4]}_{3B}(\boldsymbol{k}) =  H^B_{3B,\text{deg}}[1/2,0,-3/2,1](\boldsymbol{k}) - \dfrac{1}{4} \Lambda_3\,.
\end{equation}
We show the linked nodal structure in Fig.\;\ref{fig_3B_NL}(c), that again exhibits eight stable nodal points at $k_z=0$, but now with two adjacent nodal rings leading to a doubled linking number $\vert\mathsf{Ln}_I(\mathcal{L}_{(I,II)})\vert = 4$.

\subsection{Four-band $PT$-models and their linked nodal structure}\label{sec_PT_4B}

\begin{figure*}[t!]
\centering
\begin{tabular}{ll}
    (a)~$\Delta[\boldsymbol{\chi}]=[0,0]-[1,1]$ &
    (b)~$\Delta[\boldsymbol{\chi}]=[1,-1]-[1,1]$ \\
	\includegraphics[width=0.31\linewidth]{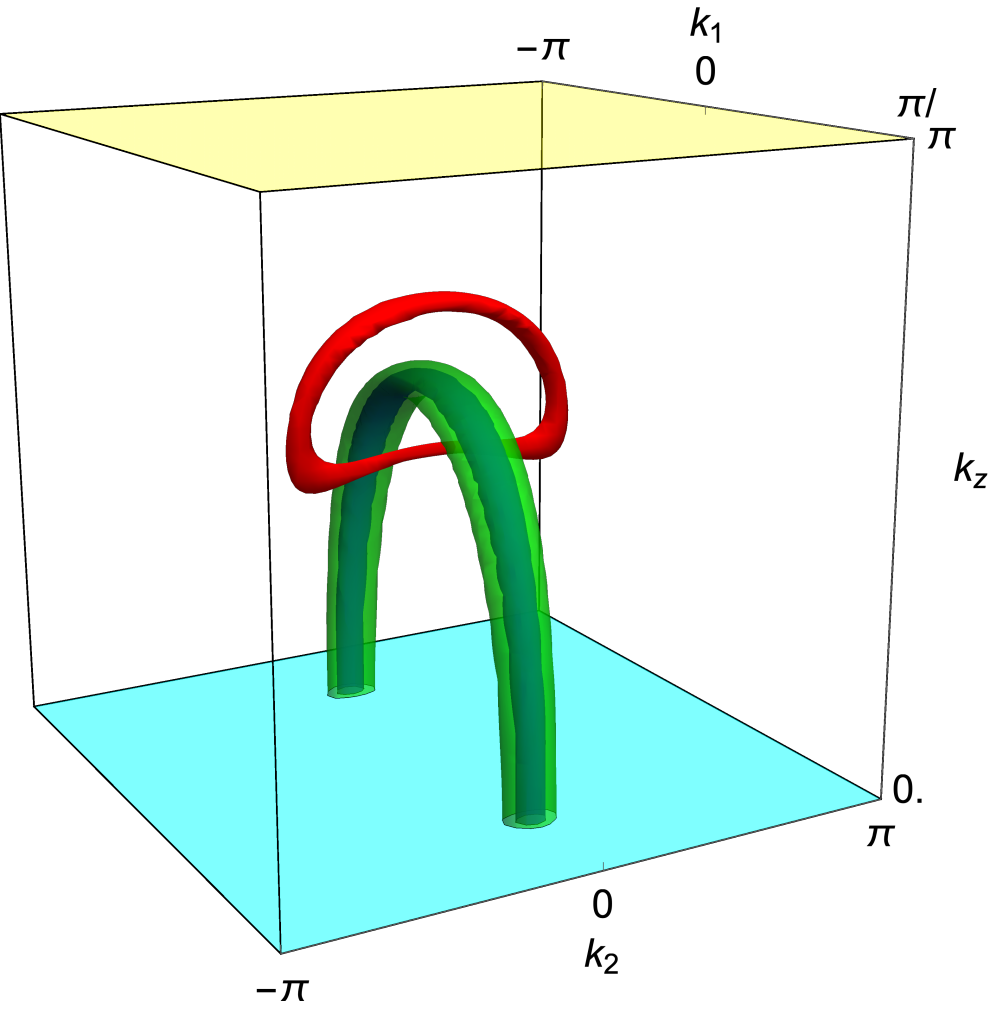}&
	\includegraphics[width=0.31\linewidth]{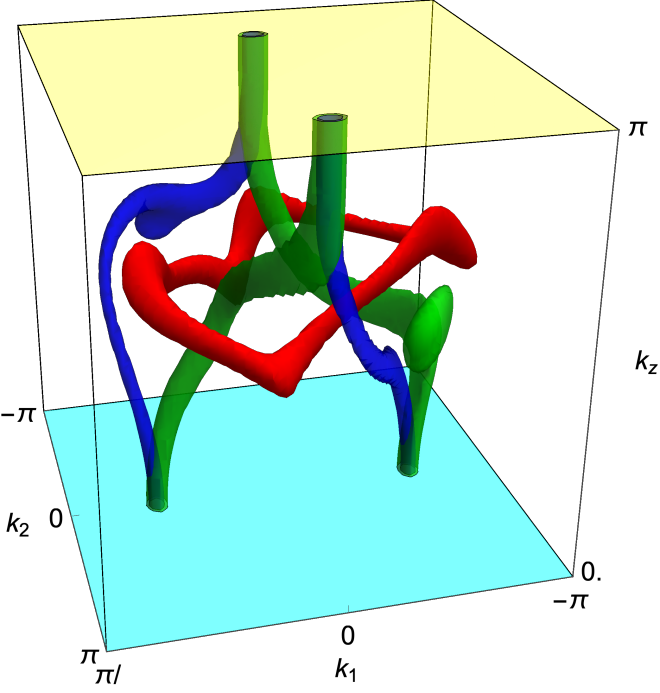} \\
    (c)~$\Delta[\boldsymbol{\chi}]=[0,0]-[2,2]$  &
    (d)~$\Delta[\boldsymbol{\chi}]=[2,-2]-[2,2]$ \\
    \includegraphics[width=0.31\linewidth]{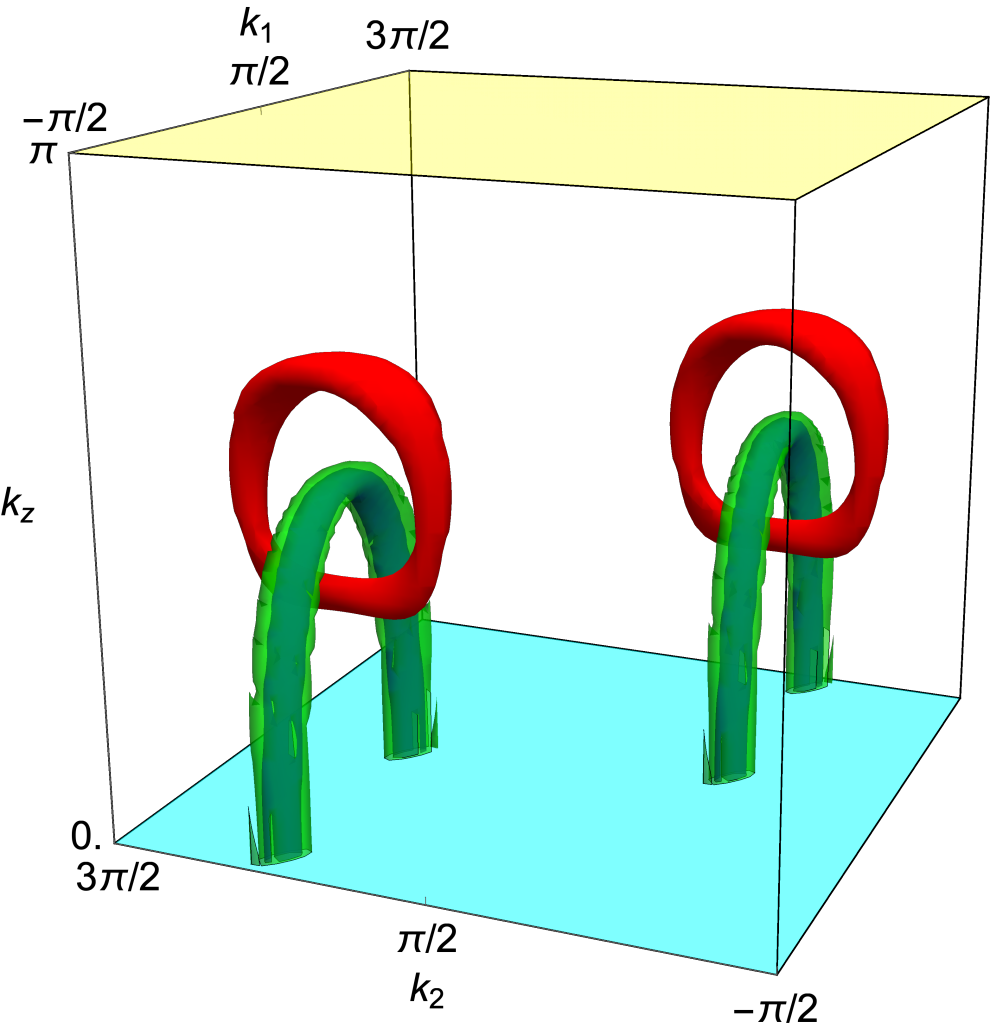} &
	\includegraphics[width=0.31\linewidth]{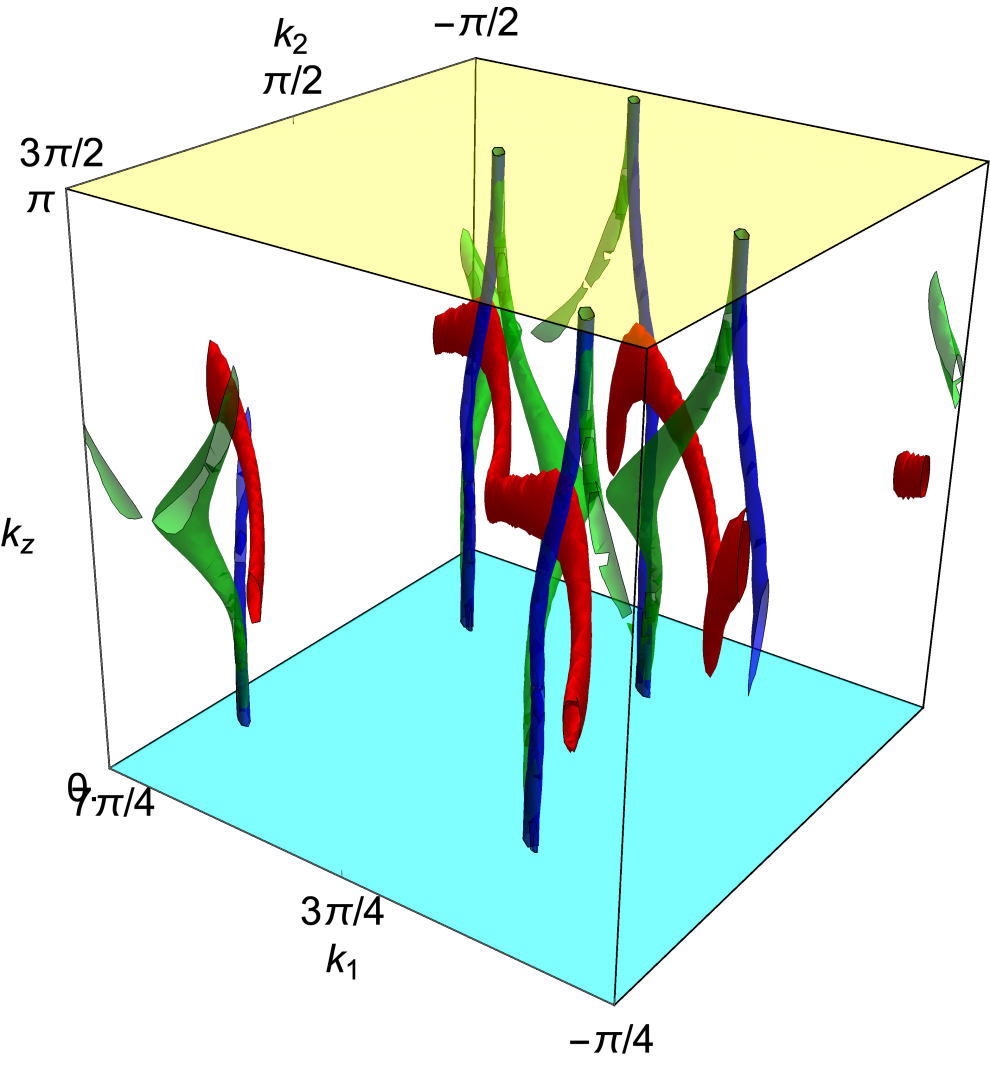} 
\end{tabular}
\caption{\label{fig_4B_balanced_NL} $PT$-symmetry protected linked nodal structures in 3D generated by a pair of 2D Euler phases, located at $k_z=0$ (cyan plane) and $ k_z=\pi$ (yellow plane), in the balanced four-band case. The $I$-th ($II$-th) nodal braids are drawn in green (blue), and the adjacent nodal rings $\mathcal{L}_{(I,II)}$ (\ie within the $(I,II)$-gap) are in red. Similarly to Fig.\;(\ref{fig_3B_NL}), the transition between two inequivalent Euler phases enforces the linking of the adjacent nodal rings with the $I$-th and $II$-th nodal braids, leading to the linking numbers $\boldsymbol{\boldsymbol{Ln}}=(\mathsf{Ln}_I,\mathsf{Ln}_{II})$ (note that the signed linking numbers require the global fixing of the gauge, see Section \ref{sec_3D_NL}). (a) $\boldsymbol{\boldsymbol{Ln}}[\mathcal{L}_{(I,II)}]=(1,1)$. (b) $\boldsymbol{\boldsymbol{Ln}}[\mathcal{L}_{(I,II)}]=(1,0)$. (c) $\boldsymbol{\boldsymbol{Ln}}[\mathcal{L}_{(I,II)}]=(2,2)$. (d) $\boldsymbol{\boldsymbol{Ln}}[\mathcal{L}_{(I,II)}]=(0,4)$.
}
\end{figure*}

\begin{figure*}[t!]
\centering
\begin{tabular}{ll}
    (a)~$\Delta[\boldsymbol{\chi}]=[0,0]-[2,0]$  
    &
    (b)~$\Delta[\boldsymbol{\chi}]=[0,0]-[4,0]$ 
        \\
	\includegraphics[width=0.31\linewidth]{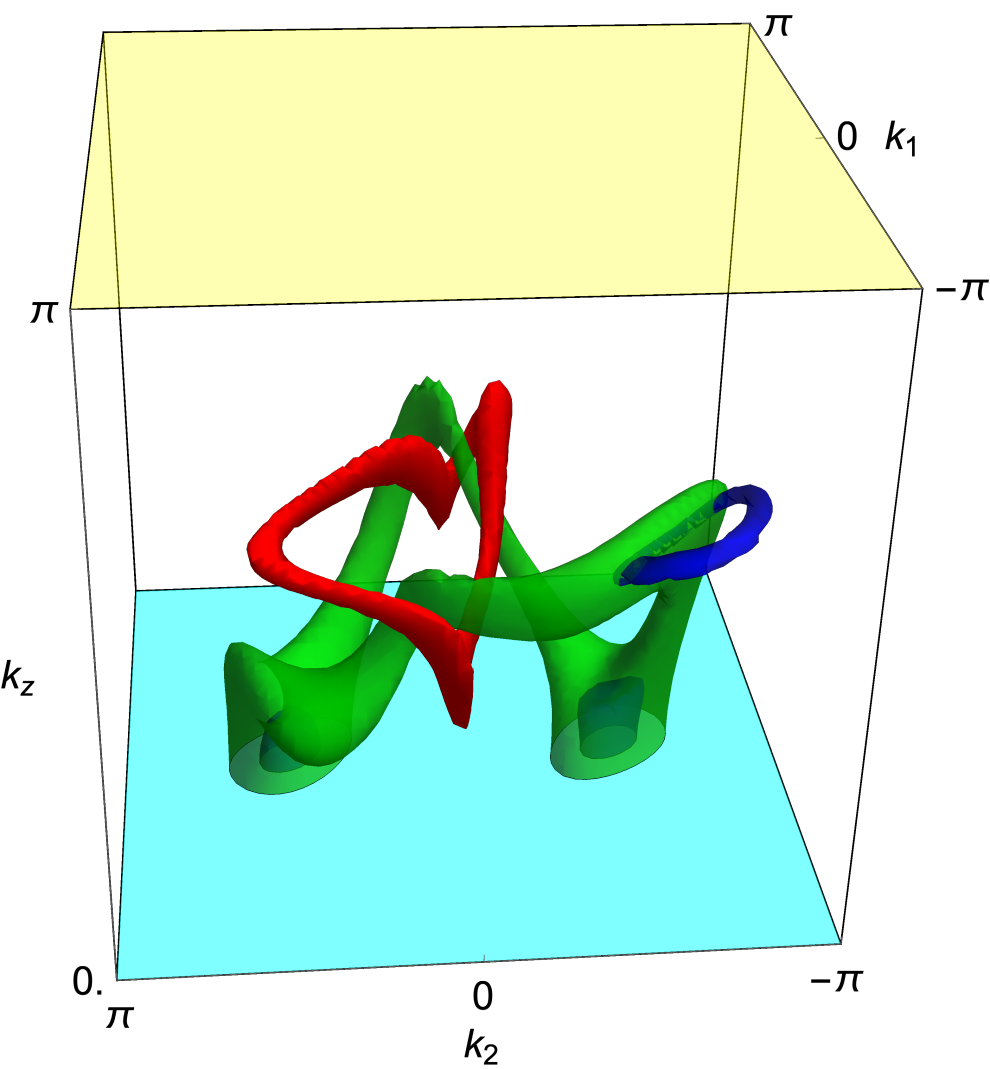} &
	\includegraphics[width=0.31\linewidth]{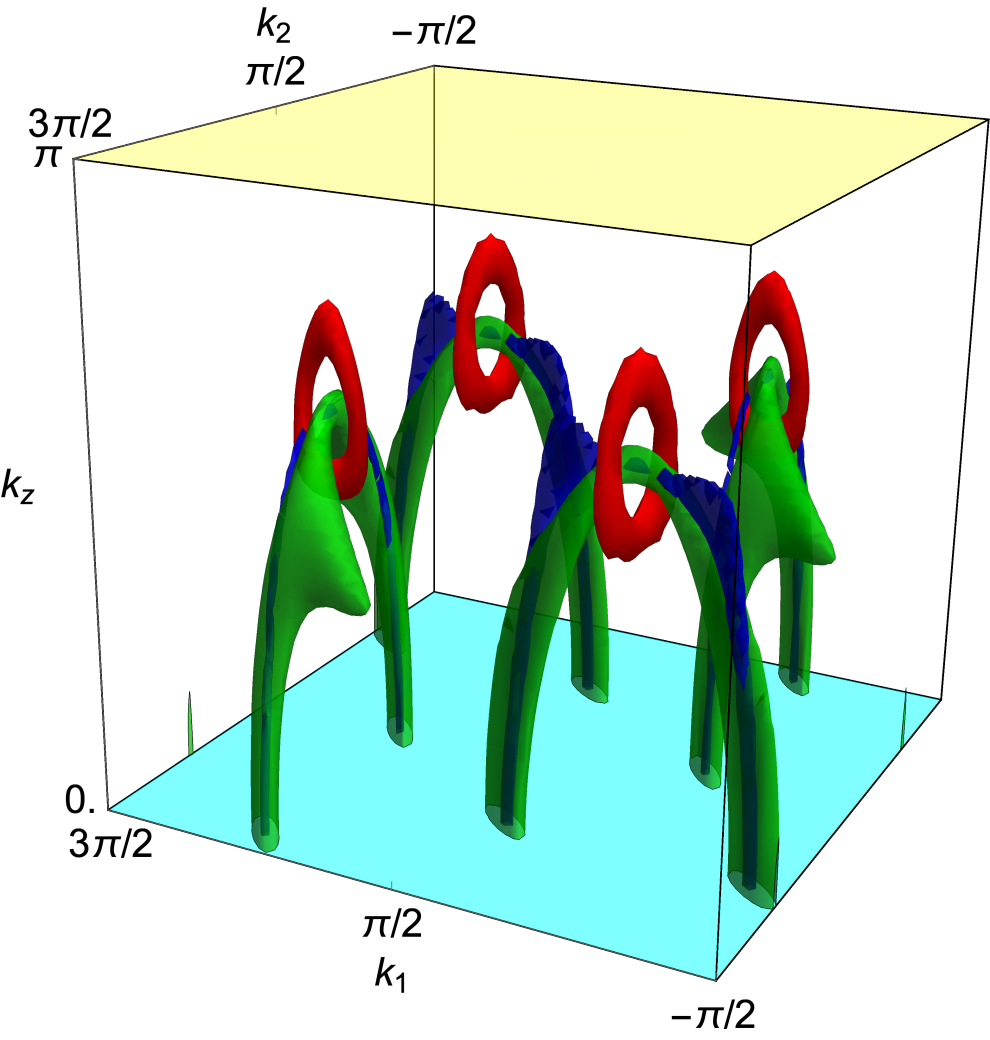} 
	\\
	(c)~$\Delta[\boldsymbol{\chi}]=[0,0]-[3,1]$  &
    (d)~$\Delta[\boldsymbol{\chi}]=[3,-1]-[3,1]$  \\
	\includegraphics[width=0.31\linewidth]{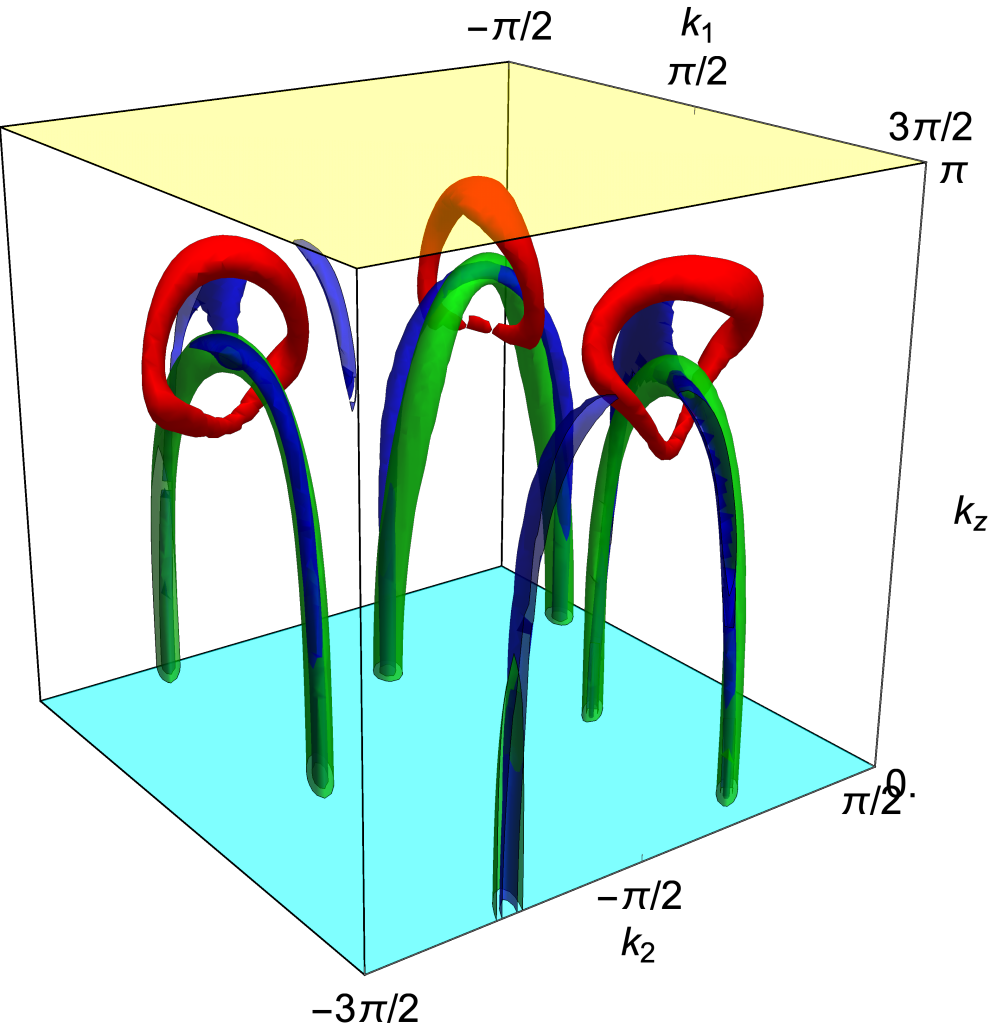}  &
	\includegraphics[width=0.31\linewidth]{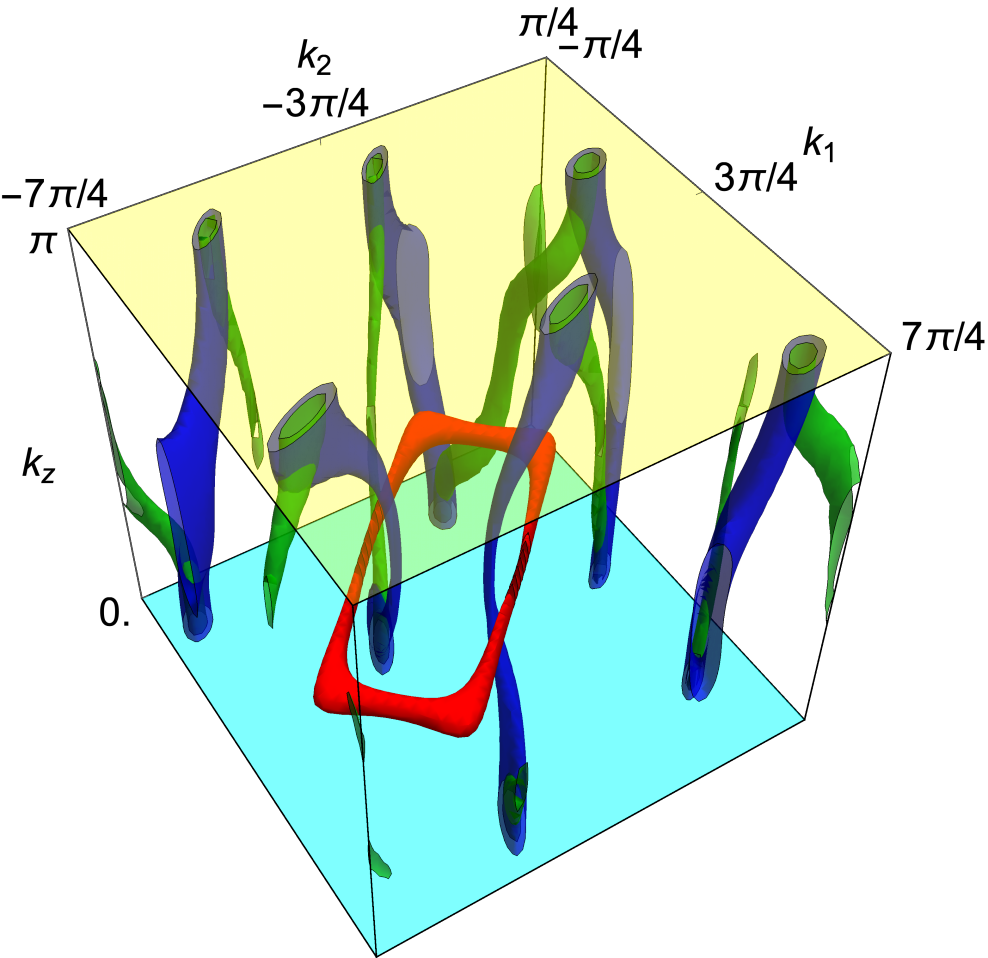} 
\end{tabular}
\caption{\label{fig_4B_imbalanced_NL} Same data as those presented in Fig.\;\ref{fig_4B_balanced_NL} for the imbalanced four-band case. (a) $\boldsymbol{\mathsf{Ln}}[\mathcal{L}_{(I,II)}]=(2,0)$. 
(b) $\boldsymbol{\mathsf{Ln}}[\mathcal{L}_{(I,II)}]=(4,0)$. 
(c) $\boldsymbol{\mathsf{Ln}}[\mathcal{L}_{(I,II)}]=(3,1)$. 
(d) $\boldsymbol{\mathsf{Ln}}[\mathcal{L}_{(I,II)}]=(0,2)$. 
}
\end{figure*}

\subsection{$\Delta[\boldsymbol{\chi}] = [0,0] - [1,1]$}

We take 
\begin{equation}
        H^{[\boldsymbol{\chi}^{0}]} (\boldsymbol{k}) = H^{[1,1]}_{4B}(\boldsymbol{k})\,,~
        H^{[\boldsymbol{\chi}^{\pi}]} (\boldsymbol{k}) = H^{[0,0]}_{4B}(\boldsymbol{k})\,,
\end{equation}
with $H^{[1,1]}_{4B}(\boldsymbol{k})$ given by Eq.\;(\ref{eq_min_H_E11}) with $(m,t_1,\delta)=(1,3/2,1/2)$, and for the trivial phase, we take Eq.\;(\ref{eq_4B_deg}) with $(m,t_1,t_2)=(1,0,0,1/2)$ and $(m',t_1',t_2')=(1,0,0,1/2)$, \ie
\begin{equation}
    H^{[0,0]}_{4B}(\boldsymbol{k}) = H^{AA}_{\text{4B,deg}}
    \left[\begin{smallmatrix}
        1 & 0 & 0 & 0 \\
        1 & 0 & 0 & 0
    \end{smallmatrix}\right]
    (\boldsymbol{k}) 
    + \dfrac{1}{2}\, \Gamma_{13}\,.
\end{equation}
We show the linked nodal structure in Fig.\;\ref{fig_4B_balanced_NL}(a), where the $I$-th nodal braid is down in green, the $II$-th nodal braid is blue (here visible by transparency below the green line), and the adjacent linked nodal ring in red. There is a pair of stable nodes at $k_z=0$ in both the $I$-th and $II$-th subspaces. Each pair is connected via a braid that is lined with the red nodal ring. The linking numbers of the later are $\boldsymbol{\mathsf{Ln}}=(1,1)$. 

\subsection{$\Delta[\boldsymbol{\chi}] = [1,-1] - [1,1]$}

We take 
\begin{equation}
        H^{[\boldsymbol{\chi}^{0}]} (\boldsymbol{k}) = H^{[1,1]}_{4B}(\boldsymbol{k})\,,~
        H^{[\boldsymbol{\chi}^{\pi}]} (\boldsymbol{k}) = H^{[1,-1]}_{4B}(\boldsymbol{k})\,,
\end{equation}
with $H^{[1,1]}_{4B}(\boldsymbol{k})$ the same as above, and for the $[1,-1]$ phase, we take Eq.\;(\ref{eq_min_H_E11m}) with $(m',t_1',\delta)=(1,0,1/2)$. We show the linked nodal structure in Fig.\;\ref{fig_4B_balanced_NL}(b). The adjacent nodal ring is now linked with the $I$-th nodal braids only, given the linking numbers $\boldsymbol{\mathsf{Ln}}=(2,0)$.

\subsection{$\Delta[\boldsymbol{\chi}] = [0,0] - [2,2]$}

We take 
\begin{equation}
        H^{[\boldsymbol{\chi}^{0}]} (\boldsymbol{k}) = H^{[2,2]}_{4B}(\boldsymbol{k})\,,~
        H^{[\boldsymbol{\chi}^{\pi}]} (\boldsymbol{k}) = H^{[0,0]}_{4B}(\boldsymbol{k})\,,
\end{equation}
with $H^{[2,2]}_{4B}(\boldsymbol{k})$ given by Eq.\;(\ref{eq_min_H_E22}) with $(m,t_1,t_2,\delta)=(1/2,0,-3/2,1/2)$, and $H^{[0,0]}_{4B}(\boldsymbol{k})$ the same as above. We show the linked nodal structure in Fig.\;\ref{fig_4B_balanced_NL}(c), where two adjacent nodal rings are each linked with both one $I$-th and one $II$-th nodal braid, leading to the linking numbers $\boldsymbol{\mathsf{Ln}}=(2,2)$.

\subsection{$\Delta[\boldsymbol{\chi}] = [2,-2] - [2,2]$}

We take 
\begin{equation}
        H^{[\boldsymbol{\chi}^{0}]} (\boldsymbol{k}) = H^{[2,2]}_{4B}(\boldsymbol{k})\,,~
        H^{[\boldsymbol{\chi}^{\pi}]} (\boldsymbol{k}) = H^{[2,-2]}_{4B}(\boldsymbol{k})\,,
\end{equation}
with $H^{[2,2]}_{4B}(\boldsymbol{k})$ the same as above, and with $H^{[2,-2]}_{4B}(\boldsymbol{k})$ given by Eq.\;(\ref{eq_min_H_E22m}) with $(m',t_1',t_2',\delta')=(1/2,0,-3/2,1/2)$. We show the linked nodal structure in Fig.\;\ref{fig_4B_balanced_NL}(d), where one adjacent nodal ring is linked with four $II$-th nodal braids (blue), leading to the linking numbers $\boldsymbol{\mathsf{Ln}}=(0,4)$.

\subsection{$\Delta[\boldsymbol{\chi}] = [0,0] - [2,0]$}

We take 
\begin{equation}
        H^{[\boldsymbol{\chi}^{0}]} (\boldsymbol{k}) = H^{[2,0]}_{4B}(\boldsymbol{k})\,,~
        H^{[\boldsymbol{\chi}^{\pi}]} (\boldsymbol{k}) = H^{[0,0]}_{4B}(\boldsymbol{k})\,,
\end{equation}
with $H^{[2,0]}_{4B}(\boldsymbol{k})$ given by Eq.\;(\ref{eq_min_H_E20}), and with
\begin{equation}
    H^{[0,0]}_{4B}(\boldsymbol{k}) = H^{CA}_{\text{4B,deg}}
    \left[\begin{smallmatrix}
        1 & 0 & 0 & 1 \\
        -1 & 0 & 0 & 1
    \end{smallmatrix}\right]
    (\boldsymbol{k}) 
    + \dfrac{1}{2}\, \Gamma_{11}\,.
\end{equation}
We show the linked nodal structure in Fig.\;\ref{fig_4B_imbalanced_NL}(a), where one adjacent nodal ring is linked with two $I$-th nodal braids (green), leading to the linking numbers $\boldsymbol{\mathsf{Ln}}=(2,0)$.

\subsection{$\Delta[\boldsymbol{\chi}] = [0,0] - [4,0]$}

We take 
\begin{equation}
        H^{[\boldsymbol{\chi}^{0}]} (\boldsymbol{k}) = H^{[4,0]}_{4B}(\boldsymbol{k})\,,~
        H^{[\boldsymbol{\chi}^{\pi}]} (\boldsymbol{k}) = H^{[0,0]}_{4B}(\boldsymbol{k})\,,
\end{equation}
with $H^{[4,0]}_{4B}(\boldsymbol{k})$ given by Eq.\;(\ref{eq_min_H_E40}), and with
\begin{equation}
    H^{[0,0]}_{4B}(\boldsymbol{k}) = H^{AA}_{\text{4B,deg}}
    \left[\begin{smallmatrix}
        1 & 0 & 0 & 1 \\
        1 & 0 & 0 & -1
    \end{smallmatrix}\right]
    (\boldsymbol{k}) 
    + \dfrac{1}{2}\, \Gamma_{13}\,.
\end{equation}
We show the linked nodal structure in Fig.\;\ref{fig_4B_imbalanced_NL}(b), where four adjacent nodal rings are each linked one time with both one $I$-th and one $II$-th nodal braid. Each adjacent nodal ring that is linked one time with a single $I$-th braid, must also be linked one time with a $II$-th braid, otherwise its non-Abelian charge would be ill defined \cite{Tiwari:2019}. This has the consequence that, even though the $II$-th braids (blue) are not stable (since $\chi^0_{II}=0$), they must be linked with the adjacent ring. The instability of the $II$-th braids tells us that Fig.\;\ref{fig_4B_imbalanced_NL}(b) can be deformed adiabatically such that the $II$-th braids annihilate while leaving the linked adjacent rings paired two-by-two. The linking numbers here are $\boldsymbol{\mathsf{Ln}}=(4,0)$.

\subsection{$\Delta[\boldsymbol{\chi}] = [0,0] - [3,1]$}

We take 
\begin{equation}
        H^{[\boldsymbol{\chi}^{0}]} (\boldsymbol{k}) = H^{[3,1]}_{4B}(\boldsymbol{k})\,,~
        H^{[\boldsymbol{\chi}^{\pi}]} (\boldsymbol{k}) = H^{[0,0]}_{4B}(\boldsymbol{k})\,,
\end{equation}
with $H^{[3,1]}_{4B}(\boldsymbol{k})$ given by Eq.\;(\ref{eq_min_H_E31}), and with
\begin{equation}
    H^{[0,0]}_{4B}(\boldsymbol{k}) = \dfrac{1}{4} H^{AA}_{\text{4B,deg}}
    \left[\begin{smallmatrix}
        -3 & 0 & 0 & 0 \\
        -1 & 0 & 0 & 0
    \end{smallmatrix}\right]
    (\boldsymbol{k}) 
    + \dfrac{1}{2}\, \Gamma_{13}\,.
\end{equation}
We show the linked nodal structure in Fig.\;\ref{fig_4B_imbalanced_NL}(c), where three adjacent nodal rings are each linked one time with both one $I$-th and one $II$-th nodal braid. Here again, the $II$-th braids (blue) tend to follow the the $I$-th braids (green), in order to satisfy the consistency of their non-Abelian charges. However, only one $II$-th braid is stable, implying that the two others can be removed upon the pairing of two adjacent nodal rings. The linking numbers are $\boldsymbol{\mathsf{Ln}}=(3,1)$.

\subsection{$\Delta[\boldsymbol{\chi}] = [3,-1] - [3,1]$}

We take 
\begin{equation}
        H^{[\boldsymbol{\chi}^{0}]} (\boldsymbol{k}) = H^{[3,1]}_{4B}(\boldsymbol{k})\,,~
        H^{[\boldsymbol{\chi}^{\pi}]} (\boldsymbol{k}) = H^{[3,-1]}_{4B}(\boldsymbol{k})\,,
\end{equation}
with $H^{[3,1]}_{4B}(\boldsymbol{k})$ the same as above, and with $H^{[3,-1]}_{4B}(\boldsymbol{k})$ given by Eq.\;(\ref{eq_min_H_E31m}). We show the linked nodal structure in Fig.\;\ref{fig_4B_imbalanced_NL}(d), where one adjacent nodal ring is lined with two $II$-th nodal braids. This is compatible with the linking numbers $\boldsymbol{\mathsf{Ln}}=(0,2)$.

\section{2D Chern phases and 3D chiral phases}\label{sec_chiral}

By breaking $C_2T$ (or by breaking $PT$ and taking $0<\vert k_z\vert<\pi$, in the 3D context), the stable nodes of the Euler phases may be gapped. When starting from a phase that hosts $C_{2z}T$ symmetry only, the breaking of this symmetry readily converts the Euler topology to the Chern topology. In this section, we first want to identify the minimal terms $H_{\text{ch}}(\boldsymbol{k})$ that bring every nontrivial Euler phase discussed in Section \ref{sec_min_models} to a nontrivial Chern phase upon the breaking of $C_{2z}T$ perturbatively, \ie taking
\begin{equation}
\label{eq_c2t_breaking}
    H^{c_1}(\boldsymbol{k}) = H^{[\chi]}(\boldsymbol{k}) +\lambda_{\text{ch}} H_{\text{ch}}(\boldsymbol{k})
\end{equation}
with $0<\vert \lambda_{\text{ch}}\vert$, and such that the Chern number is directly determined by the Euler class of the symmetric model. More precisely, the gaping of the $\nu$-th connected two-band subspace with an Euler class $\chi_{\nu}$ gives rise to two separated bands, each carrying a finite Chern number $c_{1,(\nu,1)} = -c_{1,(\nu,2)}$ with $\vert c_{1,\nu}\vert = \vert \chi_{\nu}\vert$, where the sign of the Chern number of each band is determined by the sign of $\lambda_{\text{ch}}$. In the three-band case we find $\vert c_1 \vert = \vert\chi_I \vert$. In the four-band case, we find $\vert c_{1,I}\vert =\vert c_{1,II}\vert = \vert \chi_{I}\vert = \vert \chi_{II}\vert$ for the balanced Euler phases, while there is more freedom for the imbalanced phases with $\vert c_{1,I(II)} \vert \in \{\vert \chi_{I}\vert ,\vert \chi_{II}\vert\}$. We have listed in Table \ref{tab_2} all the terms that gap the connected two-band subspaces of every Euler model of Section \ref{sec_min_models}, distinguishing those that lead to the nontrivial Chern phases. Importantly, the results of Table \ref{tab_2} are conditioned by the additional symmetries carried by the models of Section \ref{sec_min_models}. Below, we discuss separately and in more details each three-band and four-band phase, together with the effect of their additional symmetries on the Chern phases. 

From the identification of the symmetry-breaking terms $\lambda_{\text{ch}} H_{\text{ch}}(\boldsymbol{k})$ leading to the nontrivial Chern phases, we can then formulate a systematic route for the building of 3D chiral models from pairs of 2D Euler phases. Our choice is to take 
\begin{multline}
\label{eq_3D_chiral_ansatz}
    H^{\Delta c_1}_{\text{Weyl}} (\boldsymbol{k}_{\parallel},k_z) = H^{\Delta[\chi]}_{PT}(\boldsymbol{k}_{\parallel},k_z) + \\
    \lambda_{\text{ch}} \sin k_z 
    \left(\dfrac{1+\cos k_z}{2} H^{0}_{\text{ch}}(\boldsymbol{k}_    {\parallel})    
    +\dfrac{1-\cos k_z}{2} H^{\pi}_{\text{ch}}(\boldsymbol{k}_{\parallel}) 
    \right)\,, 
\end{multline}
with $H^{\Delta[\chi]}_{PT}(\boldsymbol{k}_{\parallel},k_z)$ defined in section \ref{sec_3D_NL} and given explicitly for all pairs of the Euler phases in Section \ref{sec_3D_NL_numerics}, and where $\{H^{k_z}_{\text{ch}}(\boldsymbol{k}_{\parallel})\}_{k_z=0,\pi}$ are minimal terms of Table \ref{tab_2}. These 3D chiral phases are characterized by two $C_{2}T$ planes, at $k_z=0$ and $k_z=\pi$, on which the Euler topology is preserved and where the associated stable nodal points are pinned. Then, for $0<\vert k_z\vert <\pi$, $C_2T$ symmetry is broken ($PT$ symmetry is broken all together) and the nodes of the $C_2T$-symmetric planes become gapped. In other words, the nodes pinned at $k_z=0,\pi$ by $C_{2}T$ constitute Weyl points in the 3D Brillouin zone. We show below, for the three-band and the four-band systems, that the transition between two inequivalent Euler phases, from $k_z=0$ to $k_z=\pi$, is mediated by the presence of simple Weyl points between the two symmetric planes (\ie $0< \vert k_z\vert <\pi$), the number of which matches the difference in the Chern numbers of the 2D planes at $\vert k_z\vert = \varepsilon$ and at $\vert k_z\vert=\pi- \varepsilon$, for a small deviation $\varepsilon $. 

In the following, first for the three-band case, then for the four-band case, we start with a discussion of the terms that give 2D Chern phases upon breaking $C_2T$ symmetry, and we show one example of embedding of a pair of 2D Euler phases into one 3D chiral phase. 


\begin{table}[t!]
\begin{equation*}
\arraycolsep=2.4pt\def\arraystretch{1.4}
\begin{array}{ l | c | c }
		 & \vert c_{1,I(II)}\vert =0 & \vert c_{1,I(II)}\vert \in \{\vert\chi_{I}\vert,\vert\chi_{II}\vert\}    \\
	\hline
	\hline
	{[}\chi_I{]}={[}2{]} & \Lambda_7 & \sin k_2\Lambda_5 \\
	{[}\chi_I{]}={[}4{]} & \Lambda_2 & \sin k_2\Lambda_7 \\
	\hline
	[\boldsymbol{\chi}]=[1,1] & \Gamma_{21},\Gamma_{32} &  \Gamma_{20} \\
	\hline
	[\boldsymbol{\chi}]=[1,-1] & \Gamma_{21},\Gamma_{32} &  \Gamma_{23} \\
	\hline
	[\boldsymbol{\chi}]=[2,2] & \Gamma_{21},\Gamma_{32} &  \Gamma_{20} \\
	\hline
	[\boldsymbol{\chi}]=[2,-2] & \Gamma_{21},\Gamma_{32} &  \Gamma_{23} \\
	\hline
	[\boldsymbol{\chi}]=[2,0] & \Gamma_{23},\Gamma_{32} & \sin k_2 \{\Gamma_{23},\Gamma_{32}\} \\
	\hline
	[\boldsymbol{\chi}]=[4,0] & \Gamma_{21}, \Gamma_{32} &  \sin k_1 \Gamma_{21}+\sin k_2 \Gamma_{32} \\
	\hline
	[\boldsymbol{\chi}]=[3,1] & \Gamma_{21},\Gamma_{32} & 
	\sin k_1 \Gamma_{21}+\sin k_2 \Gamma_{32}\\
	\hline
	[\boldsymbol{\chi}]=[3,-1] & \Gamma_{21},\Gamma_{32} & 
		\sin k_1 \Gamma_{21}+\sin k_2 \Gamma_{32}
\end{array}
\end{equation*}
\caption{List of the $C_2T$-breaking terms $H_{\text{ch}}(\boldsymbol{k})$ that gap perturbatively (\ie for $0<\vert \lambda_{\text{ch}}\vert$ in Eq.\;(\ref{eq_c2t_breaking})) the connected two-band subspaces of the Euler phases of Section \ref{sec_min_models}. The second column lists the terms that gap the two-band subspaces and convert them into nontrivial Chern bands.}
\label{tab_2}
\end{table}

\subsection{Three-band Chern and chiral phases}

Since the three-band models span all the Gell-Mann matrices, see Section \ref{sec_min_3B}, there is no extra symmetry beyond $C_{2z}T$ (with the caveat of an accidental symmetry discussed below). Given the representation of $C_{2z}T$ in the Bloch orbital basis, \ie $U_{C_{2z}T}\mathcal{K} = \mathbb{1}_3 \mathcal{K} $, $C_{2z}T$ is broken whenever we add a term with a complex Gell-Mann matrix, \ie 
\begin{equation}
\begin{array}{l}
    \begin{array}{lll}
        \Lambda_2 \,=\, 
        \begin{rsmallmatrix}
            0 & -\imi & 0 \\ \imi & 0 & 0 \\ 0 & 0 & 0
        \end{rsmallmatrix},&
            \Lambda_5 \,=\, \begin{rsmallmatrix}
            0 & 0 & -\imi \\ 0 & 0 & 0 \\ \imi & 0 & 0
            \end{rsmallmatrix},&
            \Lambda_7 \,=\, \begin{rsmallmatrix}
            0 & 0 & 0 \\ 0 & 0 & -\imi \\ 0 & \imi & 0
            \end{rsmallmatrix}.
        \end{array}
    \end{array}
\end{equation}
We list in Table \ref{tab_2} the minimal terms that open a gap perturbatively (\ie with a small prefactor $\vert\lambda_{\text{ch}}\vert = \varepsilon > 0$), distinguishing the terms that lead to nontrivial Chern phase (third column). We note that the models $H^{[2]}_{\text{3B}}(\boldsymbol{k})$ and $H^{[4]}_{\text{3B}}(\boldsymbol{k})$, defined in Section \ref{sec_min_3B}, do possess an accidental symmetry upon adding the term $\lambda_{\text{ch}} \Lambda_2$, and $\lambda_{\text{ch}} \Lambda_5$, respectively, which maintains the nodes over a finite range of $\lambda_{\text{ch}}$. This explains why these terms do not appear in Table \ref{tab_2}.

\begin{figure}[t!]
\centering
\begin{tabular}{ll}
    (a) & (b) \\
	\includegraphics[width=0.5\linewidth]{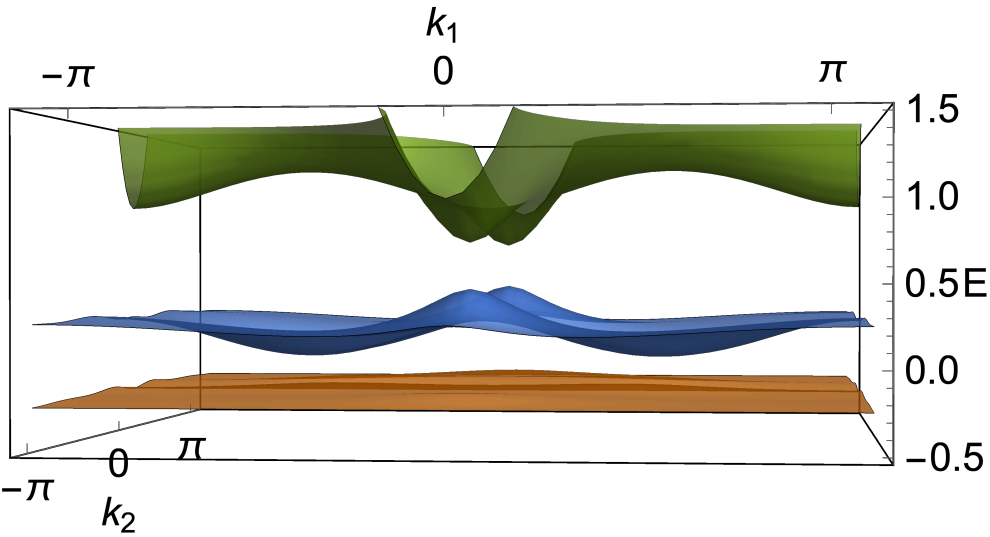} &
	\includegraphics[width=0.5\linewidth]{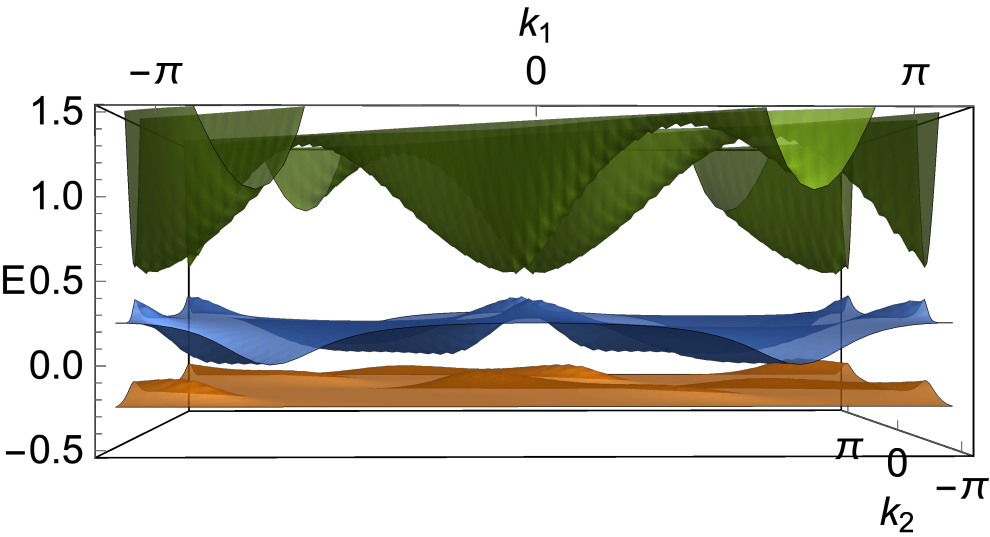}    \\
	(c) & (d) \\
	\includegraphics[width=0.5\linewidth]{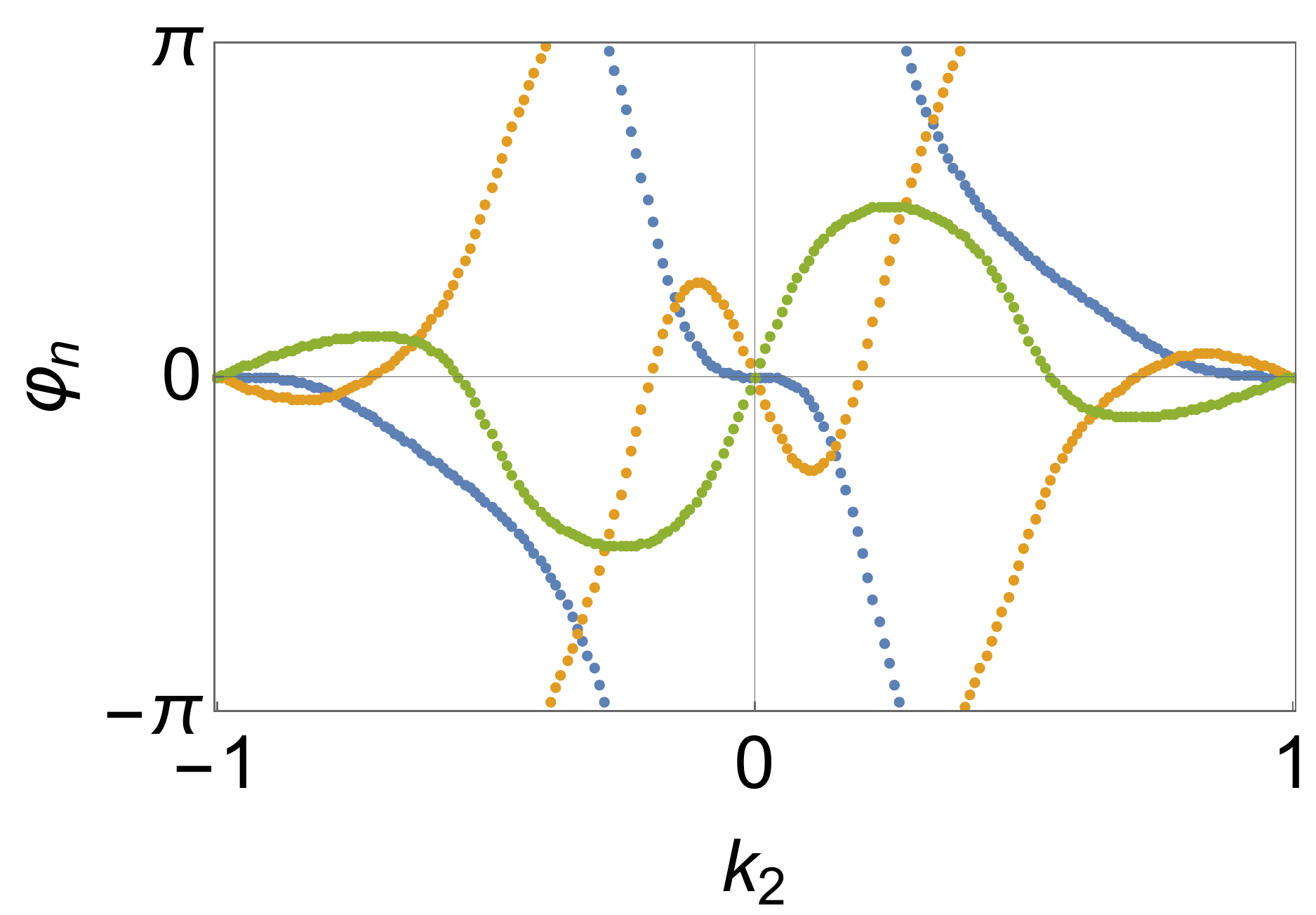}  &
	\includegraphics[width=0.5\linewidth]{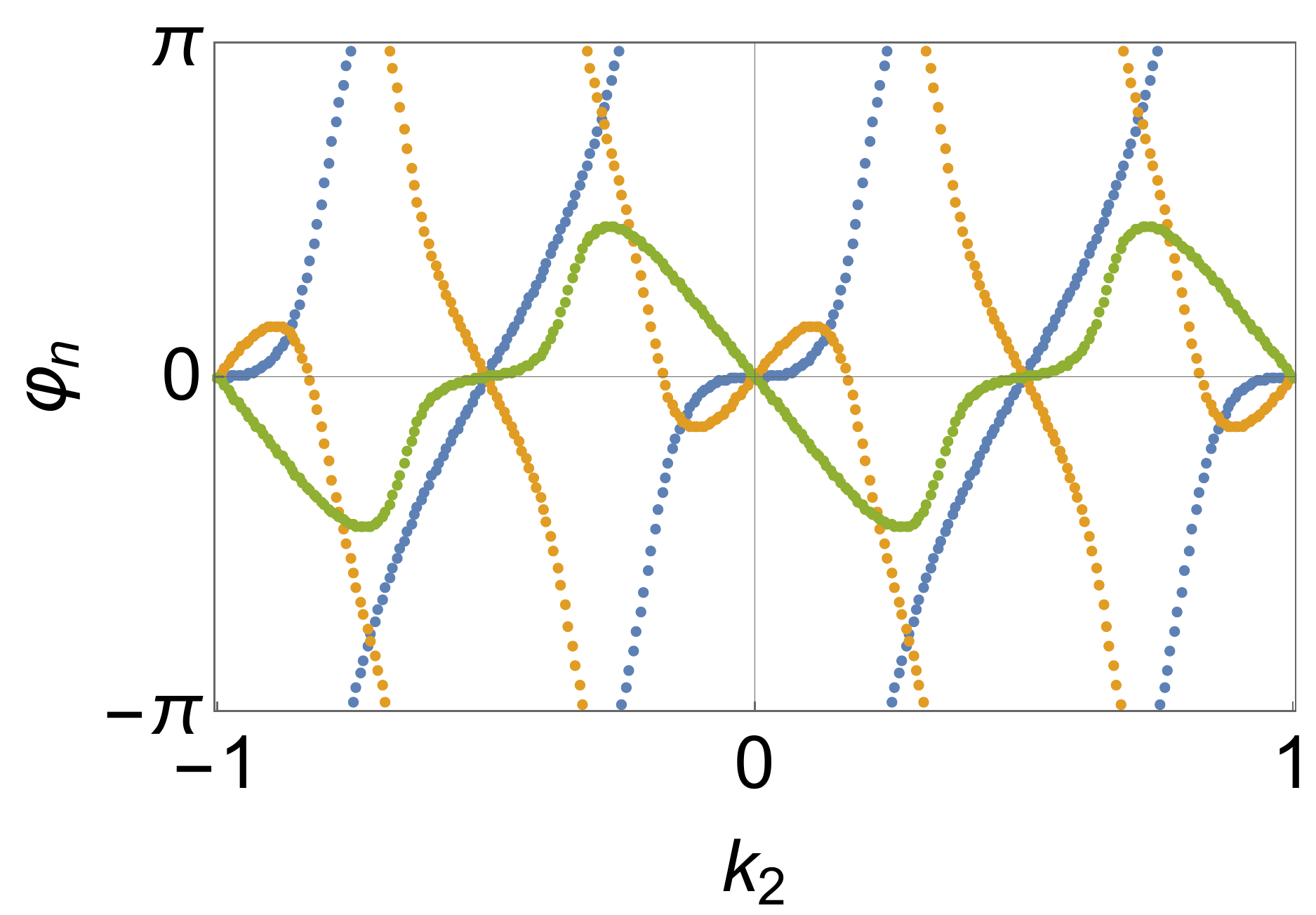} 
\end{tabular}
\caption{\label{fig_3B_chern} (a,b) Gaped band structure and (c,d) flow of Berry phase per band of the Chern phases obtained upon breaking $C_{2}T$ symmetry in the three-band phases. (a,c) Chern phase descending from the Euler phase $\vert\chi_I\vert=2$, with the Chern numbers $ c_{1,(I,1)} = -c_{1,(I,2)} = \vert\chi_I\vert=2$, and $c_{1,II}=0$. (b,d) Chern phase descending from the Euler phase $\vert\chi_I\vert=4$, with the Chern numbers $ c_{1,(I,1)} = -c_{1,(I,2)} =- \vert\chi_I\vert=-4$, and $c_{1,II}=0$.
}
\end{figure}

In Figure \ref{fig_3B_chern} we show the band structure and the flow of Berry phase per band of the nontrivial Chern phases obtained from the Euler phases (a,c) $[\chi_I]=[2]$, and (b,d) $[\chi_I]=[4]$, after adding the symmetry-breaking term of Table \ref{tab_2} (third column). We find that the minimal perturbative breaking of symmetry gives $\vert c_{1,I} \vert = \vert \chi_I \vert$ in both cases. 

\begin{figure}[t!]
\centering
\begin{tabular}{ll}
    (a)~$[\chi_I^0]=[2]$,$[\chi_I^{\pi}]=[0]$ & (b)~$[\boldsymbol{\chi}^0]=[1,1]$,$[\boldsymbol{\chi}^{\pi}]=[0,0]$ \\
	\includegraphics[width=0.5\linewidth]{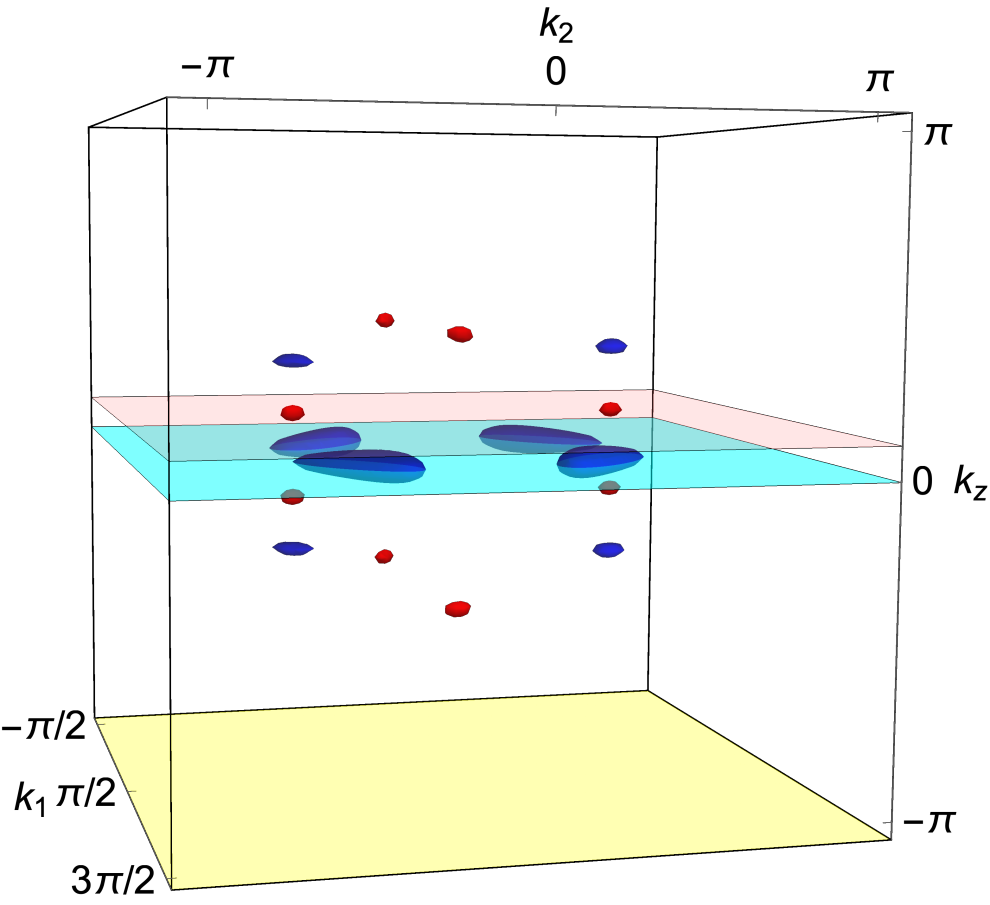} &
	\includegraphics[width=0.5\linewidth]{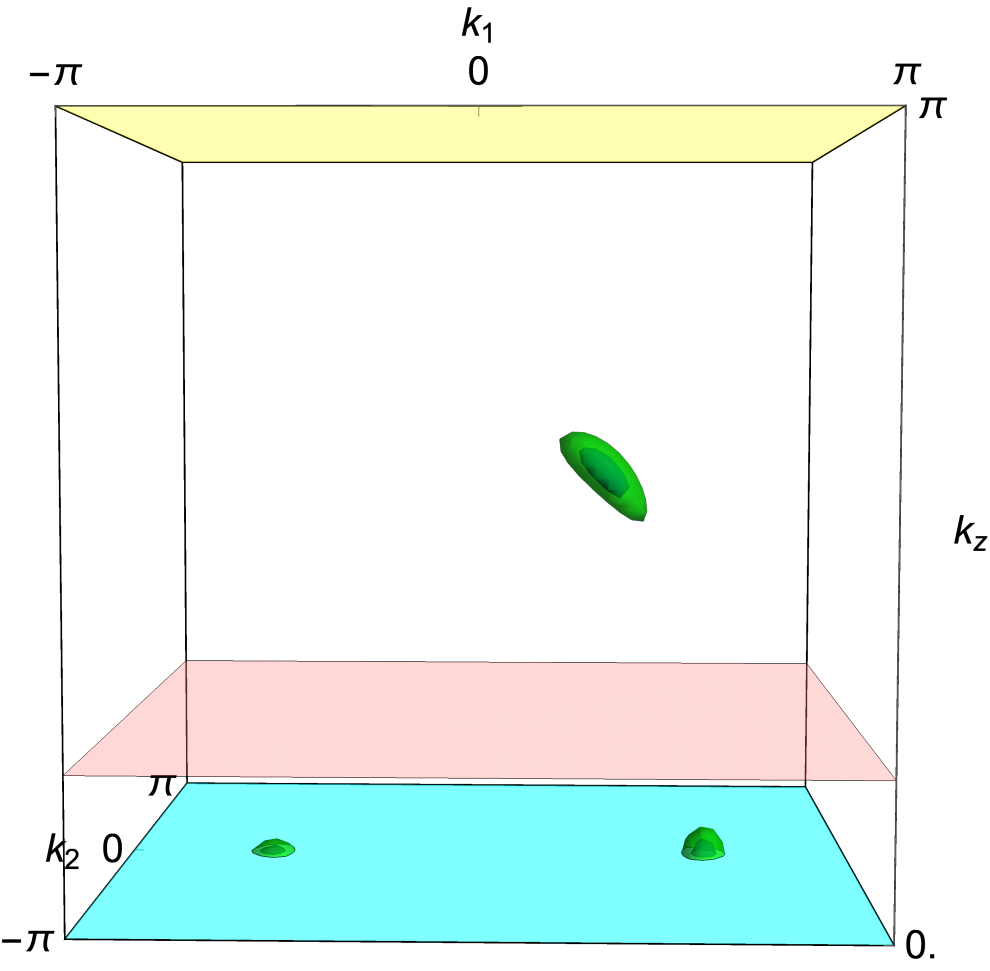} \\
	(c) & (d) \\
	\includegraphics[width=0.5\linewidth]{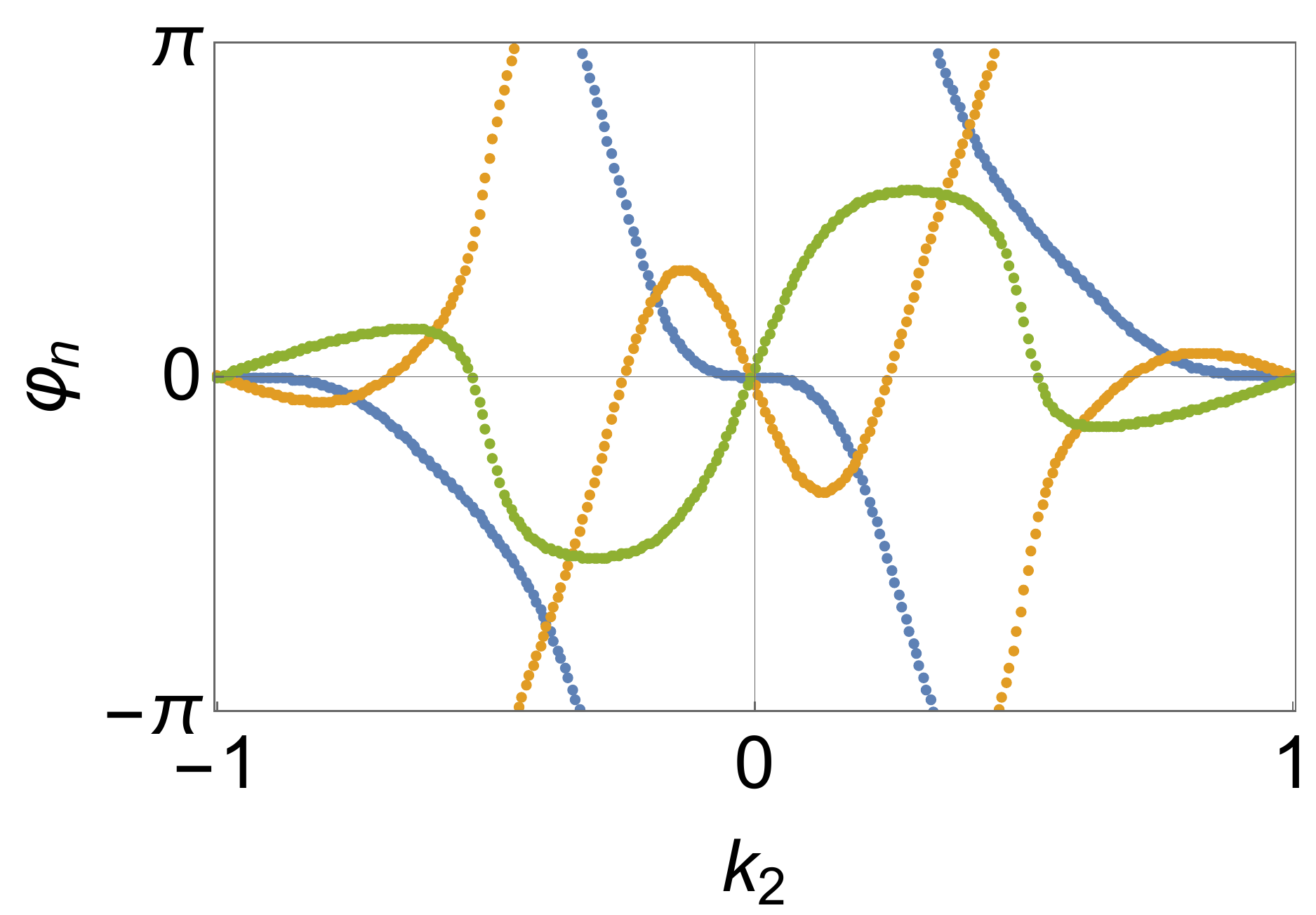} &
	\includegraphics[width=0.5\linewidth]{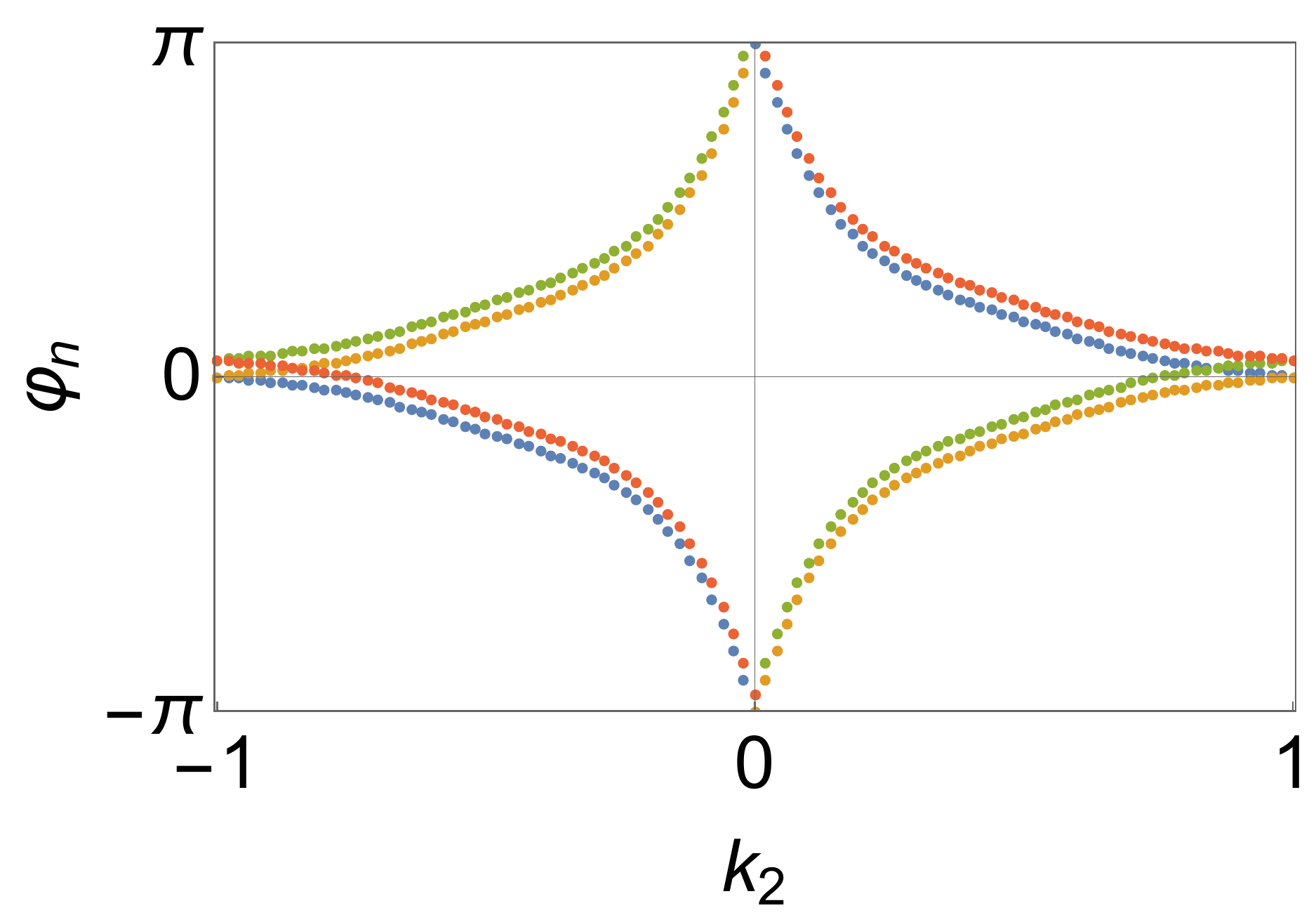}
\end{tabular}
\caption{\label{fig_3D_chiral} 3D chiral phases obtained from the embedding of a pair of Euler phases, at $k_z=0$ (cyan plane) and $\vert k_z\vert=\pi$ (yellow plane), obtained by breaking $C_{2T}$ symmetry. (a,c) Three-band example for the transition $\Delta[\chi_I] = [\chi^{\pi}_I] - [\chi^0_I] = [0] - [2]$. (b,d) Four-band example for the transition $\Delta[\boldsymbol{\chi}] = [0,0] - [1,1]$. The panels (a,b) show the Weyl points inherited from the Euler phases (blue, green) on the cyan planes, and the new Weyl points between the two Euler phases (\ie at $0<k_z<\pi$) that mediate the phase transition between the two different inherited Chern phases at $k_z=\varepsilon$ and $k_z=\varepsilon$. (c,d) Flow of Berry phase per band computed over the pink plane, \ie at $k_z=\varepsilon$, indicating a Chern phase with $\vert c_{1,I}\vert = 2$ in (c), and $\vert c_{1,I}\vert = \vert c_{1,II}\vert =  1$ in (d). Panel (a) contains adjacent Weyl points, \ie in the $(I,II)$-gap, which are accidental (red), in the sense that there can be removed without closing the energy gap of the Euler phases at $k_z=0$ and $k_z=\pi$.
}
\end{figure}

We now show one example of embedding of two Euler phases within one 3D chiral phase. Starting from the first example of linked nodal structure in Section \ref{sec_PT_3B}, \ie with $H^{[\chi_I^0]} = H_{\text{3B}}^{[2]}(\boldsymbol{k}_{\parallel})$ and $H^{[\chi_I^{\pi}]} = H_{\text{3B}}^{[0]}(\boldsymbol{k}_{\parallel})$, we then substitute the following terms in the ansatz Eq.\;(\ref{eq_3D_chiral_ansatz})
\begin{equation}
    H^0_{\text{ch}}(\boldsymbol{k}_{\parallel}) = \sin k_2 \Lambda_5 \,,~ 
    H^{\pi}_{\text{ch}}(\boldsymbol{k}_{\parallel}) = \Lambda_5 \,,
\end{equation}
and take $\lambda_{\text{ch}}=1$. We plot in Figure \ref{fig_3D_chiral}(a) the Weyl points of the 3D chiral phase inherited from the transition between the Euler phase $\chi^0_I = 2$ at $k_z=0$ (cyan plane), and the Euler trivial phase $\chi_I=0$ at $k_z=\pi$ (yellow plane). The blue dots at $k_z=0$ indicate the four $I$-th Weyl points corresponding to the four stable nodal points of the Euler phase $[\chi^0_I] = [2]$. The pink plane corresponds to a section at a small $k_z$ above zero, over which we have computed the flow of Berry phase shown in Figure \ref{fig_3D_chiral}(c), indicating a Chern phase with $\vert c_{1,I}\vert =2$. This phase actually directly corresponds to the 2D Chern phase obtained by breaking $C_2T$ discussed above. In Figure \ref{fig_3D_chiral}, beyond the presence of accidental adjacent Weyl points (\ie in the $(I,II)$-gap, eight nodes colored in red), there are two additional $I$-th Weyl points (blue) located between the two Euler planes (\ie for $0<k_z<\pi$). These Weyl points are required to make the transiton from the nontrivial Chern phase at $k_z = \varepsilon >0$ and the trivial Chern phase at $k_z = \pi - \varepsilon$.

\subsection{Four-band Chern and chiral phases}

Given the representation of the $C_{2z}T$ symmetry ($U_{C_{2z}T}\mathcal{K}=\mathbb{1}_4 \mathcal{K}$) for the models of the four-band Euler phases written in the real gauge, this symmetry is broken by adding a term with any of the complex Dirac matrices, \ie among $\{\Gamma_{2j},\Gamma_{j2}\}_{j=0,1,3}$. However, some care must be taken when considering the Chern phases descending from the four-band Euler phases introduced in Section \ref{sec_4B_models} upon breaking the $C_2T$ symmetry. Indeed, as we have analyzed in detail in Section \ref{sec_sym}, these models carry additional symmetries due to their simplicity. It tuns out that several of these symmetries interact with the stability of the nodal points, and the cancellation of Chern numbers. As in Section \ref{sec_sym}, we take $\boldsymbol{k}=(k_1,k_2) = (k_x,k_y)$ in the following, \ie we assume a rectangular lattice. 

On one hand, the vertical mirror symmetries $\{m_y,m_x\}$ (or equivalently the horizontal rotational symmetries $\{C_{2y},C_{2x\}}$) and the chiral symmetries $\{Sm_y,Sm_x\}$ interact with the stability of the nodal points of the Euler phases. Indeed, the vertical mirror symmetry $m_i:k_i\rightarrow -k_i$, for $i=x,y$, protects the band crossings on the $(k_i=0)$-plane happening between two Bloch eigenstates with distinct mirror eigenvalues (\ie belonging to distinct irreducible representations of $m_i$). Then, the chiral symmetry $Sm_i$, for $i=x,y$, protects the band crossings on the $(k_i=0)$-plane between any pair of bands whenever their chiral winding number is finite (or, equivalently, when they carry a $\pi$-Berry phase \cite{Bouhon_HHL}). As a consequence, the gaping of the connected two-band subspaces of the Euler phases also requires the breaking of the mirror and chiral symmetries, on top of breaking $C_{2}T$. This explains why several complex Dirac matrices are absent from Table \ref{tab_2}.  

On the other hand, the vertical mirror symmetries $\{m_y,m_x\}$ and time reversal symmetry $T$ enforce the Chern number to vanish. As a consequence, the mirror and time reversal symmetries must be broken, together with the breaking of $C_{2}T$, to generate a nontrivial Chern phase. This explains (partially \footnote{There actually remain some accidental symmetries that explains more completely the structure of Table \ref{tab_2}, especially for the imbalanced Euler phases. We call them ``accidental'' because these symmetries are due to the very specific form of the models in Section \ref{sec_min_models}. Since removing these symmetries amounts to depart from the simplicity of the models, which was our primary aim, we do not address these further here, and will give a more detailed treatment elsewhere.}) the entries of the last column in Table \ref{tab_2}. 

\begin{figure*}[t!]
\centering
\begin{tabular}{llll}
    (a) ~$[\boldsymbol{\chi}]=[1,1]$ & (b) ~$[\boldsymbol{\chi}]=[1,-1]$ &
    (c)~$[\boldsymbol{\chi}]=[2,2]$ & (d)~$[\boldsymbol{\chi}]=[2,-2]$  \\
	\includegraphics[width=0.25\linewidth]{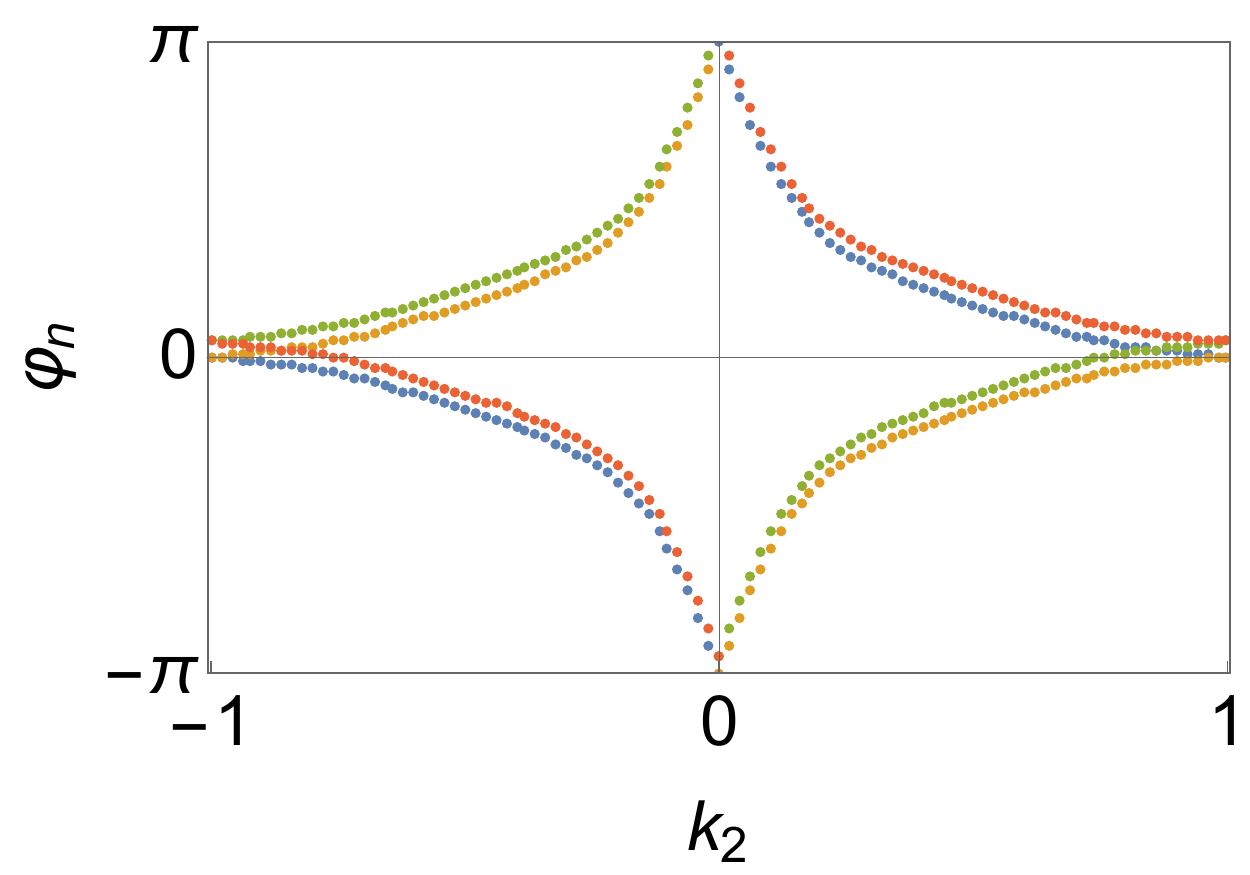} &
	\includegraphics[width=0.25\linewidth]{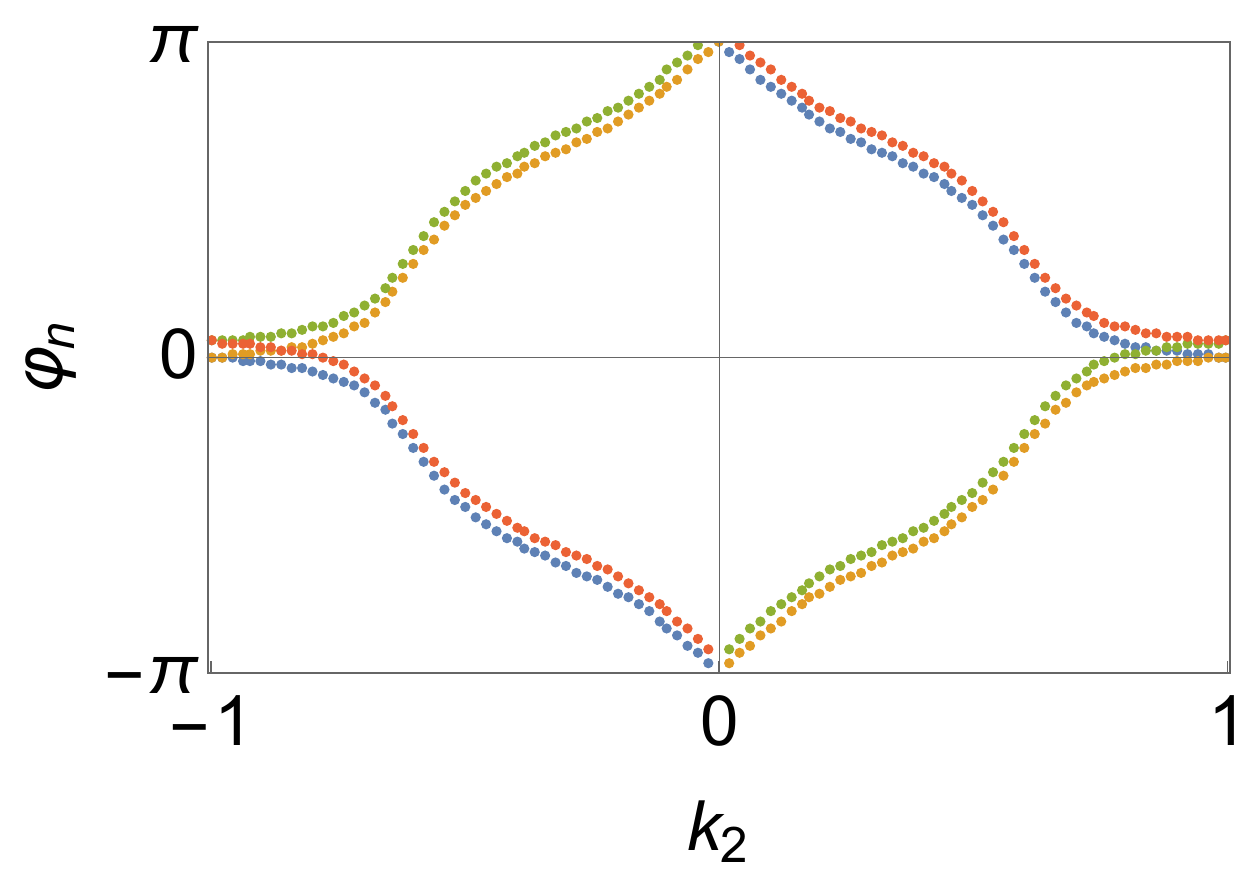}  &
	\includegraphics[width=0.25\linewidth]{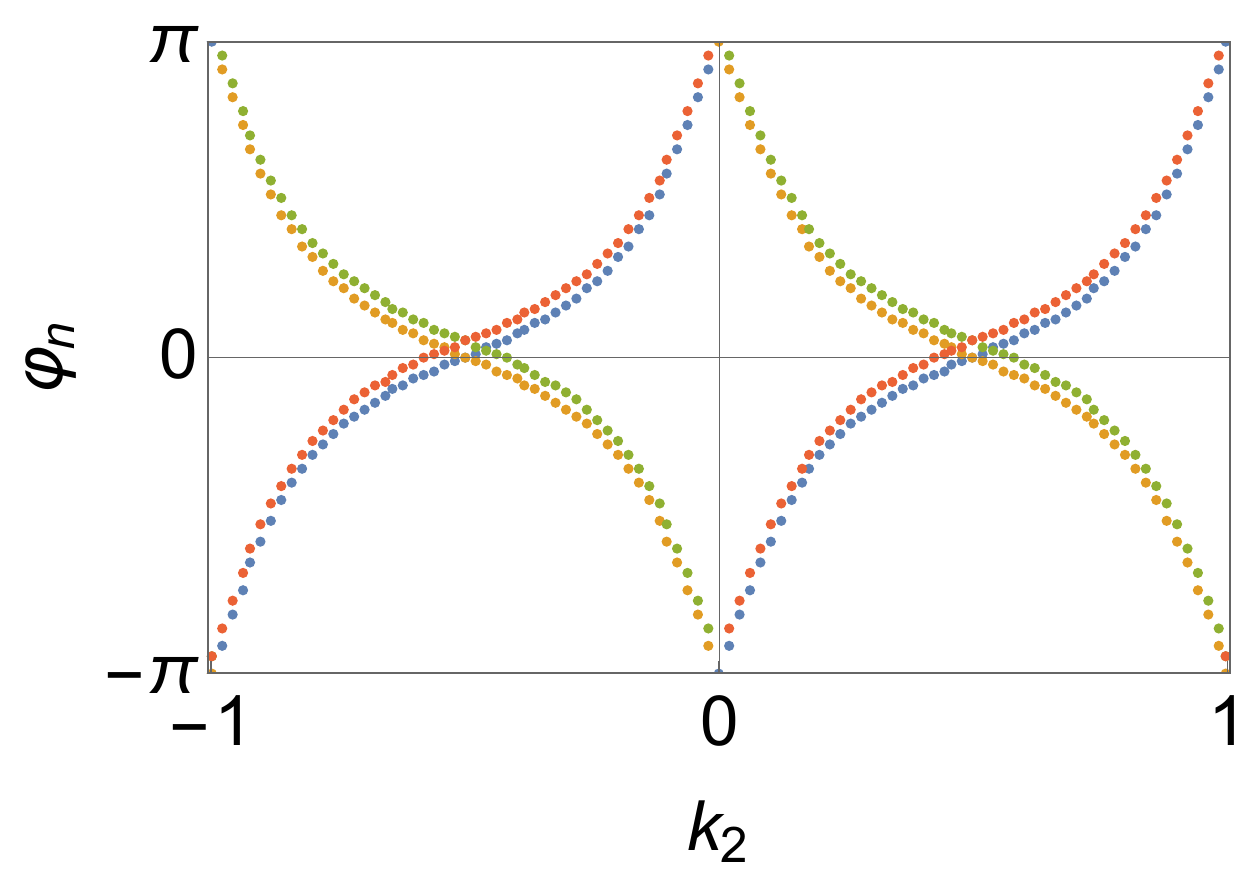}  &
	\includegraphics[width=0.25\linewidth]{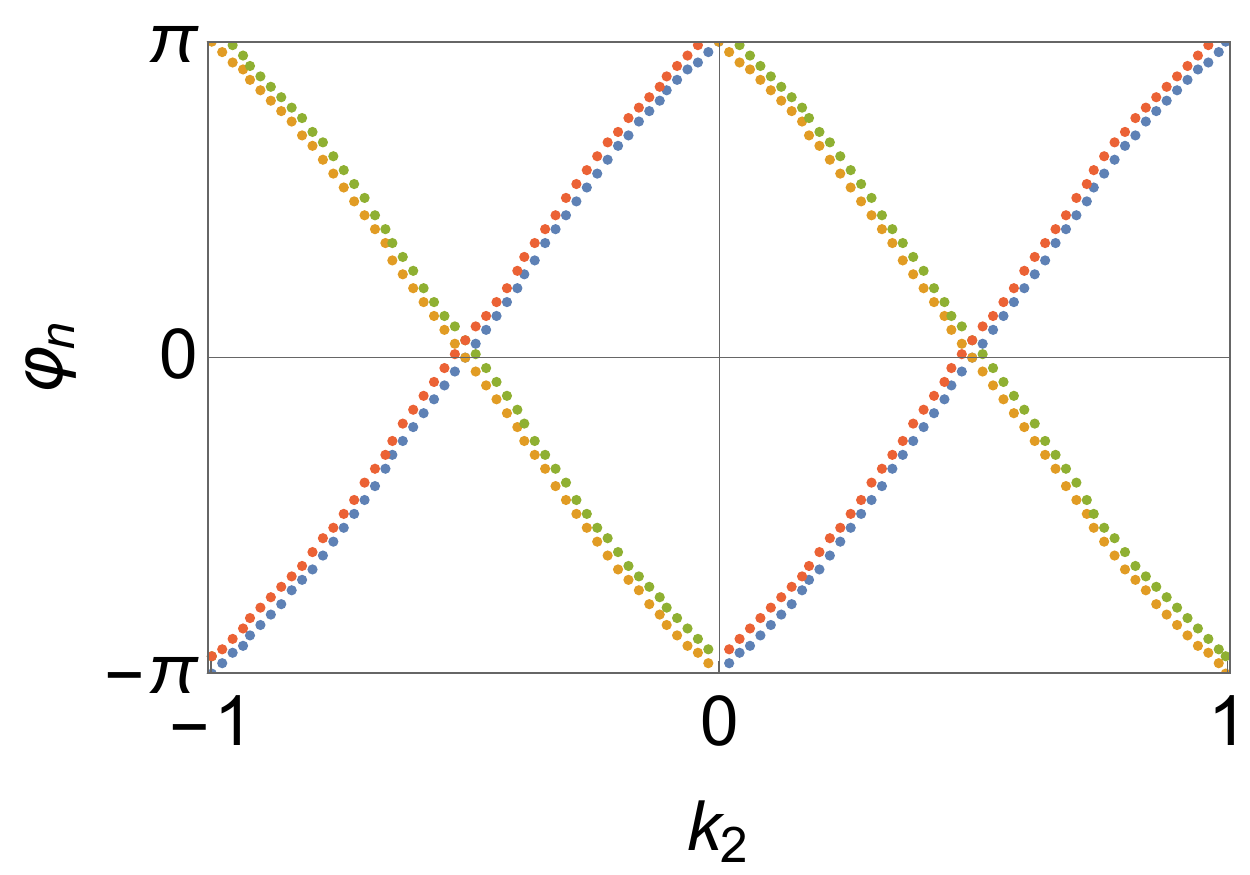}\\
	(e)~$[\boldsymbol{\chi}]=[2,0]$ & (f) ~$[\boldsymbol{\chi}]=[4,0]$ &
	(g)~$[\boldsymbol{\chi}]=[3,1]$ & (h) ~$[\boldsymbol{\chi}]=[3,-1]$ \\
	\includegraphics[width=0.25\linewidth]{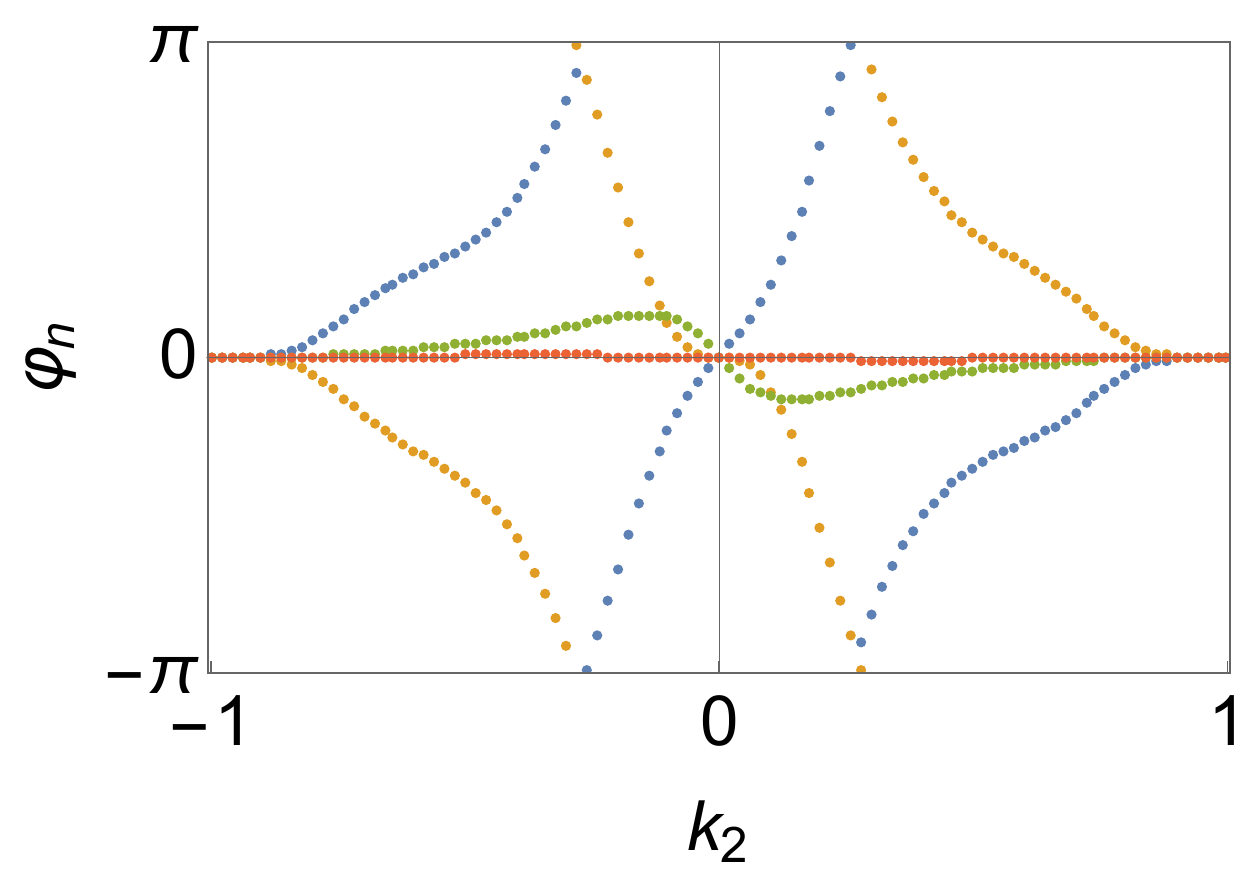}  &
	\includegraphics[width=0.25\linewidth]{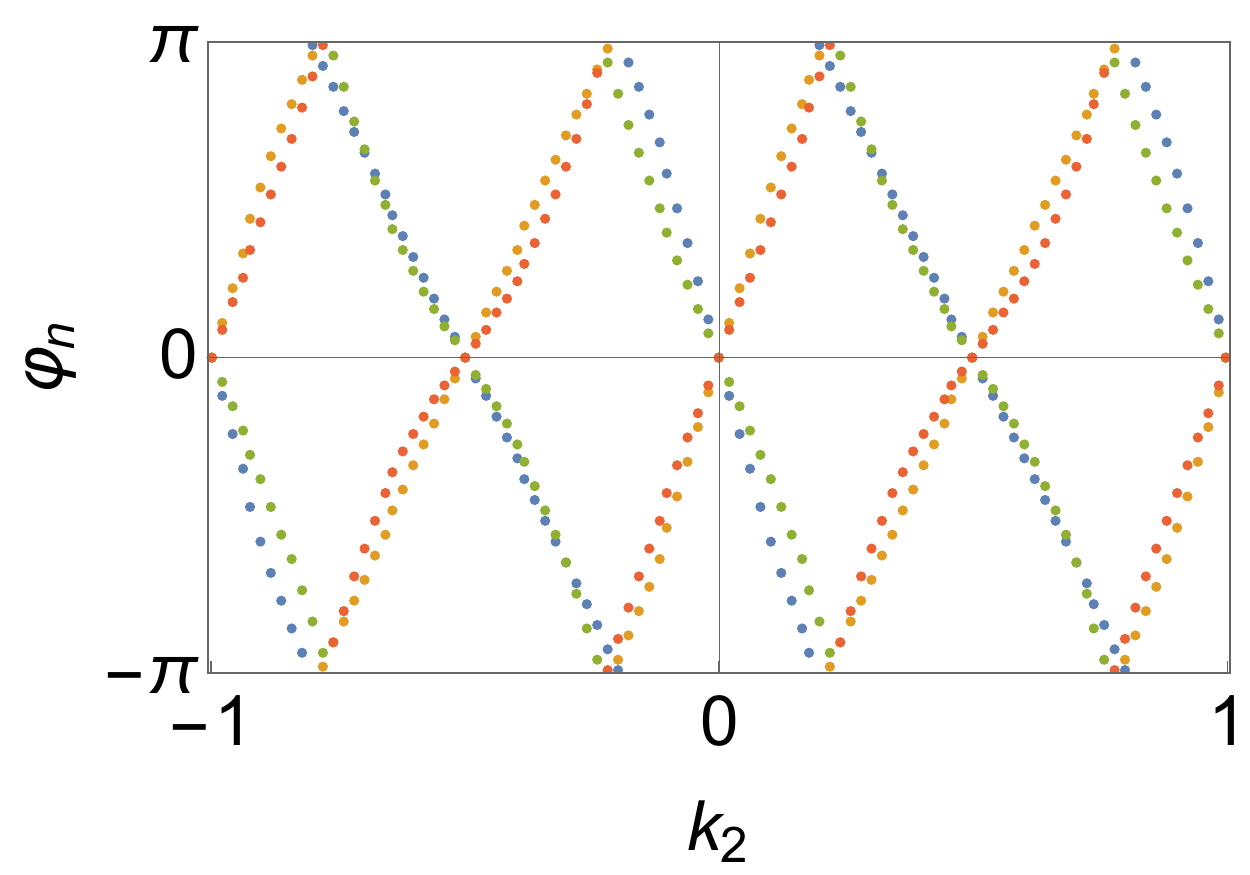} &
	\includegraphics[width=0.25\linewidth]{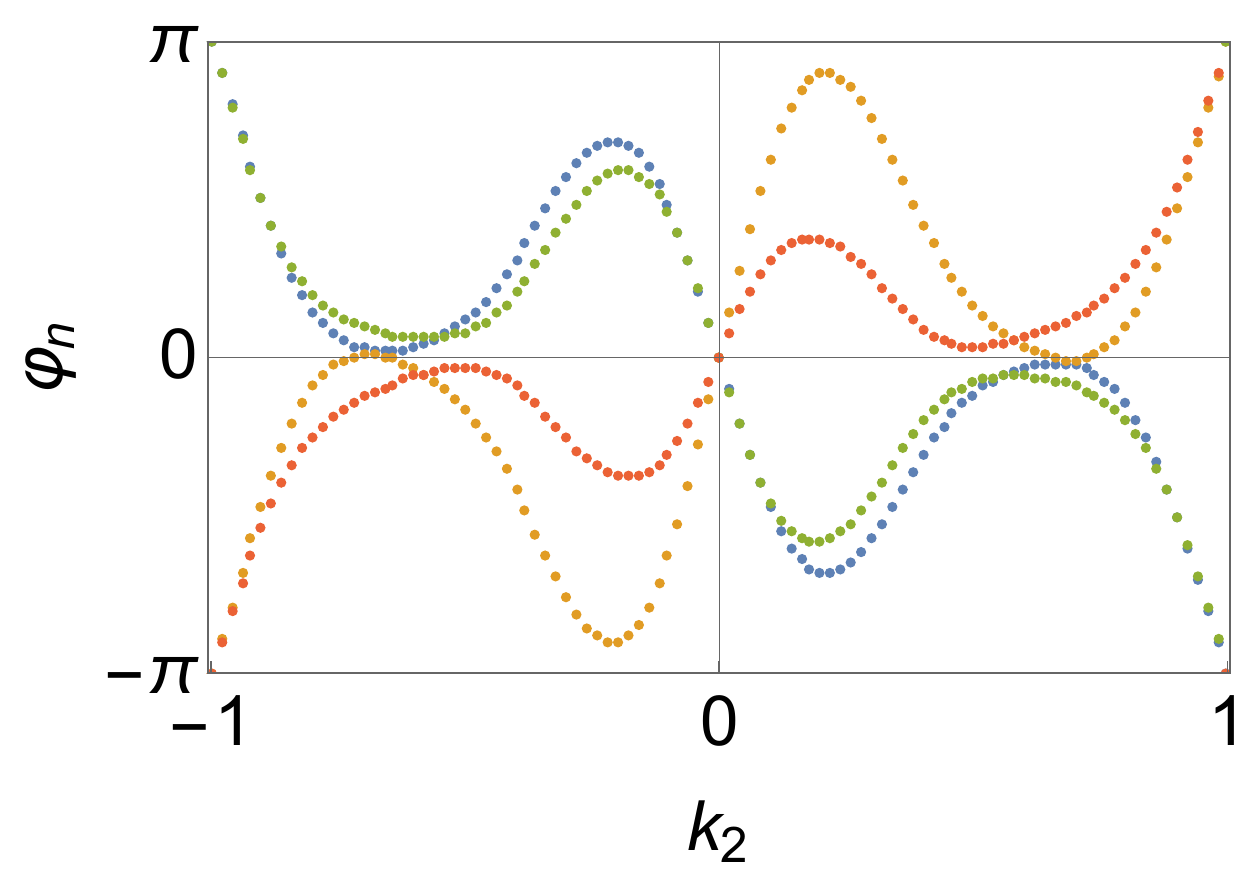}  &
	\includegraphics[width=0.25\linewidth]{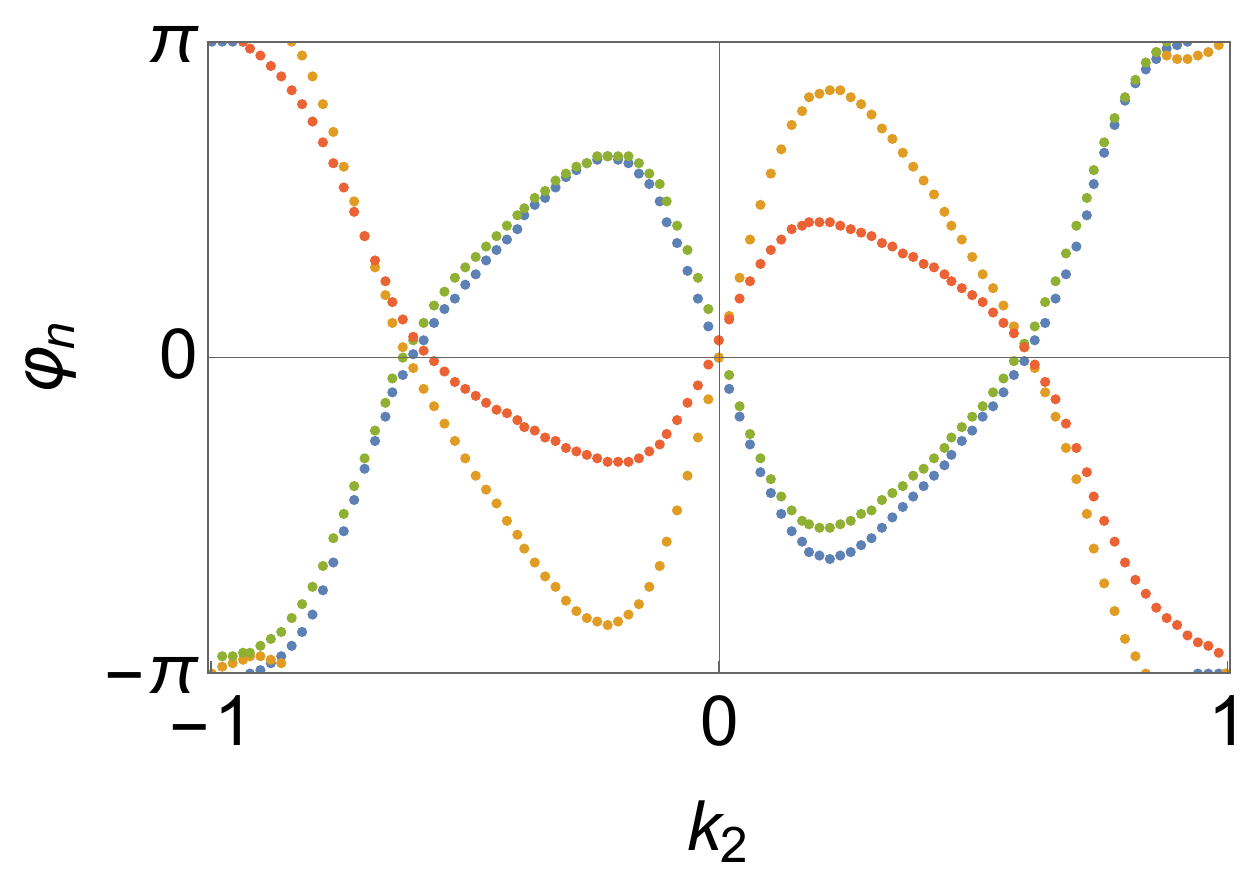}
\end{tabular}
\caption{\label{fig_4B_chern} Flow of Berry phase per band of the gapped Chern phases obtained from the models of Section \ref{sec_4B_models}, upon breaking the $C_{2z}T$ symmetry, as well as the additional mirror, chiral, and time reversal symmetries of Section \ref{sec_sym}, obtained by adding the terms of Table \ref{tab_2}. The color code of orange, blue, green, red follows an ordering of the bands with increasing energies.  
}
\end{figure*}

We show in Figure \ref{fig_4B_chern} the flows of Berry phase per band for the Chern phases generated from all the models of Section \ref{sec_4B_models} by adding the terms of Table \ref{tab_2}. We verify that all balanced Euler phases generate Chern bands with $\vert c_{1,\nu} \vert = \vert \chi_{\nu}\vert$ for both subspaces $\nu=I,II$. In the case of the imbalanced Euler phases, we get $\vert c_{1,\nu} \vert \in \{ \vert \chi_{I}\vert,\vert \chi_{II}\vert  \}$ for $\nu=I,II$. We also find that there is no qualitative difference between one phase $[\chi_I,\chi_{II}]$ and its partner $[\chi_I,-\chi_{II}]$.

We end this section with an example of 3D chiral phase obtained from the embedding of two distinct four-band Euler phases. For this we chose $H^{[\boldsymbol{\chi}^0]} = H^{[1,1]}_{\text{4B}}(\boldsymbol{k}_{\parallel})$ and $H^{[\boldsymbol{\chi}^{\pi}]} = H^{[1,1]}_{\text{4B}}(\boldsymbol{k}_{\parallel})$, which we take the same as in the first four-band linked nodal structure of Section \ref{sec_PT_4B}, and we substitute the following terms in Eq.\;(\ref{eq_3D_chiral_ansatz}),
\begin{equation}
    H^0_{\text{ch}}(\boldsymbol{k}_{\parallel}) = 2\Gamma_{20}\,,~H^{\pi}_{\text{ch}}(\boldsymbol{k}_{\parallel}) = \dfrac{3}{2}\Gamma_{20}\,.
\end{equation}
We show the resulting Weyl phase in Figure \ref{fig_3D_chiral}(b,d). We find that two intermediary Weyl points, one in the gap of each two-band subspace $\nu=I,II$, are necessary for the transition between the Euler phases $[1,1]$ and $[0,0]$. The panel (d) shows the flow of Berry phase on the pink plane [drawn in (b)], indicating a Chern phase with $c_{1,\nu} =\pm 1$ for both subspaces $\nu=I,II$. The Weyl points at an intermediary position ($0<\vert k_z\vert \pi$), are then necessary within both the $I$-th and $II$-th two-band subspaces to annihilate the Chern numbers on the pink plane, and connect with the trivial phase at $k_z=\pi$ (yellow plane).

\section{Conclusions and discussion}\label{sec:conclusions}
We present a general modeling formulation encapsulating multi-gap topologies quantified by Euler class invariants. Utilizing previous, albeit slightly technically involved,  work~\cite{bouhon2020geometric} that addresses multi-gap topological parametrizations using homotopy perspectives, we here derive explicit models that can be readily used as a benchmark for experimental and theoretical pursuits. Recent interest on both these fronts, exemplified by trapped-ion experiments~\cite{zhao2022observation} that verified predicted multi-gap topological signatures~\cite{Unal_quenched_Euler} and an ever increasing interest in theoretical predictions and characterizations~\cite{Peng2021,chen2021manipulation, ezawa2021euler, Lange2022,Eulersc,guan2021landau}, suggests that these results may be anticipated to be of general interest as well as of use to further progress this nascent research field, for example by considering different additional symmetries or (non-Hermitian) extensions. We here take an illustrative first step in uncovering this rich panorama by discussing possible descendant Chern-valued phases upon including specific symmetry-breaking terms, which could for example also flourish in the context of magnetism~\cite{magnetic}.

These pursuits are moreover not only limited to the presented specific models {\it per se}. Indeed,  we uncover that inequivalent 2D Euler phases directly relate to $PT$-symmetric nodal structures in three spatial dimensions, when the effective interpolation parameter is interpreted as the extra dimension. As the transitions from one Euler phase to another are generically mediated by the presence of adjacent nodal rings linked with sub-gap nodal lines, forming the trajectories that correspond to the braiding or debraiding of nodal points, the stability of
these adjacent nodal rings then ties to specific monople charges that root in the Euler invariant. This represents another case in point in showing the mentioned potential of the presented models and the rather rich interplay with several theoretical concepts.

Given these results and their  potential for new directions, we anticipate that our results will contribute in the pursuit of fully harvesting of this upcoming field.

\section{Acknowledgements}
A.~B. has been partly funded by a Marie-Sklodowska-Curie fellowship, grant no. 101025315. R.~J.~S acknowledges funding from a New Investigator Award, EPSRC grant EP/W00187X/1, as well as Trinity college, Cambridge.

\bibliography{bib}{}

\begin{thebibliography}{79}%
\makeatletter
\providecommand \@ifxundefined [1]{%
 \@ifx{#1\undefined}
}%
\providecommand \@ifnum [1]{%
 \ifnum #1\expandafter \@firstoftwo
 \else \expandafter \@secondoftwo
 \fi
}%
\providecommand \@ifx [1]{%
 \ifx #1\expandafter \@firstoftwo
 \else \expandafter \@secondoftwo
 \fi
}%
\providecommand \natexlab [1]{#1}%
\providecommand \enquote  [1]{``#1''}%
\providecommand \bibnamefont  [1]{#1}%
\providecommand \bibfnamefont [1]{#1}%
\providecommand \citenamefont [1]{#1}%
\providecommand \href@noop [0]{\@secondoftwo}%
\providecommand \href [0]{\begingroup \@sanitize@url \@href}%
\providecommand \@href[1]{\@@startlink{#1}\@@href}%
\providecommand \@@href[1]{\endgroup#1\@@endlink}%
\providecommand \@sanitize@url [0]{\catcode `\\12\catcode `\$12\catcode
  `\&12\catcode `\#12\catcode `\^12\catcode `\_12\catcode `\%12\relax}%
\providecommand \@@startlink[1]{}%
\providecommand \@@endlink[0]{}%
\providecommand \url  [0]{\begingroup\@sanitize@url \@url }%
\providecommand \@url [1]{\endgroup\@href {#1}{\urlprefix }}%
\providecommand \urlprefix  [0]{URL }%
\providecommand \Eprint [0]{\href }%
\providecommand \doibase [0]{http://dx.doi.org/}%
\providecommand \selectlanguage [0]{\@gobble}%
\providecommand \bibinfo  [0]{\@secondoftwo}%
\providecommand \bibfield  [0]{\@secondoftwo}%
\providecommand \translation [1]{[#1]}%
\providecommand \BibitemOpen [0]{}%
\providecommand \bibitemStop [0]{}%
\providecommand \bibitemNoStop [0]{.\EOS\space}%
\providecommand \EOS [0]{\spacefactor3000\relax}%
\providecommand \BibitemShut  [1]{\csname bibitem#1\endcsname}%
\let\auto@bib@innerbib\@empty
\bibitem [{\citenamefont {Qi}\ and\ \citenamefont {Zhang}(2011)}]{Rmp1}%
  \BibitemOpen
  \bibfield  {author} {\bibinfo {author} {\bibfnamefont {Xiao-Liang}\
  \bibnamefont {Qi}}\ and\ \bibinfo {author} {\bibfnamefont {Shou-Cheng}\
  \bibnamefont {Zhang}},\ }\bibfield  {title} {\enquote {\bibinfo {title}
  {Topological insulators and superconductors},}\ }\href {\doibase
  10.1103/RevModPhys.83.1057} {\bibfield  {journal} {\bibinfo  {journal} {Rev.
  Mod. Phys.}\ }\textbf {\bibinfo {volume} {83}},\ \bibinfo {pages}
  {1057--1110} (\bibinfo {year} {2011})}\BibitemShut {NoStop}%
\bibitem [{\citenamefont {Hasan}\ and\ \citenamefont {Kane}(2010)}]{Rmp2}%
  \BibitemOpen
  \bibfield  {author} {\bibinfo {author} {\bibfnamefont {M.~Z.}\ \bibnamefont
  {Hasan}}\ and\ \bibinfo {author} {\bibfnamefont {C.~L.}\ \bibnamefont
  {Kane}},\ }\bibfield  {title} {\enquote {\bibinfo {title} {Colloquium:
  {T}opological {I}nsulators},}\ }\href {\doibase 10.1103/RevModPhys.82.3045}
  {\bibfield  {journal} {\bibinfo  {journal} {Rev. Mod. Phys.}\ }\textbf
  {\bibinfo {volume} {82}},\ \bibinfo {pages} {3045--3067} (\bibinfo {year}
  {2010})}\BibitemShut {NoStop}%
\bibitem [{\citenamefont {Armitage}\ \emph {et~al.}(2018)\citenamefont
  {Armitage}, \citenamefont {Mele},\ and\ \citenamefont
  {Vishwanath}}]{Weylrmp}%
  \BibitemOpen
  \bibfield  {author} {\bibinfo {author} {\bibfnamefont {N.~P.}\ \bibnamefont
  {Armitage}}, \bibinfo {author} {\bibfnamefont {E.~J.}\ \bibnamefont {Mele}},
  \ and\ \bibinfo {author} {\bibfnamefont {Ashvin}\ \bibnamefont
  {Vishwanath}},\ }\bibfield  {title} {\enquote {\bibinfo {title} {Weyl and
  dirac semimetals in three-dimensional solids},}\ }\href {\doibase
  10.1103/RevModPhys.90.015001} {\bibfield  {journal} {\bibinfo  {journal}
  {Rev. Mod. Phys.}\ }\textbf {\bibinfo {volume} {90}},\ \bibinfo {pages}
  {015001} (\bibinfo {year} {2018})}\BibitemShut {NoStop}%
\bibitem [{\citenamefont {Volovik}(2003)}]{volovik2003universe}%
  \BibitemOpen
  \bibfield  {author} {\bibinfo {author} {\bibfnamefont {Grigory~E}\
  \bibnamefont {Volovik}},\ }\href@noop {} {\emph {\bibinfo {title} {The
  universe in a helium droplet}}},\ Vol.\ \bibinfo {volume} {117}\ (\bibinfo
  {publisher} {OUP Oxford},\ \bibinfo {year} {2003})\BibitemShut {NoStop}%
\bibitem [{\citenamefont {Kruthoff}\ \emph {et~al.}(2017)\citenamefont
  {Kruthoff}, \citenamefont {de~Boer}, \citenamefont {van Wezel}, \citenamefont
  {Kane},\ and\ \citenamefont {Slager}}]{Clas3}%
  \BibitemOpen
  \bibfield  {author} {\bibinfo {author} {\bibfnamefont {Jorrit}\ \bibnamefont
  {Kruthoff}}, \bibinfo {author} {\bibfnamefont {Jan}\ \bibnamefont {de~Boer}},
  \bibinfo {author} {\bibfnamefont {Jasper}\ \bibnamefont {van Wezel}},
  \bibinfo {author} {\bibfnamefont {Charles~L.}\ \bibnamefont {Kane}}, \ and\
  \bibinfo {author} {\bibfnamefont {Robert-Jan}\ \bibnamefont {Slager}},\
  }\bibfield  {title} {\enquote {\bibinfo {title} {Topological {C}lassification
  of {C}rystalline {I}nsulators through {B}and {S}tructure {C}ombinatorics},}\
  }\href {\doibase 10.1103/PhysRevX.7.041069} {\bibfield  {journal} {\bibinfo
  {journal} {Phys. Rev. X}\ }\textbf {\bibinfo {volume} {7}},\ \bibinfo {pages}
  {041069} (\bibinfo {year} {2017})}\BibitemShut {NoStop}%
\bibitem [{\citenamefont {Bouhon}\ and\ \citenamefont
  {Black-Schaffer}(2017)}]{Wi2}%
  \BibitemOpen
  \bibfield  {author} {\bibinfo {author} {\bibfnamefont {Adrien}\ \bibnamefont
  {Bouhon}}\ and\ \bibinfo {author} {\bibfnamefont {Annica~M.}\ \bibnamefont
  {Black-Schaffer}},\ }\bibfield  {title} {\enquote {\bibinfo {title} {Global
  band topology of simple and double {D}irac-point semimetals},}\ }\href
  {\doibase 10.1103/PhysRevB.95.241101} {\bibfield  {journal} {\bibinfo
  {journal} {Phys. Rev. B}\ }\textbf {\bibinfo {volume} {95}},\ \bibinfo
  {pages} {241101} (\bibinfo {year} {2017})}\BibitemShut {NoStop}%
\bibitem [{\citenamefont {Hughes}\ \emph {et~al.}(2011)\citenamefont {Hughes},
  \citenamefont {Prodan},\ and\ \citenamefont {Bernevig}}]{InvTIBernevig}%
  \BibitemOpen
  \bibfield  {author} {\bibinfo {author} {\bibfnamefont {Taylor~L.}\
  \bibnamefont {Hughes}}, \bibinfo {author} {\bibfnamefont {Emil}\ \bibnamefont
  {Prodan}}, \ and\ \bibinfo {author} {\bibfnamefont {B.~Andrei}\ \bibnamefont
  {Bernevig}},\ }\bibfield  {title} {\enquote {\bibinfo {title}
  {Inversion-symmetric topological insulators},}\ }\href {\doibase
  10.1103/PhysRevB.83.245132} {\bibfield  {journal} {\bibinfo  {journal} {Phys.
  Rev. B}\ }\textbf {\bibinfo {volume} {83}},\ \bibinfo {pages} {245132}
  (\bibinfo {year} {2011})}\BibitemShut {NoStop}%
\bibitem [{\citenamefont {Fu}(2011)}]{Clas1}%
  \BibitemOpen
  \bibfield  {author} {\bibinfo {author} {\bibfnamefont {Liang}\ \bibnamefont
  {Fu}},\ }\bibfield  {title} {\enquote {\bibinfo {title} {Topological
  {C}rystalline {I}nsulators},}\ }\href {\doibase
  10.1103/PhysRevLett.106.106802} {\bibfield  {journal} {\bibinfo  {journal}
  {Phys. Rev. Lett.}\ }\textbf {\bibinfo {volume} {106}},\ \bibinfo {pages}
  {106802} (\bibinfo {year} {2011})}\BibitemShut {NoStop}%
\bibitem [{\citenamefont {Turner}\ \emph {et~al.}(2012)\citenamefont {Turner},
  \citenamefont {Zhang}, \citenamefont {Mong},\ and\ \citenamefont
  {Vishwanath}}]{InvTIVish}%
  \BibitemOpen
  \bibfield  {author} {\bibinfo {author} {\bibfnamefont {Ari~M.}\ \bibnamefont
  {Turner}}, \bibinfo {author} {\bibfnamefont {Yi}~\bibnamefont {Zhang}},
  \bibinfo {author} {\bibfnamefont {Roger S.~K.}\ \bibnamefont {Mong}}, \ and\
  \bibinfo {author} {\bibfnamefont {Ashvin}\ \bibnamefont {Vishwanath}},\
  }\bibfield  {title} {\enquote {\bibinfo {title} {Quantized response and
  topology of magnetic insulators with inversion symmetry},}\ }\href {\doibase
  10.1103/PhysRevB.85.165120} {\bibfield  {journal} {\bibinfo  {journal} {Phys.
  Rev. B}\ }\textbf {\bibinfo {volume} {85}},\ \bibinfo {pages} {165120}
  (\bibinfo {year} {2012})}\BibitemShut {NoStop}%
\bibitem [{\citenamefont {Slager}\ \emph {et~al.}(2013)\citenamefont {Slager},
  \citenamefont {Mesaros}, \citenamefont {Juri{\v c}i{\'c}},\ and\
  \citenamefont {Zaanen}}]{Clas2}%
  \BibitemOpen
  \bibfield  {author} {\bibinfo {author} {\bibfnamefont {Robert-Jan}\
  \bibnamefont {Slager}}, \bibinfo {author} {\bibfnamefont {Andrej}\
  \bibnamefont {Mesaros}}, \bibinfo {author} {\bibfnamefont {Vladimir}\
  \bibnamefont {Juri{\v c}i{\'c}}}, \ and\ \bibinfo {author} {\bibfnamefont
  {Jan}\ \bibnamefont {Zaanen}},\ }\bibfield  {title} {\enquote {\bibinfo
  {title} {The space group classification of topological band-insulators},}\
  }\href {http://dx.doi.org/10.1038/nphys2513} {\bibfield  {journal} {\bibinfo
  {journal} {Nat. Phys.}\ }\textbf {\bibinfo {volume} {9}},\ \bibinfo {pages}
  {98} (\bibinfo {year} {2013})}\BibitemShut {NoStop}%
\bibitem [{\citenamefont {Juri\ifmmode \check{c}\else
  \v{c}\fi{}i\ifmmode~\acute{c}\else \'{c}\fi{}}\ \emph
  {et~al.}(2012)\citenamefont {Juri\ifmmode \check{c}\else
  \v{c}\fi{}i\ifmmode~\acute{c}\else \'{c}\fi{}}, \citenamefont {Mesaros},
  \citenamefont {Slager},\ and\ \citenamefont {Zaanen}}]{probes_2D}%
  \BibitemOpen
  \bibfield  {author} {\bibinfo {author} {\bibfnamefont {Vladimir}\
  \bibnamefont {Juri\ifmmode \check{c}\else \v{c}\fi{}i\ifmmode~\acute{c}\else
  \'{c}\fi{}}}, \bibinfo {author} {\bibfnamefont {Andrej}\ \bibnamefont
  {Mesaros}}, \bibinfo {author} {\bibfnamefont {Robert-Jan}\ \bibnamefont
  {Slager}}, \ and\ \bibinfo {author} {\bibfnamefont {Jan}\ \bibnamefont
  {Zaanen}},\ }\bibfield  {title} {\enquote {\bibinfo {title} {Universal
  {P}robes of {T}wo-{D}imensional {T}opological {I}nsulators: {D}islocation and
  $\ensuremath{\pi}$ {F}lux},}\ }\href {\doibase
  10.1103/PhysRevLett.108.106403} {\bibfield  {journal} {\bibinfo  {journal}
  {Phys. Rev. Lett.}\ }\textbf {\bibinfo {volume} {108}},\ \bibinfo {pages}
  {106403} (\bibinfo {year} {2012})}\BibitemShut {NoStop}%
\bibitem [{\citenamefont {Shiozaki}\ and\ \citenamefont
  {Sato}(2014)}]{Shiozaki14}%
  \BibitemOpen
  \bibfield  {author} {\bibinfo {author} {\bibfnamefont {Ken}\ \bibnamefont
  {Shiozaki}}\ and\ \bibinfo {author} {\bibfnamefont {Masatoshi}\ \bibnamefont
  {Sato}},\ }\bibfield  {title} {\enquote {\bibinfo {title} {Topology of
  crystalline insulators and superconductors},}\ }\href {\doibase
  10.1103/PhysRevB.90.165114} {\bibfield  {journal} {\bibinfo  {journal} {Phys.
  Rev. B}\ }\textbf {\bibinfo {volume} {90}},\ \bibinfo {pages} {165114}
  (\bibinfo {year} {2014})}\BibitemShut {NoStop}%
\bibitem [{\citenamefont {Slager}(2019)}]{Codefects2}%
  \BibitemOpen
  \bibfield  {author} {\bibinfo {author} {\bibfnamefont {Robert-Jan}\
  \bibnamefont {Slager}},\ }\bibfield  {title} {\enquote {\bibinfo {title} {The
  translational side of topological band insulators},}\ }\href {\doibase
  https://doi.org/10.1016/j.jpcs.2018.01.023} {\bibfield  {journal} {\bibinfo
  {journal} {J. Phys. Chem. Solids}\ }\textbf {\bibinfo {volume} {128}},\
  \bibinfo {pages} {24 -- 38} (\bibinfo {year} {2019})},\ \bibinfo {note}
  {spin-Orbit Coupled Materials}\BibitemShut {NoStop}%
\bibitem [{\citenamefont {Chiu}\ \emph {et~al.}(2016)\citenamefont {Chiu},
  \citenamefont {Teo}, \citenamefont {Schnyder},\ and\ \citenamefont
  {Ryu}}]{SchnyderClass}%
  \BibitemOpen
  \bibfield  {author} {\bibinfo {author} {\bibfnamefont {Ching-Kai}\
  \bibnamefont {Chiu}}, \bibinfo {author} {\bibfnamefont {Jeffrey C.~Y.}\
  \bibnamefont {Teo}}, \bibinfo {author} {\bibfnamefont {Andreas~P.}\
  \bibnamefont {Schnyder}}, \ and\ \bibinfo {author} {\bibfnamefont {Shinsei}\
  \bibnamefont {Ryu}},\ }\bibfield  {title} {\enquote {\bibinfo {title}
  {Classification of topological quantum matter with symmetries},}\ }\href
  {\doibase 10.1103/RevModPhys.88.035005} {\bibfield  {journal} {\bibinfo
  {journal} {Rev. Mod. Phys.}\ }\textbf {\bibinfo {volume} {88}},\ \bibinfo
  {pages} {035005} (\bibinfo {year} {2016})}\BibitemShut {NoStop}%
\bibitem [{\citenamefont {Alexandradinata}\ \emph {et~al.}(2014)\citenamefont
  {Alexandradinata}, \citenamefont {Dai},\ and\ \citenamefont
  {Bernevig}}]{Wi1}%
  \BibitemOpen
  \bibfield  {author} {\bibinfo {author} {\bibfnamefont {A.}~\bibnamefont
  {Alexandradinata}}, \bibinfo {author} {\bibfnamefont {Xi}~\bibnamefont
  {Dai}}, \ and\ \bibinfo {author} {\bibfnamefont {B.~Andrei}\ \bibnamefont
  {Bernevig}},\ }\bibfield  {title} {\enquote {\bibinfo {title} {Wilson-loop
  characterization of inversion-symmetric topological insulators},}\ }\href
  {\doibase 10.1103/PhysRevB.89.155114} {\bibfield  {journal} {\bibinfo
  {journal} {Phys. Rev. B}\ }\textbf {\bibinfo {volume} {89}},\ \bibinfo
  {pages} {155114} (\bibinfo {year} {2014})}\BibitemShut {NoStop}%
\bibitem [{\citenamefont {Alexandradinata}\ \emph {et~al.}(2016)\citenamefont
  {Alexandradinata}, \citenamefont {Wang},\ and\ \citenamefont
  {Bernevig}}]{Wi3}%
  \BibitemOpen
  \bibfield  {author} {\bibinfo {author} {\bibfnamefont {A.}~\bibnamefont
  {Alexandradinata}}, \bibinfo {author} {\bibfnamefont {Zhijun}\ \bibnamefont
  {Wang}}, \ and\ \bibinfo {author} {\bibfnamefont {B.~Andrei}\ \bibnamefont
  {Bernevig}},\ }\bibfield  {title} {\enquote {\bibinfo {title} {Topological
  {I}nsulators from {G}roup {C}ohomology},}\ }\href {\doibase
  10.1103/PhysRevX.6.021008} {\bibfield  {journal} {\bibinfo  {journal} {Phys.
  Rev. X}\ }\textbf {\bibinfo {volume} {6}},\ \bibinfo {pages} {021008}
  (\bibinfo {year} {2016})}\BibitemShut {NoStop}%
\bibitem [{\citenamefont {Scheurer}\ and\ \citenamefont
  {Slager}(2020)}]{UnsupMach}%
  \BibitemOpen
  \bibfield  {author} {\bibinfo {author} {\bibfnamefont {Mathias~S.}\
  \bibnamefont {Scheurer}}\ and\ \bibinfo {author} {\bibfnamefont {Robert-Jan}\
  \bibnamefont {Slager}},\ }\bibfield  {title} {\enquote {\bibinfo {title}
  {Unsupervised machine learning and band topology},}\ }\href {\doibase
  10.1103/PhysRevLett.124.226401} {\bibfield  {journal} {\bibinfo  {journal}
  {Phys. Rev. Lett.}\ }\textbf {\bibinfo {volume} {124}},\ \bibinfo {pages}
  {226401} (\bibinfo {year} {2020})}\BibitemShut {NoStop}%
\bibitem [{\citenamefont {Shiozaki}\ \emph {et~al.}(2017)\citenamefont
  {Shiozaki}, \citenamefont {Sato},\ and\ \citenamefont
  {Gomi}}]{ShiozakiSatoGomiK}%
  \BibitemOpen
  \bibfield  {author} {\bibinfo {author} {\bibfnamefont {Ken}\ \bibnamefont
  {Shiozaki}}, \bibinfo {author} {\bibfnamefont {Masatoshi}\ \bibnamefont
  {Sato}}, \ and\ \bibinfo {author} {\bibfnamefont {Kiyonori}\ \bibnamefont
  {Gomi}},\ }\bibfield  {title} {\enquote {\bibinfo {title} {Topological
  crystalline materials: {G}eneral formulation, module structure, and wallpaper
  groups},}\ }\href {\doibase 10.1103/PhysRevB.95.235425} {\bibfield  {journal}
  {\bibinfo  {journal} {Phys. Rev. B}\ }\textbf {\bibinfo {volume} {95}},\
  \bibinfo {pages} {235425} (\bibinfo {year} {2017})}\BibitemShut {NoStop}%
\bibitem [{\citenamefont {Po}\ \emph {et~al.}(2017)\citenamefont {Po},
  \citenamefont {Vishwanath},\ and\ \citenamefont {Watanabe}}]{Clas4}%
  \BibitemOpen
  \bibfield  {author} {\bibinfo {author} {\bibfnamefont {Hoi~Chun}\
  \bibnamefont {Po}}, \bibinfo {author} {\bibfnamefont {Ashvin}\ \bibnamefont
  {Vishwanath}}, \ and\ \bibinfo {author} {\bibfnamefont {Haruki}\ \bibnamefont
  {Watanabe}},\ }\bibfield  {title} {\enquote {\bibinfo {title} {Symmetry-based
  indicators of band topology in the 230 space groups},}\ }\href {\doibase
  10.1038/s41467-017-00133-2} {\bibfield  {journal} {\bibinfo  {journal} {Nat.
  Commun.}\ }\textbf {\bibinfo {volume} {8}},\ \bibinfo {pages} {50} (\bibinfo
  {year} {2017})}\BibitemShut {NoStop}%
\bibitem [{\citenamefont {{Ran}}\ \emph {et~al.}(2009)\citenamefont {{Ran}},
  \citenamefont {{Zhang}},\ and\ \citenamefont {{Vishwanath}}}]{ran2009one}%
  \BibitemOpen
  \bibfield  {author} {\bibinfo {author} {\bibfnamefont {Ying}\ \bibnamefont
  {{Ran}}}, \bibinfo {author} {\bibfnamefont {Yi}~\bibnamefont {{Zhang}}}, \
  and\ \bibinfo {author} {\bibfnamefont {Ashvin}\ \bibnamefont
  {{Vishwanath}}},\ }\bibfield  {title} {\enquote {\bibinfo {title}
  {{One-dimensional topologically protected modes in topological insulators
  with lattice dislocations}},}\ }\href {\doibase 10.1038/nphys1220} {\bibfield
   {journal} {\bibinfo  {journal} {Nature Physics}\ }\textbf {\bibinfo {volume}
  {5}},\ \bibinfo {pages} {298--303} (\bibinfo {year} {2009})}\BibitemShut
  {NoStop}%
\bibitem [{\citenamefont {Rhim}\ \emph {et~al.}(2018)\citenamefont {Rhim},
  \citenamefont {Bardarson},\ and\ \citenamefont {Slager}}]{UnifiedBBc}%
  \BibitemOpen
  \bibfield  {author} {\bibinfo {author} {\bibfnamefont {Jun-Won}\ \bibnamefont
  {Rhim}}, \bibinfo {author} {\bibfnamefont {Jens~H.}\ \bibnamefont
  {Bardarson}}, \ and\ \bibinfo {author} {\bibfnamefont {Robert-Jan}\
  \bibnamefont {Slager}},\ }\bibfield  {title} {\enquote {\bibinfo {title}
  {Unified bulk-boundary correspondence for band insulators},}\ }\href
  {\doibase 10.1103/PhysRevB.97.115143} {\bibfield  {journal} {\bibinfo
  {journal} {Phys. Rev. B}\ }\textbf {\bibinfo {volume} {97}},\ \bibinfo
  {pages} {115143} (\bibinfo {year} {2018})}\BibitemShut {NoStop}%
\bibitem [{\citenamefont {Teo}\ and\ \citenamefont {Kane}(2010)}]{teodef}%
  \BibitemOpen
  \bibfield  {author} {\bibinfo {author} {\bibfnamefont {Jeffrey C.~Y.}\
  \bibnamefont {Teo}}\ and\ \bibinfo {author} {\bibfnamefont {C.~L.}\
  \bibnamefont {Kane}},\ }\bibfield  {title} {\enquote {\bibinfo {title}
  {Topological defects and gapless modes in insulators and superconductors},}\
  }\href {\doibase 10.1103/physrevb.82.115120} {\bibfield  {journal} {\bibinfo
  {journal} {Physical Review B}\ }\textbf {\bibinfo {volume} {82}} (\bibinfo
  {year} {2010}),\ 10.1103/physrevb.82.115120}\BibitemShut {NoStop}%
\bibitem [{\citenamefont {Bradlyn}\ \emph {et~al.}(2017)\citenamefont
  {Bradlyn}, \citenamefont {Elcoro}, \citenamefont {Cano}, \citenamefont
  {Vergniory}, \citenamefont {Wang}, \citenamefont {Felser}, \citenamefont
  {Aroyo},\ and\ \citenamefont {Bernevig}}]{Clas5}%
  \BibitemOpen
  \bibfield  {author} {\bibinfo {author} {\bibfnamefont {Barry}\ \bibnamefont
  {Bradlyn}}, \bibinfo {author} {\bibfnamefont {L.}~\bibnamefont {Elcoro}},
  \bibinfo {author} {\bibfnamefont {Jennifer}\ \bibnamefont {Cano}}, \bibinfo
  {author} {\bibfnamefont {M.~G.}\ \bibnamefont {Vergniory}}, \bibinfo {author}
  {\bibfnamefont {Zhijun}\ \bibnamefont {Wang}}, \bibinfo {author}
  {\bibfnamefont {C.}~\bibnamefont {Felser}}, \bibinfo {author} {\bibfnamefont
  {M.~I.}\ \bibnamefont {Aroyo}}, \ and\ \bibinfo {author} {\bibfnamefont
  {B.~Andrei}\ \bibnamefont {Bernevig}},\ }\bibfield  {title} {\enquote
  {\bibinfo {title} {Topological quantum chemistry},}\ }\href
  {http://dx.doi.org/10.1038/nature23268} {\bibfield  {journal} {\bibinfo
  {journal} {Nature}\ }\textbf {\bibinfo {volume} {547}},\ \bibinfo {pages}
  {298} (\bibinfo {year} {2017})}\BibitemShut {NoStop}%
\bibitem [{\citenamefont {Slager}\ \emph {et~al.}(2015)\citenamefont {Slager},
  \citenamefont {Rademaker}, \citenamefont {Zaanen},\ and\ \citenamefont
  {Balents}}]{Codefects1}%
  \BibitemOpen
  \bibfield  {author} {\bibinfo {author} {\bibfnamefont {Robert-Jan}\
  \bibnamefont {Slager}}, \bibinfo {author} {\bibfnamefont {Louk}\ \bibnamefont
  {Rademaker}}, \bibinfo {author} {\bibfnamefont {Jan}\ \bibnamefont {Zaanen}},
  \ and\ \bibinfo {author} {\bibfnamefont {Leon}\ \bibnamefont {Balents}},\
  }\bibfield  {title} {\enquote {\bibinfo {title} {Impurity-bound states and
  {G}reen's function zeros as local signatures of topology},}\ }\href {\doibase
  10.1103/PhysRevB.92.085126} {\bibfield  {journal} {\bibinfo  {journal} {Phys.
  Rev. B}\ }\textbf {\bibinfo {volume} {92}},\ \bibinfo {pages} {085126}
  (\bibinfo {year} {2015})}\BibitemShut {NoStop}%
\bibitem [{\citenamefont {Bouhon}\ \emph {et~al.}(2018)\citenamefont {Bouhon},
  \citenamefont {Schmidt},\ and\ \citenamefont {Black-Schaffer}}]{Bouhon_HHL}%
  \BibitemOpen
  \bibfield  {author} {\bibinfo {author} {\bibfnamefont {Adrien}\ \bibnamefont
  {Bouhon}}, \bibinfo {author} {\bibfnamefont {Johann}\ \bibnamefont
  {Schmidt}}, \ and\ \bibinfo {author} {\bibfnamefont {Annica~M.}\ \bibnamefont
  {Black-Schaffer}},\ }\bibfield  {title} {\enquote {\bibinfo {title}
  {Topological nodal superconducting phases and topological phase transition in
  the hyperhoneycomb lattice},}\ }\href {\doibase 10.1103/PhysRevB.97.104508}
  {\bibfield  {journal} {\bibinfo  {journal} {Phys. Rev. B}\ }\textbf {\bibinfo
  {volume} {97}},\ \bibinfo {pages} {104508} (\bibinfo {year}
  {2018})}\BibitemShut {NoStop}%
\bibitem [{\citenamefont {Slager}\ \emph {et~al.}(2017)\citenamefont {Slager},
  \citenamefont {Juri\ifmmode \check{c}\else \v{c}\fi{}i\ifmmode~\acute{c}\else
  \'{c}\fi{}},\ and\ \citenamefont {Roy}}]{BbcWeyl}%
  \BibitemOpen
  \bibfield  {author} {\bibinfo {author} {\bibfnamefont {Robert-Jan}\
  \bibnamefont {Slager}}, \bibinfo {author} {\bibfnamefont {Vladimir}\
  \bibnamefont {Juri\ifmmode \check{c}\else \v{c}\fi{}i\ifmmode~\acute{c}\else
  \'{c}\fi{}}}, \ and\ \bibinfo {author} {\bibfnamefont {Bitan}\ \bibnamefont
  {Roy}},\ }\bibfield  {title} {\enquote {\bibinfo {title} {Dissolution of
  topological {F}ermi arcs in a dirty {W}eyl semimetal},}\ }\href {\doibase
  10.1103/PhysRevB.96.201401} {\bibfield  {journal} {\bibinfo  {journal} {Phys.
  Rev. B}\ }\textbf {\bibinfo {volume} {96}},\ \bibinfo {pages} {201401}
  (\bibinfo {year} {2017})}\BibitemShut {NoStop}%
\bibitem [{\citenamefont {Alexandradinata}\ \emph {et~al.}(2020)\citenamefont
  {Alexandradinata}, \citenamefont {H\"oller}, \citenamefont {Wang},
  \citenamefont {Cheng},\ and\ \citenamefont {Lu}}]{alex2019crystallographic}%
  \BibitemOpen
  \bibfield  {author} {\bibinfo {author} {\bibfnamefont {A.}~\bibnamefont
  {Alexandradinata}}, \bibinfo {author} {\bibfnamefont {J.}~\bibnamefont
  {H\"oller}}, \bibinfo {author} {\bibfnamefont {Chong}\ \bibnamefont {Wang}},
  \bibinfo {author} {\bibfnamefont {Hengbin}\ \bibnamefont {Cheng}}, \ and\
  \bibinfo {author} {\bibfnamefont {Ling}\ \bibnamefont {Lu}},\ }\bibfield
  {title} {\enquote {\bibinfo {title} {Crystallographic splitting theorem for
  band representations and fragile topological photonic crystals},}\ }\href
  {\doibase 10.1103/PhysRevB.102.115117} {\bibfield  {journal} {\bibinfo
  {journal} {Phys. Rev. B}\ }\textbf {\bibinfo {volume} {102}},\ \bibinfo
  {pages} {115117} (\bibinfo {year} {2020})}\BibitemShut {NoStop}%
\bibitem [{\citenamefont {Slager}\ \emph {et~al.}(2014)\citenamefont {Slager},
  \citenamefont {Mesaros}, \citenamefont {Juri\ifmmode \check{c}\else
  \v{c}\fi{}i\ifmmode~\acute{c}\else \'{c}\fi{}},\ and\ \citenamefont
  {Zaanen}}]{Mode2}%
  \BibitemOpen
  \bibfield  {author} {\bibinfo {author} {\bibfnamefont {Robert-Jan}\
  \bibnamefont {Slager}}, \bibinfo {author} {\bibfnamefont {Andrej}\
  \bibnamefont {Mesaros}}, \bibinfo {author} {\bibfnamefont {Vladimir}\
  \bibnamefont {Juri\ifmmode \check{c}\else \v{c}\fi{}i\ifmmode~\acute{c}\else
  \'{c}\fi{}}}, \ and\ \bibinfo {author} {\bibfnamefont {Jan}\ \bibnamefont
  {Zaanen}},\ }\bibfield  {title} {\enquote {\bibinfo {title} {Interplay
  between electronic topology and crystal symmetry: {D}islocation-line modes in
  topological band insulators},}\ }\href {\doibase 10.1103/PhysRevB.90.241403}
  {\bibfield  {journal} {\bibinfo  {journal} {Phys. Rev. B}\ }\textbf {\bibinfo
  {volume} {90}},\ \bibinfo {pages} {241403} (\bibinfo {year}
  {2014})}\BibitemShut {NoStop}%
\bibitem [{\citenamefont {Fang}\ \emph {et~al.}(2012)\citenamefont {Fang},
  \citenamefont {Gilbert},\ and\ \citenamefont {Bernevig}}]{Chenprb2012}%
  \BibitemOpen
  \bibfield  {author} {\bibinfo {author} {\bibfnamefont {Chen}\ \bibnamefont
  {Fang}}, \bibinfo {author} {\bibfnamefont {Matthew~J.}\ \bibnamefont
  {Gilbert}}, \ and\ \bibinfo {author} {\bibfnamefont {B.~Andrei}\ \bibnamefont
  {Bernevig}},\ }\bibfield  {title} {\enquote {\bibinfo {title} {Bulk
  topological invariants in noninteracting point group symmetric insulators},}\
  }\href {\doibase 10.1103/PhysRevB.86.115112} {\bibfield  {journal} {\bibinfo
  {journal} {Phys. Rev. B}\ }\textbf {\bibinfo {volume} {86}},\ \bibinfo
  {pages} {115112} (\bibinfo {year} {2012})}\BibitemShut {NoStop}%
\bibitem [{\citenamefont {\"Unal}\ \emph {et~al.}(2019)\citenamefont {\"Unal},
  \citenamefont {Eckardt},\ and\ \citenamefont {Slager}}]{Unal2019}%
  \BibitemOpen
  \bibfield  {author} {\bibinfo {author} {\bibfnamefont {F.~Nur}\ \bibnamefont
  {\"Unal}}, \bibinfo {author} {\bibfnamefont {Andr\'e}\ \bibnamefont
  {Eckardt}}, \ and\ \bibinfo {author} {\bibfnamefont {Robert-Jan}\
  \bibnamefont {Slager}},\ }\bibfield  {title} {\enquote {\bibinfo {title}
  {Hopf characterization of two-dimensional {F}loquet topological
  insulators},}\ }\href {\doibase 10.1103/PhysRevResearch.1.022003} {\bibfield
  {journal} {\bibinfo  {journal} {Phys. Rev. Research}\ }\textbf {\bibinfo
  {volume} {1}},\ \bibinfo {pages} {022003} (\bibinfo {year}
  {2019})}\BibitemShut {NoStop}%
\bibitem [{\citenamefont {Cornfeld}\ and\ \citenamefont
  {Carmeli}(2021)}]{Cornfeld_2021}%
  \BibitemOpen
  \bibfield  {author} {\bibinfo {author} {\bibfnamefont {Eyal}\ \bibnamefont
  {Cornfeld}}\ and\ \bibinfo {author} {\bibfnamefont {Shachar}\ \bibnamefont
  {Carmeli}},\ }\bibfield  {title} {\enquote {\bibinfo {title} {Tenfold
  topology of crystals: Unified classification of crystalline topological
  insulators and superconductors},}\ }\href {\doibase
  10.1103/PhysRevResearch.3.013052} {\bibfield  {journal} {\bibinfo  {journal}
  {Phys. Rev. Research}\ }\textbf {\bibinfo {volume} {3}},\ \bibinfo {pages}
  {013052} (\bibinfo {year} {2021})}\BibitemShut {NoStop}%
\bibitem [{\citenamefont {Po}\ \emph {et~al.}(2018)\citenamefont {Po},
  \citenamefont {Watanabe},\ and\ \citenamefont {Vishwanath}}]{Ft1}%
  \BibitemOpen
  \bibfield  {author} {\bibinfo {author} {\bibfnamefont {Hoi~Chun}\
  \bibnamefont {Po}}, \bibinfo {author} {\bibfnamefont {Haruki}\ \bibnamefont
  {Watanabe}}, \ and\ \bibinfo {author} {\bibfnamefont {Ashvin}\ \bibnamefont
  {Vishwanath}},\ }\bibfield  {title} {\enquote {\bibinfo {title} {Fragile
  {T}opology and {W}annier {O}bstructions},}\ }\href {\doibase
  10.1103/PhysRevLett.121.126402} {\bibfield  {journal} {\bibinfo  {journal}
  {Phys. Rev. Lett.}\ }\textbf {\bibinfo {volume} {121}},\ \bibinfo {pages}
  {126402} (\bibinfo {year} {2018})}\BibitemShut {NoStop}%
\bibitem [{\citenamefont {Bouhon}\ \emph {et~al.}(2019)\citenamefont {Bouhon},
  \citenamefont {Black-Schaffer},\ and\ \citenamefont
  {Slager}}]{bouhon2018wilson}%
  \BibitemOpen
  \bibfield  {author} {\bibinfo {author} {\bibfnamefont {Adrien}\ \bibnamefont
  {Bouhon}}, \bibinfo {author} {\bibfnamefont {Annica~M.}\ \bibnamefont
  {Black-Schaffer}}, \ and\ \bibinfo {author} {\bibfnamefont {Robert-Jan}\
  \bibnamefont {Slager}},\ }\bibfield  {title} {\enquote {\bibinfo {title}
  {Wilson loop approach to fragile topology of split elementary band
  representations and topological crystalline insulators with time-reversal
  symmetry},}\ }\href {\doibase 10.1103/PhysRevB.100.195135} {\bibfield
  {journal} {\bibinfo  {journal} {Phys. Rev. B}\ }\textbf {\bibinfo {volume}
  {100}},\ \bibinfo {pages} {195135} (\bibinfo {year} {2019})}\BibitemShut
  {NoStop}%
\bibitem [{\citenamefont {Bradlyn}\ \emph {et~al.}(2019)\citenamefont
  {Bradlyn}, \citenamefont {Wang}, \citenamefont {Cano},\ and\ \citenamefont
  {Bernevig}}]{Bradlyn_fragile}%
  \BibitemOpen
  \bibfield  {author} {\bibinfo {author} {\bibfnamefont {Barry}\ \bibnamefont
  {Bradlyn}}, \bibinfo {author} {\bibfnamefont {Zhijun}\ \bibnamefont {Wang}},
  \bibinfo {author} {\bibfnamefont {Jennifer}\ \bibnamefont {Cano}}, \ and\
  \bibinfo {author} {\bibfnamefont {B.~Andrei}\ \bibnamefont {Bernevig}},\
  }\bibfield  {title} {\enquote {\bibinfo {title} {Disconnected elementary band
  representations, fragile topology, and wilson loops as topological indices:
  {A}n example on the triangular lattice},}\ }\href {\doibase
  10.1103/PhysRevB.99.045140} {\bibfield  {journal} {\bibinfo  {journal} {Phys.
  Rev. B}\ }\textbf {\bibinfo {volume} {99}},\ \bibinfo {pages} {045140}
  (\bibinfo {year} {2019})}\BibitemShut {NoStop}%
\bibitem [{\citenamefont {Hwang}\ \emph {et~al.}(2019)\citenamefont {Hwang},
  \citenamefont {Ahn},\ and\ \citenamefont {Yang}}]{Hwang_inversion_fragile}%
  \BibitemOpen
  \bibfield  {author} {\bibinfo {author} {\bibfnamefont {Yoonseok}\
  \bibnamefont {Hwang}}, \bibinfo {author} {\bibfnamefont {Junyeong}\
  \bibnamefont {Ahn}}, \ and\ \bibinfo {author} {\bibfnamefont {Bohm-Jung}\
  \bibnamefont {Yang}},\ }\bibfield  {title} {\enquote {\bibinfo {title}
  {Fragile topology protected by inversion symmetry: {D}iagnosis, bulk-boundary
  correspondence, and {W}ilson loop},}\ }\href {\doibase
  10.1103/PhysRevB.100.205126} {\bibfield  {journal} {\bibinfo  {journal}
  {Phys. Rev. B}\ }\textbf {\bibinfo {volume} {100}},\ \bibinfo {pages}
  {205126} (\bibinfo {year} {2019})}\BibitemShut {NoStop}%
\bibitem [{\citenamefont {Song}\ \emph {et~al.}(2020)\citenamefont {Song},
  \citenamefont {Elcoro},\ and\ \citenamefont {Bernevig}}]{Song794}%
  \BibitemOpen
  \bibfield  {author} {\bibinfo {author} {\bibfnamefont {Zhi-Da}\ \bibnamefont
  {Song}}, \bibinfo {author} {\bibfnamefont {Luis}\ \bibnamefont {Elcoro}}, \
  and\ \bibinfo {author} {\bibfnamefont {B.~Andrei}\ \bibnamefont {Bernevig}},\
  }\bibfield  {title} {\enquote {\bibinfo {title} {Twisted bulk-boundary
  correspondence of fragile topology},}\ }\href {\doibase
  10.1126/science.aaz7650} {\bibfield  {journal} {\bibinfo  {journal}
  {Science}\ }\textbf {\bibinfo {volume} {367}},\ \bibinfo {pages} {794--797}
  (\bibinfo {year} {2020})}\BibitemShut {NoStop}%
\bibitem [{\citenamefont {Palumbo}(2021)}]{paulmbo_metricw}%
  \BibitemOpen
  \bibfield  {author} {\bibinfo {author} {\bibfnamefont {Giandomenico}\
  \bibnamefont {Palumbo}},\ }\bibfield  {title} {\enquote {\bibinfo {title}
  {Non-abelian tensor berry connections in multiband topological systems},}\
  }\href {\doibase 10.1103/PhysRevLett.126.246801} {\bibfield  {journal}
  {\bibinfo  {journal} {Phys. Rev. Lett.}\ }\textbf {\bibinfo {volume} {126}},\
  \bibinfo {pages} {246801} (\bibinfo {year} {2021})}\BibitemShut {NoStop}%
\bibitem [{\citenamefont {Lange}\ \emph {et~al.}(2021)\citenamefont {Lange},
  \citenamefont {Bouhon},\ and\ \citenamefont {Slager}}]{SubD_Gunnar}%
  \BibitemOpen
  \bibfield  {author} {\bibinfo {author} {\bibfnamefont {Gunnar~F.}\
  \bibnamefont {Lange}}, \bibinfo {author} {\bibfnamefont {Adrien}\
  \bibnamefont {Bouhon}}, \ and\ \bibinfo {author} {\bibfnamefont {Robert-Jan}\
  \bibnamefont {Slager}},\ }\bibfield  {title} {\enquote {\bibinfo {title}
  {Subdimensional topologies, indicators, and higher order boundary effects},}\
  }\href {\doibase 10.1103/PhysRevB.103.195145} {\bibfield  {journal} {\bibinfo
   {journal} {Phys. Rev. B}\ }\textbf {\bibinfo {volume} {103}},\ \bibinfo
  {pages} {195145} (\bibinfo {year} {2021})}\BibitemShut {NoStop}%
\bibitem [{\citenamefont {Wieder}\ and\ \citenamefont
  {Bernevig}(2018)}]{Wieder_axion}%
  \BibitemOpen
  \bibfield  {author} {\bibinfo {author} {\bibfnamefont {Benjamin~J.}\
  \bibnamefont {Wieder}}\ and\ \bibinfo {author} {\bibfnamefont {B.~Andrei}\
  \bibnamefont {Bernevig}},\ }\href@noop {} {\enquote {\bibinfo {title} {The
  axion insulator as a pump of fragile topology},}\ } (\bibinfo {year}
  {2018}),\ \Eprint {http://arxiv.org/abs/arXiv:1810.02373} {arXiv:1810.02373}
  \BibitemShut {NoStop}%
\bibitem [{\citenamefont {Peri}\ \emph {et~al.}(2020)\citenamefont {Peri},
  \citenamefont {Song}, \citenamefont {Serra-Garcia}, \citenamefont {Engeler},
  \citenamefont {Queiroz}, \citenamefont {Huang}, \citenamefont {Deng},
  \citenamefont {Liu}, \citenamefont {Bernevig},\ and\ \citenamefont
  {Huber}}]{Peri797}%
  \BibitemOpen
  \bibfield  {author} {\bibinfo {author} {\bibfnamefont {Valerio}\ \bibnamefont
  {Peri}}, \bibinfo {author} {\bibfnamefont {Zhi-Da}\ \bibnamefont {Song}},
  \bibinfo {author} {\bibfnamefont {Marc}\ \bibnamefont {Serra-Garcia}},
  \bibinfo {author} {\bibfnamefont {Pascal}\ \bibnamefont {Engeler}}, \bibinfo
  {author} {\bibfnamefont {Raquel}\ \bibnamefont {Queiroz}}, \bibinfo {author}
  {\bibfnamefont {Xueqin}\ \bibnamefont {Huang}}, \bibinfo {author}
  {\bibfnamefont {Weiyin}\ \bibnamefont {Deng}}, \bibinfo {author}
  {\bibfnamefont {Zhengyou}\ \bibnamefont {Liu}}, \bibinfo {author}
  {\bibfnamefont {B.~Andrei}\ \bibnamefont {Bernevig}}, \ and\ \bibinfo
  {author} {\bibfnamefont {Sebastian~D.}\ \bibnamefont {Huber}},\ }\bibfield
  {title} {\enquote {\bibinfo {title} {Experimental characterization of fragile
  topology in an acoustic metamaterial},}\ }\href {\doibase
  10.1126/science.aaz7654} {\bibfield  {journal} {\bibinfo  {journal}
  {Science}\ }\textbf {\bibinfo {volume} {367}},\ \bibinfo {pages} {797--800}
  (\bibinfo {year} {2020})}\BibitemShut {NoStop}%
\bibitem [{\citenamefont {Wu}\ \emph {et~al.}(2019)\citenamefont {Wu},
  \citenamefont {Soluyanov},\ and\ \citenamefont {Bzdu{\v s}ek}}]{Wu1273}%
  \BibitemOpen
  \bibfield  {author} {\bibinfo {author} {\bibfnamefont {QuanSheng}\
  \bibnamefont {Wu}}, \bibinfo {author} {\bibfnamefont {Alexey~A.}\
  \bibnamefont {Soluyanov}}, \ and\ \bibinfo {author} {\bibfnamefont
  {Tom{\'a}{\v s}}\ \bibnamefont {Bzdu{\v s}ek}},\ }\bibfield  {title}
  {\enquote {\bibinfo {title} {Non-{A}belian band topology in noninteracting
  metals},}\ }\href {\doibase 10.1126/science.aau8740} {\bibfield  {journal}
  {\bibinfo  {journal} {Science}\ }\textbf {\bibinfo {volume} {365}},\ \bibinfo
  {pages} {1273--1277} (\bibinfo {year} {2019})}\BibitemShut {NoStop}%
\bibitem [{\citenamefont {Tiwari}\ and\ \citenamefont
  {Bzdu\v{s}ek}(2020)}]{Tiwari:2019}%
  \BibitemOpen
  \bibfield  {author} {\bibinfo {author} {\bibfnamefont {Apoorv}\ \bibnamefont
  {Tiwari}}\ and\ \bibinfo {author} {\bibfnamefont {Tom\'{a}\v{s}}\
  \bibnamefont {Bzdu\v{s}ek}},\ }\bibfield  {title} {\enquote {\bibinfo {title}
  {Non-{A}belian topology of nodal-line rings in $\mathcal{PT}$-symmetric
  systems},}\ }\href {\doibase 10.1103/PhysRevB.101.195130} {\bibfield
  {journal} {\bibinfo  {journal} {Phys. Rev. B}\ }\textbf {\bibinfo {volume}
  {101}},\ \bibinfo {pages} {195130} (\bibinfo {year} {2020})}\BibitemShut
  {NoStop}%
\bibitem [{\citenamefont {Ahn}\ \emph {et~al.}(2019)\citenamefont {Ahn},
  \citenamefont {Park},\ and\ \citenamefont {Yang}}]{BJY_nielsen}%
  \BibitemOpen
  \bibfield  {author} {\bibinfo {author} {\bibfnamefont {Junyeong}\
  \bibnamefont {Ahn}}, \bibinfo {author} {\bibfnamefont {Sungjoon}\
  \bibnamefont {Park}}, \ and\ \bibinfo {author} {\bibfnamefont {Bohm-Jung}\
  \bibnamefont {Yang}},\ }\bibfield  {title} {\enquote {\bibinfo {title}
  {Failure of {N}ielsen-{N}inomiya {T}heorem and {F}ragile {T}opology in
  {T}wo-{D}imensional {S}ystems with {S}pace-{T}ime {I}nversion {S}ymmetry:
  {A}pplication to {T}wisted {B}ilayer {G}raphene at {M}agic {A}ngle},}\ }\href
  {\doibase 10.1103/PhysRevX.9.021013} {\bibfield  {journal} {\bibinfo
  {journal} {Phys. Rev. X}\ }\textbf {\bibinfo {volume} {9}},\ \bibinfo {pages}
  {021013} (\bibinfo {year} {2019})}\BibitemShut {NoStop}%
\bibitem [{\citenamefont {Bouhon}\ \emph
  {et~al.}(2020{\natexlab{a}})\citenamefont {Bouhon}, \citenamefont {Wu},
  \citenamefont {Slager}, \citenamefont {Weng}, \citenamefont {Yazyev},\ and\
  \citenamefont {Bzdu{\v s}ek}}]{bouhon2019nonabelian}%
  \BibitemOpen
  \bibfield  {author} {\bibinfo {author} {\bibfnamefont {Adrien}\ \bibnamefont
  {Bouhon}}, \bibinfo {author} {\bibfnamefont {QuanSheng}\ \bibnamefont {Wu}},
  \bibinfo {author} {\bibfnamefont {Robert-Jan}\ \bibnamefont {Slager}},
  \bibinfo {author} {\bibfnamefont {Hongming}\ \bibnamefont {Weng}}, \bibinfo
  {author} {\bibfnamefont {Oleg~V.}\ \bibnamefont {Yazyev}}, \ and\ \bibinfo
  {author} {\bibfnamefont {Tom{\'a}{\v s}}\ \bibnamefont {Bzdu{\v s}ek}},\
  }\bibfield  {title} {\enquote {\bibinfo {title} {Non-abelian reciprocal
  braiding of weyl points and its manifestation in zrte},}\ }\href {\doibase
  10.1038/s41567-020-0967-9} {\bibfield  {journal} {\bibinfo  {journal} {Nature
  Physics}\ }\textbf {\bibinfo {volume} {16}},\ \bibinfo {pages} {1137--1143}
  (\bibinfo {year} {2020}{\natexlab{a}})}\BibitemShut {NoStop}%
\bibitem [{\citenamefont {Alexander}\ \emph {et~al.}(2012)\citenamefont
  {Alexander}, \citenamefont {Chen}, \citenamefont {Matsumoto},\ and\
  \citenamefont {Kamien}}]{Kamienrmp}%
  \BibitemOpen
  \bibfield  {author} {\bibinfo {author} {\bibfnamefont {Gareth~P.}\
  \bibnamefont {Alexander}}, \bibinfo {author} {\bibfnamefont {Bryan Gin-ge}\
  \bibnamefont {Chen}}, \bibinfo {author} {\bibfnamefont {Elisabetta~A.}\
  \bibnamefont {Matsumoto}}, \ and\ \bibinfo {author} {\bibfnamefont
  {Randall~D.}\ \bibnamefont {Kamien}},\ }\bibfield  {title} {\enquote
  {\bibinfo {title} {{C}olloquium: {D}isclination loops, point defects, and all
  that in nematic liquid crystals},}\ }\href {\doibase
  10.1103/RevModPhys.84.497} {\bibfield  {journal} {\bibinfo  {journal} {Rev.
  Mod. Phys.}\ }\textbf {\bibinfo {volume} {84}},\ \bibinfo {pages} {497--514}
  (\bibinfo {year} {2012})}\BibitemShut {NoStop}%
\bibitem [{\citenamefont {Liu}\ \emph {et~al.}(2016)\citenamefont {Liu},
  \citenamefont {Nissinen}, \citenamefont {Slager}, \citenamefont {Wu},\ and\
  \citenamefont {Zaanen}}]{Genqcs2016}%
  \BibitemOpen
  \bibfield  {author} {\bibinfo {author} {\bibfnamefont {Ke}~\bibnamefont
  {Liu}}, \bibinfo {author} {\bibfnamefont {Jaakko}\ \bibnamefont {Nissinen}},
  \bibinfo {author} {\bibfnamefont {Robert-Jan}\ \bibnamefont {Slager}},
  \bibinfo {author} {\bibfnamefont {Kai}\ \bibnamefont {Wu}}, \ and\ \bibinfo
  {author} {\bibfnamefont {Jan}\ \bibnamefont {Zaanen}},\ }\bibfield  {title}
  {\enquote {\bibinfo {title} {Generalized {L}iquid {C}rystals: {G}iant
  {F}luctuations and the {V}estigial {C}hiral {O}rder of ${I}$, ${O}$, and
  ${T}$ {M}atter},}\ }\href {\doibase 10.1103/PhysRevX.6.041025} {\bibfield
  {journal} {\bibinfo  {journal} {Phys. Rev. X}\ }\textbf {\bibinfo {volume}
  {6}},\ \bibinfo {pages} {041025} (\bibinfo {year} {2016})}\BibitemShut
  {NoStop}%
\bibitem [{\citenamefont {Volovik}\ and\ \citenamefont
  {Mineev}(2018)}]{volovik2018investigation}%
  \BibitemOpen
  \bibfield  {author} {\bibinfo {author} {\bibfnamefont {G.~E.}\ \bibnamefont
  {Volovik}}\ and\ \bibinfo {author} {\bibfnamefont {V.~P.}\ \bibnamefont
  {Mineev}},\ }\bibfield  {title} {\enquote {\bibinfo {title} {Investigation of
  singularities in superfluid {H}e$^3$ in liquid crystals by the homotopic
  topology methods},}\ }in\ \href@noop {} {\emph {\bibinfo {booktitle} {Basic
  Notions Of Condensed Matter Physics}}}\ (\bibinfo  {publisher} {CRC Press},\
  \bibinfo {year} {2018})\ pp.\ \bibinfo {pages} {392--401}\BibitemShut
  {NoStop}%
\bibitem [{\citenamefont {Beekman}\ \emph {et~al.}(2017)\citenamefont
  {Beekman}, \citenamefont {Nissinen}, \citenamefont {Wu}, \citenamefont {Liu},
  \citenamefont {Slager}, \citenamefont {Nussinov}, \citenamefont {Cvetkovic},\
  and\ \citenamefont {Zaanen}}]{Beekman20171}%
  \BibitemOpen
  \bibfield  {author} {\bibinfo {author} {\bibfnamefont {Aron~J.}\ \bibnamefont
  {Beekman}}, \bibinfo {author} {\bibfnamefont {Jaakko}\ \bibnamefont
  {Nissinen}}, \bibinfo {author} {\bibfnamefont {Kai}\ \bibnamefont {Wu}},
  \bibinfo {author} {\bibfnamefont {Ke}~\bibnamefont {Liu}}, \bibinfo {author}
  {\bibfnamefont {Robert-Jan}\ \bibnamefont {Slager}}, \bibinfo {author}
  {\bibfnamefont {Zohar}\ \bibnamefont {Nussinov}}, \bibinfo {author}
  {\bibfnamefont {Vladimir}\ \bibnamefont {Cvetkovic}}, \ and\ \bibinfo
  {author} {\bibfnamefont {Jan}\ \bibnamefont {Zaanen}},\ }\bibfield  {title}
  {\enquote {\bibinfo {title} {Dual gauge field theory of quantum liquid
  crystals in two dimensions},}\ }\href {\doibase
  https://doi.org/10.1016/j.physrep.2017.03.004} {\bibfield  {journal}
  {\bibinfo  {journal} {Phys. Rep.}\ }\textbf {\bibinfo {volume} {683}},\
  \bibinfo {pages} {1 -- 110} (\bibinfo {year} {2017})},\ \bibinfo {note} {dual
  gauge field theory of quantum liquid crystals in two dimensions}\BibitemShut
  {NoStop}%
\bibitem [{\citenamefont {Ahn}\ \emph {et~al.}(2018)\citenamefont {Ahn},
  \citenamefont {Kim}, \citenamefont {Kim},\ and\ \citenamefont
  {Yang}}]{BJY_linking}%
  \BibitemOpen
  \bibfield  {author} {\bibinfo {author} {\bibfnamefont {Junyeong}\
  \bibnamefont {Ahn}}, \bibinfo {author} {\bibfnamefont {Dongwook}\
  \bibnamefont {Kim}}, \bibinfo {author} {\bibfnamefont {Youngkuk}\
  \bibnamefont {Kim}}, \ and\ \bibinfo {author} {\bibfnamefont {Bohm-Jung}\
  \bibnamefont {Yang}},\ }\bibfield  {title} {\enquote {\bibinfo {title} {Band
  topology and linking structure of nodal line semimetals with ${Z}_{2}$
  monopole charges},}\ }\href {\doibase 10.1103/PhysRevLett.121.106403}
  {\bibfield  {journal} {\bibinfo  {journal} {Phys. Rev. Lett.}\ }\textbf
  {\bibinfo {volume} {121}},\ \bibinfo {pages} {106403} (\bibinfo {year}
  {2018})}\BibitemShut {NoStop}%
\bibitem [{\citenamefont {Bouhon}\ \emph
  {et~al.}(2020{\natexlab{b}})\citenamefont {Bouhon}, \citenamefont {Bzdusek},\
  and\ \citenamefont {Slager}}]{bouhon2020geometric}%
  \BibitemOpen
  \bibfield  {author} {\bibinfo {author} {\bibfnamefont {Adrien}\ \bibnamefont
  {Bouhon}}, \bibinfo {author} {\bibfnamefont {Tomas}\ \bibnamefont {Bzdusek}},
  \ and\ \bibinfo {author} {\bibfnamefont {Robert-Jan}\ \bibnamefont
  {Slager}},\ }\bibfield  {title} {\enquote {\bibinfo {title} {Geometric
  approach to fragile topology beyond symmetry indicators},}\ }\href {\doibase
  10.1103/PhysRevB.102.115135} {\bibfield  {journal} {\bibinfo  {journal}
  {Phys. Rev. B}\ }\textbf {\bibinfo {volume} {102}},\ \bibinfo {pages}
  {115135} (\bibinfo {year} {2020}{\natexlab{b}})}\BibitemShut {NoStop}%
\bibitem [{\citenamefont {\"Unal}\ \emph {et~al.}(2020)\citenamefont {\"Unal},
  \citenamefont {Bouhon},\ and\ \citenamefont {Slager}}]{Unal_quenched_Euler}%
  \BibitemOpen
  \bibfield  {author} {\bibinfo {author} {\bibfnamefont {F.~Nur}\ \bibnamefont
  {\"Unal}}, \bibinfo {author} {\bibfnamefont {Adrien}\ \bibnamefont {Bouhon}},
  \ and\ \bibinfo {author} {\bibfnamefont {Robert-Jan}\ \bibnamefont
  {Slager}},\ }\bibfield  {title} {\enquote {\bibinfo {title} {Topological
  euler class as a dynamical observable in optical lattices},}\ }\href
  {\doibase 10.1103/PhysRevLett.125.053601} {\bibfield  {journal} {\bibinfo
  {journal} {Phys. Rev. Lett.}\ }\textbf {\bibinfo {volume} {125}},\ \bibinfo
  {pages} {053601} (\bibinfo {year} {2020})}\BibitemShut {NoStop}%
\bibitem [{\citenamefont {Zhao}\ \emph {et~al.}(2022)\citenamefont {Zhao},
  \citenamefont {Yang}, \citenamefont {Jiang}, \citenamefont {Mao},
  \citenamefont {Guo}, \citenamefont {Qiu}, \citenamefont {Wang}, \citenamefont
  {Yao}, \citenamefont {He}, \citenamefont {Zhou}, \citenamefont {Xu},\ and\
  \citenamefont {Duan}}]{zhao2022observation}%
  \BibitemOpen
  \bibfield  {author} {\bibinfo {author} {\bibfnamefont {W.~D.}\ \bibnamefont
  {Zhao}}, \bibinfo {author} {\bibfnamefont {Y.~B.}\ \bibnamefont {Yang}},
  \bibinfo {author} {\bibfnamefont {Y.}~\bibnamefont {Jiang}}, \bibinfo
  {author} {\bibfnamefont {Z.~C.}\ \bibnamefont {Mao}}, \bibinfo {author}
  {\bibfnamefont {W.~X.}\ \bibnamefont {Guo}}, \bibinfo {author} {\bibfnamefont
  {L.~Y.}\ \bibnamefont {Qiu}}, \bibinfo {author} {\bibfnamefont {G.~X.}\
  \bibnamefont {Wang}}, \bibinfo {author} {\bibfnamefont {L.}~\bibnamefont
  {Yao}}, \bibinfo {author} {\bibfnamefont {L.}~\bibnamefont {He}}, \bibinfo
  {author} {\bibfnamefont {Z.~C.}\ \bibnamefont {Zhou}}, \bibinfo {author}
  {\bibfnamefont {Y.}~\bibnamefont {Xu}}, \ and\ \bibinfo {author}
  {\bibfnamefont {L.~M.}\ \bibnamefont {Duan}},\ }\href@noop {} {\enquote
  {\bibinfo {title} {Observation of topological euler insulators with a
  trapped-ion quantum simulator},}\ } (\bibinfo {year} {2022}),\ \Eprint
  {http://arxiv.org/abs/2201.09234} {arXiv:2201.09234 [quant-ph]} \BibitemShut
  {NoStop}%
\bibitem [{\citenamefont {Park}\ \emph {et~al.}(2021)\citenamefont {Park},
  \citenamefont {Hwang}, \citenamefont {Choi},\ and\ \citenamefont
  {Yang}}]{park2021}%
  \BibitemOpen
  \bibfield  {author} {\bibinfo {author} {\bibfnamefont {Sungjoon}\
  \bibnamefont {Park}}, \bibinfo {author} {\bibfnamefont {Yoonseok}\
  \bibnamefont {Hwang}}, \bibinfo {author} {\bibfnamefont {Hong~Chul}\
  \bibnamefont {Choi}}, \ and\ \bibinfo {author} {\bibfnamefont {Bohm~Jung}\
  \bibnamefont {Yang}},\ }\bibfield  {title} {\enquote {\bibinfo {title}
  {{Topological acoustic triple point}},}\ }\href {\doibase
  10.1038/s41467-021-27158-y} {\bibfield  {journal} {\bibinfo  {journal}
  {Nature Communications}\ }\textbf {\bibinfo {volume} {12}},\ \bibinfo {pages}
  {1--9} (\bibinfo {year} {2021})}\BibitemShut {NoStop}%
\bibitem [{\citenamefont {Lange}\ \emph {et~al.}(2022)\citenamefont {Lange},
  \citenamefont {Bouhon}, \citenamefont {Monserrat},\ and\ \citenamefont
  {Slager}}]{Lange2022}%
  \BibitemOpen
  \bibfield  {author} {\bibinfo {author} {\bibfnamefont {Gunnar~F.}\
  \bibnamefont {Lange}}, \bibinfo {author} {\bibfnamefont {Adrien}\
  \bibnamefont {Bouhon}}, \bibinfo {author} {\bibfnamefont {Bartomeu}\
  \bibnamefont {Monserrat}}, \ and\ \bibinfo {author} {\bibfnamefont
  {Robert-Jan}\ \bibnamefont {Slager}},\ }\bibfield  {title} {\enquote
  {\bibinfo {title} {Topological continuum charges of acoustic phonons in two
  dimensions and the nambu-goldstone theorem},}\ }\href {\doibase
  10.1103/PhysRevB.105.064301} {\bibfield  {journal} {\bibinfo  {journal}
  {Phys. Rev. B}\ }\textbf {\bibinfo {volume} {105}},\ \bibinfo {pages}
  {064301} (\bibinfo {year} {2022})}\BibitemShut {NoStop}%
\bibitem [{\citenamefont {Peng}\ \emph
  {et~al.}(2022{\natexlab{a}})\citenamefont {Peng}, \citenamefont {Bouhon},
  \citenamefont {Monserrat},\ and\ \citenamefont {Slager}}]{Peng2021}%
  \BibitemOpen
  \bibfield  {author} {\bibinfo {author} {\bibfnamefont {Bo}~\bibnamefont
  {Peng}}, \bibinfo {author} {\bibfnamefont {Adrien}\ \bibnamefont {Bouhon}},
  \bibinfo {author} {\bibfnamefont {Bartomeu}\ \bibnamefont {Monserrat}}, \
  and\ \bibinfo {author} {\bibfnamefont {Robert-Jan}\ \bibnamefont {Slager}},\
  }\bibfield  {title} {\enquote {\bibinfo {title} {Phonons as a platform for
  non-abelian braiding and its manifestation in layered silicates},}\ }\href
  {\doibase 10.1038/s41467-022-28046-9} {\bibfield  {journal} {\bibinfo
  {journal} {Nature Communications}\ }\textbf {\bibinfo {volume} {13}},\
  \bibinfo {pages} {423} (\bibinfo {year} {2022}{\natexlab{a}})}\BibitemShut
  {NoStop}%
\bibitem [{\citenamefont {Peng}\ \emph
  {et~al.}(2022{\natexlab{b}})\citenamefont {Peng}, \citenamefont {Bouhon},
  \citenamefont {Slager},\ and\ \citenamefont {Monserrat}}]{peng2022multi}%
  \BibitemOpen
  \bibfield  {author} {\bibinfo {author} {\bibfnamefont {Bo}~\bibnamefont
  {Peng}}, \bibinfo {author} {\bibfnamefont {Adrien}\ \bibnamefont {Bouhon}},
  \bibinfo {author} {\bibfnamefont {Robert-Jan}\ \bibnamefont {Slager}}, \ and\
  \bibinfo {author} {\bibfnamefont {Bartomeu}\ \bibnamefont {Monserrat}},\
  }\bibfield  {title} {\enquote {\bibinfo {title} {Multigap topology and
  non-abelian braiding of phonons from first principles},}\ }\href {\doibase
  10.1103/PhysRevB.105.085115} {\bibfield  {journal} {\bibinfo  {journal}
  {Phys. Rev. B}\ }\textbf {\bibinfo {volume} {105}},\ \bibinfo {pages}
  {085115} (\bibinfo {year} {2022}{\natexlab{b}})}\BibitemShut {NoStop}%
\bibitem [{\citenamefont {Chen}\ \emph {et~al.}(2022)\citenamefont {Chen},
  \citenamefont {Bouhon}, \citenamefont {Slager},\ and\ \citenamefont
  {Monserrat}}]{chen2021manipulation}%
  \BibitemOpen
  \bibfield  {author} {\bibinfo {author} {\bibfnamefont {Siyu}\ \bibnamefont
  {Chen}}, \bibinfo {author} {\bibfnamefont {Adrien}\ \bibnamefont {Bouhon}},
  \bibinfo {author} {\bibfnamefont {Robert-Jan}\ \bibnamefont {Slager}}, \ and\
  \bibinfo {author} {\bibfnamefont {Bartomeu}\ \bibnamefont {Monserrat}},\
  }\bibfield  {title} {\enquote {\bibinfo {title} {Non-abelian braiding of weyl
  nodes via symmetry-constrained phase transitions},}\ }\href {\doibase
  10.1103/PhysRevB.105.L081117} {\bibfield  {journal} {\bibinfo  {journal}
  {Phys. Rev. B}\ }\textbf {\bibinfo {volume} {105}},\ \bibinfo {pages}
  {L081117} (\bibinfo {year} {2022})}\BibitemShut {NoStop}%
\bibitem [{\citenamefont {Bouhon}\ \emph {et~al.}(2021)\citenamefont {Bouhon},
  \citenamefont {Lange},\ and\ \citenamefont {Slager}}]{magnetic}%
  \BibitemOpen
  \bibfield  {author} {\bibinfo {author} {\bibfnamefont {Adrien}\ \bibnamefont
  {Bouhon}}, \bibinfo {author} {\bibfnamefont {Gunnar~F.}\ \bibnamefont
  {Lange}}, \ and\ \bibinfo {author} {\bibfnamefont {Robert-Jan}\ \bibnamefont
  {Slager}},\ }\bibfield  {title} {\enquote {\bibinfo {title} {Topological
  correspondence between magnetic space group representations and
  subdimensions},}\ }\href {\doibase 10.1103/PhysRevB.103.245127} {\bibfield
  {journal} {\bibinfo  {journal} {Phys. Rev. B}\ }\textbf {\bibinfo {volume}
  {103}},\ \bibinfo {pages} {245127} (\bibinfo {year} {2021})}\BibitemShut
  {NoStop}%
\bibitem [{\citenamefont {K\"onye}\ \emph {et~al.}(2021)\citenamefont
  {K\"onye}, \citenamefont {Bouhon}, \citenamefont {Fulga}, \citenamefont
  {Slager}, \citenamefont {van~den Brink},\ and\ \citenamefont
  {Facio}}]{Koneye2021}%
  \BibitemOpen
  \bibfield  {author} {\bibinfo {author} {\bibfnamefont {Viktor}\ \bibnamefont
  {K\"onye}}, \bibinfo {author} {\bibfnamefont {Adrien}\ \bibnamefont
  {Bouhon}}, \bibinfo {author} {\bibfnamefont {Ion~Cosma}\ \bibnamefont
  {Fulga}}, \bibinfo {author} {\bibfnamefont {Robert-Jan}\ \bibnamefont
  {Slager}}, \bibinfo {author} {\bibfnamefont {Jeroen}\ \bibnamefont {van~den
  Brink}}, \ and\ \bibinfo {author} {\bibfnamefont {Jorge~I.}\ \bibnamefont
  {Facio}},\ }\bibfield  {title} {\enquote {\bibinfo {title} {Chirality flip of
  weyl nodes and its manifestation in strained ${\mathrm{mote}}_{2}$},}\ }\href
  {\doibase 10.1103/PhysRevResearch.3.L042017} {\bibfield  {journal} {\bibinfo
  {journal} {Phys. Rev. Research}\ }\textbf {\bibinfo {volume} {3}},\ \bibinfo
  {pages} {L042017} (\bibinfo {year} {2021})}\BibitemShut {NoStop}%
\bibitem [{\citenamefont {Yu}\ \emph {et~al.}(2021)\citenamefont {Yu},
  \citenamefont {Chen},\ and\ \citenamefont {Sarma}}]{Eulersc}%
  \BibitemOpen
  \bibfield  {author} {\bibinfo {author} {\bibfnamefont {Jiabin}\ \bibnamefont
  {Yu}}, \bibinfo {author} {\bibfnamefont {Yu-An}\ \bibnamefont {Chen}}, \ and\
  \bibinfo {author} {\bibfnamefont {Sankar~Das}\ \bibnamefont {Sarma}},\
  }\bibfield  {title} {\enquote {\bibinfo {title} {Euler obstructed cooper
  pairing: Nodal superconductivity and hinge majorana zero modes},}\ }\href
  {\doibase 10.48550/ARXIV.2109.02685} {\  (\bibinfo {year} {2021}),\
  10.48550/ARXIV.2109.02685}\BibitemShut {NoStop}%
\bibitem [{\citenamefont {Lian}\ \emph {et~al.}(2020)\citenamefont {Lian},
  \citenamefont {Xie},\ and\ \citenamefont {Bernevig}}]{biao2020landau}%
  \BibitemOpen
  \bibfield  {author} {\bibinfo {author} {\bibfnamefont {Biao}\ \bibnamefont
  {Lian}}, \bibinfo {author} {\bibfnamefont {Fang}\ \bibnamefont {Xie}}, \ and\
  \bibinfo {author} {\bibfnamefont {B.~Andrei}\ \bibnamefont {Bernevig}},\
  }\bibfield  {title} {\enquote {\bibinfo {title} {Landau level of fragile
  topology},}\ }\href {\doibase 10.1103/PhysRevB.102.041402} {\bibfield
  {journal} {\bibinfo  {journal} {Phys. Rev. B}\ }\textbf {\bibinfo {volume}
  {102}},\ \bibinfo {pages} {041402} (\bibinfo {year} {2020})}\BibitemShut
  {NoStop}%
\bibitem [{\citenamefont {Guan}\ \emph {et~al.}(2021)\citenamefont {Guan},
  \citenamefont {Bouhon},\ and\ \citenamefont {Yazyev}}]{guan2021landau}%
  \BibitemOpen
  \bibfield  {author} {\bibinfo {author} {\bibfnamefont {Yifei}\ \bibnamefont
  {Guan}}, \bibinfo {author} {\bibfnamefont {Adrien}\ \bibnamefont {Bouhon}}, \
  and\ \bibinfo {author} {\bibfnamefont {Oleg~V.}\ \bibnamefont {Yazyev}},\
  }\href@noop {} {\enquote {\bibinfo {title} {Landau levels of the euler class
  topology},}\ } (\bibinfo {year} {2021}),\ \Eprint
  {http://arxiv.org/abs/2108.10353} {arXiv:2108.10353 [cond-mat.mes-hall]}
  \BibitemShut {NoStop}%
\bibitem [{\citenamefont {Guo}\ \emph {et~al.}(2021)\citenamefont {Guo},
  \citenamefont {Jiang}, \citenamefont {Zhang}, \citenamefont {Zhang},
  \citenamefont {Zhang}, \citenamefont {Yang}, \citenamefont {Zhang},\ and\
  \citenamefont {Chan}}]{Guo1Dexp}%
  \BibitemOpen
  \bibfield  {author} {\bibinfo {author} {\bibfnamefont {Qinghua}\ \bibnamefont
  {Guo}}, \bibinfo {author} {\bibfnamefont {Tianshu}\ \bibnamefont {Jiang}},
  \bibinfo {author} {\bibfnamefont {Ruo-Yang}\ \bibnamefont {Zhang}}, \bibinfo
  {author} {\bibfnamefont {Lei}\ \bibnamefont {Zhang}}, \bibinfo {author}
  {\bibfnamefont {Zhao-Qing}\ \bibnamefont {Zhang}}, \bibinfo {author}
  {\bibfnamefont {Biao}\ \bibnamefont {Yang}}, \bibinfo {author} {\bibfnamefont
  {Shuang}\ \bibnamefont {Zhang}}, \ and\ \bibinfo {author} {\bibfnamefont
  {C.~T.}\ \bibnamefont {Chan}},\ }\bibfield  {title} {\enquote {\bibinfo
  {title} {Experimental observation of non-abelian topological charges and edge
  states},}\ }\href {\doibase 10.1038/s41586-021-03521-3} {\bibfield  {journal}
  {\bibinfo  {journal} {Nature}\ }\textbf {\bibinfo {volume} {594}},\ \bibinfo
  {pages} {195--200} (\bibinfo {year} {2021})}\BibitemShut {NoStop}%
\bibitem [{\citenamefont {Park}\ \emph {et~al.}(2022)\citenamefont {Park},
  \citenamefont {Gao}, \citenamefont {Zhang},\ and\ \citenamefont
  {Oh}}]{park2022nodal}%
  \BibitemOpen
  \bibfield  {author} {\bibinfo {author} {\bibfnamefont {Haedong}\ \bibnamefont
  {Park}}, \bibinfo {author} {\bibfnamefont {Wenlong}\ \bibnamefont {Gao}},
  \bibinfo {author} {\bibfnamefont {Xiao}\ \bibnamefont {Zhang}}, \ and\
  \bibinfo {author} {\bibfnamefont {Sang~Soon}\ \bibnamefont {Oh}},\
  }\href@noop {} {\enquote {\bibinfo {title} {Nodal lines in momentum space:
  topological invariants and recent realizations in photonic and other
  systems},}\ } (\bibinfo {year} {2022}),\ \Eprint
  {http://arxiv.org/abs/2201.06639} {arXiv:2201.06639 [cond-mat.mtrl-sci]}
  \BibitemShut {NoStop}%
\bibitem [{\citenamefont {Jiang}\ \emph {et~al.}(2021)\citenamefont {Jiang},
  \citenamefont {Bouhon}, \citenamefont {Lin}, \citenamefont {Zhou},
  \citenamefont {Hou}, \citenamefont {Li}, \citenamefont {Slager},\ and\
  \citenamefont {Jiang}}]{Jiang2021}%
  \BibitemOpen
  \bibfield  {author} {\bibinfo {author} {\bibfnamefont {Bin}\ \bibnamefont
  {Jiang}}, \bibinfo {author} {\bibfnamefont {Adrien}\ \bibnamefont {Bouhon}},
  \bibinfo {author} {\bibfnamefont {Zhi-Kang}\ \bibnamefont {Lin}}, \bibinfo
  {author} {\bibfnamefont {Xiaoxi}\ \bibnamefont {Zhou}}, \bibinfo {author}
  {\bibfnamefont {Bo}~\bibnamefont {Hou}}, \bibinfo {author} {\bibfnamefont
  {Feng}\ \bibnamefont {Li}}, \bibinfo {author} {\bibfnamefont {Robert-Jan}\
  \bibnamefont {Slager}}, \ and\ \bibinfo {author} {\bibfnamefont {Jian-Hua}\
  \bibnamefont {Jiang}},\ }\bibfield  {title} {\enquote {\bibinfo {title}
  {Experimental observation of non-abelian topological acoustic semimetals and
  their phase transitions},}\ }\href {\doibase 10.1038/s41567-021-01340-x}
  {\bibfield  {journal} {\bibinfo  {journal} {Nature Physics}\ }\textbf
  {\bibinfo {volume} {17}},\ \bibinfo {pages} {1239--1246} (\bibinfo {year}
  {2021})}\BibitemShut {NoStop}%
\bibitem [{\citenamefont {Qiu}\ \emph {et~al.}(2022)\citenamefont {Qiu},
  \citenamefont {Zhang}, \citenamefont {Liu}, \citenamefont {Fan},
  \citenamefont {Zhang},\ and\ \citenamefont {Qiu}}]{qiu2022minimal}%
  \BibitemOpen
  \bibfield  {author} {\bibinfo {author} {\bibfnamefont {Huahui}\ \bibnamefont
  {Qiu}}, \bibinfo {author} {\bibfnamefont {Qicheng}\ \bibnamefont {Zhang}},
  \bibinfo {author} {\bibfnamefont {Tingzhi}\ \bibnamefont {Liu}}, \bibinfo
  {author} {\bibfnamefont {Xiying}\ \bibnamefont {Fan}}, \bibinfo {author}
  {\bibfnamefont {Fan}\ \bibnamefont {Zhang}}, \ and\ \bibinfo {author}
  {\bibfnamefont {Chunyin}\ \bibnamefont {Qiu}},\ }\bibfield  {title} {\enquote
  {\bibinfo {title} {Minimal non-abelian nodal braiding in ideal
  metamaterials},}\ }\href@noop {} {\  (\bibinfo {year} {2022})},\ \Eprint
  {http://arxiv.org/abs/2202.01467} {arXiv:2202.01467 [cond-mat.other]}
  \BibitemShut {NoStop}%
\bibitem [{\citenamefont {Ezawa}(2021)}]{ezawa2021euler}%
  \BibitemOpen
  \bibfield  {author} {\bibinfo {author} {\bibfnamefont {Motohiko}\
  \bibnamefont {Ezawa}},\ }\bibfield  {title} {\enquote {\bibinfo {title}
  {Topological euler insulators and their electric circuit realization},}\
  }\href {\doibase 10.1103/PhysRevB.103.205303} {\bibfield  {journal} {\bibinfo
   {journal} {Phys. Rev. B}\ }\textbf {\bibinfo {volume} {103}},\ \bibinfo
  {pages} {205303} (\bibinfo {year} {2021})}\BibitemShut {NoStop}%
\bibitem [{\citenamefont {{Bzdu{\v s}ek}}\ and\ \citenamefont
  {Sigrist}(2017)}]{BzduSigristRobust}%
  \BibitemOpen
  \bibfield  {author} {\bibinfo {author} {\bibfnamefont {Tom\'{a}\v{s}}\
  \bibnamefont {{Bzdu{\v s}ek}}}\ and\ \bibinfo {author} {\bibfnamefont
  {Manfred}\ \bibnamefont {Sigrist}},\ }\bibfield  {title} {\enquote {\bibinfo
  {title} {Robust doubly charged nodal lines and nodal surfaces in
  centrosymmetric systems},}\ }\href {\doibase 10.1103/PhysRevB.96.155105}
  {\bibfield  {journal} {\bibinfo  {journal} {Phys. Rev. B}\ }\textbf {\bibinfo
  {volume} {96}},\ \bibinfo {pages} {155105} (\bibinfo {year}
  {2017})}\BibitemShut {NoStop}%
\bibitem [{\citenamefont {Sticlet}\ \emph {et~al.}(2012)\citenamefont
  {Sticlet}, \citenamefont {Pi\'echon}, \citenamefont {Fuchs}, \citenamefont
  {Kalugin},\ and\ \citenamefont {Simon}}]{sticlet2012engineering}%
  \BibitemOpen
  \bibfield  {author} {\bibinfo {author} {\bibfnamefont {Doru}\ \bibnamefont
  {Sticlet}}, \bibinfo {author} {\bibfnamefont {Frederic}\ \bibnamefont
  {Pi\'echon}}, \bibinfo {author} {\bibfnamefont {Jean-No\"el}\ \bibnamefont
  {Fuchs}}, \bibinfo {author} {\bibfnamefont {Pavel}\ \bibnamefont {Kalugin}},
  \ and\ \bibinfo {author} {\bibfnamefont {Pascal}\ \bibnamefont {Simon}},\
  }\bibfield  {title} {\enquote {\bibinfo {title} {Geometrical engineering of a
  two-band chern insulator in two dimensions with arbitrary topological
  index},}\ }\href {\doibase 10.1103/PhysRevB.85.165456} {\bibfield  {journal}
  {\bibinfo  {journal} {Phys. Rev. B}\ }\textbf {\bibinfo {volume} {85}},\
  \bibinfo {pages} {165456} (\bibinfo {year} {2012})}\BibitemShut {NoStop}%
\bibitem [{\citenamefont {Hatcher}(2001)}]{Hatcher_1}%
  \BibitemOpen
  \bibfield  {author} {\bibinfo {author} {\bibfnamefont {A.}~\bibnamefont
  {Hatcher}},\ }\href@noop {} {\emph {\bibinfo {title} {{A}lgebraic
  {T}opology}}}\ (\bibinfo  {publisher} {Cambridge University Press},\ \bibinfo
  {year} {2001})\BibitemShut {NoStop}%
\bibitem [{\citenamefont {Wojcik}\ \emph {et~al.}(2020)\citenamefont {Wojcik},
  \citenamefont {Sun}, \citenamefont {Bzdu\ifmmode~\check{s}\else
  \v{s}\fi{}ek},\ and\ \citenamefont {Fan}}]{wojcik2020homotopy}%
  \BibitemOpen
  \bibfield  {author} {\bibinfo {author} {\bibfnamefont {Charles~C.}\
  \bibnamefont {Wojcik}}, \bibinfo {author} {\bibfnamefont {Xiao-Qi}\
  \bibnamefont {Sun}}, \bibinfo {author} {\bibfnamefont {Tom\'a\ifmmode
  \check{s}\else~\v{s}\fi{}}\ \bibnamefont {Bzdu\ifmmode~\check{s}\else
  \v{s}\fi{}ek}}, \ and\ \bibinfo {author} {\bibfnamefont {Shanhui}\
  \bibnamefont {Fan}},\ }\bibfield  {title} {\enquote {\bibinfo {title}
  {Homotopy characterization of non-hermitian hamiltonians},}\ }\href {\doibase
  10.1103/PhysRevB.101.205417} {\bibfield  {journal} {\bibinfo  {journal}
  {Phys. Rev. B}\ }\textbf {\bibinfo {volume} {101}},\ \bibinfo {pages}
  {205417} (\bibinfo {year} {2020})}\BibitemShut {NoStop}%
\bibitem [{\citenamefont {Zhao}\ and\ \citenamefont {Lu}(2017)}]{Zhao_PT}%
  \BibitemOpen
  \bibfield  {author} {\bibinfo {author} {\bibfnamefont {Y.~X.}\ \bibnamefont
  {Zhao}}\ and\ \bibinfo {author} {\bibfnamefont {Y.}~\bibnamefont {Lu}},\
  }\bibfield  {title} {\enquote {\bibinfo {title} {${P}{T}$-{S}ymmetric real
  {D}irac {F}ermions and {S}emimetals},}\ }\href {\doibase
  10.1103/PhysRevLett.118.056401} {\bibfield  {journal} {\bibinfo  {journal}
  {Phys. Rev. Lett.}\ }\textbf {\bibinfo {volume} {118}},\ \bibinfo {pages}
  {056401} (\bibinfo {year} {2017})}\BibitemShut {NoStop}%
\bibitem [{\citenamefont {Bouhon}(2020)}]{abouhon_EulerClassTightBinding}%
  \BibitemOpen
  \bibfield  {author} {\bibinfo {author} {\bibfnamefont {Adrien}\ \bibnamefont
  {Bouhon}},\ }\href {https://github.com/abouhon/EulerClassTightBinding}
  {\enquote {\bibinfo {title} {3-band and 4-band real symmetric tight-binding
  models with arbitrary {E}uler class},}\ }\bibinfo {howpublished} {GitHub}
  (\bibinfo {year} {2020}),\ \bibinfo {note} {publicly available Mathematica
  code, https://github.com/abouhon/EulerClassTightBinding}\BibitemShut
  {NoStop}%
\bibitem [{\citenamefont {Khalaf}\ \emph {et~al.}(2018)\citenamefont {Khalaf},
  \citenamefont {Po}, \citenamefont {Vishwanath},\ and\ \citenamefont
  {Watanabe}}]{Khalaf_sum_indicators}%
  \BibitemOpen
  \bibfield  {author} {\bibinfo {author} {\bibfnamefont {Eslam}\ \bibnamefont
  {Khalaf}}, \bibinfo {author} {\bibfnamefont {Hoi~Chun}\ \bibnamefont {Po}},
  \bibinfo {author} {\bibfnamefont {Ashvin}\ \bibnamefont {Vishwanath}}, \ and\
  \bibinfo {author} {\bibfnamefont {Haruki}\ \bibnamefont {Watanabe}},\
  }\bibfield  {title} {\enquote {\bibinfo {title} {Symmetry {I}ndicators and
  {A}nomalous {S}urface {S}tates of {T}opological {C}rystalline insulators},}\
  }\href {\doibase 10.1103/PhysRevX.8.031070} {\bibfield  {journal} {\bibinfo
  {journal} {Phys. Rev. X}\ }\textbf {\bibinfo {volume} {8}},\ \bibinfo {pages}
  {031070} (\bibinfo {year} {2018})}\BibitemShut {NoStop}%
\bibitem [{Note1()}]{Note1}%
  \BibitemOpen
  \bibinfo {note} {One caveat here comes from the presence of some accidental
  symmetries in the three-band models and in the imbalanced four-band models
  that merely come from the specific ansatz in Eq.\protect \tmspace
  +\thickmuskip {.2777em}(\ref {eq_3B_nonflat}). The effect of these accidental
  symmetries is further discussed in Section \ref {sec_chiral}.}\BibitemShut
  {Stop}%
\bibitem [{\citenamefont {Burkov}\ \emph {et~al.}(2011)\citenamefont {Burkov},
  \citenamefont {Hook},\ and\ \citenamefont {Balents}}]{burkov2011semimetal}%
  \BibitemOpen
  \bibfield  {author} {\bibinfo {author} {\bibfnamefont {A.~A.}\ \bibnamefont
  {Burkov}}, \bibinfo {author} {\bibfnamefont {M.~D.}\ \bibnamefont {Hook}}, \
  and\ \bibinfo {author} {\bibfnamefont {Leon}\ \bibnamefont {Balents}},\
  }\bibfield  {title} {\enquote {\bibinfo {title} {Topological nodal
  semimetals},}\ }\href {\doibase 10.1103/PhysRevB.84.235126} {\bibfield
  {journal} {\bibinfo  {journal} {Phys. Rev. B}\ }\textbf {\bibinfo {volume}
  {84}},\ \bibinfo {pages} {235126} (\bibinfo {year} {2011})}\BibitemShut
  {NoStop}%
\bibitem [{\citenamefont {Fang}\ \emph {et~al.}(2015)\citenamefont {Fang},
  \citenamefont {Chen}, \citenamefont {Kee},\ and\ \citenamefont {Fu}}]{FuC2T}%
  \BibitemOpen
  \bibfield  {author} {\bibinfo {author} {\bibfnamefont {Chen}\ \bibnamefont
  {Fang}}, \bibinfo {author} {\bibfnamefont {Yige}\ \bibnamefont {Chen}},
  \bibinfo {author} {\bibfnamefont {Hae-Young}\ \bibnamefont {Kee}}, \ and\
  \bibinfo {author} {\bibfnamefont {Liang}\ \bibnamefont {Fu}},\ }\bibfield
  {title} {\enquote {\bibinfo {title} {Topological nodal line semimetals with
  and without spin-orbital coupling},}\ }\href {\doibase
  10.1103/PhysRevB.92.081201} {\bibfield  {journal} {\bibinfo  {journal} {Phys.
  Rev. B}\ }\textbf {\bibinfo {volume} {92}},\ \bibinfo {pages} {081201}
  (\bibinfo {year} {2015})}\BibitemShut {NoStop}%
\bibitem [{\citenamefont {{Bouhon}}\ and\ \citenamefont
  {{Black-Schaffer}}(2017)}]{BBS_nodal_lines}%
  \BibitemOpen
  \bibfield  {author} {\bibinfo {author} {\bibfnamefont {A.}~\bibnamefont
  {{Bouhon}}}\ and\ \bibinfo {author} {\bibfnamefont {A.~M.}\ \bibnamefont
  {{Black-Schaffer}}},\ }\bibfield  {title} {\enquote {\bibinfo {title} {{Bulk
  topology of line-nodal structures protected by space group symmetries in
  class {A}{I}}},}\ }\href@noop {} {\bibfield  {journal} {\bibinfo  {journal}
  {ArXiv e-prints}\ } (\bibinfo {year} {2017})},\ \Eprint
  {http://arxiv.org/abs/1710.04871} {arXiv:1710.04871 [cond-mat.mtrl-sci]}
  \BibitemShut {NoStop}%
\bibitem [{Note2()}]{Note2}%
  \BibitemOpen
  \bibinfo {note} {There actually remain some accidental symmetries that
  explains more completely the structure of Table \ref {tab_2}, especially for
  the imbalanced Euler phases. We call them ``accidental'' because these
  symmetries are due to the very specific form of the models in Section \ref
  {sec_min_models}. Since removing these symmetries amounts to depart from the
  simplicity of the models, which was our primary aim, we do not address these
  further here, and will give a more detailed treatment elsewhere.}\BibitemShut
  {Stop}%
\end{thebibliography}%


%

\newpage
\appendix

\clearpage
\section{Full expression of $R(\phi_+,\theta_+,\phi_-,\theta_-)$ in Eq.\;(\ref{eq_H_imb_gen}a)}

The representative $R\in \mathsf{SO}(4)$ from which we model the four-band Euler phases is given by (see Ref.\;\cite{bouhon2020geometric} for a derivation, and in Ref.\;\cite{abouhon_EulerClassTightBinding} two \texttt{Mathematica} notebooks can be downloaded that generate arbitrary three-band and four-band tight-binding Euler models)
\begin{widetext}
\begin{equation}
\label{eq_R_full}
    \begin{aligned}
         &R(\phi_+,\theta_+,\phi_-,\theta_-) = \\
         &\left(
\begin{array}{c}
 \sin \left(\frac{\theta _-}{2}\right) \sin
   \left(\frac{\phi _+}{2}\right) \sin \left(\frac{1}{2}
   \left(\phi _--\theta _+\right)\right)+\cos
   \left(\frac{\theta _-}{2}\right) \cos \left(\frac{\phi
   _+}{2}\right) \cos \left(\frac{1}{2} \left(\theta
   _++\phi _-\right)\right) \\
 \sin \left(\frac{\theta _-}{2}\right) \cos
   \left(\frac{\phi _+}{2}\right) \sin \left(\frac{1}{2}
   \left(\phi _--\theta _+\right)\right)-\cos
   \left(\frac{\theta _-}{2}\right) \sin \left(\frac{\phi
   _+}{2}\right) \cos \left(\frac{1}{2} \left(\theta
   _++\phi _-\right)\right) \\
 \cos \left(\frac{\theta _-}{2}\right) \cos
   \left(\frac{\phi _+}{2}\right) \sin \left(\frac{1}{2}
   \left(\theta _++\phi _-\right)\right)-\sin
   \left(\frac{\theta _-}{2}\right) \sin \left(\frac{\phi
   _+}{2}\right) \cos \left(\frac{1}{2} \left(\theta
   _+-\phi _-\right)\right) \\
 -\sin \left(\frac{\theta _-}{2}\right) \cos
   \left(\frac{\phi _+}{2}\right) \cos
   \left(\frac{1}{2} \left(\theta _+-\phi
   _-\right)\right) -\cos \left(\frac{\theta
   _-}{2}\right) \sin \left(\frac{\phi _+}{2}\right) \sin
   \left(\frac{1}{2} \left(\theta _++\phi
   _-\right)\right) \\
\end{array}
\right.\\
         & \quad\quad\quad \left.
\begin{array}{c}
 \cos \left(\frac{\theta _-}{2}\right) \sin
   \left(\frac{\phi _+}{2}\right) \cos \left(\frac{1}{2}
   \left(\theta _+-\phi _-\right)\right)-\sin
   \left(\frac{\theta _-}{2}\right) \cos \left(\frac{\phi
   _+}{2}\right) \sin \left(\frac{1}{2} \left(\theta
   _++\phi _-\right)\right) \\
 \sin \left(\frac{\theta _-}{2}\right) \sin
   \left(\frac{\phi _+}{2}\right) \sin \left(\frac{1}{2}
   \left(\theta _++\phi _-\right)\right)+\cos
   \left(\frac{\theta _-}{2}\right) \cos \left(\frac{\phi
   _+}{2}\right) \cos \left(\frac{1}{2} \left(\phi
   _--\theta _+\right)\right) \\
 \sin \left(\frac{\theta _-}{2}\right) \cos
   \left(\frac{\phi _+}{2}\right) \cos \left(\frac{1}{2}
   \left(\theta _++\phi _-\right)\right)+\cos
   \left(\frac{\theta _-}{2}\right) \sin \left(\frac{\phi
   _+}{2}\right) \sin \left(\frac{1}{2} \left(\phi
   _--\theta _+\right)\right) \\
 \cos \left(\frac{\theta _-}{2}\right) \cos
   \left(\frac{\phi _+}{2}\right) \sin \left(\frac{1}{2}
   \left(\phi _--\theta _+\right)\right)-\sin
   \left(\frac{\theta _-}{2}\right) \sin \left(\frac{\phi
   _+}{2}\right) \cos \left(\frac{1}{2} \left(\theta
   _++\phi _-\right)\right) \\
\end{array}
\right.\\
         & \quad\quad\quad\quad\quad\left.
\begin{array}{c}
 -\cos \left(\frac{\theta _-}{2}\right) \cos
   \left(\frac{\phi _+}{2}\right) \sin \left(\frac{1}{2}
   \left(\theta _++\phi _-\right)\right)-\sin
   \left(\frac{\theta _-}{2}\right) \sin \left(\frac{\phi
   _+}{2}\right) \cos \left(\frac{1}{2} \left(\theta
   _+-\phi _-\right)\right) \\
 \cos \left(\frac{\theta _-}{2}\right) \sin
   \left(\frac{\phi _+}{2}\right) \sin \left(\frac{1}{2}
   \left(\theta _++\phi _-\right)\right)-\sin
   \left(\frac{\theta _-}{2}\right) \cos \left(\frac{\phi
   _+}{2}\right) \cos \left(\frac{1}{2} \left(\theta
   _+-\phi _-\right)\right) \\
 \sin \left(\frac{\theta _-}{2}\right) \sin
   \left(\frac{\phi _+}{2}\right) \sin \left(\frac{1}{2}
   \left(\theta _+-\phi _-\right)\right)+\cos
   \left(\frac{\theta _-}{2}\right) \cos \left(\frac{\phi
   _+}{2}\right) \cos \left(\frac{1}{2} \left(\theta
   _++\phi _-\right)\right) \\
 \sin \left(\frac{\theta _-}{2}\right) \cos
   \left(\frac{\phi _+}{2}\right) \sin \left(\frac{1}{2}
   \left(\theta _+-\phi _-\right)\right)-\cos
   \left(\frac{\theta _-}{2}\right) \sin \left(\frac{\phi
   _+}{2}\right) \cos \left(\frac{1}{2} \left(\theta
   _++\phi _-\right)\right) \\
\end{array}
\right.\\
         &\quad \quad\quad \quad\quad \quad\quad \left.
\begin{array}{c}
 \sin \left(\frac{\theta _-}{2}\right) \cos
   \left(\frac{\phi _+}{2}\right) \cos \left(\frac{1}{2}
   \left(\theta _++\phi _-\right)\right)+\cos
   \left(\frac{\theta _-}{2}\right) \sin \left(\frac{\phi
   _+}{2}\right) \sin \left(\frac{1}{2} \left(\theta
   _+-\phi _-\right)\right) \\
 \cos \left(\frac{\theta _-}{2}\right) \cos
   \left(\frac{\phi _+}{2}\right) \sin \left(\frac{1}{2}
   \left(\theta _+-\phi _-\right)\right)-\sin
   \left(\frac{\theta _-}{2}\right) \sin \left(\frac{\phi
   _+}{2}\right) \cos \left(\frac{1}{2} \left(\theta
   _++\phi _-\right)\right) \\
 \sin \left(\frac{\theta _-}{2}\right) \cos
   \left(\frac{\phi _+}{2}\right) \sin \left(\frac{1}{2}
   \left(\theta _++\phi _-\right)\right)+\cos
   \left(\frac{\theta _-}{2}\right) \sin \left(\frac{\phi
   _+}{2}\right) \cos \left(\frac{1}{2} \left(\theta
   _+-\phi _-\right)\right) \\
 \cos \left(\frac{\theta _-}{2}\right) \cos
   \left(\frac{\phi _+}{2}\right) \cos \left(\frac{1}{2}
   \left(\phi _--\theta _+\right)\right)-\sin
   \left(\frac{\theta _-}{2}\right) \sin \left(\frac{\phi
   _+}{2}\right) \sin \left(\frac{1}{2} \left(\theta
   _++\phi _-\right)\right) \\
\end{array}
\right).
    \end{aligned}
\end{equation}
\end{widetext}

\section{Mirror Chern number of the balanced degenerate Euler phases}\label{ap_mirror}

We here slightly extend the argument given in Ref.\;\cite{guan2021landau}. We first note that the eigenvalues $\{\epsilon_1,\epsilon_2\}$ in Eq.\;(\ref{eq_H_imb_gen}) do not need to be constant. Assuming $\boldsymbol{k}$-dependent eigenvalues, the only condition for the definition of Eq.\;(\ref{eq_H_imb_gen}) is the two-by-two degeneracy of the eigenvalues, \ie 
\begin{equation}
    \epsilon_1(\boldsymbol{k}) = E_1(\boldsymbol{k})=E_2(\boldsymbol{k}) < \epsilon_2(\boldsymbol{k}) =
    E_3(\boldsymbol{k})=E_4(\boldsymbol{k}).
\end{equation} 
for all $\boldsymbol{k}$. In other words, the Hamiltonian $H[\boldsymbol{n},\boldsymbol{n}';\epsilon_1,\epsilon_2]$ is the most general expression of a four-band two-by-two degenerate Euler Hamiltonian, modulo any change of orbital basis, $Q(\boldsymbol{k})\mapsto O\cdot Q(\boldsymbol{k})\cdot O^T$ with $O\in\mathsf{O}(4)$. We now show that all {\it balanced} degenerate Euler phases must satisfy an effective spinful mirror symmetry, \ie there always exists an unitary matrix $U_{\sigma_h}$ with $U_{\sigma_h}^2=-\mathbb{1}_4$, such that 
\begin{equation}
\label{eq_mirror_sym}
    U_{\sigma_h} \cdot 
    H_{\text{bal}} 
    \cdot U_{\sigma_h}^{\dagger} = H_{\text{bal}},
\end{equation}
where the balanced Hamiltonian is obtained by keeping one of the unit vectors constant, \ie 
\begin{equation}
\begin{alignedat}{2}
    &H_{\text{bal}}(\phi,\theta,\phi'_c,\theta'_c) &&= H[\boldsymbol{n},\boldsymbol{n}'_0;\epsilon_1,\epsilon_2],\\
    \text{or}~
    &H_{\text{bal}}(\phi_c,\theta_c,\phi',\theta') &&=
    H[\boldsymbol{n}_0,\boldsymbol{n}';\epsilon_1,\epsilon_2],
\end{alignedat}
\end{equation}
with $\boldsymbol{n}_{q=0} = \boldsymbol{n}(\phi_c,\theta_c)$ and $\boldsymbol{n}'_{q=0} = \boldsymbol{n}'(\phi_c',\theta_c')$ constant. Our strategy is to first find the mirror operator in a special case, and then obtain the mirror operator in the general case as induced by the deformation of the Hamiltonian. Setting \eg $(\phi'_c,\theta'_c)=(0,0)$ for which $\boldsymbol{n}'_0 = (0,0,1)$, the balanced and degenerate Hamiltonian is
\begin{subequations}
\begin{multline}
    H_{\text{bal}}(\phi,\theta,0,0) = \\
    \dfrac{\epsilon_1+\epsilon_2}{2}\;\Gamma_{00} + \dfrac{-\epsilon_1+\epsilon_2}{2}  Q^{(2+2)}[\boldsymbol{n},(0,0,1)],
\end{multline}
with 
\begin{multline}
   Q^{(2+2)}[\boldsymbol{n}(\phi,\theta),(0,0,1)] = \\
   n_1(\phi,\theta) \Gamma_{01} + 
   n_2(\phi,\theta) \Gamma_{03} - 
   n_3(\phi,\theta) \Gamma_{22},
\end{multline}
such that $H_{\text{bal}}(\phi,\theta,0,0)$ satisfies Eq.\;(\ref{eq_mirror_sym}) with
\begin{equation}
\label{eq_sym_matrix}
    U_{\sigma_h}(\phi'_c=0,\theta'_c=0) = -\mathrm{i}\sigma_2\otimes \sigma_0 = -\mathrm{i}\Gamma_{20}.
\end{equation}
\end{subequations}
For the general case $\boldsymbol{n}'_{c} \in \mathbb{S}^2_c$, we first note the relation 
\begin{subequations}
\begin{multline}
    H_{\text{bal}}(\phi,\theta,\phi'_c,\theta'_c) =\\
    \Delta R(\phi'_c,\theta'_c) \cdot
    H_{\text{bal}}(\phi,\theta,0,0) \cdot 
    \Delta R(\phi'_c,\theta'_c)^T,
\end{multline}
where
\begin{equation}
\begin{aligned}
    \Delta R(\phi'_c,\theta'_c) &= R(\phi,\theta,\phi'_c,\theta'_c) \cdot R(\phi,\theta,0,0)^T,\\
    &= \left(
        \begin{array}{cccc}
            cc & -cs & 
                -sc & -ss \\
            cs & cc & 
                ss & -sc \\
            sc & -ss &
            cc & cs \\
            ss & sc & 
            -cs & cc
        \end{array}
    \right),
\end{aligned}
\end{equation}
with 
\begin{equation}
\begin{array}{rclrcl}
    cc &=& \cos (\phi'_c/2) \cos (\theta'_c/2),
    & sc &=& \sin (\phi'_c/2) \cos (\theta'_c/2), \\
    cs &=& \cos (\phi'_c/2) \sin (\theta'_c/2),
    & ss &=& \sin (\phi'_c/2) \sin (\theta'_c/2)\,.
\end{array}
\end{equation}
\end{subequations}
In the above expression, we importantly note that $\Delta R(\phi'_c,\theta'_c)$ is constant, \ie it is independent of the varying point $(\phi,\theta)$. Then, $H_{\text{bal}}(\phi,\theta,\phi'_c,\theta'_c)$ satisfies Eq.\;(\ref{eq_mirror_sym}) with the generalized (constant) mirror operator
\begin{equation}
    U_{\sigma_h}(\phi'_c,\theta'_c) = 
        \Delta R(\phi'_c,\theta'_c) \cdot U_{\sigma_h}(0,0) \cdot \Delta R(\phi'_c,\theta'_c)^T.
\end{equation}
We thus conclude that all balanced and degenerate phases are also mirror symmetric. 

On the other hand, $H_{\text{bal}}(\phi,\theta,0,0)$ can be rotated such that it decomposes into mirror-symmetry sectors, \ie rotating the Bloch orbital basis $\vert \widetilde{\phi}_{\alpha} ,\boldsymbol{k} \rangle = \vert \phi_{\beta} ,\boldsymbol{k} \rangle V_{\beta\alpha}$ 
with 
\begin{subequations}
\label{eq_mirror_rot}
\begin{equation}
    V = \dfrac{1}{\sqrt{2}} \left(\Gamma_{01} + \mathrm{i} \Gamma_{11} \right),
\end{equation}
we get the block-diagonal decomposition
\begin{multline}
   V^{\dagger} \cdot H_{\text{bal}}(\phi,\theta,0,0) \cdot V = \\
   \left[ \dfrac{\epsilon_1+\epsilon_2}{2}\mathbb{1}_2
    + \dfrac{-\epsilon_1+\epsilon_2}{2} 
     \left( n_1 \sigma_1 
    + n_3 \sigma_2
    - n_2 \sigma_3
   \right) \right] \oplus \\
   \quad \quad\left[ \dfrac{\epsilon_1+\epsilon_2}{2}\mathbb{1}_2
    + \dfrac{-\epsilon_1+\epsilon_2}{2} 
     \left( n_1 \sigma_1 
    - n_3 \sigma_2
    - n_2 \sigma_3
   \right) \right]
\end{multline}
and the simultaneous diagonalization of the mirror symmetry matrix Eq.\;(\ref{eq_sym_matrix}), 
\begin{equation}
     V^{\dagger} \cdot U_{\sigma_h}(0,0) \cdot V = \mathrm{diag}(-\mathrm{i},-\mathrm{i},\mathrm{i},\mathrm{i}).
\end{equation}
\end{subequations}
The above block diagonalization of the balanced Hamiltonian, with the pair of eigenvalues $\{1/2(\epsilon_1+\epsilon_2\pm \vert \epsilon_1-\epsilon_2\vert) \}$ for each block, and the single mirror-symmetry eigenvalue found for each block, tells us that the Bloch eigenstates of each block are characterized by one mirror eigenvalue, \ie each eigenvector is a mirror eigenstate with either the $(-\mathrm{i})$- or the $\mathrm{i}$-mirror eigenvalues. As a consequence, we can characterize the balanced and degenerate Euler phases with a mirror Chern number. From the form of Eq.\;(\ref{eq_mirror_rot}b), we readily obtain that the mirror Chern number is given by the winding number of $\boldsymbol{n}=(n_1,n_2,n_3)$ and thus relates to the Euler class. Taking the $(-\mathrm{i})$-mirror sector as a reference, we get
\begin{equation}
    C^{(-\mathrm{i})} = W[\boldsymbol{n}] = q = \chi_I = \chi_{II}.
\end{equation}
See also Ref.\;\cite{guan2021landau} for further details.

\end{document}